%% file: Verma_Thesis.tex
\documentclass[a4paper,12pt,onecolumn,twoside]{ufcthese-utinam}
\usepackage[style=listdotted,acronym,xindy,toc]{glossaries}
\makeglossaries
\usepackage[xindy]{imakeidx}
\makeindex

\usepackage{graphicx}
\usepackage{natbib}
\usepackage[utf8]{inputenc}
\usepackage{graphics}
\usepackage[subnum]{cases}
\usepackage{threeparttable}
\usepackage[titletoc,title]{appendix}
\usepackage[dotinlabels]{titletoc}
\usepackage{makecell}
\usepackage{lscape}
\usepackage[figuresright]{rotating}
\usepackage{multirow,booktabs}
\usepackage{amsmath}
\usepackage{ulem}
\usepackage{txfonts}
\usepackage{colortbl}                          
\usepackage{stfloats}
\usepackage{caption}
\usepackage{subcaption}
\usepackage{enumerate}
\usepackage[b]{esvect }
\usepackage{textcomp}
\usepackage{pdfpages}
\usepackage{pifont}
\newcommand{\myparagraph}[1]{\paragraph{#1}\mbox{}\\ \\}

\begin{document}
\include{acronyms}

\thispagestyle{titreUFC}
 \begin{sffamily}
\begin{center}
\vspace*{-1cm}
{\fufc \'Ecole Doctorale Carnot-Pasteur \\ \small (ED CP n$^\circ$ 554)}
\vspace*{2.5cm}

  {\Huge \bf PhD Thesis}
   \vspace{5mm}


  presented by
  \vspace{6mm}

  {\Large \bf Ashok Kumar VERMA}
  \vspace{1.5cm}

\shadowbox{
\begin{minipage}{13.5cm}
\huge
\bigskip
\begin{center}

Improvement of the planetary ephemerides using spacecraft navigation data and its application to fundamental physics

\end{center}
\bigskip
\end{minipage}
}

\vspace{1cm}

{\bf directed by Agnes Fienga }\\[5mm]
 $19^{th}$ September $2013$


\end{center}
\vspace{10mm}

\begin{center}
{\bf Jury :}\\[1mm]

 \hskip5mm
\begin{tabular}{l l l}
    Pr\'esident :      & Veronique Dehant     & ROB, Belgium\\
    Rapporteurs :   & Gilles Metris             & OCA, France\\
                             & Richard Biancale     & CNES, France\\
    Examinateurs : & Jacques Laskar       & IMCCE, France\\
                             & Luciano Iess            & University of Rome, Italy\\
                             & Veronique Dehant   & ROB, Belgium\\
                             & Jose Lages             & UFC, France\\
    Directeur:  & Agnes Fienga          & OCA, France \\

  \end{tabular}
  \end{center}

\end{sffamily}

\chapternonum{Acknowledgements}
Foremost, I would like to express my sincere gratitude to my advisor Dr. Agnes Fienga for the continuous support of my Ph.D study and research, for her patience, motivation, enthusiasm, and immense knowledge. Her guidance helped me in all the time of research, and writing of scientific papers and this thesis. I could not have imagined having a better advisor and mentor for my Ph.D study.

I would like to acknowledge the financial support of the French Space Agency (CNES) and Region Franche-Comte. Part of this thesis was made using the GINS software; I would like to acknowledge CNES, who provided us access to this software. I am also grateful to J.C Marty (CNES) and P. Rosenbatt (Royal Observatory of Belgium) for their support in handling the GINS software. 

I would also like to thanks Observatoire de Besancon, UTINAM for providing me a library and computer facilities of the lab to pursue this study. I am grateful to all respective faculties, staff and colleagues of the lab for their direct and indirect contributions to made my stay fruitful and pleasant.   

Needless to say, my Besancon years would not have been as much fun without the company of my friends, Arvind Rajpurohit, Eric Grux and Andre Martins. Without their support and love, I could not have imagined my successful stay at Besancon. And at last but not least, I am happy that the distance to my family and my friends back home has remained purely geographical. I wish to thank my family for their love and support which provided my inspiration and was my driving force. I owe them everything and wish I could show them just how much I love and appreciate them.

\cleardoublepage
\chapternonum{Abstract}

The planetary ephemerides play a crucial role for spacecraft navigation, mission planning, reduction and analysis of the most precise astronomical observations. The construction of such ephemerides is highly constrained by the tracking observations, in particular range, of the space probes collected by the tracking stations on the Earth. The present planetary ephemerides (DE, INPOP, EPM) are mainly based on such observations. However, the data used by the planetary ephemerides are not the direct raw tracking data, but measurements deduced after the analysis of raw data made by the space agencies and the access to such processed measurements remains difficult in terms of availability.

The goal of the thesis is to use archives of past and present space missions data independently from the space agencies, and to provide data analysis tools for the improvement of the planetary ephemerides INPOP, as well as to use improved ephemerides to perform tests of physics such as general relativity, solar corona studies, etc. 

The first part of the study deals with the analysis of the Mars Global Surveyor (MGS) tracking data as an academic case for understanding. The CNES orbit determination software GINS was used for such analysis. The tracking observations containing one-, two-, and three-way Doppler and two-way range are then used to reconstruct MGS orbit precisely and obtained results are consistent with those published in the literature. As a supplementary exploitation of MGS, we derived the solar corona model and estimated the average electron density along the line of sight separately for slow and fast wind regions. Estimated electron densities are comparable with the one found in the literature. Fitting the planetary ephemerides, including additional data which were corrected for the solar corona perturbations, noticeably improves the extrapolation capability of the planetary ephemerides and the estimation of the asteroid masses \citep{verma12}.

The second part of the thesis deals with the complete analysis of the MESSENGER tracking data. This analysis improved the Mercury ephemeris up to two order of magnitude compared to any latest ephemerides. Such high precision ephemeris, INPOP13a, is then used to perform general relativity tests of PPN-formalism. Our estimations of PPN parameters ($\beta$ and $\gamma$) are the most stringent than previous results \citep{verma14}.
\cleardoublepage

\cleardoublepage
\chapternonum{R\'esum\'e}

Les \'eph\'em\'erides plan\'etaires jouent un r$\hat{o}$le crucial pour la navigation des missions spatiales actuelles et la mise en place des missions futures ainsi que la r\'eduction et l'analyse des observations astronomiques les plus pr\'ecises. La construction de ces \'eph\'em\'erides est fortement contrainte par les observations de suivi des sondes spatiales collect\'ees par les stations de suivi sur la Terre. Les \'eph\'em\'erides plan\'etaires actuelles (DE, INPOP, EPM) sont principalement bas\'ees sur ces observations. Toutefois, les donn\'ees utilis\'ees par les \'eph\'em\'erides plan\'etaires ne sont pas issues directement des donn\'ees brutes du suivi, mais elles d\'ependent de mesures d\'eduites apr\`es l' analyse des donn\'ees brutes. Ces analyses sont faites par les agences spatiales et leur acc\`es demeure difficile en terme de disponibilit\'e. 

L'objectif de la th\`ese est d'utiliser des archives de donn\'ees de missions spatiales pass\'ees et pr\'esentes et de fournir des outils d'analyse de donn\'ees pour l'am\'elioration de l'\'eph\'em\'eride  plan\'etaire INPOP, ainsi que pour une meilleure utilisation des \'eph\'em\'erides pour effectuer des tests de la physique tels que la relativit\'e g\'en\'erale, les \'etudes de la couronne solaire, etc.

La premi\`ere partie de l'\'etude porte sur l'analyse des donn\'ees de suivi de la sonde Mars Global Surveyor (MGS) prise comme un cas d'\'ecole pour la comp\'ehension de l'observable. Le logiciel du CNES pour la d\'etermination d'orbite GINS a \'et\'e utilis\'e pour une cette analyse. Les r\'esultats obtenus sont coh\'erents avec ceux publi\'es dans la litt\'erature. Comme exploitation suppl\'ementaire des donn\'ees MGS, nous avons \'etudi\'e des mod\`eles de couronne solaire et estim\'e la densit\'e moyenne d'\'electrons le long de la ligne de vis\'ee s\'epar\'ement pour les zones de vents solaires lents et rapides. Les densit\'es \'electroniques estim\'ees sont comparables \`a celles que l'on trouve dans la litt\'erature par d'autres techniques. L'ajout dans l'ajustement des \'eph\'em\'erides plan\'etaires des donn\'ees qui ont \'et\'e corrig\'ees pour les perturbations de plasma solaire, am\'eliore sensiblement la capacit\'e d'extrapolation des \'eph\'em\'erides plan\'etaires et l'estimation des masses d'ast\'eroides \citep{verma12}.

La deuxi\`eme partie de la th\`ese traite de l'analyse compl\`ete des donn\'ees de suivi d'une sonde actuellement en orbite autour de Mercure, Messenger. Cette analyse a am\'elior\'e les \'eph\'em\'erides de Mercure jusqu'\`a deux ordres de grandeur par rapport \`a toutes les derni\`eres \'eph\'em\'erides. La nouvelle \'eph\'em\'eride de haute pr\'ecision, INPOP13a, est ensuite utilis\'ee pour effectuer des tests de la relativit\'e g\'en\'erale via le formalisme PPN. Nos estimations des param\`etres PPN ($\gamma$ et $\beta$) donnent de plus fortes contraintes que les r\'esultats ant\'erieurs \citep{verma14}.

\cleardoublepage

\tableofcontents \markboth{Contents}{Contents}
\addcontentsline{toc}{chapter}{Contents}
\cleardoublepage

\listoffigures \markboth{Table of figures}{Table of figures}
\addcontentsline{toc}{chapter}{List of figures}
\cleardoublepage

\listoftables \markboth{List of tables}{List of tables}
\addcontentsline{toc}{chapter}{List of tables}

\printglossaries
\cleardoublepage

\setlength{\parskip}{1.2em}
\mainmatter
\include{CHP0}
\include{CHP1}
\include{CHP2}
\include{CHP3}

\include{CHP4}
\include{CHP5}

\chapternonum{Publications}
\addcontentsline{toc}{chapter}{Publications}
\begin{itemize}
\item Verma, A. K., Fienga, A., Laskar, J., Manche, H., Gastineau, M., 2013b. Use of MESSENGER radioscience data to improve planetary ephemeris and to test general relativity. Astronomy $\&$ Astrophysics, submitted.

\item Verma, A. K., Fienga, A., Laskar, J., Issautier, K., Manche, H., Gastineau, M., Feb. 2013a. Electron density distribution and solar plasma correction of radio signals using MGS, MEX, and VEX spacecraft navigation data and its application to planetary ephemerides. Astronomy $\&$ Astrophysics 550, A124.

\item Fienga, A.; Manche, H.; Laskar, J.; Gastineau, M.; Verma, A. INPOP new release: INPOP10e. eprint arXiv:1301.1510, 2013.

\item Fienga, A.; Laskar, J.; Verma, A.; Manche, H.; Gastineau, M. INPOP: Evolution, applications, and perspective. SF2A-2012: Proceedings of the Annual meeting of the French Society of Astronomy and Astrophysics. Eds.: S. Boissier, P. de Laverny, N. Nardetto, R. Samadi, D. Valls-Gabaud and H. Wozniak, pp.25-33, 2012.

\item Verma, A.K., Fienga, A. New developments in spacecraft raw data direct analysis for the INPOP planetary ephemerides. IAU Joint Discussion 7: Space-Time Reference Systems for Future Research at IAU General Assembly-Beijing, id.39, August 2012.

\item Verma, A.K., Fienga. INPOP: The planetary ephemerides and its applications. 39th COSPAR Scientific Assembly, 14-22 July 2012, Mysore, India. Abstract C5.1- 15-12, p.2082.

\item Verma, A.K., Fienga, A. Electron density distribution and solar plasma correction of radio signals using MGS, MEX and VEX spacecraft navigation data and its application to planetary ephemerides. 39th COSPAR Scientific Assembly, 14-22 July 2012, Mysore, India. Abstract D3.5-5-12, p.2081.

\item Verma, A.K., Fienga, A. Re-Estimation of Solar Corona Coefficients (a, b, c) by Using MGS $\&$ Mex Spacecraft Data. EPSC-DPS Joint Meeting 2011, 2-7 October 2011, Nantes, France, 2011epsc.conf.1828V.
\end{itemize}

\bibliographystyle{elsarticle-harv}
\bibliography{reference2.bib}
\addcontentsline{toc}{chapter}{Bibliographie}

\backmatter

\end{document}

%% file: acronyms.tex
\newacronym{CNES}{CNES}{Centre National d'Etudes Spatiales}
\newacronym{DSN}{DSN}{Deep Space Network}
\newacronym{ESA}{ESA}{European Space Agency}
\newacronym{GINS}{GINS}{``G\'eod\'esie par Int\'egrations Num\'eriques Simultan\'ees$\textquotedbl$}
\newacronym{INPOP}{INPOP}{``Int\'egrateur Num\'erique Plan\'etaire de l'Observatoire de Paris$\textquotedbl$}
\newacronym{LOS}{LOS}{line of sight}
\newacronym{MDLOS}{MDLOS}{minimum distance of the line of sight}
\newacronym{MEX}{MEX}{Mars Express}
\newacronym{MGS}{MGS}{Mars Global Surveyor}
\newacronym{MNL}{MNL}{Magnetic Neutral Line}
\newacronym{NAIF}{NAIF}{Navigation and Ancillary Information Facility}
\newacronym{NASA}{NASA}{National Aeronautics and Space Administration}
\newacronym{PDS}{PDS}{Planetary Data System}
\newacronym{SEP}{SEP}{Sun-Earth-Probe}
\newacronym{VEX}{VEX}{Venus Express}
\newacronym{WSO}{WSO}{Wilcox Solar Observatory}
\newacronym{JPL}{JPL}{Jet Propulsion Laboratory} 
\newacronym{MMNAV}{MMNAV}{Multimission Navigation}
\newacronym{RMDCT}{RMDCT}{Radio Metric Data Conditioning Team}
\newacronym{ODF}{ODF}{Orbit Data File}
\newacronym{UTC}{UTC}{Coordinated Universal Time}
\newacronym{ET}{ET}{Ephemeris Time}
\newacronym{TAI}{TAI}{International Atomic Time}
\newacronym{UT1}{UT1}{Universal Time}
\newacronym{TT}{TT}{Terrestrial Time}
\newacronym{TDB}{TDB}{Barycentric Dynamical Time}
\newacronym{TCG}{TCG}{Geocentric Coordinate Time}
\newacronym{TCB}{TCB}{Barycentric Coordinate Time}
\newacronym{EOP}{EOP}{Earth Orientation Parameters}
\newacronym{IERS}{IERS}{International Earth Rotation and Reference Systems Service}
\newacronym{IAU}{IAU}{International Astronomical Union}
\newacronym{HEF}{HEF}{high efficiency}
\newacronym{BVE}{BVE}{block 5 exciter}
\newacronym{COI}{COI}{center of integration}
\newacronym{MOI}{MOI}{Mars orbit insertion}
\newacronym{HGA}{HGA}{high gain antenna}
\newacronym{AMD}{AMD}{angular momentum wheel desaturation}
\newacronym{PPN}{PPN}{Parameterized Post-Newtonian}
\newacronym{rms}{rms}{root mean square}
\newacronym{ITRF}{ITRF}{International Terrestrial Reference Frame}
\newacronym{ODP}{ODP}{Orbit Determination Program}
\newacronym{ODY}{ODY}{Odyssey}
\newacronym{MRO}{MRO}{Mars Reconnaissance Orbiter}
\newacronym{BW}{BW$_{mm}$}{Box-Wing macro-model} 
\newacronym{SP}{SP$_{mm}$}{Spherical macro-model}
\newacronym{CME}{CME}{coronal mass ejections}
\newacronym{MGR}{MESSENGER}{MErcury Surface, Space Environment, GEochemistry, and Ranging}
\newacronym{ROB}{ROB}{Royal Observatory of Belgium}
\newacronym{DSM}{DSMs}{deep-space maneuvers}
\newacronym{TCM}{TCMs}{trajectory-correction maneuvers}
\newacronym{MDM}{MDM}{Momentum Dump Maneuver}
\newacronym{OCM}{OCM}{Orbit Correction Maneuver}
\newacronym{GR}{GR}{general relativity}
\newacronym{GCRS}{GCRS}{geocentric celestial reference system}
\newacronym{BCRS}{BCRS}{barycentric celestial reference system} 
\newacronym{IMF}{IMF}{Interplanetary Magnetic Field}
\newacronym{VLBI}{VLBI}{Very-long-baseline interferometry}
\newacronym{EPM}{EPM}{Ephemerides of Planets and the Moon}
\newacronym{LLR}{LLR}{Lunar Laser Ranging}
\newacronym{BVLS}{BVLS}{Bounded Variable Least Squares}
\newacronym{Gaia}{Gaia}{Global Astrometric Interferometer for Astrophysics}

%% file: CHP0.tex
\chapter{Introduction}    
\label{CHP0}

\section{Introduction to planetary ephemerides}
\label{intro_pe}

The word {\it ephemeris} originated from the Greek language ``$\varepsilon\varphi\eta\mu\epsilon\rho o\varsigma$ $\textquotedbl$. The planetary ephemeris gives the positions and velocities of major bodies of the solar system at a given epoch. Historically, positions (right ascension and declination) were given as printed tables of values, at regular intervals of date and time. Nowadays, the modern ephemerides are often computed electronically from mathematical models of the motion.

Before 1960\textquoteright s, analytical models were used for describing the state of the solar system bodies as a function of time. At that time only optical angular measurements of solar system bodies were available. In 1964 radar measurements of the terrestrial planets have been measured. These measurements significantly improved the knowledge of the position of the objects in space. The first laser ranges to the lunar corner cube retroreflectors were then obtained in 1969 \citep{Newhall83}. With the developments of these techniques, a group at MIT, had initiated such an ephemeris program as a support of solar system observations and resulting scientific analyses. The first modern ephemerides, deduced from radar and optical observations, were then developed at MIT \citep{Ash67}. The achieved precision in the measurements, and the improvement in the dynamics of the solar system objects, gave an opportunities to tests the theory of \gls{GR} \citep{Shapiro64}.   

In the late 1970\textquoteright s, the first numerically integrated planetary ephemerides were built by \gls{JPL}, so called DE96 \citep{Standish76}. There have been many versions of the JPL DE ephemerides, from the 1960s through the present. With the beginning of the Space Age, space probes began their journeys resulting in a revolution in knowledge that is still continuing. These ephemerides have then served for spacecraft navigation, mission planning, reduction and analysis of the most precise of astronomical observations. Number of efforts were then also devoted for testing the \gls{GR} using astrometric and radiometric observations \citep{Anderson76,Anderson78}.

Improvements in the planetary ephemerides occurred simultaneously with the evolution of the space missions and the navigation of the probes. Navigation observations were included for the first time in the construction of DE102 ephemeris \citep{Newhall83}. In this ephemeris, planet orbit were constrained by the Viking range measurements along with entire historical astronomical observations. Such addition of the Viking range measurements improved the Mars position by more than 4 order of magnitude and \gls{JPL} becomes the only source of development of high precision planetary ephemerides. In the 1970\textquoteright s and early 1980\textquoteright s, a lot of work was done in the astronomical community to update the astronomical almanacs all around the word. Four major types of observations (optical measurements, radar ranging, spacecraft ranging, and lunar laser ranging) were then included in the adjustment of the ephemeris DE200 \citep{Standish90}. This ephemeris becomes a worldwide standard for several decades. In the late 1990s, a new series of the \gls{JPL} ephemerides were introduced. In particular, DE405 \citep{DE405}, which covers the period between 1600 to 2200, was widely used for the spacecraft navigation and data analysis. DE423 \citep{DE423} is the most recent documented ephemeris produced by the \gls{JPL}. 

However, almost from a decade, the \gls{ESA} is very active in the development of interplanetary missions in collaboration with the \gls{NASA}. These missions include: Giotto for the study of the comets Halley; Ulysses for charting the poles of the Sun; Huygens for Titan; Rosetta for comet; \gls{MEX} for Mars; \gls{VEX} for Venus; \gls{Gaia} for space astrometry; BebiColombo for Mercury (future mission); JUICE for Jupiter (future mission); etc. With the new era of European interplanetary missions, the \gls{INPOP} project was initiated in 2003 to built the first European planetary ephemerides independently from the \gls{JPL}. INPOP has then evolved over the years and the first official release was made on 2008, so-called INPOP06 \citep{Fienga2008}. Currently several versions of INPOP are available to the users: INPOP06 \citep{Fienga2008}; INPOP08 \citep{Fienga2009}; INPOP10a \citep{Fienga2011}; INPOP10b \citep{Fienga2011b}; and INPOP10e \citep{Fienga13}. INPOP10a was the first planetary ephemerides solving for the mass of the Sun (GM$_\odot$) for a given fixed value of Astronomical Unit (AU). Since INPOP10a, new estimations of the Sun mass together with the oblateness of the Sun (J2$_\odot$) are regularly obtained. With the website \url {www.imcce.fr/inpop}, these ephemerides are freely distributed to the users. With this users can have access to positions and velocities of the major planets of our solar system and of the moon, the libration angles of the moon but also to the differences between the terrestrial time TT (time scale used to date the observations) and the barycentric times TDB or TCB (time scale used in the equations of motion). 

With such gradual improvement, INPOP has become an international reference for space navigation. INPOP is the official ephemerides used for the \gls{Gaia} mission navigation and the analysis of the \gls{Gaia} observations. INPOP10e \citep{Fienga13} is the latest ephemerides delivered by the INPOP team to support this mission. Moreover, the INPOP team is also involved in the preparation of the Bepi-Colombo and the JUICE missions. The brief description of the INPOP construction and its evolution are given in Section \ref{inpop_evo}.

Moreover, in addition to DE and INPOP ephemerides, there is one more numerical ephemerides which were developed at the Institute of Applied Astronomy of the Russian Academy of Sciences, called \gls{EPM}. These ephemerides are based upon the same modeling as the JPL DE ephemeris. Their of the \gls{EPM} ephemerides, the most recent are EPM2004 \citep{Pitjeva05}, EPM2008 \citep{Pitjeva10}, and EPM2011 \citep{Pitjeva13}. The EPM2004 ephemerides were constructed over the 1880-2020 time interval in the TDB time scale. In this ephemerides GM of all planets, the Sun, the Moon and value of Earth-Moon mass ratio correspond to DE405 \citep{DE405}, while for EPM2008 these values are close to DE421 \citep{DE421}.

\section{INPOP}
\label{inpop_evo}
The construction of independent planetary ephemerides is a crucial point for the strategy of space development in Europe. As mentioned before, \gls{JPL} ephemerides were used as a reference for spacecraft navigation of the US and the European missions. With the delivery of INPOP the situation has changed. Since 2006, a completely autonomous planetary ephemerides has been built in Europe and became an international reference for space navigation and for scientific research in the dynamics of the Solar System objects and in fundamental physics.

\subsection{INPOP construction}
\label{inpop_con_chp0}
INPOP numerically integrates the equations of motion of the major bodies of our solar system including about 300 asteroids about the solar system barycenter and of the motion and rotation of the Moon about the Earth. Figure \ref{flow_inpop} shows the systematic diagram for the procedure of the INPOP construction. The brief descriptions of this procedure is described as follows: 
%
 %
 \begin{figure}[!ht]
\begin{center}\includegraphics[trim={0cm 2cm 0 0cm},width=15cm]{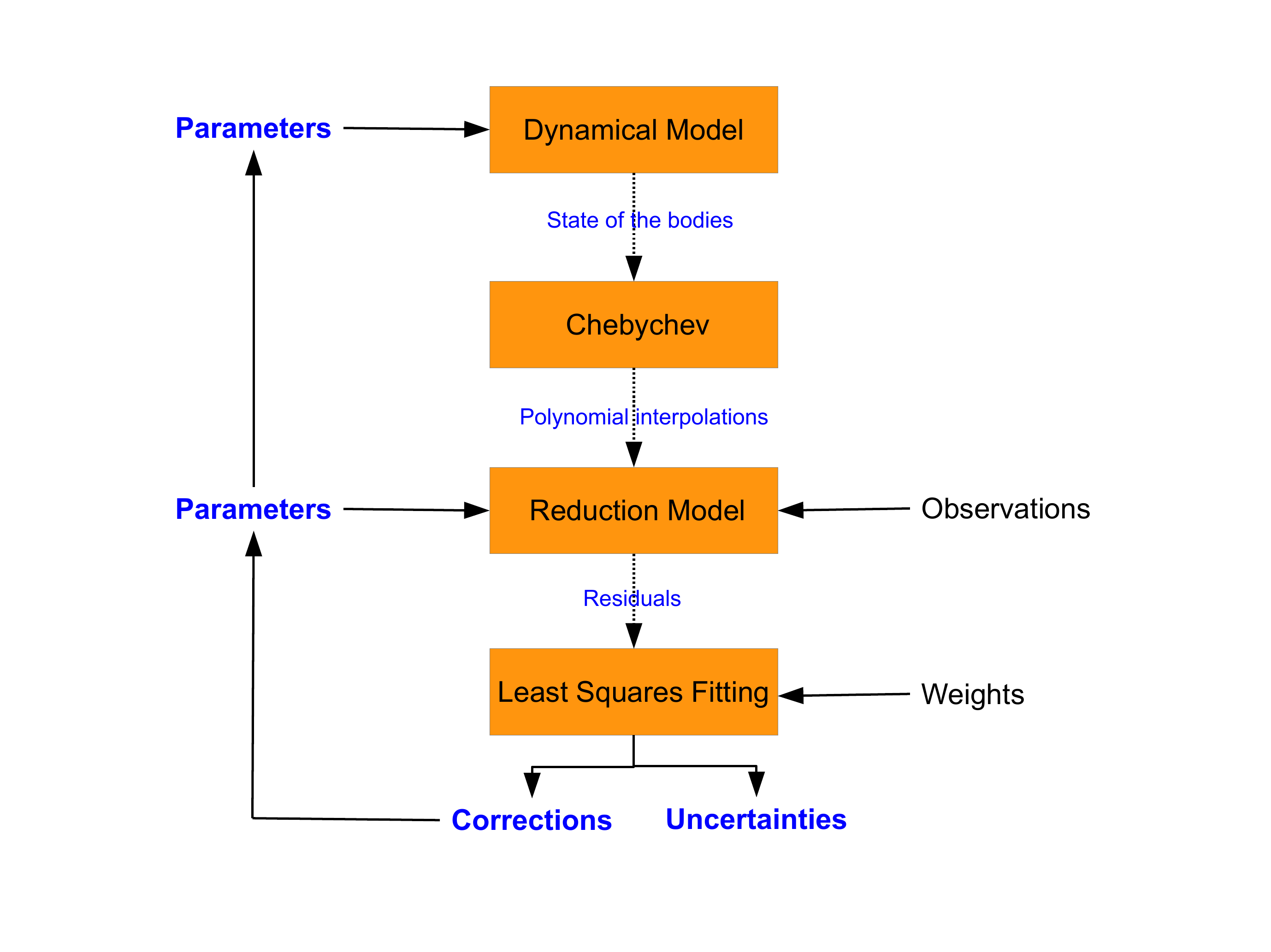}\end{center}
\caption{Schematic diagram for the procedure of the INPOP construction.}
\label{flow_inpop}
\end{figure}
\begin{itemize}
\item The dynamic model of INPOP follows the recommendations of the \gls{IAU} in terms of compatibility between time scales, \gls{TT} and \gls{TDB}, and metric in the relativistic equations of motion. It is developed in the \gls{PPN} framework, and includes the solar oblateness, the perturbations induced by the major asteroids (about 300) as well as the tidal effects of the Earth and Moon. The trajectories of the major bodies are obtained by the numerical integration of a differential equation of first order, $Y\textquotesingle = F(t,Y(t))$, where $Y$ is the parameter that describes the state vectors of the system (position/velocity of the bodies, their orientations) \citep{Manche2011}. Prior knowledge of these parameters at epoch zero (t$_0$, usually J2000) are then used to initiate the integrations with the method of Adams \citep{Hairer87}. Detailed descriptions of the INPOP dynamic modeling are given in the \cite{Fienga2008} and \cite{Manche2011}.

\item The numerical integration produces a file of positions and velocities (state vectors) of the solar system bodies at each time step of the integration. The step size of 0.055 days is usually chosen to minimize the roundoff error. Each component of the state vectors of the solar system bodies relative to the solar system barycenter and the Moon relative to the Earth are then represented by an Nth-degree expansion in Chebyshev polynomials (see \cite{Newhall89} for more details). Interpolation of these polynomials gives the access of the state vectors at any given epoch. 

%
 %
 \begin{figure}[!ht]
\begin{center}\includegraphics[width=10cm]{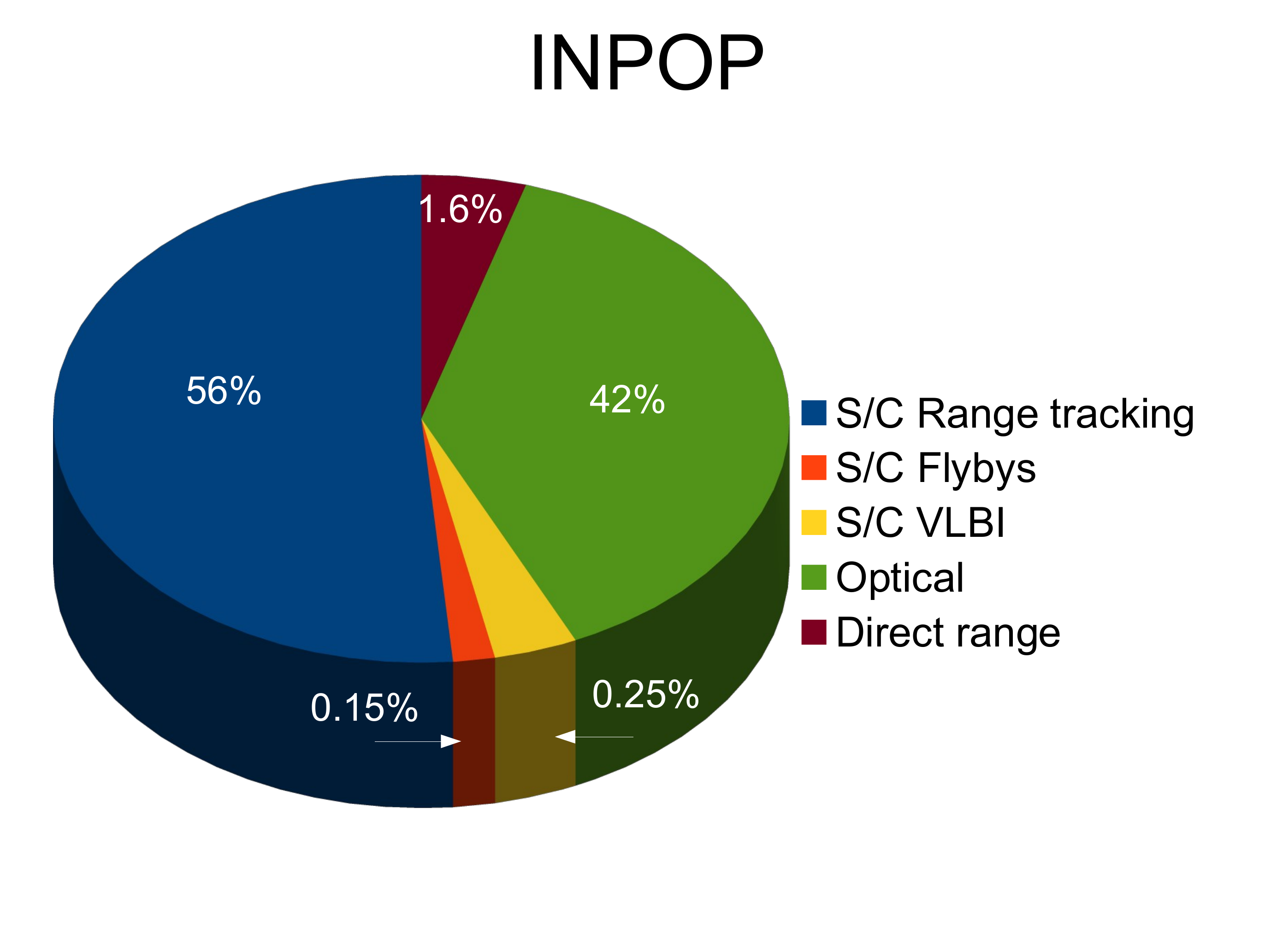}\end{center}
\caption{Percentage contribution of the data in the INPOP construction.}
\label{pie_inpop}
\end{figure}
\item Interpolated solutions of the state vectors are then used to reduce the observations. There are several types of observations that are used for the construction of planetary ephemerides \citep{Fienga2008}: direct radar observations of the planet surface (Venus, Mercury and Mars), spacecraft tracking data (radar ranging, ranging and VLBI), optical observations (transit, photographic plates and CCD observations for outer planets), and \gls{LLR} for Moon. The observations and the parameters associated with the data reduction are then induced in the data reduction models (see Figure \ref{flow_inpop}). Description of such models can be found in \cite{Fienga2008}. Figure \ref{pie_inpop} shows the contributions in percentage of the different types of data used for constraining the recent series of INPOP. More than 136,000 planetary observations are involved in this process. From Figure \ref{pie_inpop} one can noticed that, nowadays, planetary ephemerides are mainly driven by the spacecraft data. However, old astrometric data are still important especially for a better knowledge of outer planet orbits for which few or no spacecraft data are available (see Section \ref{inpop_evo_sub} for more details).

\item Data reduction process allows to compute the differences obtained between the observations and its calculation from the planetary solution, called residuals (observation-calculation). The same program can also calculate the matrix of partial derivatives with respect to the parameters that required to be adjusted.

\item The residuals and the matrix of partial derivatives are then used to adjust the parameters, associated with dynamic models and data reduction models, using least squares techniques. In addition, the file containing the weights, assigned to each observation, is also used to assist the parameter fitting. Usually, almost 400 parameters are estimated during the orbit fitting. About 70 parameters related to the Moon orbit and rotation initial conditions are fitted iteratively with the planetary parameters \citep{Manche2010,Manche2011}. The objective of the parameter fitting is to minimize the residuals using an iterative process. In this process, newly estimated parameters at $i^{th}$ iteration are then feedback to the dynamical and reduction model for initiating the $i^{th}+1$ iteration. 
\end{itemize}

\subsection{INPOP evolution}
\label{inpop_evo_sub}
\begin{figure}[!ht]
\begin{center}\includegraphics[width=7cm]{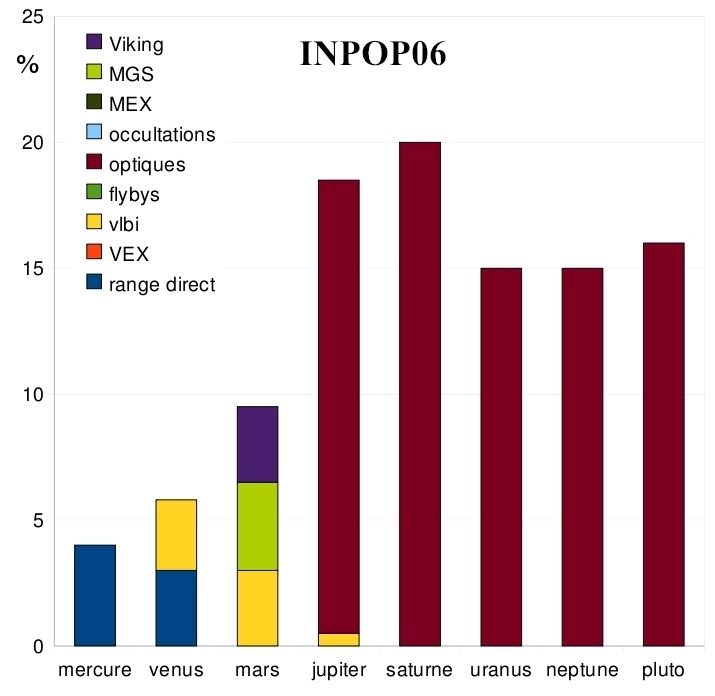}\includegraphics[width=7cm]{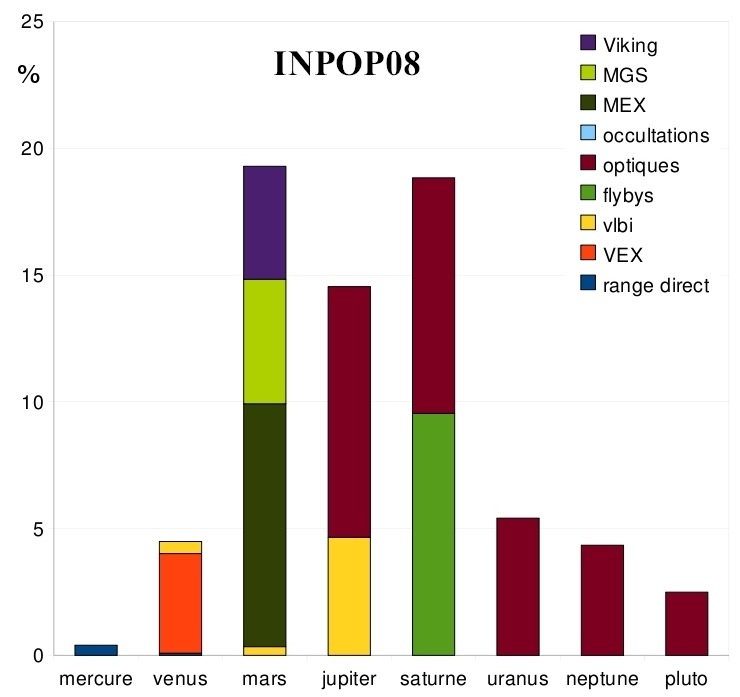} \\
\includegraphics[width=8cm]{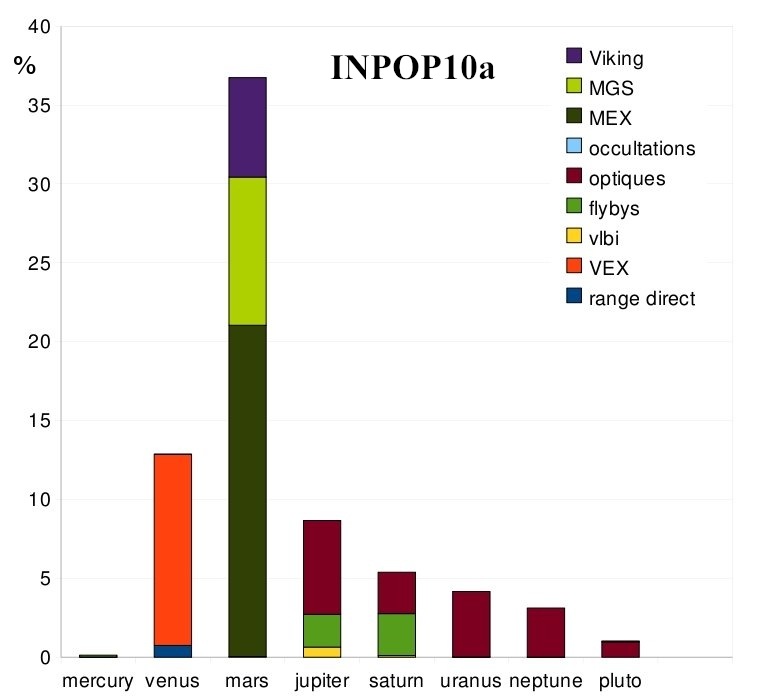}\end{center}
\caption{Percentage contribution of each data type used in the construction INPOP series \citep{FiengaHDR}.}
\label{chi2}
\end{figure}

As stated previously, since 2006, INPOP has become an international reference for space navigation and for scientific research in the dynamics of the solar system objects and in fundamental physics. Since then several versions of the INPOP have been delivered. The Figure \ref{chi2} shows the evolution of the INPOP ephemerides from INPOP06 \citep{Fienga2008} to INPOP10a \citep{Fienga2011} and Table \ref{paramajuste} gives the solve-for parameters for these ephemerides. 

%
 %

\begin{table}[!ht]
\caption{Solve-for parameters for different ephemerides. The parameters which are not modeled are denoted by \ding{55}. The parameters that are fixed during orbit computations are denoted by $F$, while fitted parameters are marked as \ding{51}, \citep{FiengaHDR}}
\begin{center}
\renewcommand{\arraystretch}{1.4}
\small
\begin{threeparttable}
\begin{tabular}{ c c c c c c c}\Xhline{2\arrayrulewidth}
\hline
{\bf Contents} & {\bf Parameters} & {\bf INPOP06} & {\bf INPOP08} & {\bf INPOP10a} & {\bf INPOP10b} & {\bf INPOP10e} \\
\Xhline{2\arrayrulewidth}
&fitted masses & 5 & 34 & 145 & 120  & 79 \\
Asteroids &fixed masses & 0 & 5 & 16 & 71   & 73 \\
&asteroid ring & \ding{51} & F & F & \ding{51} & \ding{51}  \\
&densities& 295 & 261 & \ding{55} & \ding{55}   & \ding{55} \\
\hline
&AU & F & \ding{51} & F & F & F   \\
Constants &EMRAT & F & \ding{51} & \ding{51} & F & \ding{51}  \\
&GM$_\odot$ & F & F & \ding{51} & \ding{51}   & \ding{51}  \\
&J2$_\odot$  & \ding{51} & \ding{51} & \ding{51} & F  & \ding{51} \\
\hline
Time &end of the fit & 2005.5 & 2008.5 & 2010.0 & 2010.0 & 2010.0   \\ 
\hline
Total & fitted parameters & 63 & 83 & 202 & 177 & 137\\
\hline
\end{tabular}
\end{threeparttable}
\label{paramajuste}
\end{center}
\end{table}

INPOP06 was the first version of the INPOP series, published in 2008. As one can noticed, INPOP06 was mainly driven by optical observations of outer planets and Mars tracking data, and solved for 63 parameters. The choice of the estimated parameters and the methods are very similar to DE405 (see \cite{Fienga2008} and Table \ref{paramajuste}). 

Thanks to new collaborations with ESA, \gls{VEX} and \gls{MEX} navigation data have been introduced in INPOP since INPOP08 \citep{Fienga2009}. As a result, estimation of Earth-Venus and Earth-Mars distances, based on VEX and MEX data, were improved by a factor 42 and 6, respectively, compared to INPOP06. 

With the availability of more and more processed range data, it was then possible to improved the accuracy of INPOP ephemerides significantly. Consequently, as one can see on Figure \ref{chi2}, INPOP10a is mainly driven by Mars spacecraft tracking and by the \gls{VEX} tracking data. The Mercury, Jupiter, and Saturn positions deduced from several flybys were also included in the INPOP10a adjustment. Since INPOP10a, the \gls{BVLS} algorithm \citep{Lawson95,Stark&Parker95} is used for the adjustment of parameters, especially for the mass of the asteroids. Compared to the previous versions, significant improvements were noticed in the postfit residuals and in the fitted parameters. The detailed description of these ephemerides and its comparisons are given in \cite{FiengaHDR}.

\begin{figure}[!ht]
\begin{center}\includegraphics[width=15cm]{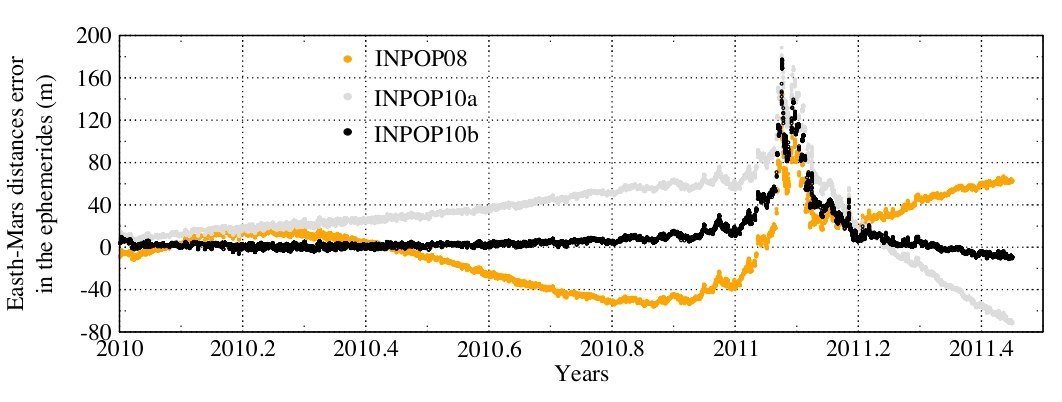}\end{center}
\caption{Extrapolation capability of the planetary ephemerides: INPOP08 \citep{Fienga2008}, INPOP10a \citep{Fienga2011}, and INPOP10b \citep{Fienga2011b}.}
\label{extra_mex}
\end{figure}

Nowadays, modern planetary ephemerides are being more and more spacecraft dependent. However, the accuracy of the planetary ephemerides is characterized by the extrapolation capability. Such capability of the ephemerides is very important for mission design and analysis. In INPOP10b therefore, the efforts were mainly devoted to improve the extrapolation capabilities of the INPOP ephemeris. On Figure \ref{extra_mex} are plotted the one-way Earth-Mars distances residuals estimated with INPOP08, INPOP10a, and INPOP10b ephemerides. The plotted \gls{MEX} data were not included in the fit of the planetary ephemerides, hence represent the extrapolated postfit residuals. By adding more informations on asteroid masses estimated (see Table \ref{paramajuste}) with other techniques (close-encounters between two asteroids, or between a spacecraft and an asteroid), INPOP10b has improved its extrapolation on the Mars-Earth distances of about a factor of 10 compared to INPOP10a. Furthermore, Figure \ref{extra_mex} also demonstrates that, it is crucial to input regularly new tracking data in order to keep the extrapolation capabilities of the planetary ephemerides below 20 meters after 2 years of extrapolation, especially for Mars.

INPOP10e is the latest INPOP version developed for the \gls{Gaia} mission. Compared to pervious versions, new sophisticated procedures related to the asteroid mass determinations have been implemented: \gls{BVLS} have been associated with a-priori sigma estimators \citep{KuchynkaPHD, Fienga2011b} and solar plasma corrections (see Chapter \ref{CHP3} and \cite{verma12}). In addition to INPOP10b data, very recent Uranus observations and the positions of Pluto deduced from Hubble Space Telescope have been also added in the construction of INPOP10e. This ephemerides further used for the analysis of the MESSENGER spacecraft radioscience data for the planetary orbits (see Chapter \ref{CHP4}). 

\section{Importances of the direct analysis of radioscience data for INPOP}
In the Figure \ref{pie_inpop}, is given the distribution of the data samples used for the construction of the INPOP planetary ephemerides. The dependency of the planetary ephemerides on the range observations of the robotic space missions (56$\%$) is obvious and will increase with the continuous addition of spacecraft and lander data like MESSENGER, Opportunity etc. However, the range observations used by the planetary ephemerides are not the direct raw tracking data, but measurements (also called range bias) deduced after the analysis of raw data (Doppler and Range) using orbit determination softwares.

Until recently, space agencies (NASA and ESA) were the only source for the access of such processed range measurements to construct INPOP. Thanks to the PDS server\footnote{\url{http://pds-geosciences.wustl.edu/}}, it is now possible to download the raw tracking observations of space missions such as MGS and MESSENGER and to use them independently for the computations of precise probe orbits and biases for planetary ephemeris construction. Furthermore, flybys of planets by spacecraft are also a good source of information. Owing to the vicinity of the spacecraft and its accurate tracking during this crucial phase of the mission, it is possible to deduce very accurate positions of the planet as the spacecraft pass by. For gaseous planets, this type of observations is the major constraint on their orbits (flybys of Jupiter, Neptune, and Saturn, mainly). Even if they are not numerous (less than 0.5$\%$ of the data sample), they provide 50$\%$ of the constraints brought to outer planet orbits. 


\begin{table}[!ht]
\caption{Sources for the processed spacecraft and lander missions data sets, used for the construction of INPOP.}
\begin{center}
\renewcommand{\arraystretch}{1.1}
\small
\begin{threeparttable}
\begin{tabular}{ c |c c c c c}\Xhline{2\arrayrulewidth}
\hline
{\bf Type} & {\bf Mission} & {\bf Planet} & {\bf Data source}   \\
\Xhline{2\arrayrulewidth}
 &VEX & Venus & ESA      \\
Orbiter &MGS & Mars & JPL/CNES/PDS     \\
 &MEX & Mars & ESA/ROB      \\
 &ODY & Mars & JPL     \\
\hline
 &Mariner 10 & Mercury & JPL   \\
 &MESSENGER & Mercury & JPL/PDS   \\
 &Pioneer 10 $\&$ 11 & Jupiter & JPL   \\
Flyby &Voyager 1 $\&$ 2 & Jupiter & JPL   \\
 &Ulysses & Jupiter & JPL   \\
 &Cassini & Jupiter & JPL   \\
 &Voyager  2 & Uranus & JPL   \\
  &Voyager  2 & Neptune & JPL   \\
\hline
Lander &Viking & Mars & JPL    \\
&pathfinder & Mars & JPL    \\
\hline
\end{tabular}
\end{threeparttable}
\label{data_source_chp0}
\end{center}
\end{table}

The goal of the thesis is therefore to analyze the radioscience data independently (see Chapter \ref{CHP1}) and then to improve INPOP. High precision ephemerides are then used for performing tests of physics such as solar corona studies (see Chapter \ref{CHP3}) and tests of \gls{GR} through the \gls{PPN} formalism (see Chapter \ref{CHP4}). In this thesis, such analysis has been performed with entire \gls{MGS} (see Chapter \ref{CHP2}) and MESSENGER (see Chapter \ref{CHP4}) radioscience data using \gls{CNES} orbit determination software \gls{GINS}. Key aspects of this thesis are:
\begin{itemize}
\item To make INPOP independent from the space agencies and to deliver most up-to-date high accurate ephemerides to the users.

\item To maintain consistency between spacecraft orbit and planet orbit constructions.

\item To perform for the first time studies of the solar corona with the ephemerides. The solar corona model derived from the range bias are then used to correct the solar corona perturbations and for the construction of INPOP ephemerides \citep{verma12}.

\item To analyze the entire MESSENGER radioscience data corresponding to the mapping phase, make INPOP the first ephemerides in the world with the high precision Mercury orbit INPOP13a of about -0.4$\pm$8.4 meters \citep{verma14}.

\item To perform one of the most sensitive \gls{GR} tests of PPN-formalism based on the Mercury improved ephemerides. Estimated \gls{PPN} parameters ($\beta$ and $\gamma$) are most stringent than previous results \citep{verma14}. 

\end{itemize}

Furthermore, the scientific progress in planetary ephemerides, radio science data modeling and orbit determinations will allow the INPOP team to better advance in the interpretation of the space data. This will propel INPOP at the forefront of planetary ephemerides and the future ephemerides will be in competition with other ephemerides ( DE ephemerides from the US for instance), which yields a better security on their validity and integrity from the checking and comparison of the series between them. \\

The outline of the thesis is as follows: \\

In Chapter \ref{CHP1}, the inherent characteristics of the radioscience data are introduced. The contents of the Orbit Determination File (ODF) that are used to measure the spacecraft motion are described. The definitions and the formulations used to index the observations and describe the spacecraft motion are given. The formulations associated with the modeling of the observables, that include precise light time solution, one-, two-, and three-way Doppler shift and two-way range, are discussed. Finally, the modeling of gravitational and non-gravitational forces used in the \gls{GINS} software to describe the spacecraft motion are also discussed briefly. 

In Chapter \ref{CHP2}, the analysis of the radioscience data of the MGS mission has been chosen as an academic case to test our understanding of the raw radiometric data and their analysis with GINS. Data processing and dynamic modeling used to reconstruct the \gls{MGS} orbit are discussed. Results obtained during the orbit computation are then compared with the estimations found in the literature. Finally, a supplementary test, that addresses the impact of the macro-model on the orbit reconstruction, and the comparison between the GINS solution and the JPL light time solution are also discussed. These results have been published in the Astronomy $\&$ Astrophysics journal, \cite{verma12}.

In Chapter \ref{CHP3}, we address issues of radio signal perturbations during the period of superior solar conjunctions of the spacecraft. Brief characteristics of the solar magnetic field, solar activity, and solar wind are given. The complete description of the solar corona models that have been derived from the range measurements of the MGS, MEX, and VEX spacecraft are discussed in details. The solar corona correction of radio signals and its impact on planetary ephemeris and on the estimation of asteroid masses is also discussed. All results corresponding to this study were published in Astronomy $\&$ Astrophysics journal, \cite{verma12}.

In Chapter \ref{CHP4}, we analyze one and half year of radioscience data of the MESSENGER mission using GINS software. Data processing and dynamic modeling used to reconstruct the MESSENGER orbit are discussed. We also discussed the construction of the first high precision Mercury ephemeris INPOP13a using the results obtained with the MESSENGER orbit determination. Finally, \gls{GR} tests of PPN formalism using updated MESSENGER and Mercury ephemerides are discussed. All these results are published in Astronomy $\&$ Astrophysics journal, \cite{verma14}.

In Chapter \ref{CHP5}, we summarize the achieved goal followed by the conclusions and prospectives of the thesis.

%% file: CHP1.tex
\chapter{The radioscience observables and their computation}    
\label{CHP1}

\section{Introduction}
\label{intro}
The radioscience study is the branch of science which usually consider the phenomenas associate with radio wave generation and propagation. In space, these radio signals could be originated from natural sources (for example: pulsars) or from artificial sources such as spacecraft. If the source of these signals is natural, then study is referred to radio astronomy. Usually the objective of radio astronomy is to perform a study of the generation and of the process of propagation of the signal.

However, if the source of the radio signals is artificial satellite, then the radioscience experiment are usually related to the phenomena that occurred along the \gls{LOS} which affect the radio waves propagation. Small changes in phase or amplitude (or both) of the radio signals, when propagating between spacecraft and the \gls{DSN} station on Earth, allow us to study, celestial mechanics, planetary atmosphere, solar corona, planetary ephemeris, planetary gravity field, test of \gls{GR}, etc.

%
 %
 \begin{figure}[!ht]
\begin{center}\includegraphics[width=10cm]{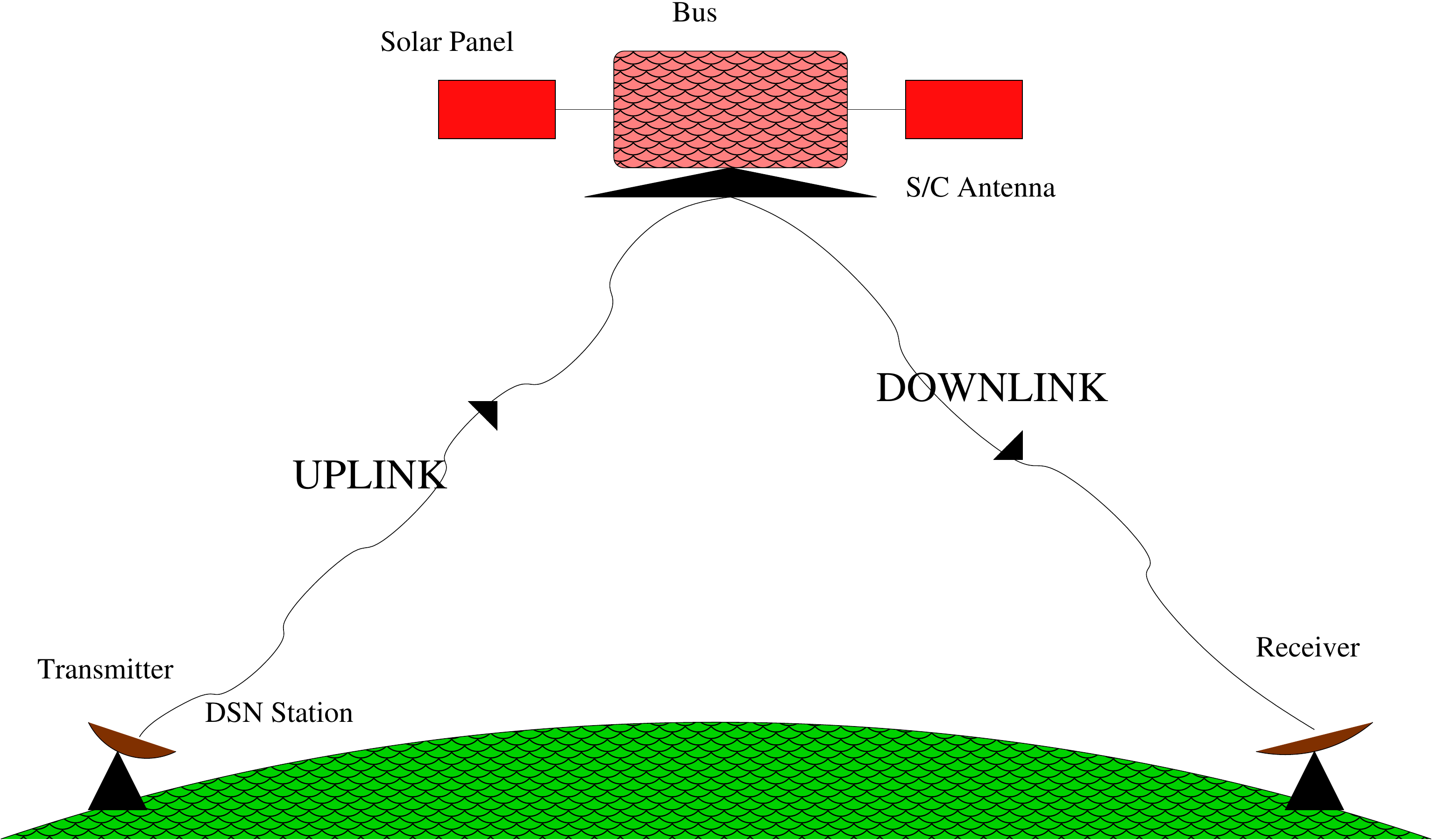}\end{center}
\caption{Two- or three-way radio wave propagation between a spacecraft and Deep Space Network (DSN) station.}
\label{fig1}
\end{figure}

The figure \ref{fig1} represents the schematic diagram of the communication between a spacecraft and the \gls{DSN} stations. These \gls{DSN} stations are used primarily for the uplink transmission of signal and downlink reception of spacecraft data. The uplink is first transmitted by the transmitter from the \gls{DSN} station at time t$_{1}$. These signals are received by the spacecraft antenna which is typically a few meter in size. The received signals are then transmitted back (downlink) by the transponder to the \gls{DSN} station at bouncing time t$_{2}$. The spacecraft transponder multiplies the uplink frequency by a transponder ratio so that downlink frequency is coherently related to the uplink frequency. The transmitted signals (downlink) are then received by the receiver at \gls{DSN} station at time t$_{3}$. 

While tracking the spacecraft, the Doppler shift is routinely measured in the frequency of the signal at the receiving \gls{DSN} station. The Doppler shift, which represents the change in the received signal frequency from the transmitted signal, may be caused by the spacecraft orbit around the planet, Earth revolution around the Sun, Earth rotation, atmospheric perturbations etc. Doppler observables, which are collected at the receiving station, are the average values of this Doppler shift over a period of time called count interval. These collected radiometric data could be one-, two-, or three-way Doppler and range observations. Time delay in terms of distance is represented by range observable and rate of change of this distance is called Doppler observables. When the \gls{DSN} stations on Earth only receive a downlink signal from a spacecraft, the communication is called one-way. The observables are called two-way if the transmitted and received antennas are the same, and three-way observables if they are different. An example of two- or three-way communication is shown in Figure \ref{fig1}).

\section{The radioscience experiments}
\label{radioscience}
The radioscience experiments are used for study the planetary environment and its physical state. Such experiments already have been performed and tested with early flight planetary missions. For example, Voyager \citep{Eshleman77,Tyler81,Tyler86} ,Ulysses \citep{Bird94,Patzold95}, Marine 10 \citep{Howard74}, Mars Global Surveyor \citep{Tyler01,Konopliv06,JMarty}, Mars Express \citep{patzold04}. Brief description of these investigation are discussed below.

\subsection{Planetary atmosphere} 
\label{platm}
%
 %
 \begin{figure}[!ht]
\begin{center}\includegraphics[width=10cm]{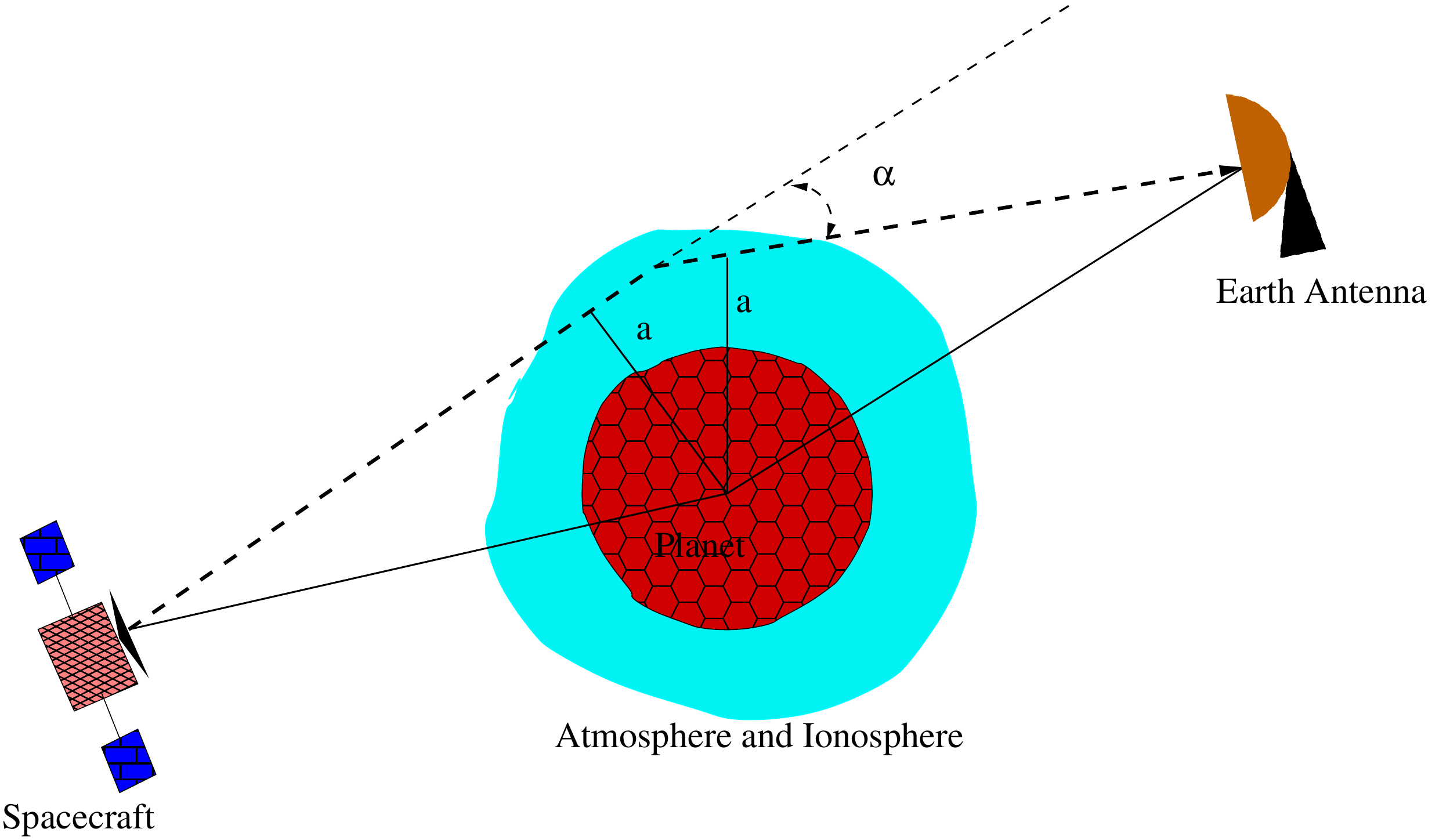}\end{center}
\caption{Radio wave bending when the spacecraft is occulted by the planet and the signal propagates through the atmosphere and ionosphere of the planet.}
\label{fig2}
\end{figure}

In order to study planetary environment, the spacecraft orbit can be arrange such that, the spacecraft passes behind the orbiting planet as seen from the \gls{DSN} stations. This phenomena known as occultation. Just before, the spacecraft is hidden by the planetary disc, signals sent between the spacecraft and the ground station will travel through the atmosphere and ionosphere of the planet. The refraction in the atmosphere and ionosphere bends the \gls{LOS}, as shown in Figure \ref{fig2}. This bending will produce a phase and frequency shift in the received signal. Analysis of this shift can be then account for investigating the atmospheric and ionospheric properties of the planet.

Measurements of the Doppler shift on a spacecraft coherent downlink determine the \gls{LOS} component of the spacecraft velocity. These Doppler and range measurements are then also useful to compute the precise orbit of the spacecraft. The geometry between the spacecraft and the Earth station are then useful to determine the refraction or bending angle, $\alpha$, as shown in Figure \ref{fig2}. The ray asymptotes, $a$ (see Figure \ref{fig2}), and the bending angle, $\alpha$, can be used to estimate the refraction profile of the atmosphere and ionosphere \citep{Fjeldbo71}. This refractivity could be then interpreted in terms of pressure and temperature by assuming the hydrostatic equilibrium \citep{patzold04}.

\subsection{Planetary gravity}
\label{plgra}
The accurate determination of the spacecraft orbit requires a precise knowledge of the gravity field and its temporal variations of the planet. Such variations in the gravity field are associated with the high and low concentration of the mass below and at the surface of the planet. They cause the slight change in the speed of the spacecraft relative to the ground station on Earth and induced small shift in the receiving frequency. After removing the Doppler shift induced by the planetary motion, spacecraft orbital motion, atmospheric friction, solar wind, it is then possible to compute the spacecraft acceleration or deceleration induced by the gravity field of the planet. 

%
 %
\begin{figure}[!ht]
\begin{center}\includegraphics[width=10cm]{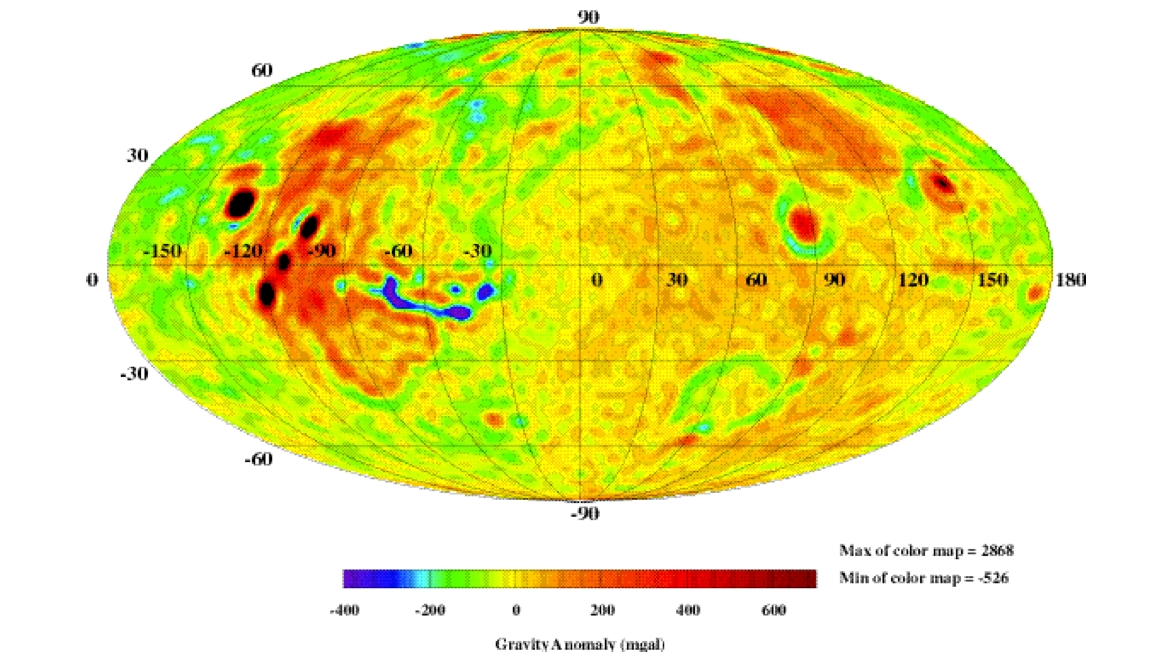}\end{center}
\caption{Mars gravity field derived from Mariner 9, Viking 1$\&$2 and Mars Global Surveyor (MGS) spacecraft \citep{patzold04}.}
\label{fig3}
\end{figure}

Gravity field mapping require the spacecraft downlink carries signal coherent with a highly stable uplink from the Earth station. The two-way radio tracking of these signals provides an accurate measurement of spacecraft velocity along the \gls{LOS} to the tracking station on Earth. Figure \ref{fig3} represents an example of the gravity field mapping of the Mars surface using such radio tracking signals of Mariner 9, Viking 1$\&$2 and Mars Global Surveyor (MGS) spacecraft \citep{patzold04}. This mapping was derived from the gravity field model \citep{Kaula66}, developed upto degree and order 75. The strong positive anomalies shown in Figure \ref{fig3} correspond to the regions of highest elevation on the Mars surface. The low circular orbiter, such as MGS, allows to mapping an accurate and complete gravity field of the planet. However, the highly eccentric orbiter, such as Mars Express (MEX) or MESSENGER, is not best suitable for investigating the global map of the gravity field of orbiting planet.

\subsection{Solar corona}
\label{socor}
When the \gls{LOS} passes close to the sun as seen from the Earth and all three bodies (Planet, Sun and Earth) approximately lies in the straight line, then such geometric configuration is called solar conjunction. During conjunction periods, strong turbulent and ionized gases of corona region severely degrade the radio wave signals when propagating between spacecraft and Earth tracking stations. Such degradations cause a delay and a greater dispersion of the radio signals. The group and phase delays induced by the Sun activity are directly proportional to the total electron contents along the \gls{LOS} and inversely with the square of carrier radio wave frequency.

%
 %
\begin{figure}[ht]
\begin{center}\includegraphics[width=10cm]{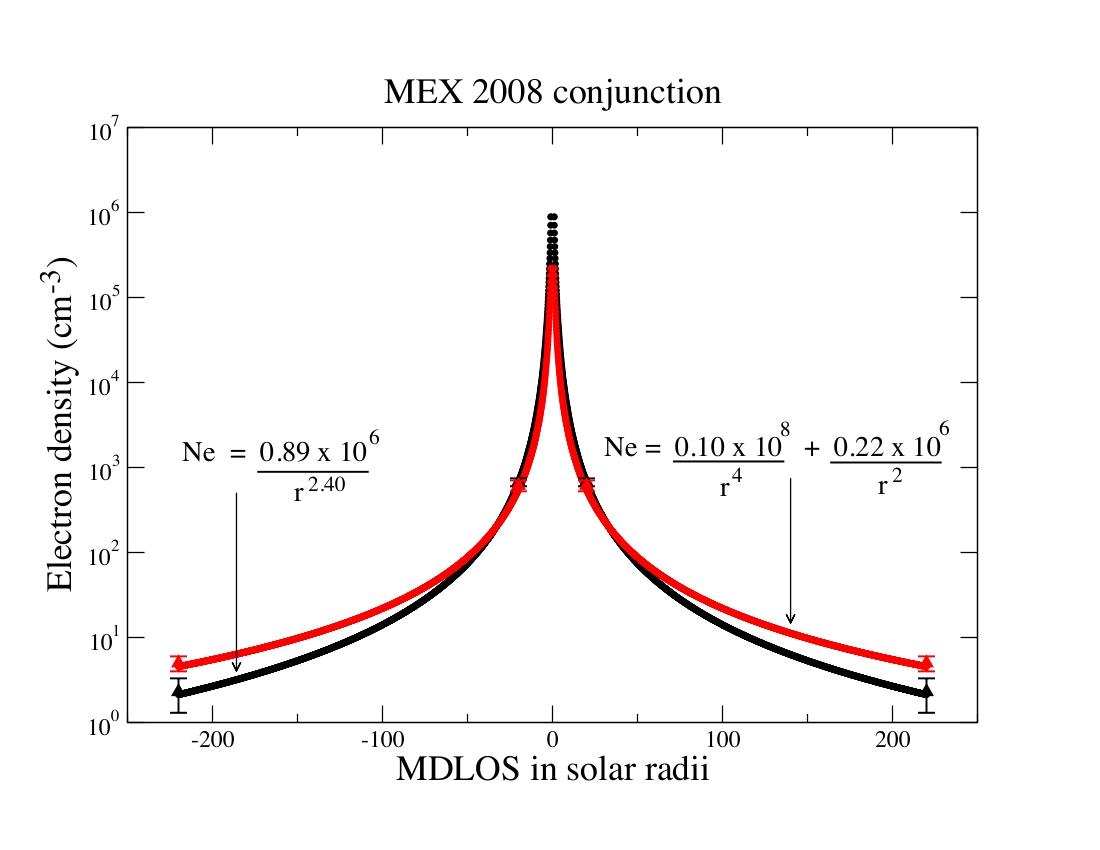}\end{center}
\caption{Electron density distribution with respect to minimum distance of the line of sight (MDLOS) from Sun \citep{verma12}. Black: profile derived from \cite{Bird96}; and Red: profile derived from \cite{Guhathakurta96}}
\label{fig4}
\end{figure}

By analyzing spacecraft radio waves which are directly intercepted by solar plasma, it is then possible to study the corona density distribution, the solar wind region and the corona mass ejection. Two viable methods which are generally used for performing these studies are \citep{Muhleman81}: (1) direct {\itshape in situ} measurements of the electron density, speed, and energies of the electron and photons (2) analysis of a single and dual frequency time delay data acquired from interplanetary spacecraft. The second method which corresponds to radioscience experiment has been already tested using radiometric data acquired at the time of solar conjunctions from interplanetary spacecraft \citep{Muhleman77,AndersonV2,Guhathakurta94,Bird94,Bird96}. 

Figure \ref{fig4} represents an example of corona density distribution computed from the Mars Express (MEX) radiometric data acquired at the time minimum solar cycle, 2008 \citep{verma12}. These electronic profiles of the density can be derived by computing the time or phase delay due to the solar corona. Computed time delay is then fitted over the radiometric data in order to estimate the solar corona model parameters and consequentially the electron density. The detailed analysis of the radioscience solar corona experiment is discussed in Chapter \ref{CHP3}.

\subsection{Celestial mechanics}
\label{cemec}
As discussed in Chapter \ref{CHP0}, the radioscience data are also useful to estimate an accurate position and velocity of the planets, and other solar system parameters from the dynamic modeling of the planetary motion, called planetary ephemerides. Accurate planetary ephemerides are necessary for the spacecraft mission design, orbitography and to perform fundamental tests of physics. However, radioscience data are not directly imposed into the planetary ephemeris software, they are instead first analyzed by the spacecraft orbit determination software. Range bias, which present the systematic error in the geometric position of the planet as seen from the Earth, can be estimated while computing the orbit of the spacecraft. These rang bias impose strong constraints on the orbits of the planet, as well as on other solar system parameters. In consequence, such data not only allow the construction of an accurate planetary ephemeris, they also contribute significantly to our knowledge of parameters such as asteroid masses. A detailed description of such analysis using MGS, and MESSENGER spacecraft radiometric data is discussed in Chapters \ref{CHP2}, and \ref{CHP4}, respectively.


\section{Radiometric data}
\label{radat}
The radiometric data which are produced by the NASA \gls{DSN} \gls{MMNAV} \gls{RMDCT} is called \gls{ODF}\footnote{\url{http://geo.pds.nasa.gov/}}. These \gls{ODF} are used to determine the spacecraft trajectories, gravity field affecting them, and radio propagation conditions. Each \gls{ODF} is in standard \gls{JPL} binary format and consists of many 36-byte logical records, which falls into 7 primary groups. In this work, we have developed  an independent software to extract the contents of these \gls{ODF}s. This software reads the binary \gls{ODF} and writes the contents in specific format, called \gls{GINS} format. \gls{GINS} is the orbit determination software, independently developed at the \gls{CNES} (see Section \ref{gins}).  

\subsection{ODF contents}
\label{odcon}
The \gls{ODF} contains several groups of informations. An \gls{ODF} usually contains most groups, but may not have all. The format of such groups are given in \cite{trk}. The brief description of the contents of these groups is given below.

\subsubsection{Group 1}
\begin{itemize}
\item This group is usually a first group among the several records. It identifies the spacecraft ID, the file creation time, the hardware, and the software associated with the \gls{ODF}. This group also provides the information about the reference date and time for \gls{ODF} time-tags. Currently the \gls{ODF} data time-tags are referenced to Earth Mean Equatorial equinox of 1950 (EME-50).
\end{itemize}

\subsubsection{Group 2}
\begin{itemize}
\item This group is usually a second group among the several groups records. It contains the string character that some time used to identify the contents of the data record, such as, TIMETAG, OBSRVBL, FREQ, ANCILLARY-DATA.
\end{itemize}

\subsubsection{Group 3}
\begin{itemize}
     \item  This is the third group that usually contains majority of the data included in the \gls{ODF}. According to the data categories, the description of this group is given below.
      \end{itemize}
     \myparagraph{Time-tags}
       \begin{itemize}
               \item{{\bf Observable time:}} First in this category is the Doppler and range observable time $TT$ measured at the receiving station. Observable time $TT$ corresponds to the time at the midpoint of the count interval, $T_c$. The integer and the fractional part of this time-tag ($TT$) is given separately in \gls{ODF}. The integer part is measured from 0 hours UTC on 1 January 1950, whereas the fractional part is given in milliseconds.
               \item{{\bf Count interval:}} Doppler observables are derived from the change in the Doppler cycle count. The time period on which these counts are accumulated is called count interval or compression time $T_c$.  Typically count times have a duration of tens of seconds to a few thousand of seconds. For example, count time could be between 1-10 $s$ when the spacecraft is near a planet or roughly 1000 $s$ for interplanetary cruise. 
               \item{{\bf Station delay:}} This gives the information corresponding to the downlink and uplink delay at the receiving and at the transmitting station respectively. It is given in nanosecond in the \gls{ODF}.
               \end{itemize}

     \myparagraph{Format IDs}
      \begin{itemize}
               \item{{\bf Spacecraft ID:}} It identities the spacecraft ID which corresponds to \gls{ODF} data. For example: 94 for MGS 
               \item{{\bf Data type ID:}} As mentioned before, the radiometric data could be one-, two-, and three-way Doppler and two-way range. The \gls{ODF} provides a specific ID associated with these data set. For example: 11, 12, and 13 integers give in \gls{ODF} correspond to one-, two-, and three-way Doppler respectively, whereas 37 stands for two-way range.
               \item{{\bf Station ID:}} This is an integer that gives the receiving and transmitting stations ID that are associated with the time period covered by the \gls{ODF}. The transmitting station ID is set to zero, if the date type is one-way Doppler.
               \item{{\bf Band ID:}} It identifies the uplink (at transmitting station), downlink (at receiving station), and exciter band (at receiving station) ID. The ID of these bands are set to 1, 2, and 3 for S, X, and Ka band.
               \item{{\bf Date Validity ID:}} It is the quality indicator of the data. It set to zero for a good quality of data and set to one for a bad data.
                \end{itemize}
        \myparagraph{Observables}
        \begin{itemize}

%
 %
                   \begin{table*}[bp]
                   \caption{Constants dependent upon transmitter or exciter band}
                   \centering
                   \renewcommand{\arraystretch}{1.2}
                   \small
                   \begin{threeparttable}
                   \begin{tabular}{cccccccccccc}\Xhline{2\arrayrulewidth}
                   \multicolumn{3}{c}{\multirow{2}{*}{Band}} &\multicolumn{7}{c}{ Transmitter Band}      \\ \cline{4-11}   
                  &        &&   T$_1$    &&     T$_2$   &&     T$_3$ &&   T$_4$ (Hz)     \\ \Xhline{2\arrayrulewidth}
                   &S       &&  240     &&   221     &&         96     &&            0     &  \\ 
                   &X       &&  240    &&    749     &&         32     &&     6.5$\times$10$^9$      \\
                   &Ku     &&  142    &&    153     &&     1000     &&     -7.0$\times$10$^9$       \\
                   &Ka     &&   14    &&       15     &&     1000     &&      1.0$\times$10$^{10}$      \\
                   \Xhline{2\arrayrulewidth}
                   \end{tabular}
                   \end{threeparttable}
                   \label{T2}
                   \end{table*}

                  \item{\bf Reference frequency:} It is the frequency measured at the reception time $t_3$ at the receiving station in UTC (see Section \ref{tisca}). This frequency can be constant or ramped. However, the given reference frequency in the \gls{ODF} could be a reference oscillator frequency $f_q$, or a transmitter frequency $f_T$, or a Doppler reference frequency $f_{REF}$. The computed values of Doppler observables are directly affected by the $f_{REF}$. Hence, the computation of the $f_{REF}$ from the reference oscillator frequency $f_q$, or from the transmitter frequency $f_T$ is discussed below.
                  \begin{enumerate}[{\bf (i)}]
                   \item When the given frequency in the \gls{ODF} is $f_q$, then it is needed to first compute the transmitter frequency, which is given by \citep{Moyer}:
                   \begin{equation}                             
                   \label{ft}
                   \mathrm{
                   f_T(t) = T_3 \times  f_q(t) + T_4       }    
                   \end{equation}
                    where T3 and T4 are the transmitter-band dependent constants as given in Table \ref{T2}. From Eq. \ref{ft}, one can compute the transmitter frequency at the receiving station and at the transmitting station by replacing the time $t$ to $t_3$ and $t_1$ respectively. Thus, the $f_{REF}$ at reception time can calculated by multiplying the spacecraft transponder ratio with $f_T$:
                   \begin{equation}
                   \label{fref}
                   \mathrm{
                   f_{REF}(t_3) = M_{2_{R}}  \times  f_T(t_3) }
                   \end{equation}
                   where M$_{2_{R}}$ is the spacecraft transponder ratio (see Table \ref{M2}). It is the function of the exciter band at the transmitting station and of the downlink band at the receiving station. Whereas, M$_{2}$ given in Table \ref{M2} is the function of uplink band at the transmitting station and the downlink band at the receiving station. Hence, the corresponding frequency $f(t_1)$ at the transmission time $t_1$ can be calculated by,
                   \begin{equation}
                   \label{f1}
                   \mathrm{
                   f(t_1) = M_{2}  \times  f_T(t_1) }
                   \end{equation}
                    In Eqs. \ref{fref} and \ref{f1}, $f_T(t_3)$ and $f_T(t_1)$ can be calculated from the Eq. \ref{ft}. If the given value of $f_q$ in the \gls{ODF} is ramped then  $f_q$ is calculated through ramp-table (see Group 4). 
                   \item When the transmitter frequency at reception time $f_T(t_3)$ is given in the \gls{ODF}, then Eq. \ref{fref} can be used to compute the $f_{REF}$. The given $f_T(t_3)$ could be the constant or ramped. The ramped $f_T(t_3)$ can be calculated from the ramped table. However, when the spacecraft is the transmitter (one-way Doppler), then $f_T$ is given by \citep{Moyer}:
                   \begin{equation}
                   \label{f1-way}
                   \mathrm{
                   f_T(t) = C_2 \times f_{S/C} }
                   \end{equation}
                    where $C_2$ is the downlink frequency multiplier. Table \ref{C2} shows the standard \gls{DSN} values of the $C_2$ for S, X, and Ka downlink bands for the data point. $f_{S/C}$ is the spacecraft transmitter frequency which is given by \citep{Moyer}:
                   \begin{equation}
                   \label{FS/C}
                   \mathrm{ 
                     f_{S/C} = f_{T_{0}} + \Delta{f_{T_{0}}}  + f_{T_{1}} (t - t_0) + f_{T_{2}} (t - t_0)^2                            }
                     \end{equation}
         where $f_{T_{0}}$ is the nominal value of $f_{S/C}$ and give in \gls{ODF}. $\Delta{f_{T_{0}}}$, $f_{T_{1}}$, and $f_{T_{2}}$ are the solve-for quadratic coefficients used to represent the departure of $f_{S/C}$. The quadratic coefficients are specified by time block with start time $t_0$.
                   
%
 %
                   \begin{table*}[tp]
                   \caption{Spacecraft transponder ratio M$_2$ (M$_{2_{R}}$)}
                   \centering
                   \renewcommand{\arraystretch}{1.9}
                   \small
                   \begin{threeparttable}
                   \begin{tabular}{cccccccccc}\Xhline{2\arrayrulewidth}
                   \multicolumn{2}{c}{\multirow{2}{*}{Uplink (Exciter)}} &\multicolumn{6}{c}{ Downlink band}      \\ \cline{3-8}   
                  & band  &   S    &&     X    &&     Ka&    \\ \Xhline{2\arrayrulewidth}
                   &S       &  \large{$\frac{240}{221}$}     &&   \large{$\frac{880}{221}$}     &&     \large{$\frac{3344}{221}$}     &  \\ 
                   &X       &  \large{$\frac{240}{749}$ }    &&   \large{$\frac{880}{749}$}     &&     \large{$\frac{3344}{749}$}     &  \\
                   &Ka     &  \large{$\frac{240}{3599}$}  &&   \large{$\frac{880}{3599}$}    &&    \large{$\frac{3344}{3599}$}    & \\
                   \Xhline{2\arrayrulewidth}
                   \end{tabular}
                   \end{threeparttable}
                   \label{M2}
                   \end{table*}

%
 %
                   \begin{table*}[tp]
                   \caption{Downlink frequency multiplier $C_2$}
                   \centering
                   \renewcommand{\arraystretch}{1.8}
                   \small
                   \begin{threeparttable}
                   \begin{tabular}{cccccccccc}\Xhline{2\arrayrulewidth}
                   \multicolumn{3}{c}{\multirow{2}{*}{Multiplier}} &\multicolumn{5}{c}{ Downlink Band}      \\ \cline{4-8}   
                  &        &&   S    &&     X   &&     Ka    \\ \Xhline{2\arrayrulewidth}
                   &$C_2$     &&  1     &&   \large{$\frac{880}{230}$}     &&         \large{$\frac{3344}{240}$}      \\ 
                   \Xhline{2\arrayrulewidth}
                   \end{tabular}
                   \end{threeparttable}
                   \label{C2}
                   \end{table*}
                   \item Finally, the given frequency in the \gls{ODF} could be a constant value of $f_{REF}$. This value is usually constant for a given pass.
                   \end{enumerate}
                  \item{{\bf Doppler observable:}} Doppler observables are derived from the change in the Doppler cycle count $N(t_3)$, which accumulates during the compression time $T_c$ at the receiving station.  These observables in the \gls{ODF} are defined as follows:
                  \begin{equation}
                   \label{dopob}
                   \mathrm{
                   Observable =  \bigg( \frac{B} {|B|} \bigg)\times\bigg[ \bigg( \frac{N_j - N_i}{t_j - t_i}\bigg) - | F_b \times K + B | \bigg] }
                   \end{equation}
                   
                   where: \\*
                   B \ \ \ = \ \ Bias placed on receiver \\*
                   N$_i$ \ \ = \ \ Doppler count at time t$_i$ \\*
                   N$_j$ \ \ = \ \ Doppler count at time t$_j$ \\*
                   t$_i$ \ \ \ \ = \ \ start time of interval \\*
                   t$_j$ \ \ \ \ = \ \ end time of interval \\*
                   F$_b$ \ \ = \ \ frequency bias \\*
                   K \ \ \ = \ \ 1 for S-band receivers \\*
                           = \ \ 11/3 for X-band receivers \\*
                        = \ \ 176/27 for Ku-band receivers \\* 
                  = \ \ 209/15 for Ka-band receivers \\*                          
                   \begin{equation}
                   \label{Fb1}
                   \mathrm{
                   F_b = ( X_1 / X_2 ) \times ( X_3 \times f_R + X_4 ) - f_{s/c} + R_3   \ \ \ \ \ \ \ \ \ \ \ \ \ \ \ \ \ \ \-  \mathrm{for} \ \ \mathrm{1-way} \ \ \mathrm{Doppler}}
                   \end{equation}                  
                   \begin{equation}
                   \label{Fb1}
                   \mathrm{
                   F_b = ( X_1 / X_2 ) \times ( X_3 \times f_{q_{R}} + X_4 ) - ( T_1 / T_2 ) \times ( T_3 \times f_{q_{T}} + T_4 )   \ \ \ \ \  \mathrm{for} \ \ \mathrm{2/3-way} \ \ }\mathrm{Doppler}
                   \end{equation}                   
                   \begin{equation}
                   \label{Tc}
                   \mathrm{
                   T_c = t_j - t_i           \ \ \ \ \ \ \ \ \ \ \ \ \ \ \ \ \ \ \ \ \ \ \ \ \ \ \ \ \ \ \ \ \ \ \ \ \ \ \ \ \ \ \ \ \ \ \ \ \ \ \ \   \mathrm{compression} \ \ \mathrm{time} }
                   \end{equation}
                   where: \\*
                   f$_{q_{R}}$ \ \ \ \ \ \ \ \ \ \ = \ \ Receiver oscillator frequency at time t$_3$ \\*
                   f$_{s/c}$ \ \ \ \ \ \ \ \ \  = \ \ Spacecraft (beacon) frequency \\*
                   f$_{q_{T}}$  \ \ \ \ \ \ \ \ \ \ = \ \ Transmitter oscillator frequency at time t$_1$ \\*
                   R$_3$ \ \ \ \ \ \ \ \ \ \ = \ \ 0 for all receiving bands  \\*
                   T$_1$ to T$_4$ \ =  \ \ Transmitters band (Table \ref{T2}) \\*
                   X$_1$ to X$_4$ \ = \ \ Exciter band ( same value as transmitter band, Table \ref{T2}) \\*

%
 %
\begin{figure}[ht]
\begin{center}\includegraphics[width=14cm]{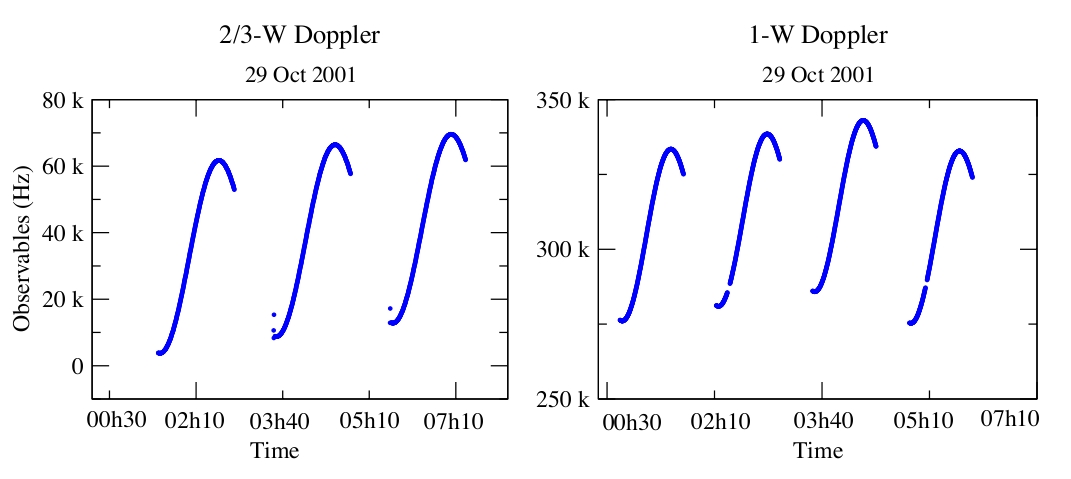}\end{center}
\caption{One- ,Two-, and three-way Doppler observables of the MGS spacecraft.}
\label{dop_data}
\end{figure}
Figure \ref{dop_data} shows an example of two and three way Doppler observables extracted from the MGS \gls{ODF}.

%
 %
\begin{figure}[ht]
\begin{center}\includegraphics[width=14cm]{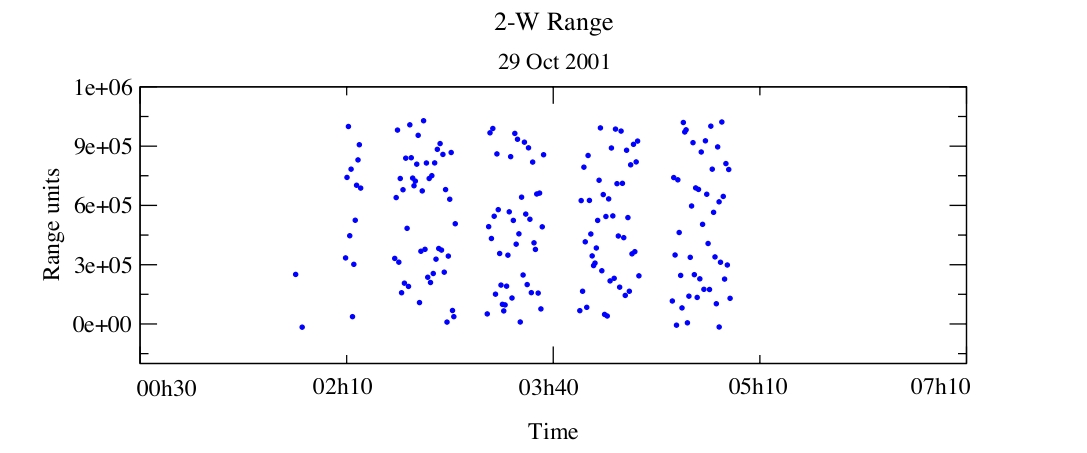}\end{center}
\caption{Two way range observables of the MGS spacecraft.}
\label{range_data}
\end{figure}
                   
                    \item{{\bf Range observable:}} Range observables are obtained from the ranging machine at the receiving station. These range observables are measured in range units (see Section \ref{drobs}) and defined in \gls{ODF} as follows:
                    \begin{equation}
                    \mathrm{
                    Observable = R - C + Z - S }
                    \end{equation}
                   where:  \\*
                    R \ \ = \ \ range measurement  \\*
                    C \ \ = \ \ station delay calibration \\* 
                    Z \ \ = \ \ Z-height correction  \\*
                    S \ \ = \ \ spacecraft delay \\*
      \end{itemize}          
Figure \ref{range_data} shows an example of two way range observables extracted from the MGS$\textquoteright$s \gls{ODF}.

\subsubsection{Group 4}
\begin{figure}[ht]
\begin{center}\includegraphics[width=14cm]{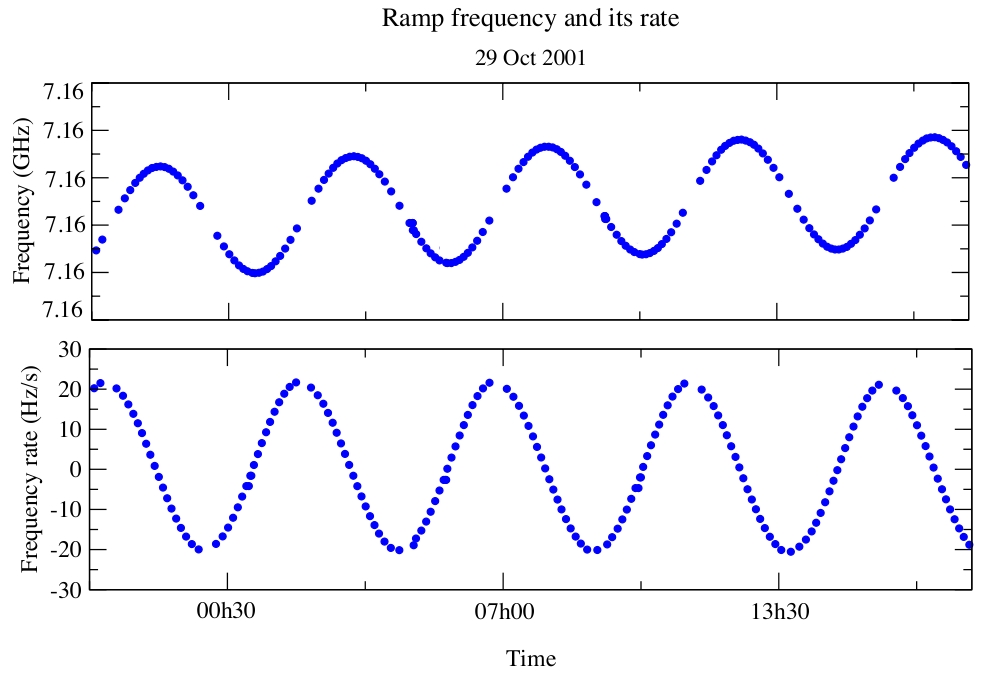}\end{center}
\caption{Ramped frequency $f_o$ and frequency rate $\dot{f}$ measured for MGS spacecraft.}
\label{ramp_data}
\end{figure}

\begin{itemize}
\item Ramp groups are usually the fourth of several groups of record in \gls{ODF}. This group contains the information about the tuning of the receiver or transmitter on the Earth station. Ramping is a technique to achieve better quality communication with spacecraft when its velocity varies with respect to ground stations and it has been implemented at the \gls{DSN}. There is usually one ramp group for each \gls{DSN} station. The contents of this group and the procedure to calculates the ramped transmitter frequency $f_T(t)$ is described below.

\myparagraph{Ramp tables} As mentioned in Group 3, the reference frequency given in \gls{ODF} can be a constant or ramped. When the given frequency is ramped then the reference frequency is computed through the ramp table. The ramp table contains the start UTC time $t_o$, end UTC time $t_f$, the values of ramped frequency $f_o$ at the start time $t_o$, the constant time derivative of frequency (ramp rate) $\dot{f}$, and the tracking station. The ramp table can be specified as the reference oscillator frequency $f_q(t)$ or the transmitter frequency $f_T(t)$. However, Eq. \ref{ft} can be used to convert reference oscillator frequency $f_q(t)$ into the transmitter frequency $f_T(t)$. The ramped frequency can be then calculated by:
               \begin{equation}
               \label{ramp}
               \mathrm{
               f_T(t) = f_o + \dot{f} ( t - t_o)}
               \end{equation}
               where $t$ is the interpolation time.  For Doppler observables, the ramp table for the receiving station gives the ramped transmitter frequency $f_T(t)$ as a function of time. This ramped frequency or a constant value of $f_T(t)$ at the receiving station can be then used to calculate the Doppler reference frequency $f_{REF}(t_3)$ at receiving station using Eq. \ref{fref}. \\*
               The Figure \ref{ramp_data} shows an example of the ramped frequency $f_o$ and the ramp rate $\dot{f}$ plotted over the start time $t_o$ of the ramp table. These ramp informations are extracted from the MGS \gls{ODF}.    
               
\end{itemize}     

%
 %

\subsubsection{Group 5}
\begin{itemize}
              \item This is a clock offset group. It is usually the fifth of several groups of record in \gls{ODF}. It contains information on clock offsets at DSN stations contributing to the \gls{ODF}. This group may be omitted from the \gls{ODF} and used only with VLBI data. It contains the start and end time of the clock offset which is measured from 0 hours UTC on 1 January 1950. It also includes the \gls{DSN} station ID and the correspond clock offset given in nanoseconds. The informations of this group are generally not useful for the radioscience studies.  
\end{itemize}

\subsubsection{Group 6}
\begin{itemize}
              \item This group is usually not include in the \gls{ODF} and omitted all the time.
\end{itemize}

\subsubsection{Group 7}
\begin{itemize}
              \item It is a data summary group which contains summary information on contents of the \gls{ODF}, such as, the first and last date of the data sample, total number of samples, used transmitting and receiving stations, band ID, and the type of data available in the \gls{ODF}. This group is optional and may be omitted from the \gls{ODF}.
               \end{itemize}


\section{Observation Model}
\label{obcom}
For given spacecraft radiometric data obtained by the \gls{DSN} are described in Section \ref{radat}. These data record for each data point contains ID information which is necessary to unambiguously identify the data point and the observed value of the observable (see Group 3 of Section \ref{odcon}). In order to better understand these radiometric data and to estimate the precise orbit of the spacecraft, it is then necessary to compute the observables. 

The computation of the observables requires the time and frequency information of the transmitted frequencies at the transmitter. The various time scales and their transformations used for these computation are described in Section \ref{tisca}. Using the time scale transformations, the reception time $t_3$ and the transmission time $t_1$ can be then derived from the light-time delay described in the Section \ref{lidel}. Using these informations it is then possible to compute the Doppler and range observables as described in Section \ref{drobs}.

\subsection{Time scales}
\label{tisca}
\begin{figure*}[tp]
\begin{center}\includegraphics[width=14cm]{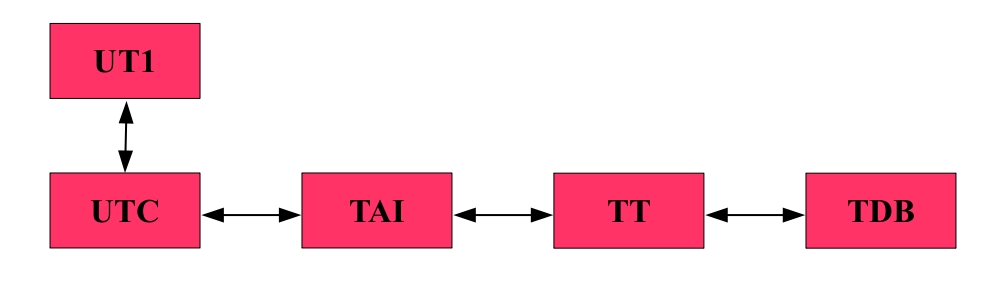}\end{center}
\caption{Transformation between the time scales.}
\label{time_scale}
\end{figure*}

As described in Group 3 of Section \ref{odcon}, the given time in the \gls{ODF} is measured in UTC from 0 hours, 1 January 1950. However, the orbital computations of the celestial body and the artificial satellite are described in \gls{TDB}. Therefore, it is necessary to transform the given UTC time into \gls{TDB}. The transformation between these time scales is give in Figure \ref{time_scale}.

%
 %

\subsubsection{Universal Time (UT or UT1)}
UT1 (or UT) is the modern equivalent of mean solar time. It is defined through the relationship with the Earth rotation angle (formerly through sidereal time), which is the Greenwich hour angle of the mean equinox of date, measured in the true equator of date. Owing the Earth rotation rate which is slightly irregular for geophysical reasons and is gradually decreasing, the UT1 is not uniform. Hence, this makes \gls{UT1} unsuitable for use as a time scale in physics applications.

\subsubsection{Coordinated Universal Time (UTC)}
\begin{figure*}[ht]
\begin{center}\includegraphics[width=12cm]{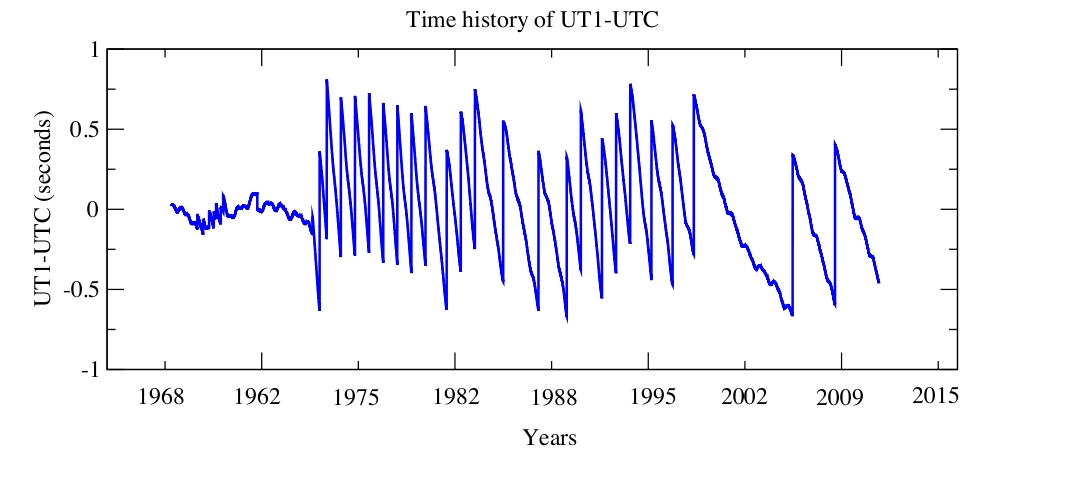}\end{center}
\caption{Time history of the $\triangle$UT1 since 1962. These values of $\triangle$UT1 are extracted from EOP file.}
\label{ut1mutc}
\end{figure*}

\gls{UTC} is the basis of civilian time which is the standard time for $0^\circ$ longitude along the Greenwich meridian. Since January 1, 1972, \gls{UTC} is given in unit of SI seconds and has been derived from the \gls{TAI}. \gls{UTC} is close to \gls{UT1} and maintained within 0.90 second of the observed \gls{UT1} by adding a positive or negative leap second to \gls{UTC}. Figure \ref{ut1mutc} shows the time history of the $\triangle$UT1 since 1962, which can be defined as the time scale difference between \gls{UT1} and \gls{UTC}:

\begin{equation}
\mathrm{
\label{ut1utc}
\triangle UT1 = UT1 - UTC}
\end{equation}

%
 %

The $\triangle$UT1 can be extracted from the \gls{EOP}\footnote{\url{http://www.iers.org/IERS/EN/DataProducts/EarthOrientationData}} file and at any given time, $\triangle$UT1 can be obtained by interpolating this file. 

\subsubsection{International Atomic Time (TAI)}
The International Atomic Time (\gls{TAI}) is measured in the unit of SI second and defined the duration of 9,192,631,770 periods of the radiation corresponding to the transition between the two hyperfine levels of the ground state of the caesium 133 atom \citep{Moyer}. \gls{TAI} is a laboratory time scale, independent of astronomical phenomena apart from having been synchronized to solar time. \gls{TAI} is obtained from a worldwide system of synchronized atomic clocks. It is calculated as a weighted average of times obtained from the individual clocks, and corrections are applied for known effects.

\begin{figure*}[ht]
\begin{center}\includegraphics[width=12cm]{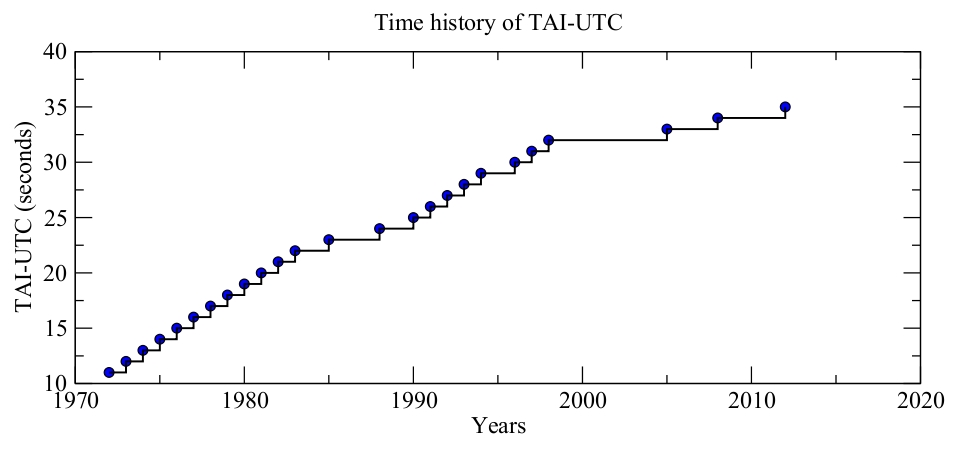}\end{center}
\caption{Time history of the $\triangle$TAI since 1973.}
\label{taimutc}
\end{figure*}

\gls{TAI} is ahead of \gls{UTC} by an integer number of seconds. The Figure \ref{taimutc} shows the time history of the difference between \gls{TAI} and \gls{UTC} time scales $\triangle$TAI since 1973. The value of the $\triangle$TAI can be extracted from the \gls{IERS}\footnote{\url{http://www.iers.org/}} and it given by

\begin{equation}
\label{taiutc}
\mathrm{
\triangle TAI = TAI - UTC}
\end{equation}
%
 %

\subsubsection{Terrestrial Time (TT)}
\gls{TT} is the theoretical time scale for clocks at sea-level. In a modern astronomical time standard, it defined by the \gls{IAU} as a measurement time for astronomical observations made from the surface of the Earth. \gls{TT} runs parallel to the atomic timescale \gls{TAI} and it is ahead of \gls{TAI} by a certain number of seconds which is given as

\begin{equation}
\label{ttmtai}
\mathrm{
TT - TAI = 32.184  \mathrm{s}}
\end{equation}

From Figure \ref{time_scale} and from Eqs. \ref{taiutc} and \ref{ttmtai}, one can transform the time scale from \gls{UTC} to \gls{TT} or from \gls{TT} to \gls{UTC}.

\subsubsection{Barycentric Dynamical Time (TDB)}
%
 %
\begin{figure}[!ht]
\begin{center}\includegraphics[width=8cm]{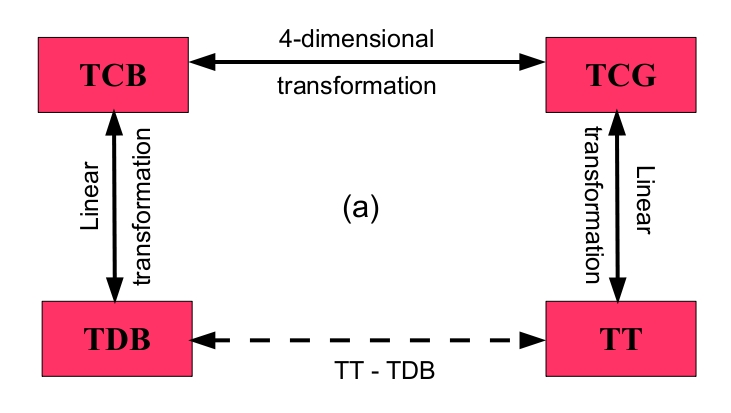}
\includegraphics[width=14cm]{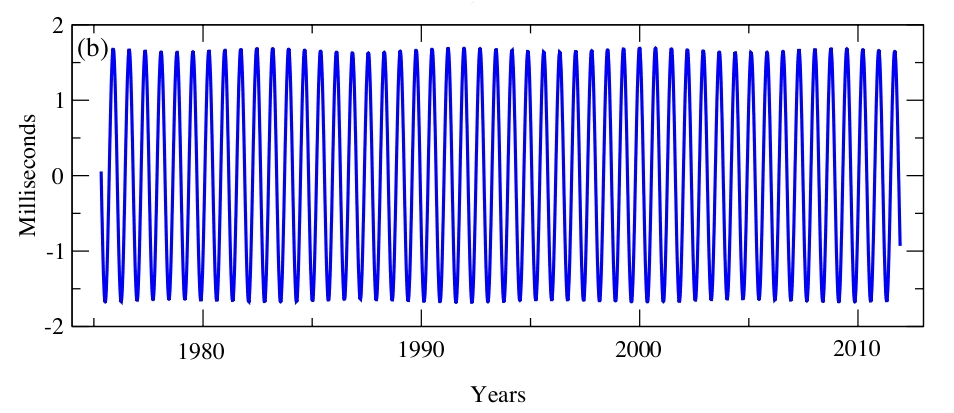}\end{center}
\caption{Panels: (a) transformations of various coordinate time scales, (b) time history of TT - TDB.}
\label{ttmtdb}
\end{figure}
\gls{TT} and \gls{TCG} are the geocenter time scales to be used in the vicinity of the Earth, while \gls{TCB} and \gls{TDB} are the solar system barycentric time scales to be used for planetary ephemerides or interplanetary spacecraft navigation. Transformation between these time scales are plotted in Figure \ref{ttmtdb}. 

The geocentric coordinate time, \gls{TCG}, is appropriate for theoretical studies of geocentric ephemerides and differ from the \gls{TT} by a constant rate with linear transformation \citep{McCarthy04}:
\begin{equation}
\label{ttmtcg}
\mathrm{
TCG - TT =  L_G \times (JD - T_0) \times 86 400}
\end{equation}
where L$_G$ = 6.969290134$\times$10$^{10}$, T$_0$ = 2443144.5003725, and JD is TAI measured in Julian days. T$_0$ is JD at 1977 January 1, 00h 00m 00s TAI. The time-scale used in the ephemerides of planetary spacecraft, as well as that of solar system bodies, is the barycentric dynamical time, \gls{TDB}, a scaled version of the barycentric coordinate time, \gls{TCB}, (the time coordinate of the IAU space-time metric BCRS) \citep{Klioner08}. The TDB stays close to TT ($\textless$ 2ms, see panel $b$ of Figure \ref{ttmtdb}) on the average by suppressing a drift in TCB due to the combined effect of the terrestrial observer orbital speed and the gravitational potential from the Sun and planets by applying a linear transformation \citep{McCarthy04}:
\begin{equation}
\label{tdbmtcb}
\mathrm{
TCB - TDB = L_B(JD - T_0) \times 86400 - TDB_0}
\end{equation}
where T$_0$ = 2443144.5003725, L$_B$ = 1.550519768$\times$10$^{-8}$, TDB$_0$ = -6.55$\times$10$^5$ s, and JD is the TCB Julian date which is T$_0$ for the event 1977 January 1, 00h 00m 00s TAI.

The barycentric coordinate time, TCB, is appropriate for applications where the observer is imagined to be stationary in the solar system so that the gravitational potential of the solar system vanishes at their location and is at rest relative to the solar system barycenter \citep{Klioner08}. The transformation from TCG to TCB thus takes account of the orbital speed of the geocenter and the gravitational potential from the Sun and planets. The difference between TCG and TCB involves a full 4-dimensional GR transformation \citep{McCarthy04}:

\begin{equation}
\label{tcgmtcb}
\mathrm{
TCB - TCG = c^{-2} \bigg\{ \int_{t_0}^{t} \bigg[ \frac{v_e^2}{2} \ + \ U_{ext}\ (\vv{x}_e)\bigg]dt \ + \ \vv{v}_e.(\vv{x} \ - \ \vv{x}_e)\bigg\} + \it{O}(c^{-4})}
\end{equation}

where $\vv{x}_e$ and $\vv{v}_e$ are the barycentric position and velocity of the geocenter, the $\vv{x}$ is the barycentric position of the observer and $U_{ext}$ is the Newtonian potential of all of the solar system bodies apart from the Earth, evaluated at the geocenter. In this formula, $t$ is TCB and $t_0$ is chosen to be consistent with 1977 January 1, 00h 00m 00s TAI. The neglected terms, $\it{O}(c^{-4})$, are of order 10$^{-16}$ in rate for terrestrial observers. $U_{ext}$ ($\vv{x}_e$) and $\vv{v}_e$ are all ephemeris-dependent, and so the resulting TCB belongs to that particular ephemeris, and the term $\vv{v}_e$.($\vv{x}$ - $\vv{x}_e$) is zero at the geocenter.

The all above set of Equations \ref{ttmtcg}-\ref{tcgmtcb} are precisely modeled in INPOP. The numerical integration has been performed to obtain a realization of Equation \ref{tcgmtcb} with a nanosecond accuracy \citep{Fienga2009}. The difference between TT-TDB therefore can be extracted at any time from the INPOP planetary ephemeris using the tool called calceph\footnote{\url{http://www.imcce.fr/inpop/calceph/}}. The spacecraft orbit determination software GINS (see Section \ref{gins}), integrates the equations of motion in the specific coordinate time called, ephemeris time (ET). In GINS, this time is also referred to as \gls{TDB}, as defined by \cite{Moyer}. As discrepancies between TT and TDB or ET are smaller than 2 ms (see panel $b$ of Figure \ref{ttmtdb}), the transformation between the time scales defined either in INPOP or GINS are analogous and show consistency between both software.

\subsection{Light time solution}
\label{lidel}
The light time solution is used to compute the one-way or round-trip light time of the signal propagating between the tracking station on the Earth and the spacecraft. In order to compute the Doppler and range observables, the first step is to obtain the light time solution. This solution can be modeled by computing the positions and velocities of the transmitter at the transmitting time $t_1(TDB)$, spacecraft at the bouncing time $t_2(TDB)$ (for round-trip) or transmitting time $t_2(TDB)$ (for one-way), and receiver at the receiving time $t_3(TDB)$.

For round-trip light time, spacecraft observations involve two tracking stations, a transmitter, and a receiver which may not be at the same location. Therefore, two light time solutions must be computed, one for up-leg of the signal (transmitter to spacecraft) and one for down-leg (spacecraft to receiver). However, one-way light time requires only single solution because the signal is transmitting by the spacecraft to the receiver. These solutions can be obtained in the Solar system barycenter space-time reference frame for a spacecraft located anywhere in the Solar system.

Since spacecraft observations are usually given at receiver time UTC (see section \ref{odcon}), the computation sequence therefore works backward in time: given the receiver time $t_3(UTC)$, bouncing or transmitting time $t_2(TDB)$ can be computed iteratively, and using this result, transmitter time $t_1(UTC)$ is also computed iteratively. The total time delay for the round-trip signal is then computed by summing the two light time solutions (up-leg and down-leg).

The procedure for modeling the spacecraft light time solution can be divided in several steps as discussed below:
\subsubsection{Time conversion}
\begin{itemize}
          \item  As discussed in the Section \ref{odcon}, the spacecraft observations are given in the receiver time $t_3(UTC)$. However, participants (transmitter, spacecraft, and receiver) state vectors (position and velocity) are must be computed in \gls{TDB}. Thus, given receiver time $t_3(UTC)$ can be transformed into receiver time $t_3(TDB)$ as described in Section \ref{tisca}.
         \end{itemize}

\subsubsection{Down-leg $\tau_U$ computation}
\begin{itemize}
         \item Figure \ref{lightime} represents the schematic diagram of the vector relationship between the participants. From this figure, the Solar system barycentric C state vector $r_3^C(t_3)$ of the Earth tracking station at receiver time $t_3(TDB)$ can be calculated by,
        \begin{equation}
        \label{r3c}
        \mathrm{
         {{r}_{3}^{\ C}(t_3)} =   {{r}_{E}^{\ C}(t_3)} \ +\  {{r}_{3}^{\ E}(t_3)} }
        \end{equation}
        where superscript and subscript are correspond to the Solar system barycenter C and the Earth geocenter E. The vector $r_E^C(t_3)$ is the state vectors of the Earth relative to Solar system barycenter C which can be obtained from the planetary ephemerides. The geocentric space-fixed state vectors $r_3^E(t_3)$ of the Earth tracking station can be calculated using proper formulation which includes Earth precession, nutation, polar motion, plate motion, ocean loading, Earth tides, and plot tide. The detail of this formulation can be find in \cite{Moyer}. 

%
 %
\begin{figure*}[ht]
\begin{center}\includegraphics[width=14cm]{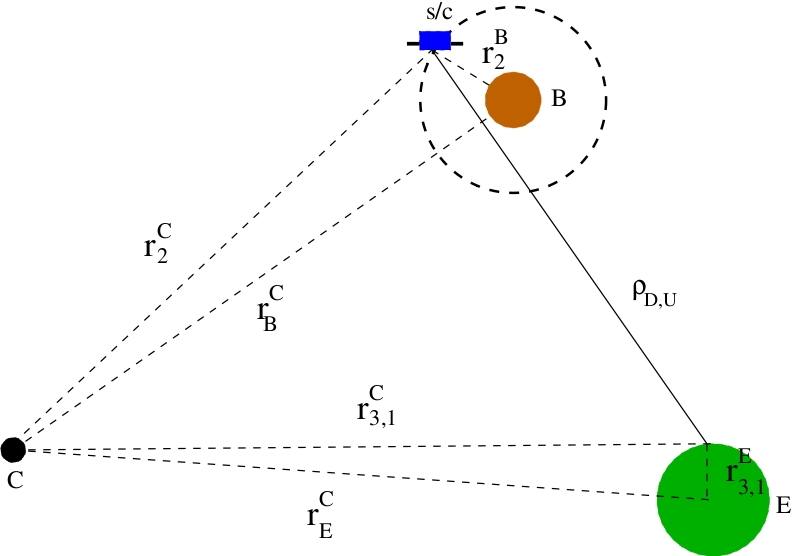}\end{center}
\caption{Geometric sketch of the vectors involved in the computation of the light time solution, where $C$ is the solar system barycentric; $E$ is the Earth geocenter; and B is the center of the central body. }
\label{lightime}
\end{figure*}
        
        \item  The transmission time $t_2(TDB)$ and the corresponding state vectors of the spacecraft has to compute through the iterative process. In order to start the iterations, first approximation of transmission time $t_2(TDB)$ can be taken as the reception time $t_3(TDB)$. Hence, using this approximation and the geometric relationship between the vectors as shown in Figure \ref{lightime}, one can compute the spacecraft state vectors relative to the Solar system barycenter $r_2^C(t_2)$ using the spacecraft and planetary ephemerides. The approximated down-leg time delay $\tau_D$ required by the signal to reach the spacecraft from the Earth receiving station can be then computed as,  
        \begin{equation}
        \label{r2c}
        \mathrm{
         {{r}_{2}^{\ C}(t_2)} =  \bigg[ {{r}_{B}^{\ C}(t_2)} \ +\  {{r}_{2}^{\ B}(t_2)} \bigg]_{t_2 = t_3} }
        \end{equation}        
        \begin{equation}
        \label{r2s}
        \mathrm{
         \tau_D \approx \frac{1}{c} \bigg[ | \ {{r}_{2}^{\ C}(t_2)} \ - \  {{r}_{3}^{\ C}(t_3)} \ | \bigg]_{t_2 = t_3} }
        \end{equation}  

         where superscript $B$ represents the central body of the orbiting spacecraft. The vector $r_B^C(t_2)$ given in Eq. \ref{r2c} is the state vector of the central body relative to the Solar system barycenter obtained from the planetary ephemerides. While, $r_2^B(t_2)$ is the spacecraft state vectors relative to center of the central body computed from the spacecraft ephemerides. In Eq. \ref{r2s}, $c$ is the speed of light and $\tau_D$ is the down-leg time delay which can be computed through the Eqs. \ref{r2c} and \ref{r3c}. An estimated value of the bouncing time $t_2$ can be then computed as,
        \begin{equation}
        \label{t2aprox}
        \mathrm{
         t_2 = t_3 - \tau_D}
        \end{equation}        
         \item Now using this result, we can then estimate the barycentric position of the spacecraft at bouncing time $t_2$. Hence, the down-leg vector $\rho_{D}$ as shown in Figure \ref{lightime} can be then obtained as,
        \begin{equation}
        \label{downleg}
        \mathrm{
         \rho_{D} =  {{r}_{2}^{\ C}(t_2)} \ - \  {{r}_{3}^{\ C}(t_3)} }
        \end{equation} 
          The improved value of the down-leg time delay $\tau_D$, in seconds, can be then estimated as, 
        \begin{equation}
        \label{delay}
        \mathrm{
         \tau_D =  \frac{1}{c} ( |\ \rho_{D} \ | )+ \delta{\tau_D} }
        \end{equation}            
         where $\delta{\tau_D}$ is a down-leg light time corrections which includes the relativistic, solar corona, and media contributions to the propagation delay. Furthermore, Eqs. \ref{t2aprox} to \ref{delay} need to iterate until the latest estimate of $\tau_D$ differs from the previous estimate by some define value such as 0.05$\mu$. 
\end{itemize}
 
\subsubsection{Up-leg $\tau_U$ computation}
\begin{itemize}
           \item For round-trip light time solution, next is to compute the up-leg time delay. A similar iterative procedure as used for down-leg solution can be used the up-leg solution. Up-leg time delay $\tau_U$ which represents the time required for signal to travel between the spacecraft and the Earth transmitting station. In order to begin the iterations, first approximation can be assumed as,
        \begin{equation}
        \label{tauDaprox}
        \mathrm{
         \tau_U \approx \tau_D }
         \end{equation}  
 
         Therefore, while using Eq. \ref{tauDaprox}, approximated transmitting time $t_1(TDB)$ can be then computed as,
        \begin{equation}
        \label{t1aprox}
        \mathrm{
         t_1 = t_2 - \tau_U }
         \end{equation}        
 
         \item The barycentric state vectors of the transmitting station $r_1^C(t_1)$ at transmitted time $t_1(TDB)$ as shown in Figure \ref{lightime} can be computed from Eq. \ref{r3c} by replacing the 3 with 1, that is,
        \begin{equation}
        \label{r1c}
        \mathrm{
         {{r}_{1}^{\ C}(t_1)} =   {{r}_{E}^{\ C}(t_1)} \ +\  {{r}_{1}^{\ E}(t_1)} }
        \end{equation}         
        Now, using Eq. \ref{r1c}, one can compute the up-leg state vector as give by,
        \begin{equation}
        \label{upleg}
        \mathrm{
         \rho_{U} =  {{r}_{2}^{\ C}(t_2)} \ - \  {{r}_{1}^{\ C}(t_1)} }
        \end{equation} 
        where $r_2^C(t_2)$ is the barycentric position of the spacecraft at bouncing time $t_2(TDB)$ and can be calculated from Eq. \ref{r2c}. Finally, the new estimation of the up-leg time delay $\tau_U$, in seconds, is given by,
        \begin{equation}
        \label{updelay}
        \mathrm{
         \tau_U =  \frac{1}{c} ( |\ \rho_{U} \ | )+ \delta{\tau_U} }
        \end{equation}
         where $\delta{\tau_U}$ is the up-leg light time correction analogous to $\delta{\tau_D}$ of Eq. \ref{delay}. Eqs. \ref{t1aprox} to \ref{updelay} are then need to iterative until the convergence is achieved.
\end{itemize}  
        
\subsubsection{Light time corrections, $\delta{\tau_D}$ and $\delta{\tau_U}$} 

\myparagraph{Relativistic correction $\delta{\tau_{RC}}$} Electromagnetic signals that are traveling between the spacecraft and the Earth tracking encounters light time delay when it passes close to the massive celestial bodies. This effects is known as $Shapiro$ $delay$ or gravitational time delay \citep{Shapiro64}. Such time delays are caused by the bending of the light path which increase the travailing path of the signal. Hence, relativistic time delays caused by the gravitational attraction of the bodies can be expressed,  in seconds, as \citep{Shapiro64, Moyer}, 
        \begin{displaymath}
        \mathrm{
         \delta{\tau_{RC_{U}}} =  \frac{(1 + \gamma)\ \mu_S}{c^3} \ ln \ \left[ \frac{ {r}_{1}^{\ S}  + {r}_{2}^{\ S} + {r}_{12}^{\ S} + \frac{(1 + \gamma)\ \mu_S}{c^2}}{{r}_{1}^{\ S}  + {r}_{2}^{\ S} - {r}_{12}^{\ S} + \frac{(1 + \gamma)\ \mu_S}{c^2}} \right]   }
        \end{displaymath}
         \begin{equation}
          \label{rc_u}
        \mathrm{
         +\sum_{B=1}^{10} \  \frac{(1 + \gamma)\ \mu_B}{c^3} \ ln \ \left[ \frac{ {r}_{1}^{\ B}  + {r}_{2}^{\ B} + {r}_{12}^{\ B}}{{r}_{1}^{\ S}  + {r}_{2}^{\ S} - {r}_{12}^{\ S} } \right] }
        \end{equation}
                          
         where superscript $S$ and $B$ correspond to the Sun and the celestial body. $r_1$,  $r_2$, and $r_{12}$ are the distance between the spacecraft and the Sun $S$ (or celestial body $B$), the Earth station and the Sun $S$ (or celestial body $B$), and the spacecraft and the Earth station, respectively. The $\mu_S$ and $\mu_B$ are the gravitational constant of the Sun and the celestial body, respectively.
         
         For round-trip signal, Eq. \ref{rc_u} represents the relativistic time delay $\delta{\tau_{RC_{U}}}$ relative to the up-leg of the signal. The corresponding down-leg relativistic time delay $\delta{\tau_{RC_{D}}}$ can be calculate using the same equation by replacing the 1 with 2 and 2 with 3. Hence, the total relativistic time delay,  in seconds, during the round-trip of the signal can be given as,   
        \begin{equation}
        \label{rc}
        \mathrm{
         \delta{\tau_{RC}} = \delta{\tau_{RC_{U}}} + \delta{\tau_{RC_{D}}}}      
         \end{equation}

\myparagraph{Solar Corona correction $\delta{\tau_{SC}}$} As mentioned in Section \ref{socor}, solar corona severely degrades the radio wave signals when propagating between spacecraft and Earth tracking stations. The delay owing to the solar corona are directly proportional to the total electron contents along the \gls{LOS} and inversely with the square of carrier radio wave frequency. Solar corona model for computing such delays for each legs are described in Chapter \ref{CHP2}. The total round-trip solar corona delay,  in seconds, can be written as,
        \begin{equation}
        \label{sc}
        \mathrm{
         \delta{\tau_{SC}} = \delta{\tau_{SC_{U}}} + \delta{\tau_{SC_{D}}} }      
         \end{equation} 
         
\myparagraph{Media corrections $\delta{\tau_{MC}}$} The media corrections consist of Earth$\textquoteright$s troposphere correction and the correction due to the charge particles of the Earth ionosphere. Such delays however relatively lesser compare to the relativistic and solar corona delays. The tropospheric model used for computing these corrections for each legs are discussed in \cite{chao, Moyer}. The total round-trip media correction,  in seconds, can be written as,
        \begin{equation}
        \label{mc}
        \mathrm{
         \delta{\tau_{MC}} = \delta{\tau_{MC_{U}}} + \delta{\tau_{MC_{D}}} }      
         \end{equation} 
         
\begin{figure*}[ht]
\begin{center}\includegraphics[width=14cm]{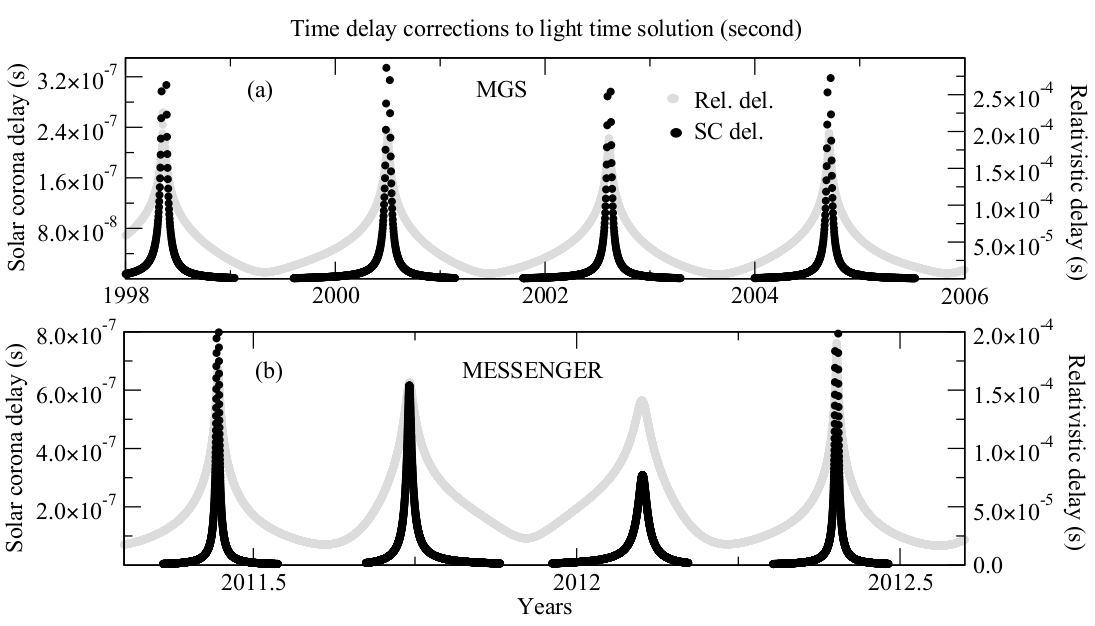}\end{center}
\caption{Relativistic and solar corona corrections to light time solution (expressed in seconds): (a) for MGS, (b) for MESSENGER.}
\label{correction_lightime}
\end{figure*}

         Figure \ref{correction_lightime} shows an example of relativistic correction and solar corona correction to light time solution for MGS and MESSENGER spacecraft.   
%
 %

\subsubsection{Total light time delay}

\myparagraph{Round-trip delay} The total round-trip delay is the sum of number of terms, that includes,
        \begin{displaymath}
        \label{rho}
        \mathrm{
         \rho =  (  \ \tau_{D} \  + \ \tau_{U} \ ) - (TDB - TAI)_{t_3}  +   (TDB - TAI)_{t_1} }      
         \end{displaymath}
         \begin{displaymath}
        \mathrm{
         - (TAI - UTC)_{t_3} +  (TAI - UTC)_{t_1}}      
         \end{displaymath}
          \begin{equation}
        \label{rho}
        \mathrm{
        + \delta{\rho_U} +  \delta{\rho_D}}      
         \end{equation}
         
         where quantities $\delta{\rho_D}$ and $\delta{\rho_U}$ are the downlink delay at receiver and uplink delay at transmitter (see Group 3 of Section \ref{odcon}), respectively. The time differences given in Eq. \ref{rho} can be obtained as described in Section \ref{tisca}, while $\tau_{D}$ and $\tau_{U}$ can be obtained from Eqs. \ref{delay} and \ref{updelay} respectively. 
%
 %
\begin{figure}[!ht]
\begin{center}\includegraphics[width=14cm]{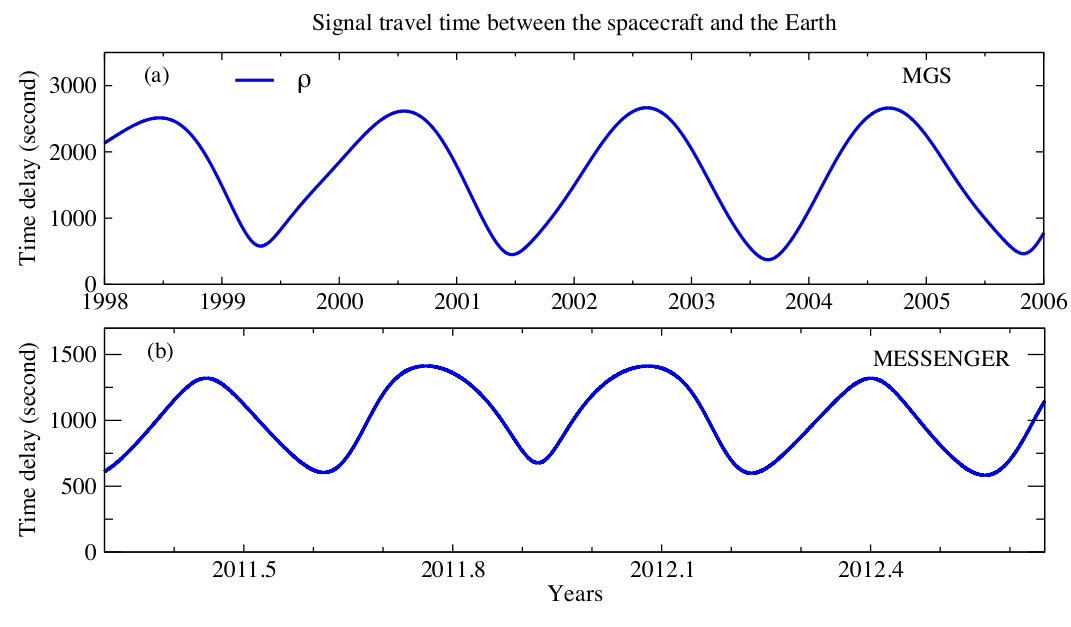}\end{center}
\caption{Round-trip light time solution of MGS (panel $a$) and MESSENGER (panel $b$) spacecraft, computed from the Eq. \ref{rho}.}
\label{lightime_sol}
\end{figure}
Figure \ref{lightime_sol} illustrates an example for the total round-trip time that required by the signal to travel from the transmitter to the spacecraft (up-leg) and then from the spacecraft to the receiver (down-leg). This time solution shown in Figure \ref{lightime_sol} corresponds to MGS (panel $a$) and MESSENGER (panel $b$) spacecraft. 
         
\myparagraph{One-way delay} One-way light,  in seconds, which is used to calculate the one-way Doppler observables can be calculated as,
        \begin{equation}
        \label{rho_1w}
        \mathrm{
         \hat{\rho_1} =   \rho_1 + (TDB - TAI)_{t_2}  }
         \end{equation}
         where $\rho_1$ is the difference between reception time $t_3(UTC)$ and the spacecraft transmitter time $t_2(TDB)$.

\subsection{Doppler and range observables}
\label{drobs}
The radiometric data obtained by the Earth tracking station (\gls{DSN}) usually consists of three kind of measurements (one-way Doppler, two/three-way Doppler, and two-way range). The detail description of these measurements and other contents of the \gls{ODF} that are useful for recognize the data are described in Section \ref{odcon}. Observation model which computes the observations requires the time history of the transmitted frequency at the transmitter. Such time history which contains transmitter time $t_1(UTC)$ and receiver time $t_3(UTC)$, can be obtained as described in Section \ref{lidel}. The corresponding transmitter frequency which can be obtained from the different forms of given frequency in \gls{ODF} is described in Group 3 of Section \ref{odcon}. 

One of the most important aspect of precise orbit determination of the spacecraft is to compute the observables. These observables require spacecraft ephemerides which can be constructed from the dynamic modeling (see Section \ref{gins}). An observation model accounts the propagation of signal and allows to compute the frequency change between received and transmitted signal, also called $Doppler$ $shift$. The difference between observed (given in \gls{ODF}) and computed values, also called residuals, are then used to adjust the dynamic model along with observation model for accounting the discrepancy in the models.

This section contains the formulations for computing the one-way Doppler, two/three-way Doppler, and two-way range observables. These formulations are based on \cite{Moyer}. The motivation for developing an observation model is to have a better understanding of the radiometric data. However, for precise computation, such as for MGS (see Chapter \ref{CHP2}) and for MESSENGER (see Chapter \ref{CHP4}), \gls{GINS} orbit determination model (see Section \ref{gins}) has been used. Moreover, \gls{GINS} observation model is also based on \cite{Moyer} formulations and the brief overview of the \gls{GINS} dynamic model is described in Section \ref{gins}.

\subsubsection{Two-way (F$_2$) and Three-way (F$_3$) Doppler} 
\label{observation_structure_para}
\myparagraph{Ramped} Doppler observable can be derived from the difference between the number of cycles received by a receiving station and the number of cycles produced by a fixed or ramped known reference frequency $f_{REF}$, during a specific count interval $T_c$. The given observables time-tag $TT$ in the \gls{ODF} is the mid-point of the count interval $T_c$ (see Group 3 of Section \ref{odcon}). To compute these observables (Eq. \ref{2-wdop}), it is thus necessary to obtain the starting time $t_{3_{s}}(UTC)$ and the ending time $t_{3_{e}}(UTC)$ of the count interval, which is given in  seconds by
        \begin{equation}
        \label{start_time}
        \mathrm{
          t_{3_{s}}(UTC) =  TT - \frac{1}{2} \ T_c  }
         \end{equation}            
        \begin{equation}
        \label{end_time}
        \mathrm{
          t_{3_{e}}(UTC) =  TT + \frac{1}{2} \ T_c }
         \end{equation}

         where $TT$ and $T_c$ can be extracted from the \gls{ODF}. Using Eqs. \ref{start_time} and \ref{end_time} the corresponding transmitting starting time $t_{1_{s}}(UTC)$ and ending time $t_{1_{e}}(UTC)$,  in seconds, can be obtained from light time solution (see Section \ref{lidel}), that is,
        \begin{equation}
        \label{start_timet1}
        \mathrm{
          t_{1_{s}}(UTC) =  t_{3_{s}}(UTC)  - \rho_s  }
         \end{equation}            
        \begin{equation}
        \label{end_timet1}
        \mathrm{
          t_{1_{e}}(UTC) =  t_{3_{e}}(UTC)  - \rho_e }
         \end{equation}  
         
         where  $\rho_s$ and  $\rho_e$ is the round-trip light time computed from Eq. \ref{rho}. Similarly, the corresponding start and end \gls{TDB} at the receiving station and at the transmission station, which are required for the light time solution, can be computed as described in Section \ref{tisca}.

          Using the time recorded history of the transmitters, the two-way Doppler $F_2$ and three-way Doppler $F_3$ can be computed as the difference in the total accumulation of the Doppler cycles, which is given as, in Hz,
 
        \begin{equation}
        \label{2-wdop1}
        \mathrm{
          F_{2,3} = \frac{1}{T_c} \ \left[  \int_{t_{3_{s}}}^{t_{3_{e}}} \ f_{REF} (t_3) \ dt_3 \ -     \int_{t_{1_{s}}}^{t_{1_{e}}} \ f(t_1) \ dt_1  \right]  }
         \end{equation}    
         where $f_{REF}(t_3)$ and $f(t_1)$ can be computed from the Eqs. \ref{fref} and \ref{f1}. The all time scales given in Eq. \ref{2-wdop1} correspond to \gls{UTC}. Now by substituting Eqs. \ref{fref} and \ref{f1} into Eq. \ref{2-wdop1} gives, in Hz:
        \begin{equation}
        \label{2-wdop}
        \mathrm{
          F_{2,3} = \frac{M_{2_{R}}}{T_c} \  \int_{t_{3_{s}}}^{t_{3_{e}}} \ f_{T} (t_3) \ dt_3 \ -     \frac{M_2}{T_c} \  \int_{t_{1_{s}}}^{t_{1_{e}}} \ f_T(t_1) \ dt_1 }
         \end{equation} 
         
         where $M_{2_{R}}$ and $M_2$ are the spacecraft turnaround ratio which is given in Table \ref{M2}. The transmitter frequency $f_T(t_1)$ at the transmitting station on Earth is ramped and can be obtained from the ramped table using Eq. \ref{ramp}. However, the transmitter frequency $f_T(t_3)$ at receiving station can be fixed or ramped. If it is ramped then it can be obtained from the ramped table using Eq. \ref{ramp} and for fixed, Eq. \ref{2-wdop} can be re-written as, in Hz:
        \begin{equation}
        \label{2-wdop2}
        \mathrm{
          F_{2,3} =  M_{2_{R}} \ f_{T} (t_3) \ -     \frac{M_2}{T_c} \  \int_{t_{1_{s}}}^{t_{1_{e}}} \ f_T(t_1) \ dt_1  }
         \end{equation} 
          
         In order to compute the observables, it is necessary to solve the integrations given in Eq. \ref{2-wdop}. Let us assume that, $W$ is the precision width of the interval of the integration,  in seconds, which is $T_c$ for reception interval and $T_c^T$ for the transmission interval, and can be expressed as, in seconds:
        \begin{equation}
        \label{2-wdop2}
        \mathrm{
        T_c^T = t_{1_{e}} - \ t_{1_{s}}}
        \end{equation}

\begin{figure*}[ht]
\begin{center}\includegraphics[width=14cm]{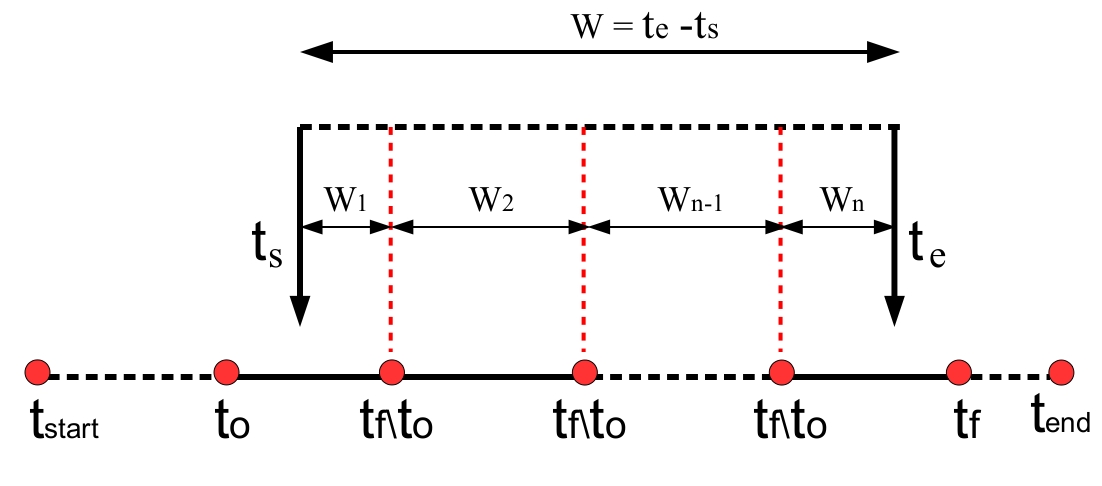}\end{center}
\caption{The technique used for computing the integration of Eq. \ref{2-wdop}. $W$ is the precision width of the interval of the integration, and t$_{start}$ and t$_{end}$ is the start and end times of the ramp table (see Section \ref{observation_structure_para}).}
\label{observation_structure}
\end{figure*}

        Furthermore, let $t_s$ be the starting time of the interval of integration which is $t_{3_{s}}(UTC)$ for the reception and $t_{1_{s}}(UTC)$ for the transmission. Similarly corresponding end time can be denoted as $t_e$. Each ramp of the ramp table given in the \gls{ODF} is specified by the start time $t_0$ and end time $t_f$ for each participating Earth stations (see Group 3 of Section \ref{odcon}). The interval of the integration can be covered by one or more ramps (let say $n$ ramps). Figure \ref{observation_structure} illustrates the above assumptions and the technique used for computing the integration of Eq. \ref{2-wdop}. 
%
 %
 In Figure \ref{observation_structure}, $t_{start}$ and $t_{end}$ is the starting and ending time of the ramp table respectively. Now using Figure \ref{observation_structure} and above made assumptions, one can compute the observables as follows: 
     \begin{enumerate}
     \item Compute the transmitter frequency at the start time $t_s$ of the integration using the first ramp (see Figure \ref{observation_structure}) transmitter frequency. It can be achieved by using the Eq. \ref{ramp}. Therefore, the new transmitter frequency can be given as, in Hz:  
        \begin{equation}
        \label{f_start}
        \mathrm{
        f_0\ (t_s) =  f_0\ (t_0) + \dot{f} \ (t_s - t_0) }
        \end{equation}
        where $\dot{f}$ is the corresponding frequency rate of first ramp expressed in Hz/s.
        
        \item If the interval of integration $W$ contains the two or more ramp as shown in Figure \ref{observation_structure}, then calculates the width of each ramp $i$ except the last ramp:
        \begin{equation}
        \label{width}
        \mathrm{
        W_i = t_f - t_0}
        \end{equation}        
       where $t_f$ and $t_0$ are the start and end times, in seconds, of each ramp (see Figure \ref{observation_structure}). Last ramp precision width $W_n$ therefore computed as:
        \begin{equation}
        \label{width_n}
        \mathrm{
        W_n = W - \sum_{i=1}^{n-1} \ W_i }
        \end{equation} 
       
         If the interval of the integration $W$ only contains the single ramp ($n$=1), then the precession width is given as:
        \begin{equation}
        \label{width_1}
        \mathrm{
        W_{n=1} = W  }
        \end{equation}         
       
        \item Calculate the average of the transmitter frequency $f_i$ for each ramp, in Hz:
        \begin{equation}
        \label{freq_avg}
        \mathrm{
        f_i = f_0 + \frac{1}{2} \ \dot{f} \ W_i }
        \end{equation}
        where $f_0$ is the transmitter frequency at the each start time $t_s$ of the ramp (see Figure \ref{observation_structure}) and $\dot{f}$ is the corresponding frequency rate. Ramp width $W_i$ for each ramp can be obtained from Eq. \ref{width}.
     
      \item The integral of the transmission frequency over the reception of transmission interval $W$ can be then obtained as: 
        \begin{equation}
        \label{freq_avg}
        \mathrm{
        \int_{t_s}^{t_e}  f_T(t)dt = \sum_{i=1}^{n} \ f_i \ W_i \hspace{20mm} cycles}
        \end{equation}
      \end{enumerate}

\myparagraph{Unramped} As mentioned earlier, the given transmitter frequency can be constant or ramped. If it is constant, then it corresponds to the unramped transmitter frequency. Let us consider that, during an interval $dt_1$, $dn$ cycles of the constant transmitter frequency $f_T(t_1)$ are transmitted. During the corresponding reception interval $dt_3$, receiving station on Earth received $M_2dn$ cycles, where $M_2$ is the spacecraft turnaround ration (see Table \ref{M2}). Therefore, the total accumulation of the constant Doppler cycles is given as, in Hz:
        \begin{equation}
        \label{2-wdop_unramp1}
        \mathrm{
          F_{2,3} = \frac{M_{2} \ f_T(t_1)}{T_c} \ \left[  \int_{t_{3_{s}}}^{t_{3_{e}}}  dt_3 \ -       \int_{t_{1_{s}}}^{t_{1_{e}}}  dt_1 \right]  }
         \end{equation} 
         where $t_{3_{s}}$ and $t_{3_{e}}$ are the start and end times of the reception time-tag $TT$ which can be computed from Eqs. \ref{start_time} and \ref{end_time} respectively. Similarly $t_{1_{s}}$ and $t_{1_{e}}$ are the corresponding transmitting times which can be computed from Eqs. \ref{start_timet1} and \ref{end_timet1} respectively. All the time given in Eq. \ref{2-wdop_unramp1} are in UTC.
         
         Now evaluating Eq. \ref{2-wdop_unramp1}:
        \begin{equation}
        \label{2-wdop_unramp}
        \mathrm{
          F_{2,3} = \frac{M_{2} \ f_T(t_1)}{T_c} \ \bigg\{ [{t_{3_{e}}} - t_{1_{e}} ] -   [ {t_{3_{s}}} - t_{1_{s}}] \bigg\}  }  
         \end{equation} 
         
         Eq. \ref{2-wdop_unramp} can be used to calculate the computed values of unramped two-way $F_2$ and three-way $F_3$ Doppler observables, in Hz.        

\subsubsection{One-way (F$_1$) Doppler} 
        When the radio signal is continuously transmitted from the spacecraft and received by the \gls{DSN} station on Earth, then the observables are referred to one-way. These observations are always unramped and can be modeled as, in Hz, \citep{Moyer}: 

         \begin{equation}
         \label{F1_1}
         \mathrm{ 
         F_{1} = C_2 \ f_{T_{0}}- \frac{1}{T_c} \int_{t_{2_{s}}(TAI)}^{t_{2_{e}}(TAI)} \bigg[ f_{T}(t_2) \bigg] \ dt_2(TAI)      }
          \end{equation} 
          where $C_2$ is the downlink frequency multiplier given in Table \ref{C2}. $f_{T_{0}}$ is the nominal value of spacecraft transmitter frequency $f_{S/C}$ (Eq. \ref{FS/C}). $f_{T}(t_2)$ is the transmitter frequency at the spacecraft at transmission time $t_2(TAI)$, given by Eq. \ref{f1-way}.

             Now by substituting Eqs. \ref{f1-way} and \ref{FS/C} in Eq. \ref{F1_1}, the one-way Doppler observables can be written as, in Hz:
             
         \begin{equation}
         \label{F1_2}
         \mathrm{ 
        F_{1} = C_2 \ f_{T_{0}} - \frac{C_2}{T_c} \int_{t_{2_{s}}(TAI)}^{t_{2_{e}}(TAI)} \bigg[ f_{T_{0}} + \Delta{f_{T_{0}}}  + f_{T_{1}} (t_2 - t_0) + f_{T_{2}} (t_2 - t_0)^2 \bigg] \ dt_2(TAI) }
          \end{equation}

          As one can see in Eq. \ref{F1_2}, the terms containing the coefficients $f_{T_{1}}$ and $f_{T_{2}}$ are functions of the spacecraft transmission time $t_2(TAI)$. Hence, the upper limit $t_{2_{e}}(TAI)$ and lower limit $t_{2_{s}}(TAI)$ of the integration are only required to evaluate these terms. Since these terms ($\textless$ 2ms) are small, the limits of the integration can be replaced with the corresponding values in coordinate time \gls{TDB} \citep{Moyer}. These coordinate times, $t_{2_{e}}(TDB)$ and $t_{2_{s}}(TDB)$, can be obtained from the light time solution (see down-leg computation of Section \ref{lidel}).

          Now by evaluating the integral of Eq. \ref{F1_2} using the above approximation it comes \citep{Moyer}, in Hz:
        \begin{displaymath}
        \mathrm{ 
          F_{1} = C_2 \ f_{T_{0}} \frac{({\rho_{1_{e}}} - {\rho_{1_{s}}} + \Delta)}{T_c}  }
         \end{displaymath} 
         \begin{equation}
         \label{F1}
         \mathrm{ 
         - C_2 \ \bigg\{  \Delta{f_{T_{0}}}  + f_{T_{1}} (t_{2_{m}} - t_0) + f_{T_{2}} (t_{2_{m}} - t_0)^2  + \frac{f_{T_{2}}}{12} \bigg(T_c^{\textquoteright}\bigg)^2 \bigg\} \ \frac{T_c^{\textquoteright}}{T_c}  }
          \end{equation}

 where ${\rho_{1_{e}}}$ and ${\rho_{1_{s}}}$ are the one-way light times at the end and at the start of the Doppler count interval $T_c$ at the receiver. These one-way light times and the time bias $\Delta$, expressed in seconds, are given by:

            \begin{equation}
            \label{rho-1waye}
            \mathrm{        
             \rho_{1_{e}}  = t_{3_{e}}(UTC)  - t_{2_{e}}(TDB)}
             \end{equation}
            \begin{equation}
            \label{rho-1ways}
            \mathrm{        
             \rho_{1_{s}}  = t_{3_{s}}(UTC)  - t_{2_{s}}(TDB) }
             \end{equation}
            \begin{equation}
            \label{rho-1way-delta}
            \mathrm{        
             \Delta  = (TDB - TAI)_{t_{2_{e}}} -  (TDB - TAI)_{t_{2_{s}}} }
             \end{equation}
 
              The quadratic coefficients $\Delta{f_{T_{0}}}$, $f_{T_{1}}$, and $f_{T_{2}}$ given in Eq. \ref{F1} are specified by time block with start a time $t_0$. In Eq. \ref{F1}, ${T}^{\textquotesingle}_c$ is the transmission interval at the spacecraft and $t_{2_{m}}$ is the average of the \gls{TDB} values of the epochs at the start and end of the transmission interval at the spacecraft. These can expressed as, in seconds:
            \begin{equation}
            \label{tc_dash}
            \mathrm{
            T_c^{\textquoteright} = T_c - ({\rho_{1_{e}}} - {\rho_{1_{s}}} + \Delta) }
            \end{equation}             
            \begin{equation}
            \label{t2m}
            \mathrm{
            t_{2_{m}} = \frac{t_{2_{e}}(TDB) + t_{2_{s}}(TDB)}{2} }
            \end{equation}

\subsubsection{Two-way ($\rho_{2,3}$) Range} 
%
 %
\begin{table}[ht] 
\caption{Constant C$_{range}$ requried for converting second to range units.}
\renewcommand{\arraystretch}{1.8}
\small
\begin{minipage}[t]{0.5\textwidth}
\vspace{0pt}
\hspace{2cm}
\begin{threeparttable}
\begin{tabular}{c c c c}\Xhline{2\arrayrulewidth}
& Type  &   C$_{range}$   &  \\ \Xhline{2\arrayrulewidth}
&S-Band       &  \large{$\frac{1}{2}$}     &  \\ 
&X-Band$^a$, HEF       &  \large{$\frac{11}{75}$ }    &  \\
&X-Band$^b$, BVE     &  \large{$\frac{221}{1496}$}  & \\
\Xhline{2\arrayrulewidth}
\end{tabular}
\end{threeparttable}
\end{minipage}
\begin{minipage}[t]{0.3\textwidth}
\vspace{0.5cm}
a: for uplink X-band at a 34-m mount high efficiency (HEF) antenna.    \\ 
b: for uplink X-band at any tracking station that as a  block 5 exciter (BVE).
\end{minipage}
\label{CRange}
\end{table}
              In addition to Doppler data, \gls{ODF} also contains the range data as described in section \ref{odcon}. These data sets usually are not included in the orbit determination process. However, processing of range data along with orbit determination are extremely useful for the improvement of the planetary ephemerides. This section will give the formulation for computing the two-way ramp range observables.
              
              The range observables given in \gls{ODF} are uniquely given in the range units. The conversion factor $F$ required to convert seconds into range units is a function of the transmitter signal frequency $f_T$ and a constant fraction depending on the uplink band \citep{Moyer}. The integral of $Fdt$ at the transmitting station gives the change in the phase of the transmitted ranging pulse (uniquely coded in range units) at its transmission time. When the spacecraft receives the ranging pulse, it returns the pulse on its downlink. The time needed by the spacecraft to turn the pulse around within its electronics is called transponder delay which is different for each spacecraft. When the pulse is received at the Earth station, the actual light time is determined including the light time delay corrections described in section \ref{lidel}. In addition it also includes the delays due to the transmitter and receiver electronics at the Earth station and the delay in the spacecraft transponder.
              
           The equation for calculating the conversion factor F at the transmitting or receiving station on Earth is a function of the uplink band at the station. For S-, and X-band transmitter frequency $f_T$, is in range units/second:
            \begin{equation}
            \label{conversion_factor}
            \mathrm{
            F = C_{range} \ f_T  }
            \end{equation}

where, C$_{range}$ is the band based constant given in Table \ref{CRange}. The classification of uplink X-band antennas, which may either be mounted as \gls{HEF} or as \gls{BVE}, can be obtained from the \gls{ODF}. The equation for calculating the computed value of 2-way ramped range observable is given by \citep{Moyer}, in modulo M expressed range units,
            \begin{equation}
            \label{2-way_range}
            \mathrm{
            \rho_2 = \int_{t_1(UTC)}^{t_3(UTC)}  F(t) \ dt   }
            \end{equation}             
            where $t_3(UTC)$ is the reception time at the receiving station and $t_1(UTC)$ is the transmission time at the transmitting station. The conversion factor $F(t)$ can be calculated from Eq. \ref{conversion_factor} for the corresponding band. Modulo $M$ is the length of the ranging pulse in range units and it is calculated from:
            \begin{equation}
            \label{modulo}
            \mathrm{
            M = 2^{n+6} }
            \end{equation} 
             
             where $n$ is the component number of the lowest frequency ranging component. The number $n$ can be obtained from the \gls{ODF}. The integral of Eq. \ref{2-way_range} can be evaluated using the same technique as described for two- and three-way ramped Doppler. 
             
             The transmitting time $t_1(UTC)$ given in Eq. \ref{2-way_range} can be obtained from the light time solution as described in Section \ref{lidel}. Now by substituting Eq. \ref{conversion_factor} in Eq. \ref{2-way_range}, it will give the time integral of ramped transmitted frequency $f_T(t)$. This integral can be then evaluated by using the algorithms given for Eq. \ref{2-wdop} by replacing $t_{3_{e}}$ or $t_{1_{e}}$ with $t_3$ and $t_{3_{s}}$ or $t_{1_{s}}$ with $t_1$.


\section{GINS: orbit determination software}
\label{gins}
An accurate orbit determination of the orbiting spacecraft involves an estimation of the positions and velocities of the spacecraft from a sequence of  observations, which are functions of the spacecraft position, and velocity. This can be accomplished by integrating the equations of motion, starting from an initial epoch to produce predicted observations. In practice, the initial state (position and velocity) of the spacecraft is never known exactly. However, some physical constants and parameters of the forces (gravitational and non-gravitational) which are required to integrate the equations of motion, are known approximately. Such limit of the knowledge of the motion, would deviate the predicted motion of the spacecraft from the actual motion. The precision of orbit determination results therefore depends on the error in spacecraft dynamic model .
%
 %
\begin{figure*}[ht]
\begin{center}\includegraphics[width=14cm]{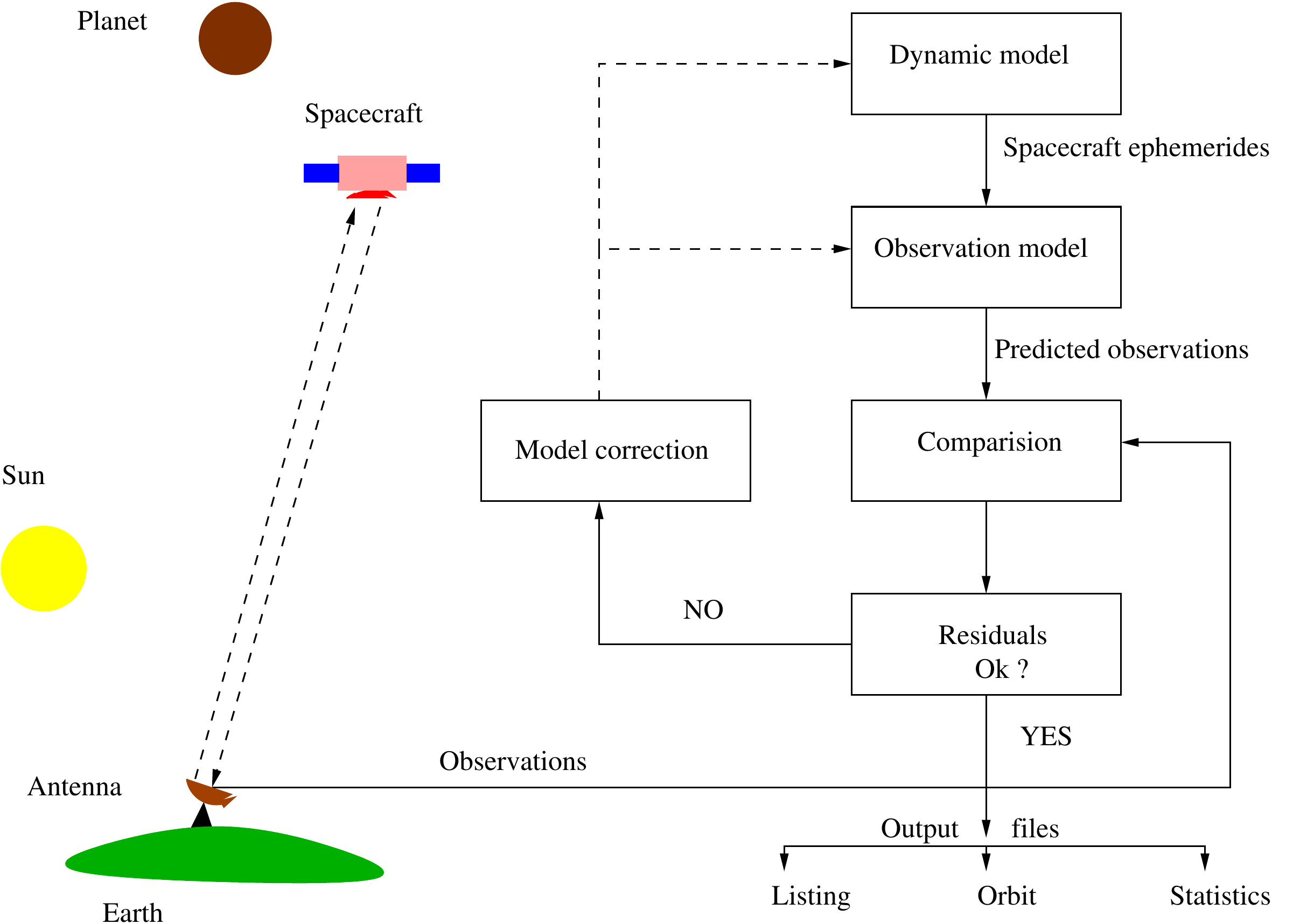}\end{center}
\caption{The process of the spacecraft orbit determination.}
\label{flow_dig}
\end{figure*}
To absorb this error, the components of the spacecraft state (position and velocity, estimated force, and measurement model parameters) at the initial epoch are then re-estimated by comparison to observations in order to minimize the observation residuals (observation - computation). The adjustment of the parameters can be achieved by using iterative least-square techniques. Hence, it can be summarized that, for the precise orbit determination, one needs a dynamic model which describes the forces acting on the spacecraft, a observational model which is a function of observed parameters and spacecraft state vector, and a least-squares estimation algorithm used to obtain the estimates.

The flow chart of orbit determination based on dynamic model in the frame of \gls{GINS} software is given Figure \ref{flow_dig}. \gls{GINS} is an orbit determination software developed by the \gls{CNES}. This software was initially developed for the Earth orbiting spacecraft and then further developed for the interplanetary spacecraft. \gls{GINS} numerically integrates the equations of motion and the associated variational equations, and simultaneously retrieves the physical parameters of the force model using an iterative least-squares technique. From Figure \ref{flow_dig}, one can see that, the first step of the orbit determination is to compute the spacecraft ephemerides from the dynamic model of the spacecraft motion. The section \ref{dymod} gives a brief description of such dynamic model used in the \gls{GINS} software for planetary spacecraft. After computing the spacecraft ephemerides, the next step is to predict the observations from the observation model. The formulation associated with the observation model, which may consist to one-, two-, and three-way Doppler and range observables, is described in Section \ref{drobs}. Finally, the difference between observed and computed observables gives the residuals, which describe the precision in the computations. These residuals are then used to adjust the dynamic and observation models by computing the solve-for parameters using least square techniques described in Section \ref{lestsq}. If the residuals meet the convergence criteria then it produce a three files output: a listing containing the estimated solve-for parameters, an ephemeris of the orbit containing the spacecraft state vectors and its acceleration at each time step of the numerical integration, and statistics of the residuals. The detailed description of \gls{GINS} software algorithms are given in the igsac-cnes webpage\footnote{\url{http://igsac-cnes.cls.fr/documents/gins/GINS_Doc_Algo_V4.html}}.

\subsection{Dynamic model}
\label{dymod}
The dynamic model which consists of equations of motion can be described in an inertial frame as follows:
            \begin{equation}
            \label{motion}
            \mathrm{
            \ddot{\vv{r}} = {\vv{a}_g} + {\vv{a}_{ng}} }
            \end{equation}

where $\vv{r}$ is the position vector between the center of mass of the spacecraft and the planet, $ {\vv{a}_g}$ is the sum of gravitational forces acting on the spacecraft, and ${\vv{a}_{ng}}$ is the sum of non-gravitational forces acting on the spacecraft. 

\subsubsection{Gravitational forces}

 When the spacecraft is orbiting the planet then the primary force acting of the spacecraft is the force of gravity, specifically, the gravitational attraction of the orbiting planet. However, there are other gravitational forces also which significantly affect the motion of the spacecraft. All these forces can be expressed as:
            \begin{equation}
            \label{gravi_force}
            \mathrm{
            \vv{a}_g = \vv{P}_{pot}  + \vv{P}_{tid} + \vv{P}_{n}  + \vv{P}_{rel} }
            \end{equation}
            
where \\

            \hspace{25mm}  $\vv{P}_{pot}$   \hspace{3mm}=\hspace{3mm}  perturbation due to the gravitational potential of the planet \\        
            
             \hspace{25mm}  $\vv{P}_{rel}$     \hspace{3mm}=\hspace{3mm}   perturbation due to solid planetary tides \\    
            
            \hspace{25mm}  $\vv{P}_{n}$       \hspace{5mm}=\hspace{3mm}   perturbation due to the Sun, Moon and planets \\ 
            
            \hspace{25mm}  $\vv{P}_{rel}$     \hspace{3mm}=\hspace{3mm}   perturbation due to \gls{GR} \\
            
          \myparagraph{Gravitational potential} 
          The perturbing forces acting of the spacecraft due to the gravitational attraction of the orbiting planet can be expressed as the gradient of the potential, $U$, which satisfies the Laplace equation, ${\nabla}^2U = 0$, that is: 
            \begin{equation}
            \label{pot_force}
            \mathrm{
            \vv{P}_{pot} = \nabla U  }
            \end{equation}           
           where $U$ is the potential due to the solid-body mass distribution. It is generally expressed in terms of a spherical harmonic expansion in a reference system fixed with respect to the planet \citep{Kaula66}:

            \begin{equation}
            \label{potential}
            \mathrm{
            U = \frac{GM}{r} + \frac{GM}{r} \sum_{l=0}^{L}\sum_{m=0}^{l} \left(\frac{R}{r}\right)^l P_{lm} (sin \ \phi) \ \bigg(C_{lm} \ cos(m\lambda) + S_{lm} \ sin(m\lambda) \bigg)     }
            \end{equation} 
where \\

            \hspace{25mm}  $GM$   \hspace{13mm}=\hspace{3mm}  the gravitational constant of the planet \\            
                            
            \hspace{25mm}  $P_{lm} (sin \ \phi)$     \hspace{3mm}=\hspace{3mm}  the Legendre function of degree $l$ and order $m$ \\
            
            \hspace{25mm}  $C_{lm}, S_{lm}$       \hspace{6mm}=\hspace{3mm} the dimensionless spherical harmonic coefficients of 
             
             \hspace{52mm} degree $l$ and order $m$ \\ 
                         
             \hspace{25mm}  $R$      \hspace{17mm}=\hspace{3mm}   the mean equatorial radius of the planet \\
            
            \hspace{25mm}  $r, \phi, \lambda$     \hspace{10mm}=\hspace{3mm} the spherical coordinates of the spacecraft in a reference 
            
             \hspace{52mm} system fixed with respect to the planet \\
            
            \hspace{25mm}  $L$     \hspace{17mm}=\hspace{3mm}  the maximum number of degree and order \\
            
            In practice, Eq. \ref{potential} is usually represented by the normalized spherical harmonic coefficients ($\bar{C}_{lm}$, $\bar{S}_{lm}$) and normalized Legendre function $\bar{P}_{lm}$. The normalized coefficients are much more uniform in magnitude than the unnormalized coefficients. These normalized coefficients are defined by:
            \begin{equation}
            \label{normal}
            \mathrm{
            \begin{Bmatrix}
            \bar{C}_{lm}\\
            \bar{S}_{lm} 
            \end{Bmatrix} 
              = \sqrt {\frac{(l+m)!}{(2-\delta_{0m})(2l+1)(l-m)!}} \ 
            \begin{Bmatrix}
            {C_{lm}} \\
            S_{lm} 
            \end{Bmatrix}  }
            \end{equation}           
           
           and 
            \begin{equation}
            \label{normal}
            \mathrm{
            \bar{P}_{lm} =  \sqrt {\frac{(2-\delta_{0m})(2l+1)(l-m)!}{(l+m)!}} \ P_{lm} }
             \end{equation}
            If one assumes that the reference system origin coincides with center of mass of the planet, then summation of Eq. \ref{potential} starts from degree ($l$) 2 or $\bar{C}_{10}$ = $\bar{C}_{11}$ = $\bar{S}_{10}$ = 0.
            
             \myparagraph{Solid planetary tides}
              Since the planet is a non-rigid elastic body, its mass distribution and the shape will be changed under the gravitational attraction of the perturbing bodies. Tides deformations of a planet or natural satellite caused by periodic variations of the local gravity acceleration as the planet or satellite rotates and revolves in the gravity field of a perturbing bodies. Tidal disturbances of a planet are primarily caused by the sun and by its satellites. The temporal variation of the free space geopotential induced from solid planet tides can be expressed as a change in the external geopotential by the following expression \citep{Wahr81,Dow88}
            \begin{equation}
            \mathrm{
            \vv{a}_{tid} = \nabla (\Delta U_{tid})}
            \end{equation}
            where,
            \begin{equation}
            \label{tides_planet}
            \mathrm{            
            \Delta U_{tid} =  \frac{GM}{R^2} \sum_{l=2}^{3}\sum_{m=0}^{l}\sum_{k(l,m)} H_k \ e^{i(\Theta_k + \chi_k)} \  k_k^0 \ \bigg [ \bigg ( \frac{R}{r}  \bigg)^{l+1} \ Y_m^l (\phi, \lambda) + k_k^+ \   \bigg( \frac{R}{r}  \bigg )^{l+3}  \ Y_m^{l+2} (\phi, \lambda)  \bigg ] }
            \end{equation}            
           
            \begin{equation}
            \mathrm{            
             Y_m^{l} (\phi, \lambda) =   (-1)^m \sqrt {\frac{(2l+1)(l-m)!}{4\pi \ (l+m)!}} \ P_{lm} (sin \ \phi)\ e^{im\lambda} }
            \end{equation}
            
             \hspace{20mm}  $P_{lm} (sin \ \phi)$   \hspace{3mm}=\hspace{3mm}  the unnormalized associated Legendre function of 
             
              \hspace{50mm} degree $l$ and order $m$ \\ 
            
            \hspace{20mm}  $H_k$   \hspace{15mm}=\hspace{3mm}  the frequency dependent tidal amplitude in meters \\
            
             \hspace{20mm}  $\Theta_k$ , $\chi_k$ \hspace{8mm}=\hspace{3mm}  Doodson argument and phase correction for constituent k 
             
              \hspace{50mm} ( $\chi_k$ = 0, if $l-m$ is even; $\chi_k$ = $\pi$/2, if $l-m$ is odd \\
             
            \hspace{20mm}  $k_k^0$, $k_k^+$  \hspace{8mm}=\hspace{3mm}  Love numbers for tidal constituent k \\
            
            \hspace{20mm}  $r$, $\phi$, $\lambda$ \hspace{8mm}=\hspace{3mm} geocentric body-fixed coordinates of the satellite \\ 
          
            \hspace{20mm} $k(l,m)$ \hspace{8mm}=\hspace{3mm} each combination of $l,m$ has a unique list of 
            
            \hspace{50mm}  tidal frequencies, k, to sum over \\

           \myparagraph{Sun, Moon and planets perturbation} 
           The gravitational perturbations of the Sun, Moon and other planets can be modeled with sufficient accuracy using point mass approximations. In the inertial coordinate system of the \gls{COI}
, the N-body accelerations can be expressed as:
            \begin{equation}
            \label{newton}
            \mathrm{
            \vv{P}_{n} =  \sum_i GM_i \ \left[ \frac{\vv{r}_{ic}}{{r_{ic}}^3} \ - \ \frac{\vv{r}_{ip}}{{r_{ip}}^3} \right] }
             \end{equation}
where \\

            \hspace{25mm}  $GM_i$      \hspace{5mm}=\hspace{3mm}   gravitational constant of body $i$ \\
                  
            \hspace{28mm}  $r_{ic}$     \hspace{5mm}=\hspace{3mm}   position of body $i$ relative to \gls{COI} \\
            
            \hspace{28mm}  $r_{ip}$     \hspace{5mm}=\hspace{3mm}  position of spacecraft relative of body $i$\\

The gravitational constant $GM$ and the position of body $i$ can be obtained from the planetary ephemerides \citep{DE405, Fienga2009, Fienga2011}.

            \myparagraph{General relativity} 
            When the massless particle moves in the field of one massive body then relativistic perturbative acceleration can be given as:
            \begin{displaymath}
            \mathrm{
            \vv{P}_{rel} = \frac{GM_c}{c^2r^3} \ \left[  \left( (2\beta + 2\gamma)  \ \frac{GM_c}{r} - \gamma (\dot{\vv{r}} . \dot{\vv{r}}) \right) \ \vv{r}   + (2+2\gamma) \ ( \vv{r} . \dot{\vv{r}}) \ \dot{\vv{r}} \right] }
             \end{displaymath} 
             
             \begin{displaymath}
            \mathrm{
                      + 2(\vv{\Omega} \times \dot{\vv{r}} )   \hspace{70mm}   }
               \end{displaymath}
               
             \begin{equation}
            \label{relativity}
            \mathrm{
                   +   L(1+\gamma) \ \frac{GM_c}{c^2r^3} \left[ \frac{3}{r^2}  \ ( \vv{r} \times \dot{\vv{r}}) \ (\vv{r} . \vv{J}) +  (\dot{\vv{r}} \times \vv{J}) \right] \hspace{10mm}   }
               \end{equation}

where \\
            \begin{displaymath}
            \mathrm{
           \vv{ \Omega} \approx \left( \frac{1+\gamma}{2} \right) \dot{\vv{R}}_{cs}  \times \left[ \frac{-GM_s \ \vv{R}_{cs}}{c^2 R^3_{cs}} \right]  }
            \end{displaymath}

            \hspace{25mm}  $GM_c$   \hspace{8mm}=\hspace{3mm}  gravitational constant of the \gls{COI} \\            
          
            \hspace{25mm}  $GM_s$      \hspace{8mm}=\hspace{3mm}   gravitational constant of Sun \\
            
            \hspace{25mm}  $\vv{r}$, $\dot{\vv{r}}$       \hspace{8mm}=\hspace{3mm}   position and velocity vectors of the spacecraft relative 
            
            \hspace{50mm} to the \gls{COI} \\ 
            
            \hspace{25mm}  $\vv{R}_{cs}$, $\dot{\vv{R}}_{cs}$     \hspace{3mm}=\hspace{3mm}   position and velocity vectors of the Sun relative
             
           \hspace{50mm}  to the \gls{COI} \\     
                   
            \hspace{25mm}  $\vv{J}$     \hspace{14mm}=\hspace{3mm}  the \gls{COI}
 angular momentum per unit mass\\
            
             \hspace{25mm}  $L$   \hspace{15mm}=\hspace{3mm}  the Lense-Thirring parameter \\            
          
            \hspace{25mm}  $\gamma$, $\beta$      \hspace{11mm}=\hspace{3mm}   the parameterized post-Newtonian (PPN) parameters \\

The first term of Eq. \ref{relativity} is the Schwarzschild motion \citep{Huang90} and describes the main effect on the spacecraft orbit with the precession of perigee. The second term of Eq. \ref{relativity} is the effect of geodesic precession, which results in a precession of the orbit plane \citep{Bertotti87}. The last term of Eq. \ref{relativity} is the Lense-Thirring precession \citep{Ciufolini86}.
              
\subsubsection{Non-Gravitational forces}
Computation of the spacecraft trajectory which relies on radio tracking is limited by the uncertainty on the spacecraft non-gravitational acceleration. There are several non-gravitational forces acting on a spacecraft, many of which must be taken into account in order to achieve high accuracy in the orbit determination. Such non-gravitational forces acting on the spacecraft can be expressed as:
            \begin{equation}
            \label{non_gravi_force}
            \mathrm{
            \vv{a}_{ng} = \vv{P}_{sr} + \vv{P}_{cd}   + \vv{P}_{th}  + \vv{P}_{rad} + \vv{P}_{mb}}
            \end{equation}
            
where \\

            \hspace{25mm}  $\vv{P}_{sr}$   \hspace{5mm}=\hspace{3mm}  perturbations due to the solar radiation pressure \\            
                      
            \hspace{25mm}  $\vv{P}_{cd}$   \hspace{5mm}=\hspace{3mm} perturbations due to the atmospheric drag \\
                       
            \hspace{25mm}  $\vv{P}_{th}$    \hspace{5mm}=\hspace{3mm}  perturbations due to the thermal radiation \\ 
            
             \hspace{25mm}  $\vv{P}_{rad}$    \hspace{4mm}=\hspace{3mm}  perturbations due to the albedo and infrared radiation \\
             
             \hspace{25mm}  $\vv{P}_{mb}$    \hspace{4mm}=\hspace{3mm}  perturbations due to the motor burn \\

\begin{figure*}[ht]
\begin{center}\includegraphics[width=14cm]{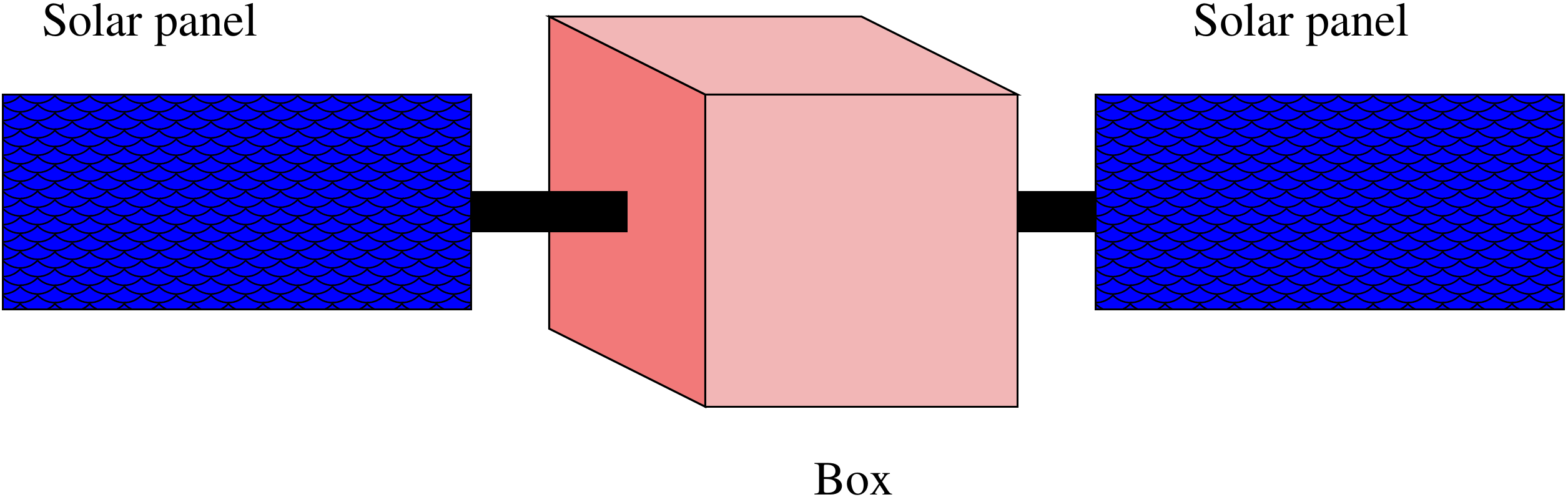}\end{center}
\caption{A simple geometric {\it macro-model} of the spacecraft. }
\label{macro}
\end{figure*}

Since the surface forces depend on the shape and orientation of the spacecraft, the force models are therefore spacecraft dependent. The $Box-Wing$ model of the spacecraft so-called {\it macro-model} \citep{Marshall92} is usually used for the modeling of non-gravitational perturbations. In the {\it macro-model}, the spacecraft main body and the solar panel are represented by a simple geometric model, a box and a wing, and the non-gravitational forces are then computed for each surface and summed over the surfaces. An example of simple {\it macro-model} is shown in Figure \ref{macro}. \\

%
 %
      
                \myparagraph{Solar radiation pressure} 
                 A spacecraft that is exposed to solar radiation experiences a small force that arise owing to the exchange of momentum between solar photons and the spacecraft, as solar photons are absorbed or reflected by the spacecraft. This force can be significant in magnitude in the vicinity of the Earth, at $\sim$ 1 AU from the Sun, especially when considering spacecraft with a large surface area, such as those with large solar panels or antennas. The solar radiation acceleration experiences by the spacecraft {\it macro-model} can be computed as \citep{Milani87}:
               
            \begin{equation}
            \label{solar_rad}
            \mathrm{
            \vv{P}_{sr} = -P\frac{F_S.\nu}{m} \sum_{i=1}^{n} A_i \ cos \ \theta_i \left[   2  \left( \frac{K_{D_i}}{3} +K_{S_i}   \ cos \ \theta_i \right) \hat{n_i} + (1-K_{S_i}) \ \hat{s} \right]}
            \end{equation}              
 where \\

            \hspace{20mm}  $P$   \hspace{5mm}=\hspace{3mm}  the momentum flux due to the Sun \\            
                      
            \hspace{20mm}  $F_S$   \hspace{5mm}=\hspace{3mm} solar radiation pressure scale factor (priori = 1)\\
                       
            \hspace{20mm}  $\nu$    \hspace{5mm}=\hspace{3mm} the eclipse factor (0 for full shadow, 1 for full Sun) \\
            
            \hspace{20mm}  $m$   \hspace{5mm}=\hspace{3mm}  mass of the spacecraft \\            
                      
            \hspace{20mm}  $A_i$   \hspace{5mm}=\hspace{3mm} surface area of the $i$-th plate \\
                       
            \hspace{20mm}  $\theta_i$    \hspace{5mm}=\hspace{3mm} angle between surface normal and spacecraft-Sun vector for $i$-th plate \\
            
            \hspace{20mm}  $\hat{n_i}$   \hspace{5mm}=\hspace{3mm}  surface normal unit vector for $i$-th plate \\            
                      
            \hspace{20mm}  $\hat{s}$   \hspace{6mm}=\hspace{3mm} spacecraft-Sun unit vector \\
                       
            \hspace{20mm}  $K_{D_i}$    \hspace{2mm}=\hspace{3mm} specular reflectivity for $i$-th plate \\ 
            
             \hspace{20mm}  $K_{S_i}$    \hspace{2mm}=\hspace{3mm} diffusive reflectivity for $i$-th plate \\
             
             \hspace{20mm}  $n$    \hspace{6mm}=\hspace{3mm} total number of plates in the model \\
               
               Typically materials used for the construction of the spacecraft, specular reflectivity $K_D$ and diffusive reflectivity $K_S$ coefficients lies in the range from 0.2 to 0.9 \citep{vanderHa77}.
               
               \myparagraph{Atmospheric drag and lift} 
               When a spacecraft is in the vicinity of a planetary body then low altitude spacecraft may experience non-gravitational perturbation due to the atmosphere of planet. This atmospheric perturbation on the spacecraft is negligible if the planet does contain an atmosphere ( for example, the Mercury and the Moon). The accurate modeling of this aerodynamic force requires the knowledge of the physical properties of the atmosphere, especially the density of the upper atmosphere. 
               
               A spacecraft of arbitrary shape moving with velocity $\vv{v}$ in an atmosphere of density $\rho$ will experience both lift and drag forces. However, the lift forces are very small as compared to the drag forces. These drag forces are directed opposite to the velocity of the spacecraft motion with respect to the atmospheric flux, hence decelerating the spacecraft. The drag and lift force can be model as:
            \begin{displaymath}
            \mathrm{
            \vv{P}_{cd} =  - \frac{1}{2} \ \rho \ \bigg [ \ F_D \sum_{i=1}^{n} C_{D_i} \ \frac{A_i}{m} \ (\vv{v}_r \ . \ \vv{n_i}) .\ \vv{v}_r }
            \end{displaymath}        
            
             \begin{equation}
             \label{drag}
             \mathrm{
            \ \ \ \  +  F_L \sum_{i=1}^{n} C_{L_i} \ \frac{A_i}{m} \ (\vv{v}_r \ . \ \vv{n_i}) .\ \vv{v}_r \bigg ]}
              \end{equation}      
                
 where \\

            \hspace{20mm}  $\rho$   \hspace{7mm}=\hspace{3mm}  the atmospheric density \\            
                      
            \hspace{20mm}  $\vv{v}_r$   \hspace{4mm}=\hspace{3mm} the spacecraft velocity relative to the atmosphere \\            
                       
            \hspace{20mm}  $m$    \hspace{5mm}=\hspace{3mm} mass of the spacecraft \\
            
            \hspace{20mm}  $n$    \hspace{6mm}=\hspace{3mm} total number of spacecraft faces directly expose to atmosphere  \\
            
            \hspace{20mm}  $\vv{n_i}$   \hspace{4mm}=\hspace{3mm} the unit vector normal to the face $i$ \\
            
            \hspace{20mm}  $A_i$   \hspace{4mm}=\hspace{3mm}  the cross-sectional area of the $i$-th face \\   
            
            \hspace{20mm}  $C_{D_i}$   \hspace{2mm}=\hspace{3mm}  the aerodynamic drag coefficient of the $i$-th face \\
            
             \hspace{20mm}  $C_{L_i}$   \hspace{2mm}=\hspace{3mm}  the aerodynamic lift coefficient of the $i$-th face \\
            
             \hspace{20mm}  $F_D$   \hspace{5mm}=\hspace{3mm}  drag force scale factor (priori = 1) \\            
             
             Density models for the upper atmosphere are partly empirical and based on the laws of static equilibrium distribution. There are a number of empirical atmospheric density models used for computing the atmospheric density. These include the Jacchia 71 \citep{Jacchia77}, the Drag Temperature Model (DTM) \citep{Barlier78}, NRLMSISE-00 \citep{Hedin96}. 
             
             The relative velocity $\vv{v}_r$ of the spacecraft with respect to the atmosphere depends on the complex atmospheric dynamics. However, a reasonable approximation of the relative velocity is obtained with the assumption that the atmosphere co-rotates with the planet. Therefore, using this assumption relative velocity can be expressed as:
            \begin{equation}
            \label{vel_rel}
            \mathrm{
            \vv{v}_r =  \vv{v} - \vv{\omega}  \ \times \vv{r}}
            \end{equation}
            where, $\vv{v}$ is the spacecraft velocity, $\vv{r}$ is the spacecraft position from the \gls{COI}, and $\vv{\omega}$ is the angular velocity of the planet.

                \myparagraph{Thermal radiation} 
                The surface temperature of a spacecraft is affected by exterior flows received from the Sun and the planet. Since the temperatures of the spacecraft surface are not uniform, there exists a force due to a net thermal radiation imbalance. This perturbation depends on the shape, the thermal property, the pattern of thermal dumping, the orbit characteristics, and the thermal environment of the spacecraft as a whole. The perturbation due to thermal radiation can be model as \citep{Afonso89}:     

            \begin{equation}
            \label{thermal}
            \mathrm{
            \vv{P}_{th} = - \frac{2}{3} \frac{\sigma}{c}  \sum_{i=1}^{n} \frac{A_i}{m} \ \varepsilon_i \ {T^4}_i  \ \vv{n_i} }
            \end{equation}
 where \\

            \hspace{20mm}  $\sigma$   \hspace{6mm}=\hspace{3mm}  the Stefan-Boltzmann constant \\            
                      
            \hspace{20mm}  $c$   \hspace{6mm}=\hspace{3mm} the speed of light \\            
                       
            \hspace{20mm}  $m$    \hspace{5mm}=\hspace{3mm} mass of the spacecraft \\
            
            \hspace{20mm}  $n$    \hspace{6mm}=\hspace{3mm} total number of spacecraft faces   \\
            
            \hspace{20mm}  $\vv{n_i}$   \hspace{4mm}=\hspace{3mm} the unit vector normal to the $i$-th face \\
            
            \hspace{20mm}  $A_i$   \hspace{4mm}=\hspace{3mm}  the cross-sectional area of the $i$-th face \\  
            
            \hspace{20mm}  $T_i$   \hspace{4mm}=\hspace{3mm}  the temperature of the $i$-th face \\     
            
             \hspace{20mm}  $\varepsilon_i$   \hspace{4mm}=\hspace{3mm}  the emissivity coefficient of the $i$-th face \\  
            
              \myparagraph{Albedo and infrared radiation } 
              In addition to the direct solar radiation pressure, the radiation emitted by the planet leads to a small pressure on the spacecraft. Such radiations can have two components, that is shortwave albedo and longwave infrared. In both cased the acceleration on the spacecraft decrease slightly with the altitude due to the inverse square law of the emitted radiation pressure. This perturbation can be model as:
               \begin{enumerate}[{\bf (i)}] 
               \item{\bf Albedo radiation} \\ 
            \begin{equation}
            \label{albedo}
            \mathrm{
            \vv{P}_{rad} = \frac{P}{\pi}\ \sum_{i=1}^{n}\  \frac{A_i}{m}  \sum_{i=1}^{n} \rho_a \ \frac{(\vv{s}.\vv{dS}) (\vv{u}.\vv{dS}) } {D^2} \vv{\beta_i}}
            \end{equation}
 \item{\bf Infrared radiation} \\ 
            \begin{equation}
            \label{infra}
            \mathrm{
            \vv{P}_{rad} = \frac{P}{\pi}\ \sum_{i=1}^{n}\  \frac{A_i}{m}  \sum_{i=1}^{n}\frac{e}{4} \ \frac{(\vv{u}.\vv{dS}) } {D^2} \vv{\beta^{\textquoteright}_i}}
            \end{equation}                                         
             \end{enumerate}

 where \\

            \hspace{20mm}  $P$   \hspace{6mm}=\hspace{3mm}  the momentum flux due to the Sun (4,5605 $\times$ 10$^{-6}$ \\
            
            \hspace{20mm}  $\vv{s}$   \hspace{6mm}=\hspace{3mm}  the unit vector in the direction \gls{COI}-Sun \\            
                      
            \hspace{20mm}  $\vv{dS}$   \hspace{4mm}=\hspace{3mm} the unit vector normal to the surface element of the orbiting planet \\      
            
             \hspace{20mm}  $\vv{n}$   \hspace{6mm}=\hspace{3mm}  the planet-spacecraft unit vector in the direction of surface element\\      
             
             \hspace{20mm}  $D$   \hspace{6mm}=\hspace{3mm} distance between the surface element of planet and the spacecraft \\
             
             \hspace{20mm}  $\rho_a$   \hspace{5mm}=\hspace{3mm}  albedo relative to the surface element\\
             
             \hspace{20mm}  $e$   \hspace{6mm}=\hspace{3mm}  emissivity relative to the surface element \\
                       
            \hspace{20mm}  $m$    \hspace{5mm}=\hspace{3mm} mass of the spacecraft \\
            
            \hspace{20mm}  $n$    \hspace{6mm}=\hspace{3mm} total number of spacecraft faces   \\
            
            \hspace{20mm}  $\vv{n_i}$   \hspace{4mm}=\hspace{3mm} the unit vector normal to the $i$-th face \\
            
            \hspace{20mm}  $A_i$   \hspace{4mm}=\hspace{3mm}  the cross-sectional area of the $i$-th face \\  
            
            \hspace{20mm}  $\vv{\beta_i}$   \hspace{4mm}=\hspace{3mm} reflectivity vector of the $i$-th face \\     
            
             \hspace{20mm}  $\vv{\beta^{\textquoteright}_i}$   \hspace{4mm}=\hspace{3mm} reflectivity infrared vector of the $i$-th face \\  
                            
             \myparagraph{Motor burn}
              The acceleration of the spacecraft due to a motor burn can be represented by \citep{Moyer1971}:
            \begin{equation}
            \label{motor_burn}
            \mathrm{
             \vv{P}_{mb} = aU \bigg [ u(t - T_0) \ + \ u(t - T_f) \bigg ] }
            \end{equation}               
             where, \\
             
              \hspace{30mm}  $a$   \hspace{6mm}=\hspace{3mm}  magnitude of  $\vv{P}_{mb}$ \\
             
              \hspace{30mm}  $U$   \hspace{6mm}=\hspace{3mm}  unit vector in direction of  $\vv{P}_{mb}$ \\ 
              
              \hspace{30mm}  $T_0$   \hspace{6mm}=\hspace{3mm} effective start time of motor \\
              
              \hspace{30mm}  $T_f$   \hspace{6mm}=\hspace{3mm}  effective end time of motor \\
              
              \hspace{30mm}  $T_f$   \hspace{6mm}=\hspace{3mm}  epoch in TDB 
              
              \hspace{30mm}  $u(t - T_0)$   \hspace{6mm}=\hspace{3mm}  1 for $t$ $\geq$ $T_0$ and 0 for $t$ $<$ $T_0$, when $T_0$ $\rightarrow$ $T_f$
              
              The acceleration magnitude $a$ is given by \citep{Moyer1971}: 
                                                                                                                             
            \begin{equation}
            \mathrm{
            a = \frac{F(t)}{m(t)} C = \frac{  F_0 + F_1\bar{t} + F_2\bar{t}^2 + F_3\bar{t}^3 + + F_4\bar{t}^4 } { m_0 + \dot{M}_0 \bar{t} + \frac{1}{2} \dot{M}_1 \bar{t}^2 + \frac{1}{3} \dot{M}_2 \bar{t}^3 + \frac{1}{4} \dot{M}_3 \bar{t}^4  } \ C    }
            \end{equation}               
             where, \\
             
             \hspace{25mm}  $F(t)$   \hspace{6mm}=\hspace{3mm}  magnitude of thrust at time $t$. The polynomial coefficients
             
             \hspace{45mm}   of $F(t)$ is solve-for paramerts   \\
             
              \hspace{34mm}  $\bar t$   \hspace{6mm}=\hspace{3mm}  t - T$_0$ \\
              
              \hspace{28mm}  $m(t)$   \hspace{6mm}=\hspace{3mm} spacecraft mass at time $t$ \\
              
              \hspace{29mm}  $\dot {M}_n$   \hspace{6mm}=\hspace{3mm} polynomial coefficients of propellant mass flow rate at time $t$ \\
            
             \hspace{32mm}  $C$   \hspace{6mm}=\hspace{3mm} 0.001 for $F$ in newtons and $m$ in $kg$
             
\subsection{Variational equations}
\label{vareq}
Variational equations are the way to describe the variations in the spacecraft state with respect to the solve-for parameters. These equations are always linear and solved simultaneously with the equations of motion. Let the differential equations of motion is given by:
            \begin{equation}
            \label{diff_motion}
            \mathrm{
             \frac{d^2\vv{r}}{dt^2}  = \vv{F} (\vv{r}, \dot{\vv{r}}, t, \vv{\varepsilon})}
            \end{equation}                                         

where $\vv{r}$ and $\dot{\vv{r}}$ are the state vector of the spacecraft relative to \gls{COI}. $\vv{\varepsilon}$ is the vector of solve-for parameter (such as, initial conditions, drag coefficient, solar radiation pressure coefficient, gravity harmonics, etc). The variational equations can be then written as:
            \begin{equation}
            \label{var_eq}
            \mathrm{
             \frac{d^2}{dt^2} {\left( \frac {\partial \vv{r}}{\partial \vv{\varepsilon}} \right)}  = \frac {\partial \vv{F}}{\partial \vv{r}} \ .  \frac {\partial \vv{r}} {\partial \vv{\varepsilon}}  +  \frac {\partial \vv{F}}{\partial \dot{\vv{r}}} \ .  \frac {\partial \dot{\vv{r}}} {\partial \vv{\varepsilon}} + \frac {\partial \vv{F}}{\partial \vv{\varepsilon}} }
            \end{equation} 
            with, at $t$ = $t_0$
            \begin{itemize}
            \item if $\vv{\varepsilon}$ is the parameter of the dynamical equations ($\vv{\varepsilon}_d$), then:
            \begin{equation}
            \label{var_eq1}
            \mathrm{
              \frac {\partial \vv{r}} {\partial \vv{\varepsilon}_d} = \frac {\partial \dot{\vv{r}}} {\partial \vv{\varepsilon}_d} = 0 }
            \end{equation}
            
            \item if $\vv{\varepsilon}$ is the initial state vectors ($\vv{\varepsilon}_0$), then:
            \begin{equation}
            \label{var_eq1}
            \mathrm{
            \frac {\partial \vv{r}} {\partial \vv{\varepsilon}_0} =             
            \begin{bmatrix}
             \dfrac {\partial {x_0}}{\partial \varepsilon_0^1} & .. & .. &  \dfrac {\partial {x_0}}{\partial \varepsilon_0^6} \\
               : & : & : & :\\
             \dfrac {\partial {z_0}}{\partial \varepsilon_0^1} & .. & .. &  \dfrac {\partial {z_0}}{\partial \varepsilon_0^6}\\
             \end{bmatrix}}
            \end{equation}
            \begin{equation}
            \label{var_eq2}
            \mathrm{
            \frac {\partial \dot{\vv{r}}} {\partial \vv{\varepsilon}_0} =             
            \begin{bmatrix}
              \dfrac {\partial \dot{x}_0}{\partial \varepsilon_0^1} & .. & .. &  \dfrac {\partial \dot{x}_0}{\partial \varepsilon_0^6} \\
               : & : & : & :\\
             \dfrac {\partial \dot{z}_0}{\partial \varepsilon_0^1} & .. & .. &  \dfrac {\partial \dot{z}_0}{\partial \varepsilon_0^6}\\
             \end{bmatrix}}
            \end{equation}           

            \end{itemize}

The matrices  ${\partial \vv{F}}/{\partial \vv{r}}$ and $ {\partial \vv{F}}/{\partial \dot{\vv{r}}}$ in Eq. \ref{var_eq} are evaluated in terms of corresponding solutions of the nonlinear equations of motion, and likewise the vector ${\partial \vv{F}}/{\partial \vv{\varepsilon}}$. These variational equations can be solved simultaneously with the nonlinear equations of motion using cowell\footnote{\url{http://igsac-cnes.cls.fr/documents/gins/Integration_Numerique/Integration_numerique.pdf}} numerical integrator.

\subsection{Parameter estimation}
\label{lestsq}
One of the most important task in the orbit determination is to estimate the solve-for parameters. To do so, the least-square technique has been used in the \gls{GINS} software to estimate such parameters. The objective of the least-square is to adjust the solve-for parameters of a model function to best fit the data set. These parameters are then refined iteratively by using the values that are obtained by successive approximation. The brief description of least-square technique is described below.

Let $\vv{\varepsilon}$ is the vector of $p$ number of solve-for parameters and $k$ is the iteration number. Therefore, the solve-for parameters can be given as:

            \begin{equation}
            \label{para}
            \mathrm{
            \vv{\varepsilon}^{k+1} = \vv{\varepsilon}^{k} + \Delta{\vv{\varepsilon}}         }
            \end{equation}
where $\Delta{\vv{\varepsilon}}$ is the correction in the adjusted solve-for parameter. By using first-order linearized approximation of a Taylor series expansion about $\vv{\varepsilon}^{k}$:
            \begin{equation}
            \label{Qc}
            \mathrm{
            Q_{c}(\vv{\varepsilon}) =  Q_{c}^k(\vv{\varepsilon})  + \sum_{i=1}^p {\frac{\partial Q_{c}^k(\vv{\varepsilon})} {\partial \varepsilon_i} } \Delta{\varepsilon_i}         }
             \end{equation}
where $Q_c$ is the theoretical values computed from the models which is a function of solve-for parameters $\vv{\varepsilon}$. If, $Q_o$ defines the real observations, then the residuals (observation - computation) can be given as:
             \begin{displaymath}
            \mathrm{
          \hspace{-1cm}  r =   Q_{o} - Q_{c} }
            \end{displaymath}
            \begin{displaymath}
            \mathrm{
            \hspace{3cm} =  Q_{o} - Q_{c}^k(\vv{\varepsilon})  + \sum_{i=1}^p {\frac{\partial Q_{c}^k(\vv{\varepsilon})} {\partial \varepsilon_i} } \Delta{\varepsilon_i}      }
             \end{displaymath}
              \begin{equation}
            \label{res}
            \mathrm{
           \hspace{1.5cm} =  \Delta{Q}  - \sum_{i=1}^p {\frac{\partial Q_{c}^k(\vv{\varepsilon})} {\partial \varepsilon_i} } \Delta{\varepsilon_i}      }
             \end{equation}
             
             If $m$ be the number of observations, then first and second terms Eq. \ref{res} can be written as:
             \begin{equation}
             \label{delQ}
            \mathrm{
            \vv{\Delta{Q}}  =   \begin{bmatrix}
              \Delta{Q_1}  \\
               : \\
               : \\
              \Delta{Q_m}\\
             \end{bmatrix} = B}
            \end{equation}                         
             \begin{equation}
             \label{delQe}
            \mathrm{
            \dfrac{\partial \vv{Q}_{c}^k} {\partial \vv {\varepsilon}}\  \Delta{\vv {\varepsilon}} =   \begin{bmatrix}
            \dfrac{\partial Q_{c_1}^k} {\partial \varepsilon_1} & .. & .. & \dfrac{\partial Q_{c_1}^k} {\partial \varepsilon_p} \\
               :  & .. & .. & :\\
               : & .. & .. & : \\
             \dfrac{\partial Q_{c_m}^k} {\partial \varepsilon_1} & .. & .. & \dfrac{\partial Q_{c_m}^k} {\partial \varepsilon_p}\\
             \end{bmatrix}} \ . \ \begin{bmatrix}
            \Delta{\varepsilon_1}  \\
             \\
               : \\
               : \\
               \\
              \Delta{\varepsilon_p}\\
             \end{bmatrix} = A . \Delta{\vv {\varepsilon}}
            \end{equation}             
             
             In order to estimate the best solve-for parameters, according to least-square technique, sum $S$ of the square of the residuals should be minimum. Where sum $S$ is given by:
             \begin{equation}
             \label{sumr2}
            \mathrm{             
             S = \sum_{j=1}^m { \bigg(B_j - A_j \ . \ \Delta{\vv {\varepsilon}\bigg)^2 }  }}
            \end{equation} 
            where $B_j$ is the $j^{th}$ value of $B$ and $A_j$ is the $j^{th}$ row of matrix $A$. For weighted least-square, the expression of sum $S$ is given as:
             \begin{equation}
             \label{wsumr2}
            \mathrm{             
             S = \sum_{j=1}^m { W_j \ . \ \bigg(B_j - A_j \ . \ \Delta{\vv {\varepsilon}\bigg)^2 }  }}
            \end{equation}             
            where $W_j$ is the $j^{th}$ value of diagonal weight matrix $W$, which is given by:
             \begin{equation}
             \label{wgt}
            \mathrm{
             W  =   \begin{bmatrix}
              W_1 &  &  & 0 \\
                 & . &  & \\
                &  & . &  \\ 
             0 &  &  & W_m \\
             \end{bmatrix} }
            \end{equation}       
          In order to achieve the minimum square of the sum $S$ of the residuals, the gradient of $S$ with respect of solve-for parameter should equal to zero. Which is given as:
             \begin{displaymath}
            \mathrm{             
            A^T \ . \ W \ . \ \bigg(B - A \ . \ \Delta{\varepsilon}\bigg)     = 0}
            \end{displaymath}          
          \begin{equation}
             \label{gradS}
            \mathrm{             
            A^T \ . \ W \ . \ A \ . \ \Delta{\varepsilon}     =   A^T \ . \ W \ . \ B }
            \end{equation}
            
           For example, if $W$ is the identity matrix and number of solve-for parameters $P$ equal to 2, then Eq. \ref{gradS} can be written as:
           
          \begin{equation}
             \label{exe}
            \mathrm{             
            \begin{bmatrix}      
            \sum_{j=1}^m { \bigg( \dfrac{\partial Q_{c_j}} {\partial \varepsilon_1} \bigg)^2 } &  \sum_{j=1}^m { \bigg( \dfrac{\partial Q_{c_j}} {\partial \varepsilon_1} \bigg) \ . \ \bigg( \dfrac{\partial Q_{c_j}} {\partial \varepsilon_2} \bigg)}  \\      \\        
             \sum_{j=1}^m { \bigg( \dfrac{\partial Q_{c_j}} {\partial \varepsilon_1} \bigg) \ . \ \bigg( \dfrac{\partial Q_{c_j}} {\partial \varepsilon_2} \bigg)}  & \sum_{j=1}^m { \bigg( \dfrac{\partial Q_{c_j}} {\partial \varepsilon_2} \bigg)^2 } \\            
            \end{bmatrix} \ . \ \begin{bmatrix}
             \Delta{\varepsilon_1}  \\ \\  \\ \\
              \Delta{\varepsilon_2}\\
             \end{bmatrix}    =   \begin{bmatrix}
              \sum_{j=1}^m { \Delta{Q} \ . \ \bigg( \dfrac{\partial Q_{c_j}} {\partial \varepsilon_1} \bigg) } \\ \\ 
              \sum_{j=1}^m { \Delta{Q} \ . \ \bigg( \dfrac{\partial Q_{c_j}} {\partial \varepsilon_2} \bigg) } 
              \end{bmatrix} }
            \end{equation}           
            
            The computed value of the observations $Q_c$ can be calculated from the state vectors of the spacecraft and the position of the stations at time of measurement (see Section \ref{drobs}). The gradient of the $Q_c$ with respect to the solve-for parameters $\varepsilon$ given in Eq. \ref{exe} can be written as:
          \begin{equation}
             \label{dQde}
            \mathrm{    
            \dfrac{\partial \vv {Q}_{c_j}} {\partial \vv {\varepsilon}} =  \dfrac{\partial \vv {Q}_{c_j}} {\partial \vv {r}} \ . \ \dfrac{\partial \vv r} {\partial \vv {\varepsilon}} +  \dfrac{\partial \vv {Q}_{c_j}} {\partial \dot{\vv {r}}} \ . \ \dfrac{\partial \dot {\vv r}} {\partial \vv {\varepsilon}}}
            \end{equation} 
            
            In Eq. \ref{dQde}, $\vv{r}$ and $\dot{\vv {r}}$,  and ${\partial \vv r} / {\partial \vv {\varepsilon}}$ and ${\partial \dot {\vv r}} / {\partial \vv {\varepsilon}}$, can be computed from the numerical integration of equations of motion and variational equations (Eqs. \ref{var_eq1} and \ref{var_eq2}) respectively. Whereas, ${\partial \vv {Q}_{c_j}} / {\partial \vv {r}}$ and ${\partial \vv {Q}_{c_j}}/ {\partial \dot{\vv {r}}}$ can be computed analytically. \\
            
            The steps used for estimating the solve-for parameters are (see Figure \ref{flow_dig}):
            \begin{itemize}
            \item to provide initial conditions and planetary ephemerides to the dynamical model,
            \item integrate numerically the equations of motion (Eq. \ref{motion}) along with the variational equations (Eq. \ref{var_eq}) in order to calculate the state vectors of the spacecraft and the partial derivative of the state vector with respect to solve-for parameters (Eqs. \ref{var_eq1} and \ref{var_eq2}),
            \item to calculate the theoretical observations $Q_c$ using the formulation given in section \ref{drobs},
            \item to calculate the partial derivative of the theoretical observations with respect to solve-for parameters and then to compute the residuals (Eq. \ref{res}),
            \item to calculate the normal matrix and its inverse and then to calculate the corrections in the solve-for parameters (Eq. \ref{gradS}),
            \item these correction are then used to refine the parameters using iterative processes (Eq. \ref{para}) until the residuals have met the specified convergence criterion.
            \end{itemize}

%% file: CHP2.tex
\chapter{Mars Global Surveyor: Radioscience data analysis}    
\label{CHP2}
\section{Introduction}
\label{intro}
Mars is the most explored terrestrial planet in the solar system. Many space missions have been attempted to Mars than to any other place in the solar system except the Moon. These missions include flyby missions (Mariner, Rosetta), orbiter missions (Mars Global Surveyor, Mars Odyssey, Mars Express, Mars Reconnaissance Orbiter ), and lander missions (Viking, Mars Exploration Rovers, Mars Science Laboratory). Obtaining scientific information is the primary reason for launching and operating such deep-space missions. Usually, science objectives of such missions involve: high resolution imaging of the planet surface, studies of the gravity and topography, studies of the atmosphere and the interior of the Mars, studies of biological, geological, and geochemical processes, etc. To achieve the mission objectives, some of these investigations are performed by using a dedicate science instrument aboard the spacecraft, which measure a particular physical phenomenon, for example:
\begin{itemize}
\item The thermal emission spectrometer: is used to measure the infrared spectrum of energy emitted by the planet. This information is used to study the composition of rock, soil, atmospheric dust, clouds, etc. 
\item The orbital laser altimeter: is used to measure the time, takes for a transmitted laser beam to reach the surface, reflect, and return. This information provides the topographic maps of Mars. 
\item The magnetometer: is used to determine a magnetic field, and its strength and orientation.
\end{itemize}
On the other hand, some objectives are undertaken as opportunities arise to take advantage of a spacecraft special capabilities or unique location or other circumstance. For example:
\begin{itemize}
\item With the radioscience experiment, which measures the Doppler shift of radio signals, it is then possible to determine the gravity field by computing the change in the speed of the spacecraft, associated with the high and low concentration of the mass below and at the surface of the Mars. Moreover, unique location such as spacecraft occultation allows the radio signals pass through the Martian atmosphere on their way to Earth. Hence, perturbations in the signals induced by the atmosphere allows to derive the atmospheric and ionospheric characteristics of the Mars. Brief description of such investigations are discussed in Chapter \ref{CHP1}.
\end{itemize}

This chapter deals with the radioscience data analyses to precisely construct the \gls{MGS} orbit. Such analyses has been already performed by several authors, such as \cite{Yuan01,Lemoine01,Konopliv06,JMarty}. We have therefore chosen \gls{MGS} as an academic case to test our understanding of the raw radiometric data (ODF, see Chapter \ref{CHP1}) and their analysis with GINS by comparing our results with the one found in the above literature. Moreover, these analyses also allowed us to derive solar corona model and to perform corona physic studies. Derivation of such models and their application to planetary ephemerides are discussed in Chapter \ref{CHP3} and in \cite{verma12}.

The outline of this chapter is as follows: in the Section \ref{overview} an overview of the \gls{MGS} mission is discussed. Data processing and dynamic modeling used for the orbit construction is described in the Section \ref{mgs_orbit}. Results obtained during the orbit computation and their comparison with the estimations found in the literature are discussed in the Section \ref{results}. The supplementary tests, which include: (a) comparison between the GINS solution and the JPL light time solution, and (b) the impact of the \textit{macro-model} on the orbit reconstruction, are discussed in Section \ref{comps}. Conclusion and prospectives of these results are reported in Section \ref{Conclusion}.

\section{Mission overview}
\label{overview}
\subsection{Mission design}
\label{mis_des}
\begin{figure}[!ht]
\begin{center}\includegraphics[width=15cm]{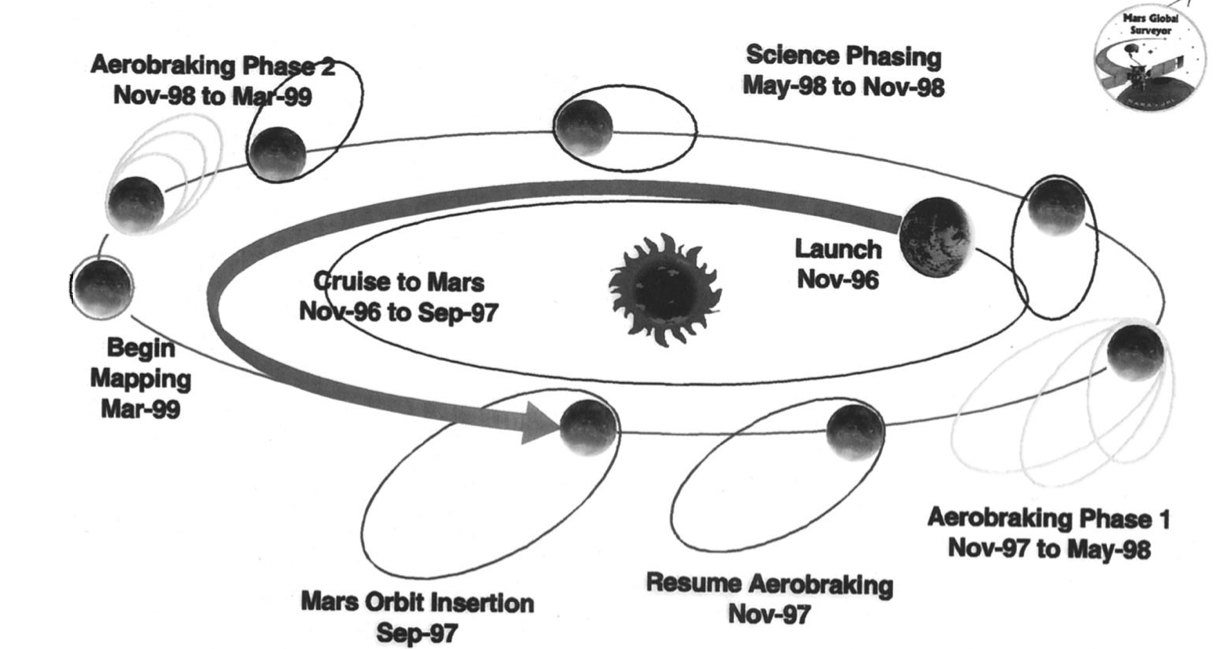}\end{center}
\caption{Summary of MGS mission phases from launch to mapping period \citep{albee01}.}
\label{mgs_phase}
\end{figure}

\gls{MGS} was a NASA\textquoteright s space mission to Mars. It was launched from the Cape Canaveral Air Station in Florida on 7$^{th}$ November 1996 aboard a Delta II rocket. The MGS began its Mars orbit insertion on 12$^{th}$ September 1997. Figure \ref{mgs_phase} illustrates the summary of the \gls{MGS} mission from launch to the \gls{MOI} in an elliptical orbit, the initial areobraking period, the science-phase period, the aerobraking resumption, and mapping in the circular orbit \citep{albee01}.
%
 %

The \gls{MOI} represents an extremely crucial maneuver, because any failure would result in a fly-by mission. The \gls{MOI} slows down the spacecraft and allows Mars to capture the probe into an elliptical orbit. Near the point of closest approach on spacecraft in-bound trajectory, the main engines fired for approximately 20 to 25 minutes to slow down the spacecraft. This burn allowed the spacecraft to run off of the hyperbolic approach trajectory and to approach the planet onto a highly elliptical orbit. Initially, the \gls{MGS} orbit had a periapsis of 262 km above the northern hemisphere, and an apoapsis of 54,026 km above the southern hemisphere and took 45 hours to complete one orbit. 

In order to attain the mapping orbit, \gls{MGS} spacecraft was designed to facilitate the use of aerobraking. Aerobraking is the utilization of atmospheric drag on the spacecraft to reduce the energy of the orbit. The friction caused by the passage of the spacecraft through the atmosphere provides a velocity change at periapsis, which results in the lowering of the apoapsis altitude. After almost sixteen months of orbit insertion, the aerobraking events were utilized to convert the elliptical orbit into an almost circular 2-hour low altitude sun synchronous polar orbit with an average altitude of 378 km. Thus, \gls{MGS} started its low altitude mapping orbital phase in March 1999 and lost communication with the ground station on 2$^{nd}$ November 2006.

\subsection{Spacecraft geometry}
\label{spa_geo}
\gls{MGS} was designed to carry science payloads to Mars, to maintain proper pointing and orbit as a three-axis stabilized platform for acquiring mapping data and return to the Earth.

%
%
 \begin{figure}[!ht]
\begin{center}\includegraphics[width=15cm]{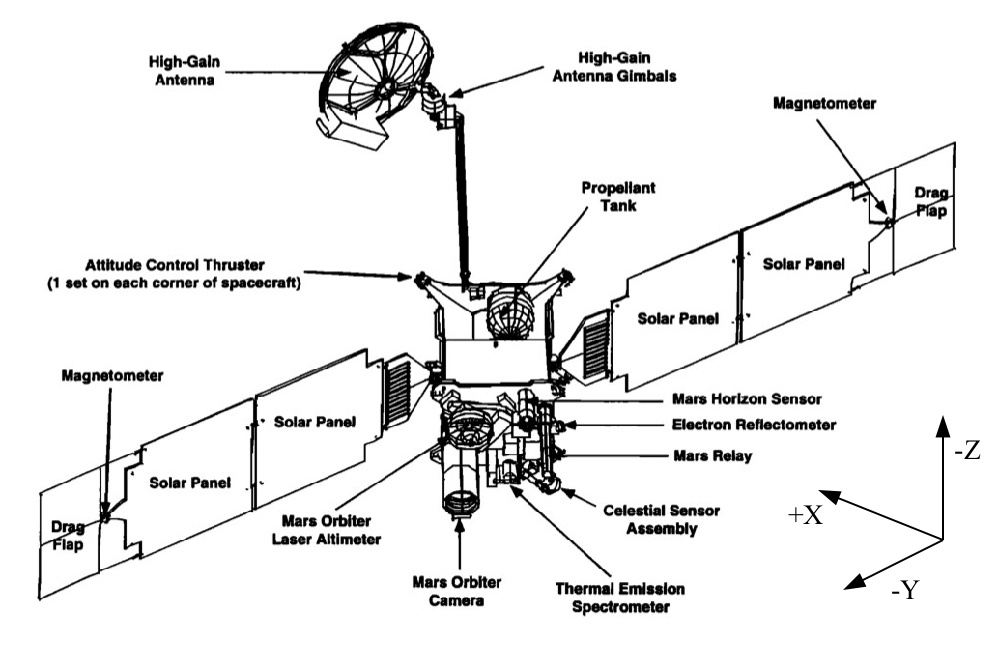}\end{center}
\caption{View of the MGS spacecraft \citep{albee01}.}
\label{mgs}
\end{figure}

In order to meet the strength mass requirements, the spacecraft structure was constructed of lightweight composite materials. It was divided into four sub-assemblies (Figure \ref{mgs}, extracted from \cite{albee01}): 
\begin{itemize}
\item {\bf Equipment module:} it houses the avionics system and the science instruments. \gls{MGS} carried six on-board instruments, that are Magnetometer$/$Electron Reflectometer, Mars orbiter camera, Mars orbiter laser altimeter, Mars relay, Thermal Emissions Spectrometer, and ultra-stable oscillator. All science instruments are bolted to the nadir equipment deck, mounted above the equipment module on the +Z panel.
\item {\bf Propulsion module:} it serves as the adapter between the launch vehicle and contains the propellant tanks, main engines, and attitude control thrusters. This module bolts beneath the equipment module on the -Z panel.
\item {\bf Solar arrays:} two solar arrays provide power for the MGS spacecraft. Each array mounts close to the top of the propulsion module on the +Y and -Y panels, near the interface between the propulsion and equipment modules. Rectangular shaped, metal drag $flaps$ mounted onto the ends of both arrays. These $flaps$ were only used to increase the total surface area of the array structure to increase the spacecraft's ballistic coefficient during aerobraking.
\item {\bf High Gain Antenna:} a parabolic \gls{HGA} structure was also bolt to the outside of the propulsion module. It was used to make high rate communication with the Earth. When fully deployed, the 1.5 meter diameter \gls{HGA} sits at the end of a 2.0 meter long boom, mounted to the +X panel of the propulsion module.
\end{itemize}
  
\begin{table}[!ht]
\caption{MGS spacecraft macro-model characteristics \citep{Jcmarty10}}
\centering
\begin{center}
\renewcommand{\arraystretch}{1.4}
\small
\begin{tabular}{ccccc}\Xhline{2\arrayrulewidth}
\hline \hline
{\bf Spacecraft body} & {\bf Components} & {\bf Area (m$^2$)}  & {\bf Diffuse ref} & {\bf Specular ref.}\\
\Xhline{2\arrayrulewidth}

Solar Arrays & Composite (front) & 7.85 & 0.049 & 0.198 \\
& Composite (back) & 7.85 & 0.079 & 0.282  \\
\Xhline{2\arrayrulewidth}
& +X, -X & 3.30 & 0.130 & 0.520 \\
Spacecraft Bus & +Y, -Y & 3.56 & 0.130 & 0.520 \\ 
& +Z, -Z & 2.31 & 0.130 & 0.520 \\
\Xhline{2\arrayrulewidth}
HGA & +X  & 1.94 & 0.100 & 0.400 \\ 
\Xhline{2\arrayrulewidth}
\end{tabular}
\end{center}
\label{macro}
\end{table}

As described in Chapter \ref{CHP1} (Section \ref{dymod}), the precise computation of the spacecraft trajectory relies on the non-gravitational accelerations that are acting in the spacecraft. These non-gravitational forces depend on the shape, size, surface properties, and orientations of the spacecraft. Thus, an accurate spacecraft geometry is an essential information for modeling such forces precisely. The characteristics of the \gls{MGS} macro-model are given in Table \ref{macro} \citep{lemoine99}.

\subsection{Radioscience data}
\label{dat_cov}
\gls{MGS} is the first operational planetary mission to employ exclusively X-band technology for radioscience observations, tracking, and spacecraft command and communication \citep{Tyler01}. The radioscience instrument used for this purpose is the spacecraft telecommunications subsystem, augmented by an ultra-stable oscillator, and the normal \gls{MGS} transmitter and receiver. The ultra-stable oscillator typically have frequency stabilities on the order of 1$\times$10$^{-13}$ for time intervals of 1 to 100 seconds \citep{Cash08}. The oscillator provides the reference frequency for the radioscience experiments and operates on the X-band 7164.624 MHz uplink and 8416.368 MHz downlink frequency. The radioscience data are then collected by the \gls{DSN} and consist of one-way Doppler, two- and three-way ramped Doppler, and two-way range observations (see Chapter \ref{CHP1} for full details of radioscience data).
%
 %
 \begin{figure}[!ht]
\begin{center}\includegraphics[width=10cm]{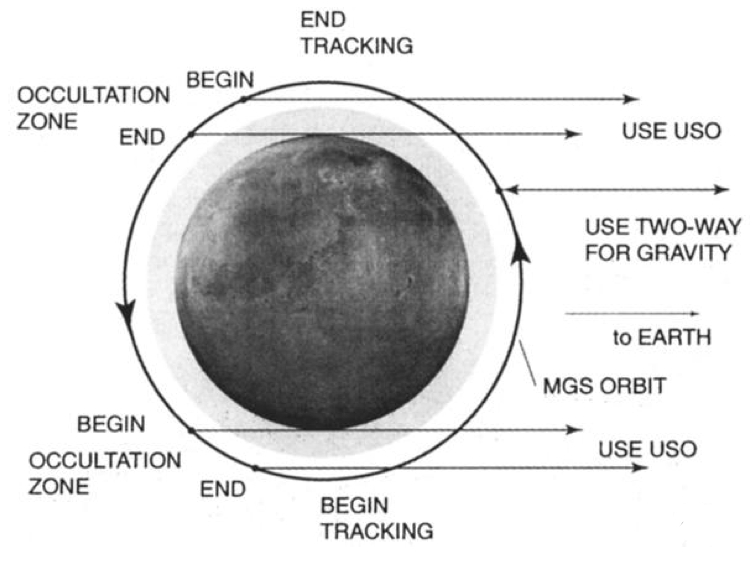}\end{center}
\caption{Pole-to-Pole tracking of the MGS spacecraft during mapping period \citep{Tyler01}.}
\label{data}
\end{figure}

Figure \ref{data} illustrates the \gls{MGS} tracking strategy during the mapping period for sharing the orbit between: i) the occultation studies when \gls{MGS} was near the limb of the planet, and ii) the gravity studies during the \gls{MGS} Earth side pass \citep{Tyler01}. Pole-to-Pole two- and three-way Doppler observations provide the primary information of Mars gravity field. Whereas, one-way Doppler observations obtained during the occultation period provide an information about the Martian atmosphere and ionosphere. Table \ref{data_mgs} gives the summary of the data coverage obtained during the mapping period and used to construct the \gls{MGS} orbit.

\begin{table}[!ht]
\caption{Year wise summary of the Doppler and range tracking data used for orbit solution.}
\centering
\begin{center}
\renewcommand{\arraystretch}{1.2}
\small
\begin{tabular}{cccccc}\Xhline{2\arrayrulewidth}
\hline \hline
 Year       & Number of        & Number of        & Number of        & Number of        & Number of  \\      
 & 1-way Dop. & 2-way Dop. & 3-way Dop. &  Range & Ramp \\
\Xhline{2\arrayrulewidth}
1999   &    133,188  & 1,060,416 &   120,931 &    46,277  &   202,188 \\
2000   & 1,472,366  &    536,142 &     74,950 &      5,192  &   192,992 \\
2001   & 1,389,279  &    877,315 &     86,482 &     54,582 &   192,624 \\
2002   & 1,772,226  &    641,729 &     86,984 &     30,533 &   203,309 \\
2003   &    937,566  &    552,073 &     64,260 &     21,784 &   155,255 \\
2004   &    660,020  &    446,934 &     52,836 &     17,022 &     78,991 \\
2005   &    205,832  &      98,130 &       2,028 &       6,200 &     23,626 \\
2006   & 1,105,734  &    452,655 &     33,486 &     23,148 &   120,846 \\
\Xhline{2\arrayrulewidth}
\end{tabular}
\end{center}
\label{data_mgs}
\end{table}

\section{Orbit determination}
\label{mgs_orbit}     
We have used Doppler- and range-tracking observations to compute the MGS orbits precisely. However as stated before, this computation was mainly performed as an academic case to understand the dynamic model and the radiometric data. With this computation, we were then able to compare our results with those obtained by \cite{Yuan01,Lemoine01,Konopliv06,JMarty} (see Section \ref{post-res}). Tests were also performed and reported in Section \ref{macro_test}, in order to understand the sensitivity of the data processing with the accuracy of non-gravitational forces by assuming different shapes of \gls{MGS} macro-model.

\subsection{Data processing and dynamic modeling}
\label{data_anyl}         
The radioscience observations, used for computing the \gls{MGS} orbit, are available on the \gls{NASA} \gls{PDS} Geoscience website\footnote{\url{http://geo.pds.nasa.gov/missions/mgs/rsraw.html}}. These observations are analyzed with the help of the \gls{GINS} software. As described in Section \ref{gins} of Chapter \ref{CHP1}, \gls{GINS} numerically integrates the equations of motion (Eqs. \ref{motion}) and the associated variational equations (Eqs. \ref{var_eq}). It simultaneously retrieves the physical parameters of the force model using an iterative least-squares technique (see Section \ref{lestsq}). The modeling of the \gls{MGS} orbit includes gravitational and non-gravitational forces that are acting on the spacecraft (see Section \ref{dymod}). In addition to these forces, third body perturbations due to Phobos and Deimos are also included. \\

The data processing and the dynamic modeling are done as follows:
\begin{itemize}

\item In order to have access to the planet positions and velocities, planetary ephemeris has been used for both measurements and force models (e.g,  DE405, INPOP10b).

\item The Mars geopotential is modeled in terms of spherical harmonic. This model is given by Eq. \ref{potential} as described in section \ref{dymod}. Fully normalized spherical harmonic coefficients of MGS95J solution\footnote{\url{http://pds-geosciences.wustl.edu/mgs/mgs-m-rss-5-sdp-v1/mors\_1033/sha/}} has been used. MGS95J is a 95x95 spherical harmonics model which was derived from 6 years of \gls{MGS} tracking data plus 3 years of measurements on the Mars \gls{ODY} spacecraft \citep{Konopliv06}. 

\item The rotation model which defines the orientation of the Mars is taken from the \cite{Konopliv06}.

\item Earth kinematics and polar motion effects are taken according to the IERS standards \citep{McCarthyIERS32}.

\item The Phobos and Deimos ephemerides are taken as developed by \cite{Lainey07}.

\item The complex geometry of the spacecraft is treated as a combination of flat plates arrange in the shape of box, with attached solar arrays and drag flaps, and attached \gls{HGA}. This \textit{Box-Wing} model includes six plates for the spacecraft bus, four plates to represent the front and back side of the +Y and -Y solar arrays, and a parabolic \gls{HGA}. The surface area and reflectivity of this macro-model are given in Table \ref{macro}. Moreover, in addition to this macro-model, a \textit{Spherical} macro-model has been also used to reconstruct the \gls{MGS} orbit and to understand the impact of macro-model over the orbit reconstruction (see Section \ref{comps}).

\item Using such configurations of the \gls{MGS} \textit{macro-models}, solar radiation pressure (Eq. \ref{solar_rad}), atmospheric drag (Eq. \ref{drag}), and Mars radiation pressure (Eqs. \ref{albedo} and \ref{infra}) forces are computed separately for each plate and \gls{HGA}. Vectorial sum of all these components are then compute to calculate the total force acting on the spacecraft. 

\item In addition to the macro-model characteristics, orientations of the spacecraft are also taken in account. The attitude of spacecraft, and of its articulated panels and antenna in inertial frame are defined in terms of quaternions. These quaternions are extracted from the SPICE \gls{NAIF} C-Kernels\footnote{\url{http://naif.jpl.nasa.gov/naif/}}.

\item \gls{MGS} periodically fires its thruster to desaturate the reactions wheels, which absorb angular momentum from disturbance torque acting on the spacecraft. Thus, empirical accelerations are modeled over the duration of each \gls{AMD} event \citep{Jcmarty10}. Constant radial, along-track, and cross-track accelerations are applied over the duration of each \gls{AMD} event, and are estimated as part of each orbit determination solution.

\item The relativistic effects on the measurements and on the spacecraft dynamics are modeled based on the \gls{PPN} formulation as described in Sections \ref{obcom} and \ref{dymod} of Chapter \ref{CHP1}.

\item The tropospheric delay corrections to the measurements are also included. Computation of this delay uses meteorological data (pressure, temperature and humidity) collected every half-hour at the \gls{DSN} sites.

\end{itemize}

\subsection{Solve-for parameters}
\label{solve_4_para_mgs}  
For the orbit computation and for the parameter estimation, a multi-arc approach is used to get independent estimates of the \gls{MGS} accelerations. In this method, orbital fits are obtained from short data-arcs of two days with two hours (approx. one orbital period of \gls{MGS}) of overlapping period. The short day data-arcs are used to accounting the model imperfections (see Section \ref{comps}) and overlapping period are used to estimate the quality of the spacecraft orbit determination by taking orbit overlap differences between two successive data-arcs (see Section \ref{results}). In order to account the effect of shortest wavelengths of Mars gravity field on the \gls{MGS} motion, we integrate the equations of motion using the time-step of 20s. An iterative least-squares fit is then performed on the complete set of Doppler- and range-tracking data-arcs. \\ 

Solve-for parameters which have been estimated during the \gls{MGS} orbit determination are:

\begin{itemize}

\item The initial state vector components at the epoch for each arc. Prior values of these vectors are taken from the SPICE \gls{NAIF} kernels.

\item Scale factor, $F_D$, for the drag force. One $F_D$ per arc is computed for accounting mis-modeling of the drag force.  

\item Scale factor, $F_S$, for the solar radiation pressure force. One $F_S$ per arc is also computed for accounting mis-modeling in the solar radiation pressure.

\item Empirical delta accelerations, radial, along-track, and cross-track, are computed at the \gls{AMD} epochs. The information of these epochs are given by \cite{Jcmarty10}.

\item For each arc, one offset per \gls{DSN} station for two or three-way Doppler measurements.

\item One offset per arc for one-way Doppler measurements accounting for the ultra stable oscillator stability uncertainty.

\item One bias per \gls{DSN} station for accounting the uncertainties on the \gls{DSN} antenna center position or the instrumental delays (such as the range group delay of the transponder)

\item One range bias per arc for ranging measurements to account the systematic geometric positions error (ephemerides error) between the Earth and the Mars.

\end{itemize}

\section{Orbit computation results}
\label{results}

\subsection{Acceleration budget}
\label{acc_bug}
Accurate orbit determination of planetary spacecraft requires good knowledge of gravitational and non-gravitational forces which act on the spacecraft. These forces are precisely modeled in the \gls{GINS} software as described in Chapter \ref{CHP1}. Figure \ref{acc_mgs} illustrates an average of various accelerations experienced by the \gls{MGS} spacecraft, that are:

\begin{itemize}
\item {\bf Accelerations due to gravitational potential:} the first two columns of the Figure \ref{acc_mgs} represent the accelerations owing to the gravitational attraction of Mars. These are the most dominating forces that are acting on the \gls{MGS} spacecraft. $GM/r + J_2$ in Figure \ref{acc_mgs} represents the mean and zonal coefficients contribution in the accelerations, which are related to $C_{lm}$ by the relation $J_l$ = -$C_{l,0}$. However, $gravity$ in the same figure represents the tesseral ($l$ $\ne$ $m$) and sectoral ($l$ = $m$) coefficients contribution in the accelerations (see Eq. \ref{potential}). An average acceleration for potentials $GM/r + J_2$ and $gravity$ is estimated as 3.0 and $8.9\times10^{-4}$ m/s$^2$ respectively. 

\item {\bf Accelerations due to third body attractions:} the third and fourth columns of the Figure \ref{acc_mgs} represent the accelerations owing to the Sun and the Moon, and planets-satellite attractions (see Eq. \ref{newton}), respectively. The Sun-Moon attraction is the third most dominating gravitational acceleration, whereas planets-satellite causes a smaller perturbation in the \gls{MGS} orbit. Average accelerations estimated for Sun-Moon and planets-satellite attractions are $6.0\times10^{-8}$ and $2.7\times10^{-12}$ m/s$^2$, respectively.

%
 %
 \begin{figure}[!ht]
\begin{center}\includegraphics[width=15cm]{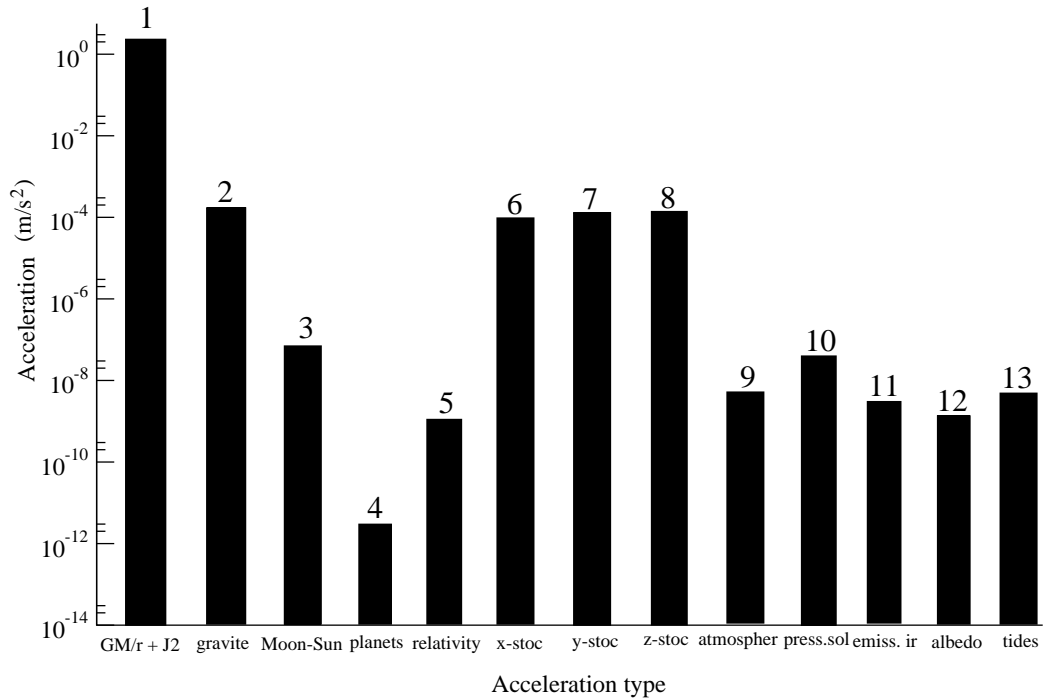}\end{center}
\caption{Gravitational and non-gravitational accelerations acting on the MGS spacecraft. See text for the explanation of each column.}
\label{acc_mgs}
\end{figure}

\item {\bf Accelerations due to general relativity:} the fifth column of the Figure \ref{acc_mgs} represents the accelerations owing to the contribution of general relativity (see Eq. \ref{relativity}). An average value of this acceleration is estimated as  $1.1\times10^{-9}$ m/s$^2$. 

\item {\bf Accelerations due to maneuvers:} the sixth, seventh, and eighth columns of the Figure \ref{acc_mgs} represent empirical accelerations that are estimated over the duration of each \gls{AMD} (see Eq. \ref{motor_burn}). Average value of radial (x-stoc), along-track (y-stoc), and cross-track (z-stoc) accelerations are estimated as $1.0\times10^{-4}$, $1.2\times10^{-4}$, and $1.8\times10^{-4}$ m/s$^2$ respectively.
 
\item {\bf Accelerations due to atmospheric drag:} the ninth column of the Figure \ref{acc_mgs} represents the acceleration owing to the resistance of the Mars atmosphere (see Eq. \ref{drag}). It is one of the largest non-gravitational accelerations acting on the low altitude spacecraft. For \gls{MGS}, an average value of this acceleration is estimated as $4.8\times10^{-9}$ m/s$^2$.

\item {\bf Accelerations due to solar radiation pressure:} the tenth column of the Figure \ref{acc_mgs} represents the acceleration due to the solar radiation pressure (see Eq. \ref{solar_rad}). It is the largest non-gravitational acceleration acting on the \gls{MGS} spacecraft with an average value of $4.5\times10^{-8}$ m/s$^2$.

\item {\bf Accelerations due to Mars radiation:} the eleventh and twelfth columns represent the accelerations due to the Infra-Red radiation (see Eq. \ref{infra}) and Albedo (see Eq.  \ref{albedo}) of the Mars. These are the smallest non-gravitational accelerations acting on the \gls{MGS} spacecraft with average values of $1.1\times10^{-9}$ and $3.0\times10^{-9}$ m/s$^2$ respectively.  

\item {\bf Accelerations due to solid planetary tides:} the thirteen column represents the accelerations owing to the contribution of solid planetary tides (see Eq. \ref{tides_planet}).  An average value of this acceleration is estimated as  $4.18\times10^{-9}$ m/s$^2$.

\end{itemize} 

\subsection{Doppler and range postfit residuals}
\label{post-res}
%
 %
 \begin{figure}[!ht]
\begin{center}\includegraphics[width=15cm]{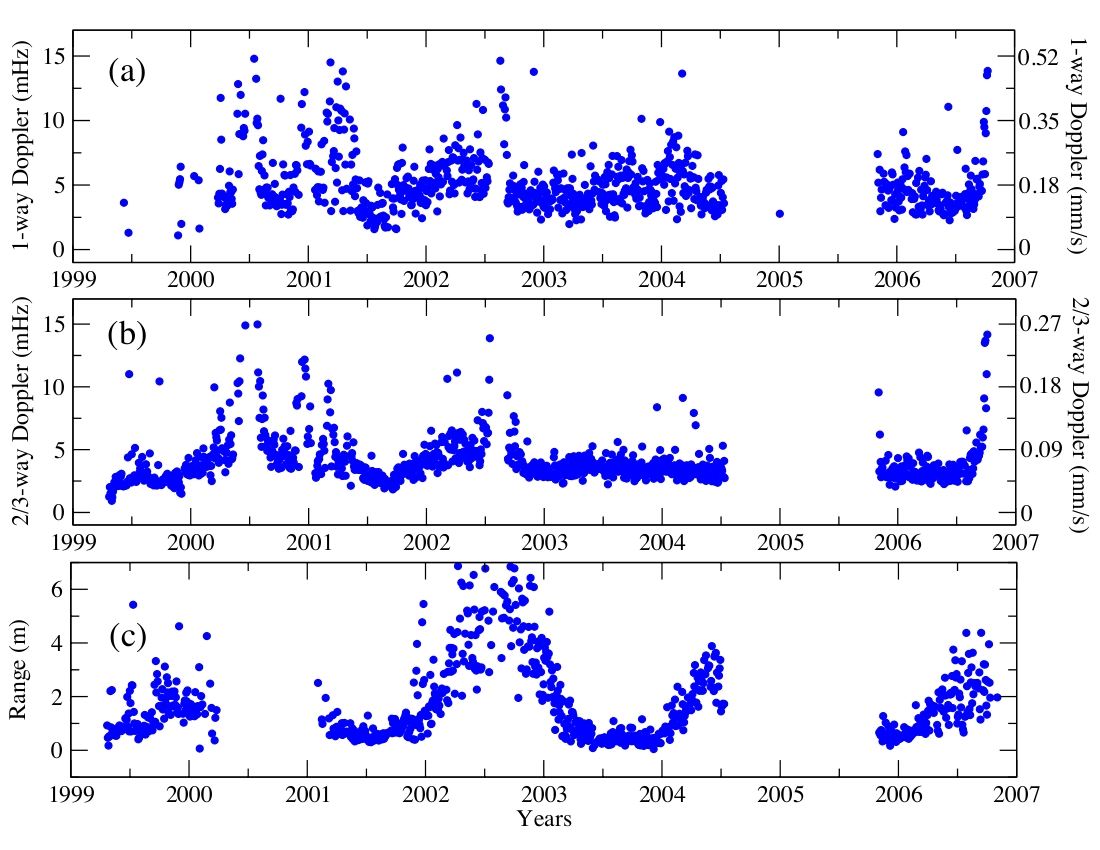}\end{center}
\caption{Quality of the MGS orbit in terms of rms values of the postfit residuals for each one-day data-arc: (a) one-way Doppler given in millihertz (1-way: 1 mHz = 0.035 mm/s =  speed of light / X-band frequency); (b) two- and three-way Doppler given in millihertz (2/3-way: 1 mHz = 0.0178 mm/s =  0.5$\times$speed of light / X-band frequency); and (c) two-way range given in meter. The peaks and gaps in residuals correspond to solar conjunction periods of MGS.}
\label{post-fit-mgs}
\end{figure}
In general, the Doppler data are mainly used for the computation of spacecraft orbit. They are sensitive to the modeling of the spacecraft dynamics and provide strong constraints on the orbit construction. However, range data are also used to assist the orbit computation. Unlike Doppler data, range data are more sensitive to the positions of the planet in the solar system and provide strong constraints to the planetary ephemerides.

Figure \ref{post-fit-mgs} illustrates the \gls{rms} values of the Doppler- and range postfit residuals estimated for each data-arc. Doppler residuals represent the accuracy in the computation of Doppler shift and in the dynamic modeling of the \gls{MGS} spacecraft, whereas range residuals represents the accuracy in the computation of range measurements. To plot, we did not consider 19\% of the data-arcs during which: i) the \gls{SEP} angle $\textless$12$^\circ$ and \gls{rms} value of the postfit Doppler and range residuals are above 15mHz and 7 m respectively, and ii) the drag coefficients and solar radiation pressures have unrealistic values. 

In Figure \ref{post-fit-mgs}, the peaks and the gaps in the postfit residuals correspond to solar conjunction periods. Excluding these periods, the \gls{rms} value of the postfit Doppler- and two- way range residuals for each data-arc is varying from1.8 to 5.8 mHz\footnote{1-way: 1 mHz = 0.035 mm/s =  speed of light / X-band frequency}$^,$\footnote{2/3-way: 1 mHz = 0.0178 mm/s =  0.5$\times$speed of light / X-band frequency} and 0.4 to 1.2m, respectively. These estimations are comparable with \cite{Yuan01,Lemoine01,JMarty}, see Table \ref{comp_mgs}. The mean value of the estimated Doppler offset for each \gls{DSN} station tracking pass is of the order of a few tenths of mHz, which is lower than the Doppler postfit residuals for each data-arc. This implies that there is no large offset in the modeling of the Doppler shift measurements at each tracking \gls{DSN} station.

\begin{table}[!ht]
\caption{Comparison of postfit Doppler and range residuals, and overlapped periods, between different authors.}
\centering
\renewcommand{\arraystretch}{1.4}
\small
\begin{threeparttable}
\begin{tabular}{ccccccccc}\Xhline{2\arrayrulewidth}
\multicolumn{1}{c}{\multirow{2}{*}{{\bf Authors}}} &\multicolumn{4}{c}{{\bf Residuals\tnote{a}}}     & \multicolumn{4}{c}{{\bf Overlap\tnote{b}}} \\ \cline{3-4} \cline{6-9}
&& {\bf Doppler (mHz)} &{\bf Range (m) }&&{\bf R (m)} &{\bf T (m)} &{\bf N (m)} &{\bf 3D (m)} \\ \Xhline{2\arrayrulewidth}
\cite{Yuan01}\tnote{c} && 3.4-4.9 &-&&1.6 & 2.5 &10.6 & -\\ 
\cite{Lemoine01}\tnote{c} && 3.4-5.6&-&&1.5 &2.55 & 8.75 & - \\ 
\cite{Konopliv06} && - &-&&{0.15} &{1.5} & 1.6 & - \\ 
\cite{JMarty} && 2-5 &0.5-1&& - &-&-& 2.1\\
This chapter && 1.8-5.8& 0.4-1.2 && 0.33$\pm$0.27\tnote{*} & 2.5$\pm$2.1\tnote{*} & 3.0$\pm$2\tnote{*} & 2.7$\pm$2.0\tnote{*} \\
\Xhline{2\arrayrulewidth}
\end{tabular}
\begin{tablenotes}
\item[a] range of the mean values estimated for each data-arc.
\item[b] mean value estimated for entire data.
\item[c] authors gave the values in mm/s, we therefore divided the values by a factor 0.0178 to obtain approximate values in mHz.
\item[*] mean$\pm$1-$\sigma$ dispersion of the rms values
\end{tablenotes}
\end{threeparttable}
\label{comp_mgs}
\end{table}

\subsection{Orbit overlap}
\label{overlap}
%
 %
 \begin{figure}[!hb]
\begin{center}\includegraphics[width=15cm]{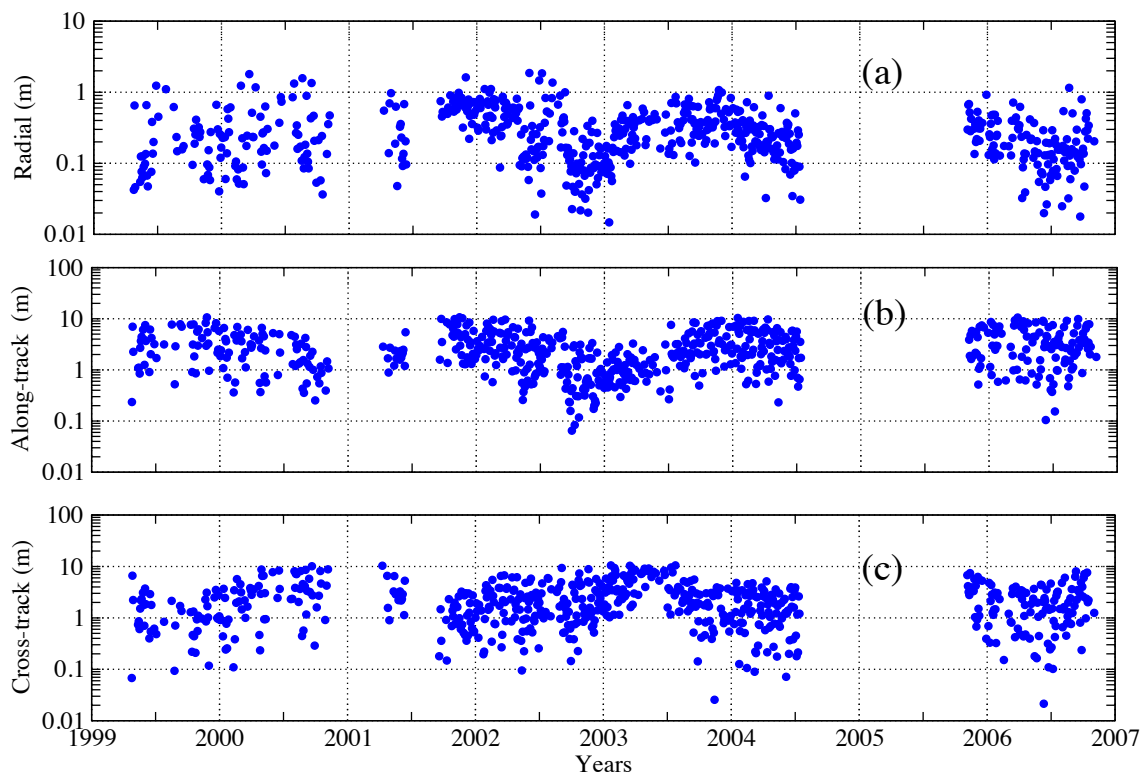}\end{center}
\caption{Orbit overlap differences for the entire mapping period of MGS mission for the (a) radial direction, (b) along-track direction, and (c) cross-track or normal to the orbit plane.}
\label{overlap_mgs}
\end{figure}
The quality of the orbit fits can also be investigated from the differences in \gls{MGS} positions between overlapping data-arcs. Such investigations are shown in Figure \ref{overlap_mgs}, which represents the \gls{rms} values of each overlap differences. These \gls{rms} values correspond to the \gls{MGS} position differences between successive two-days data-arcs over overlap duration of two-hours (one revolution of \gls{MGS} around Mars). As one can see from Figure \ref{overlap_mgs}, the radial component of the orbit error is less scattered after September 2001. This is due to the reduce number of AMDs after this period. However, mean and 1-$\sigma$ values of the radial, along-track, and cross-track components of the orbit error are 0.33$\pm$0.27\footnote{\label{ovl_mena}mean$\pm$1-$\sigma$ dispersion of the rms values}m, 2.5$\pm$2.1\textsuperscript{\ref{ovl_mena}}m, and 3.0$\pm$\textsuperscript{\ref{ovl_mena}}m respectively. The comparison of these values with the estimations of other authors are presented in Table \ref{comp_mgs}. As one can notice, statistics of our results are compatible within 1-$\sigma$ with the latest \cite{Konopliv06,JMarty}. However, overlap differences estimated by \cite{Yuan01,Lemoine01} are approximately $\sim$3 times higher in the normal direction and approximately $\sim$5 times higher in the radial-direction. It can be explained by the different gravity fields which were used for the computations. For example, \cite{Yuan01} had used a spherical harmonics model developed upto degree and order 75, while \cite{Lemoine01} updated the model upto degree and order 80. Moreover, both authors used only one year of mapping data for their computations. On the other hand, \cite{Konopliv06,JMarty} and on this chapter, the entire mapping period data since 1999 to the end of mission (late 2006) has been analyzed with 95$\times$95 spherical harmonics model. Because of the short circular orbit of the \gls{MGS}, such high order tesseral ($l$ $\ne$ $m$) and sectoral ($l$ = $m$) coefficients are important to model gravitation forces precisely.

\subsection{Estimated parameters}
\label{est_para}

\subsubsection{$F_S$ and $F_D$ scale factors}
\label{scale_factor}
The Solar radiation pressure (Eqs. \ref{solar_rad})  and the atmospheric drag (Eqs. \ref{drag}) are the most dominating non-gravitational forces that are acting on the \gls{MGS} spacecraft. As mentioned earlier, these non-gravitational forces depend upon the characteristic of the spacecraft model (macro-model) and its orientation. Thus, in order to accounting the inaccuracy in the spacecraft modeling and its orientations, an overall scale factors ($F_S$ and $F_D$) are estimated for each data-arc. However, for an accurate modeling, one can expect these values approximately equal to one. 

%
 %
 \begin{figure}[!hb]
\begin{center}\includegraphics[trim={0cm 5cm 0 0cm},width=15cm]{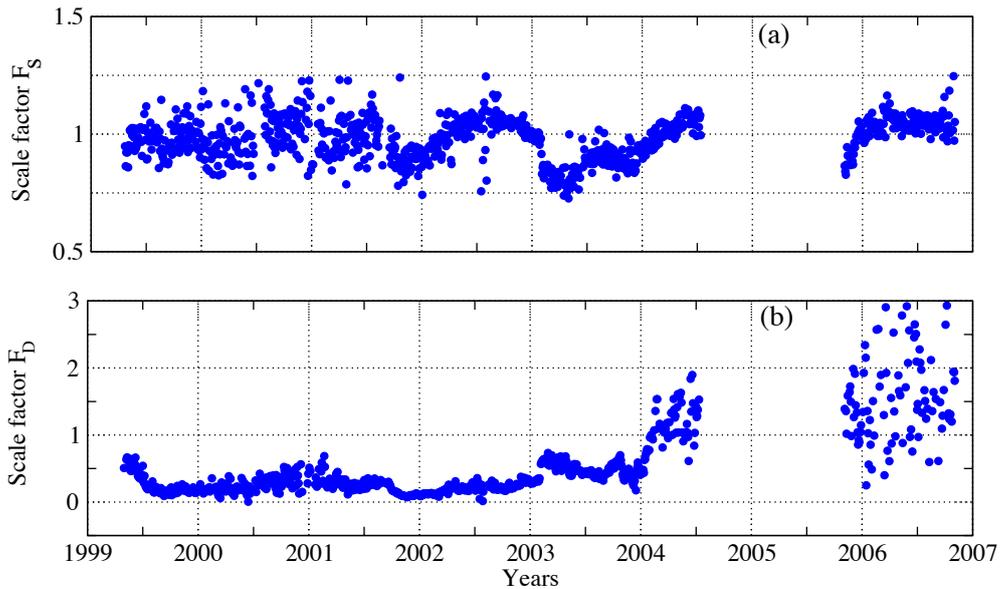}\end{center}
\caption{Scale factors: a) atmospheric drag and b) solar radiation pressure. }
\label{FDnFS_plot}
\end{figure}
 
A priori values of 1$\pm$1 were adopted for computing the two coefficients. Figure \ref{FDnFS_plot} shows the variation with time of these coefficients. As one can see on the panel $a$ of this figure, the computed values of the solar radiation pressure coefficient $F_S$ is stay around the nominal value. An average value of $F_S$ is estimated as 0.97$\pm$0.09. 
 
However, the drag coefficient (panel $b$ of same figure) stay rather below the nominal value up to the beginning of 2004, and then exhibit large variations which have been unexplained yet. An average value of $F_D$ is estimated as 0.50$\pm$0.40. This small value of $F_D$ may be a sign of lack of decorrelation with other parameters, enhanced over this time period by the weakness of the solar activity and an inadequate parameterization of the solar flux effect in the DTM-Mars model \citep{JMarty}.

\subsubsection{DSN station position and ephemeris bias}
\label{bias_eph_dsn}
\begin{figure}[!ht]
\begin{center}\includegraphics[width=15cm]{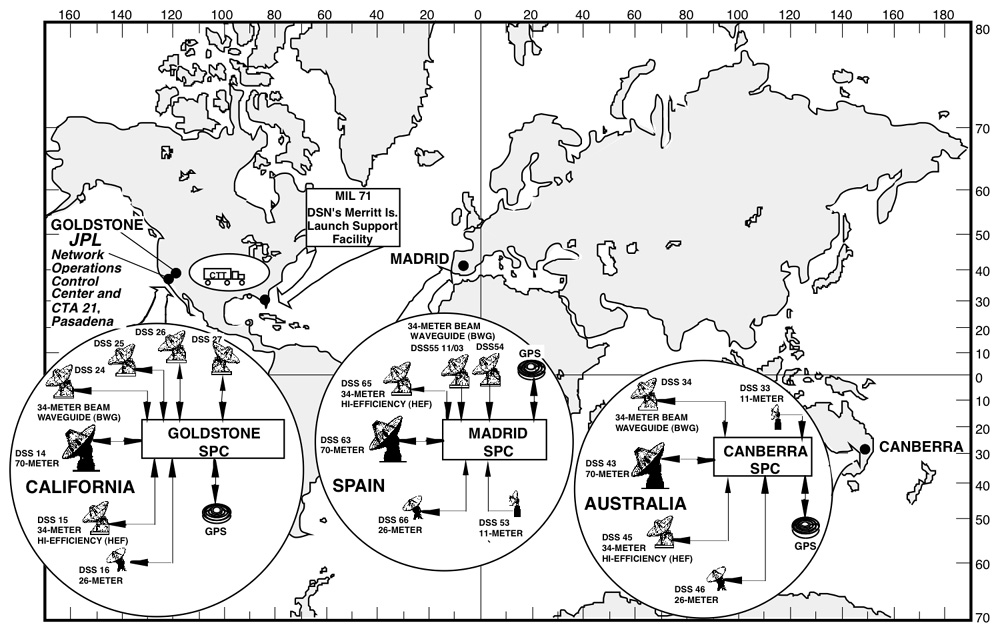}\end{center}
\caption{The continental map of DSN stations. \it{Image credit: NASA}}
\label{dsn_map}
\end{figure}

To account for the uncertainties on the \gls{DSN} antenna center position or in the instrumental delays, one bias per station is adjusted on the range measurements for each data-arc. For computing this bias, \gls{GINS} used station coordinates that are given in the 2008 \gls{ITRF}. These coordinates are then corrected from the continental drift, tides and then projected into an inertial frame through the \gls{EOP}. The continental map of \gls{DSN} stations is given in Figure \ref{dsn_map}

%
 %
%
 %
 \begin{figure}[!ht]
\begin{center}\includegraphics[trim={0cm 5cm 0 0cm},width=15cm]{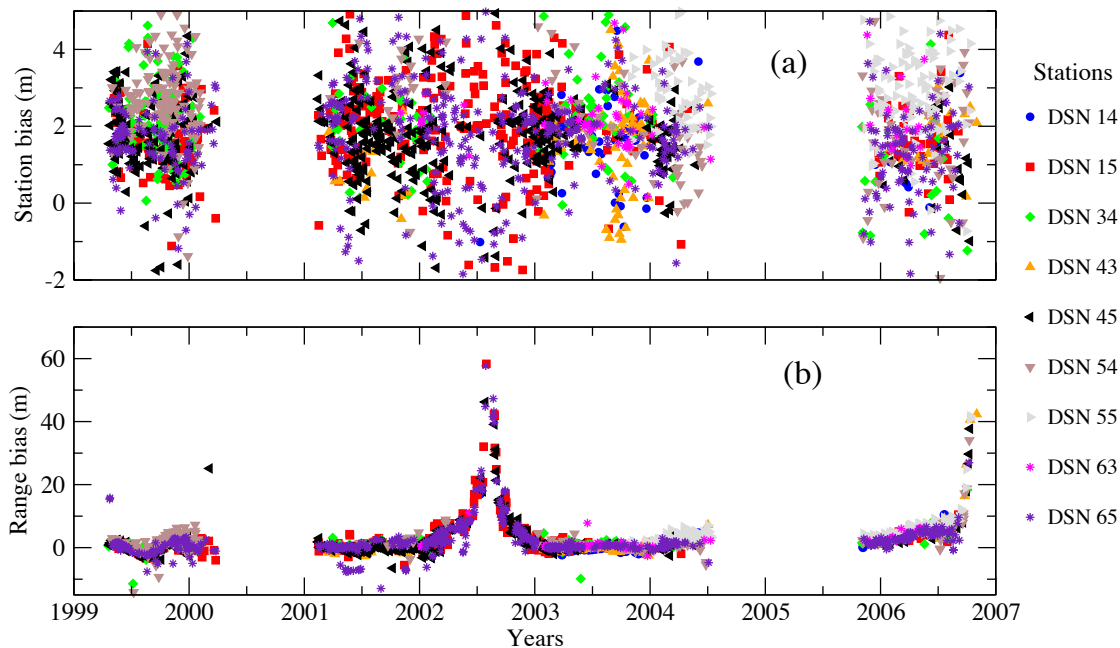}\end{center}
\caption{Distance bias estimated for each data-arc and for each participating station: a) station position bias and b) range bias corrections to the INPOP10b ephemeris.}
\label{range_bias_mgs}
\end{figure}

The adjustment of the station bias has been done simultaneously with the orbit fit. It is done independently for each station which are participating in the data-arc. This adjustment then absorbs the error at each station like uncalibrated delay in the wires. The panel $a$ of Figure \ref{range_bias_mgs} shows the variations with time of the computed station bias for each station. A mean and 1-$\sigma$ value of the station bias is estimated as 2$\pm$1 meter, which is compatible with the \cite{Konopliv06}.

In addition to the station bias, one ephemeris bias (so called range bias) is computed for each data-arc. The range bias represents the systematic error in the geometric positions between the Earth and Mars. Similar to the station bias, the range bias is also estimated from the range measurements of each data-arc. The panel $b$ of Figure \ref{range_bias_mgs} shows the variations with time of the estimated range bias per station compared to distances estimated with INPOP10b ephemeris. The peaks and gaps shown in Figure \ref{range_bias_mgs} demonstrate the effect of the solar conjunction on the range bias.

Moreover, such estimation of range bias are very crucial for the construction of planetary ephemeris and also to perform solar corona studies \citep{verma12}. The complete description of solar corona investigations (that are performed with these range bias) and its impact on the planetary ephemeris and the asteroids mass determination are discussed in Chapter \ref{CHP3}.

\section{Supplementary investigations}
\label{comps}

\subsection{GINS solution vs JPL Light time solutions} 
\label{ginsvsjpl}
One of the important information brought by the radioscience analysis is the range bias measurements between the Earth and the planet. These measurements are important for the estimation of the planet orbit. However, the range measurement accuracy is limited by the calibration of the radio signal delays at the tracking antennas and by the accuracy of the spacecraft orbit reconstruction. In order to check the accuracy of our estimations of the range bias, we computed the range bias (separately from the \gls{GINS}) from the light time data\footnote{\url{http://iau-comm4.jpl.nasa.gov/plan-eph-data/}} provided by the \gls{JPL} and compared them with the one obtained from the \gls{GINS}.  

The \gls{JPL} data represent a round-trip light time for each range measurement of \gls{MGS} spacecraft made by the \gls{DSN} relative to Mars system barycenter. Unlike to the radioscience data, the light-time is a processed data from the JPL \gls{ODP}. \gls{ODP} estimated the \gls{MGS} orbit from the Doppler tracking data and then measured the Mars position relative to an Earth station by adjusting the spacecraft range measurements for the position of the \gls{MGS} relative to the Mars center-of-mass \citep{Konopliv06}. In the \gls{ODP} software, a calibration for the tropospheric and ionospheric path delay at the \gls{DSN} stations has been applied based on calibration data specific to the time of each measurement. Moreover, a calibration for the electronic delay in the spacecraft transponder and a calibration for the \gls{DSN} tracking station measured for each tracking pass are also applied in the \gls{ODP} \citep{Konopliv06}. However, no calibration or model for solar plasma has been applied for this JPL release of the MGS light-time.

%
 %
 \begin{figure}[!ht]
\begin{center}\includegraphics[width=15cm]{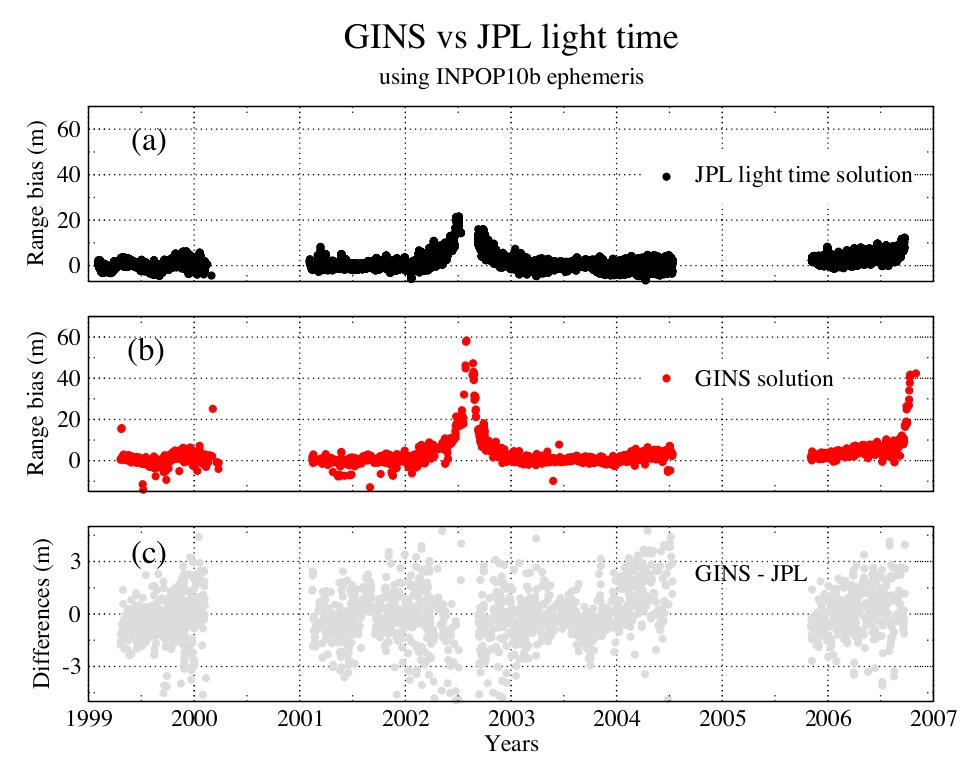}\end{center}
\caption{Range bias comparison between GINS solution and Light time solution using INPOP10b ephemeris: (a) range bias computed from the JPL light time solution, (b) range bias corresponding to the GINS solution, (c) difference between the GINS and JPL light time solutions.}
\label{range_bias_comp}
\end{figure}

In addition to \gls{MGS}, JPL also provides light time data for Odyssey and MRO missions on an irregular time basis. In order to use these light time in the INPOP construction, we therefore modeled a precise light time solution (based on the algorithms given in Section \ref{lidel} of Chapter \ref{CHP1}) to compute the round-trip time delay from Earth station to planet barycenter (in this case Mars) using INPOP planetary ephemerides. Except solar corona, all corrections which introduced perturbations in the radio signal have been taken in account. With this configuration of light time solution, we are then able to analyze the effect of the solar corona over the ranging data. The solar corona model derived from this analysis is discussed in Chapter \ref{CHP3}. 

The range bias obtained from the light time solution using INPOP10b ephemeris are shown in panel $a$ of Figure \ref{range_bias_comp}. Panel $b$ of the same figure represents the range bias obtained from the \gls{GINS} software. As one can see in Figure \ref{range_bias_comp}, both range bias show a similar behavior. The major difference between both solutions is the density of the data. JPL light time data sets are denser than the range bias obtained with \gls{GINS} as this latest estimated one bias over each two days data-arc when the JPL provides one light time measurement for each range measurement. To plot the approximated differences between both range measurements, we computed an average values of the JPL light time solution over each two days. The differences between \gls{GINS} range bias and averaged \gls{JPL} light time are plotted in panel $c$. In this panel, one can notice meter-level fluctuations in the differences, especially during 2004. Such fluctuations may be explained, by the degradations in the computation of the atmospheric drag forces (see Figure \ref{FDnFS_plot} and Section \ref{scale_factor}), and by the different approaches and softwares that have been used for the analysis of the \gls{MGS} radiometric data. Although, the average differences between both solutions was estimated as -0.08$\pm$1.2\footnote{mean$\pm$1-$\sigma$ dispersion}m, which is less than the current accuracy of the planetary ephemerides.

\subsection{\textit{Box-Wing} macro-model vs  \textit{Spherical} macro-model}
\label{macro_test}
As mentioned earlier, non-gravitational forces which are acting on the spacecraft are function of spacecraft model characteristics. These forces are however not as important in amplitude as the gravitational forces as shown in Figure \ref{acc_mgs}. Although, despite of their smaller contributions, these forces are extremely important for the precise computation of spacecraft orbit and the detection of geophysical signatures. However, in practice, the complete information of the spacecraft shape (also called macro-model) is either not precisely known or not publicly available. Therefore, the motivation of this study is to understand the impact of in-perfect macro-model over the spacecraft orbit and estimated parameters.

%
 %
 \begin{figure}[!ht]
\begin{center}\includegraphics[height=9cm,width=15cm]{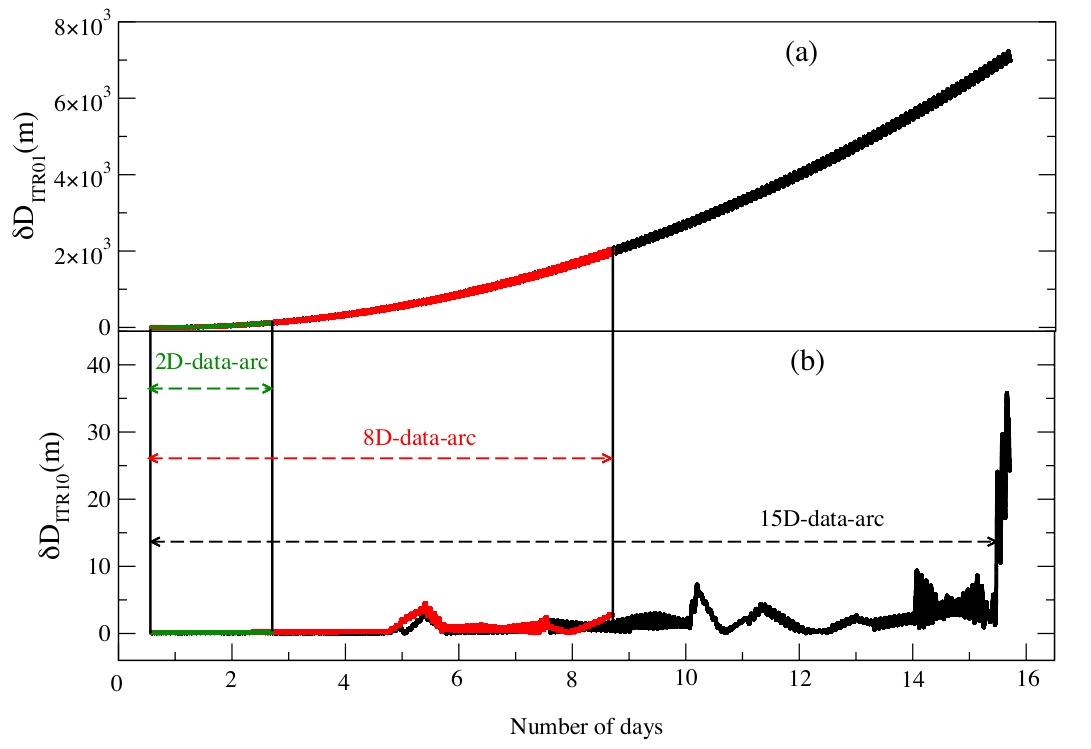}\end{center}
\caption{Evolution of the orbit change with respect to number of days in the data -arc and with the number of iterations. Panels of the figure are corresponding to a) Iteration-01 and b) Iterations-10. The green, red, and black colors in the figure are corresponding to 2-, 8-, and 15-days data-arc respectively. }
\label{macro_test}
\end{figure}

To perform this test, we have chosen two kinds of macro-model representing the \gls{MGS} spacecraft: 1)  the \textit{\gls{BW}} and 2) the \textit{\gls{SP}}. The characteristics of \gls{BW} are approximately close to the original one and are given in Table \ref{macro}, whereas \gls{SP} represents the spherical shape macro-model whose characteristics have been chosen randomly. These two models allowed us to clearly distinguish the impact of the macro-model over the orbit determination and the related parameters.

The most common method to minimize the impact of unperfect spacecraft modeling is to shorten the data-arc. We therefore test the evolution of the orbit change with respect to length of the data-arc. Figure \ref{macro_test} illustrate such changes using 2- (green), 8- (red), and 15-days (black) data-arcs. The panel $a$ of this figure represents the differences, $\delta$D$_{ITR01}$, between the integrated orbits using \gls{BW} and \gls{SP}. These orbits were not fitted to the measurements. As one can see, without the orbit fit $\delta$D$_{ITR01}$ is significantly propagating with time. Hence, for 2-, 8-, and 15-days data-arcs, the maximum differences in the orbits are found as 140m, 2000m, and 7200m respectively. 

%
 %
 \begin{figure}[!ht]
\begin{center}\includegraphics[width=15cm]{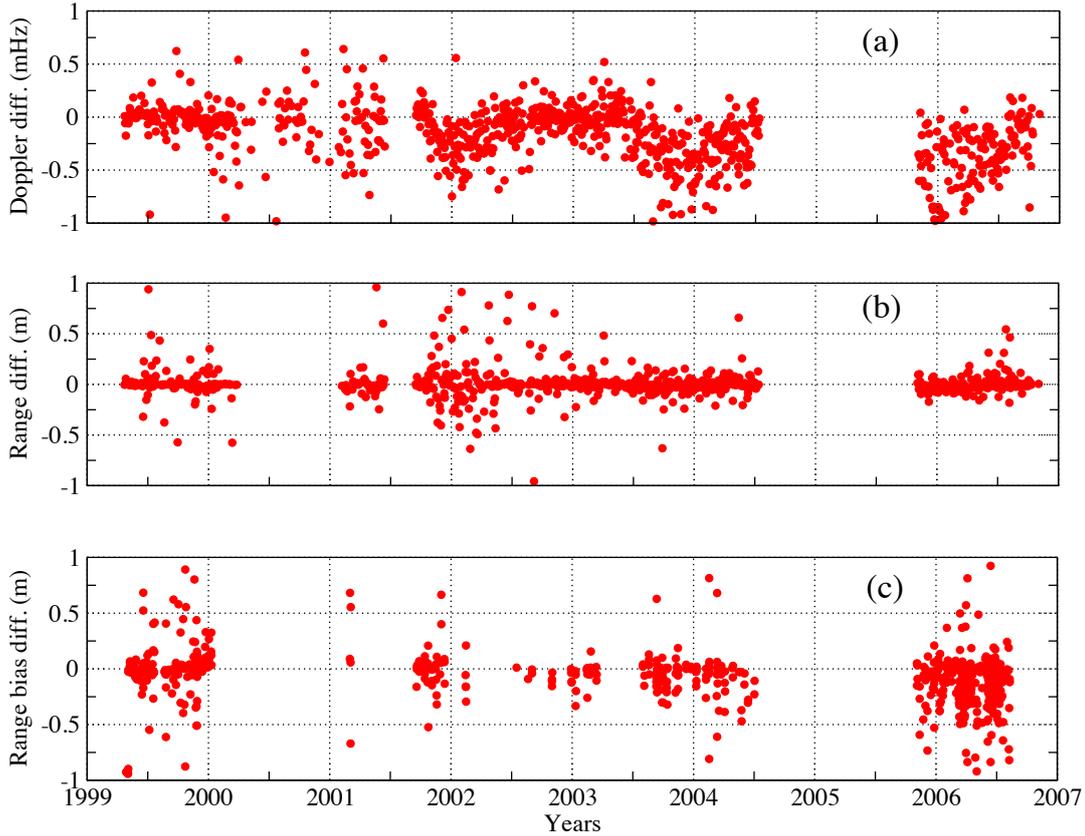}\end{center}
\caption{Difference between \textit{Box-Wing} and \textit{Spherical} macro-models: (a) Doppler \gls{rms} residuals, (b) range \gls{rms} residuals, and (c) range bias.}
\label{resd_diff}
\end{figure}

However, during the orbit determination, least-squares fitting of an orbit using an iterative process may absorb the orbit perturbation induced by the selection of macro-model. Hence, the panel $b$ of Figure \ref{macro_test} demonstrates the differences in the orbits, $\delta$D$_{ITR10}$, after ten iterations. From this figure, one can see that, the value of $\delta$D$_{ITR10}$ is significantly smaller than $\delta$D$_{ITR01}$. The maximum values of $\delta$D$_{ITR10}$ for 2-, 8-, and 15-days data-arcs are found as 0.22m, 4.5m, and 36m respectively. In particular, for 2-days data-arc, the maximum value of $\delta$D$_{ITR10}$ is less than the 1-$\sigma$ dispersion found in the overlap differences using \gls{BW} (see Section \ref{post-res}). Thus, short days data-arc would be the best choice for compensating the macro-model impact during the orbit determination. However, these statistics may vary during solar conjunction periods where most of the perturbations in the radio-signals and in the computed orbit are due to the solar corona.  

Furthermore based on this analysis, using 2-days data-arc, we re-analyzed the entire \gls{MGS} data to understand the impact of \gls{SP} on the forces (that are acting on the \gls{MGS} spacecraft) and on the estimated parameters. This analysis has been done in the same way as for \gls{BW} (see Section \ref{orbit}).

%
 %
 \begin{figure}[!ht]
\begin{center}\includegraphics[width=15cm]{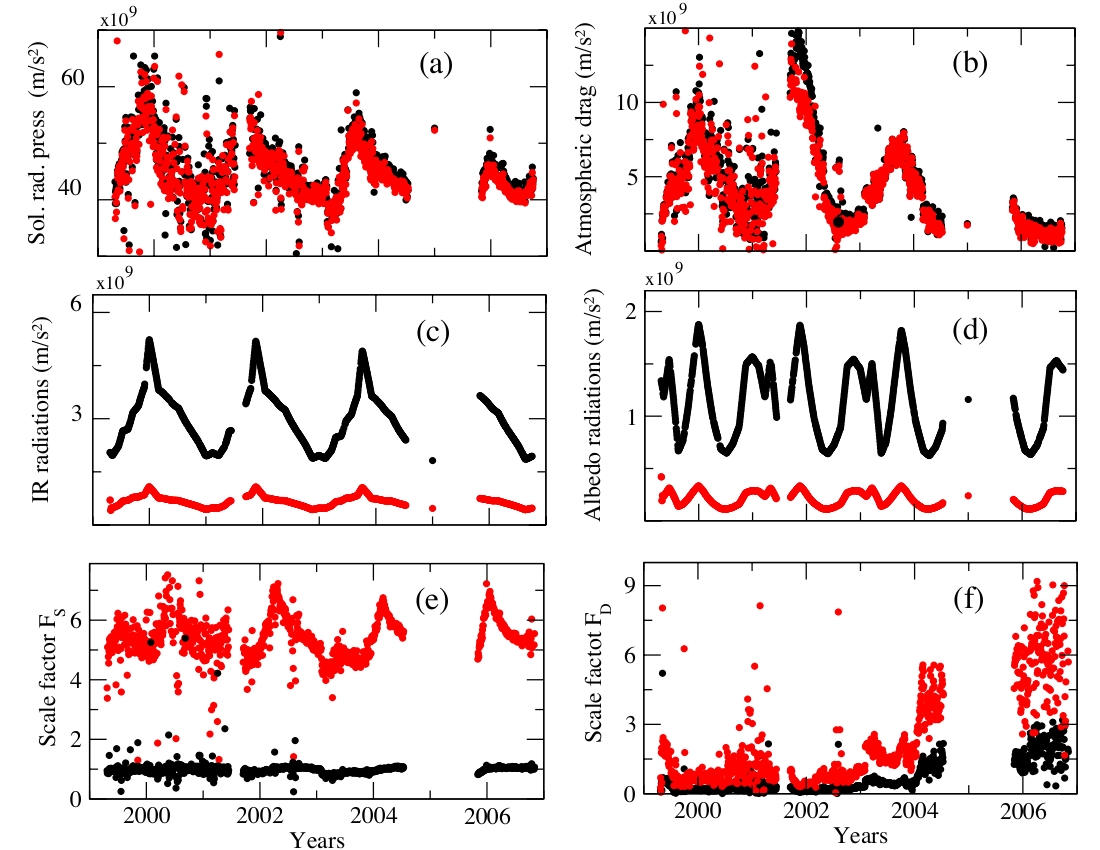}\end{center}
\caption{Non-gravitational accelerations and scale factors estimated using \textit{Box-Wing} macro-model (Black) and \textit{Spherical} macro-model (Red).}
\label{macro-acc-test}
\end{figure}

Figure \ref{resd_diff} illustrates the differences in the residuals (Doppler and range) and the range bias that are obtained during the orbit determination using \gls{BW} and \gls{SP}. Average values of these differences are estimated as -0.16$\pm$0.26mHz (-2.8$\pm$4.5 $\mu$m/s), 0.003$\pm$0.17m, and -0.0154$\pm$0.22m for Doppler residuals, range residuals, and range bias respectively. These differences are below 1-$\sigma$ of dispersion found for \gls{BW} (see Section \ref{post-res}). 

Furthermore, as mentioned earlier, characteristics of the spacecraft model only influence the non-gravitational accelerations. These accelerations are changing with one macro-model to another (see Eqs. \ref{solar_rad} to \ref{infra}). We have computed the non-gravitational accelerations which are acting on the \gls{MGS} spacecraft using \gls{BW} and \gls{SP}. Summary of these accelerations are plotted in Figure \ref{macro-acc-test}. 

In this figure, black dots correspond to \gls{BW} and red dots correspond to \gls{SP}. As one can see that, scale factor estimations for the solar radiation pressure $F_S$ (panel $e$) and for the atmospheric drag $F_D$ (panel $f$) are much higher for \gls{SP} than \gls{BW}. These scale-factors are estimated to account the mis-modeling in the corresponding accelerations. As a result, both \gls{BW} and \gls{SP} experienced approximately a similar accelerations due to the solar radiation pressure (panel $a$) and the atmospheric drag (panel $b$). 

However, such scale factors have not been estimated for accelerations due to the Mars radiations (Infra-Red radiation (panel $c$) and Albedo (panel $d$)). Owing to the different physical properties of both \textit{macro-models}, as expected, \gls{BW} experienced approximately five times greater Mars radiation accelerations than \gls{SP}. Due to the small contribution of these accelerations over the spacecraft motion, they did not bring much perturbations in the overall spacecraft orbit (Figure \ref{macro_test}) and postfit residuals (Figure \ref{resd_diff}).

\section{Conclusion and prospectives}
\label{Conclusion}

The radioscience data analysis of the \gls{MGS} spacecraft was chosen as an academic case to test our understanding of the raw radiometric data and their analysis with GINS by comparing our results with the literature. In this chapter, we have analyzed the entire radioscience data of Mars Global Surveyor since 1999 to 2006. This analysis has been done independently by using \gls{GINS} orbit determination software. For accounting the non-gravitational forces precisely, a ten-plate with parabolic high gain antenna \textit{Box-Wing} macro-model was used. In addition, orientations of the spacecraft and its articulated panels were also taken in account and modeled in terms of quaternions. Two-day data-arcs have been used to perform the numerical integration using updated 95$\times$95 MGS95J Mars gravity model.  

The estimated accuracy of the orbit and the parameters are consistent with the results found in the literature (\cite{Yuan01,Lemoine01,Konopliv06,JMarty}, see Table \ref{comp_mgs}). Moreover, we also compared range bias that were computed from our analysis with reduced light time data provided by JPL. An independent light time solution software has been developed to treat the JPL light time data. The range bias computed from both softwares are consistent with each other, and hence confirm the validity of our analysis with respect to JPL \gls{ODP} software.

 To understand the impact of the macro-model over the orbit perturbations and estimated parameters, we developed a new \textit{Spherical} macro-model. We then re-analyzed the entire radioscience data and compared the outcomes with the one obtained from the \textit{Box-Wing} macro-model. With this comparison, we confirmed that, in the absence of precise knowledge of the spacecraft characteristics, short data-arc can be preferable to accounting the mis-modeling in the spacecraft model without costing the orbit accuracy. However, this analysis may be not preferable for extracting an accurate geophysical signals from the Doppler measurements, such as the estimation of gravity field coefficients. In that case, an accurate modeling of the non-gravitational forces which depend upon the quality of macro-model are highly essential.

The part of these analysis, which consists of an accurate orbit determination of the \gls{MGS} with \textit{Box-Wing} macro-model, is published in \cite{verma12}. Moreover as a prospective of these analysis, we performed solar physics studies using data acquired at the time of solar conjunction periods. The Chapter \ref{CHP3} deals with the deduction of a solar corona model from the range bias that are computed at the time of the solar conjunctions. This model is then used to correct the range bias from the solar corona perturbations and also to estimate the corresponding averaged electron densities. The corrected or improved range bias are then used for the construction of planetary ephemerides \citep{verma12}. 

Furthermore, the hypothesis based on the supplementary tests of macro-model and the choice of data-arc, have been successfully used in Chapter \ref{CHP4} for the analysis of MESSENGER radioscience data and for the precise computations of MESSENGER orbit.

%% file: CHP3.tex
\chapter{Solar corona correction of radio signals and its application to planetary ephemeris}    
\label{CHP3}
\section{Introduction}
\label{intro}
The corona is a high temperature portion of the Sun outer atmosphere, beginning slightly above the visible surface and extending hundreds of thousands of kilometers, or further, into interplanetary space. It has a temperature of millions of degrees, but it is 10 billion times less dense than the atmosphere of the Earth at the sea level. The solar corona is the result of highly dense and strongly turbulent ionized gases that are ejected from the Sun. The particular combination of temperature and particle density in the corona leads to treat this ionized gas as a $plasma$. The term plasma represents the state of matter in which the neutral atoms are separated into charged components and with relatively strong electromagnetic forces between them.

In deep space navigation, the superior conjunction of a probe refers to the situation where the spacecraft, the Earth and the Sun lie in the same line with the spacecraft located on the opposite side of the sun with respect to the Earth. During this occasion, radio signals sent out by the spacecraft pass through the solar corona regions as they travel towards the Earth. Due to strongly turbulent and inhomogeneous plasma in the solar corona regions, radio frequency signals suffer severe degradation in their amplitude, frequency and phase.

The plasma effect on the radio signal propagation may change with the solar wind (slow or fast) and with the solar activity (minimum or maximum). Sections \ref{solcy} and \ref{solwd} give brief descriptions of these solar activity and solar wind characteristics, respectively. Moreover, the solar flares and \gls{CME} events that cause the perturbation in the radio signals during the solar conjunction are described in Section \ref{rad_sig_per}. A brief description of solar corona correction of radio signals and its application to planetary ephemeris are given in Section \ref{summary_paper}. Conclusions of the analysis are then discussed in Section \ref{chp3_con}.

All results described in this chapter are published in the Astronomy $\&$ Astrophysics journal. We therefore represent Section \ref{paper} by \cite{verma12}. This section gives the complete description of the solar corona model that has been derived from the range measurements of the \gls{MGS}, \gls{MEX}, and \gls{VEX}, spacecraft. The improvement in the extrapolation capability of the planetary ephemeris and the estimation of the asteroid masses are also discussed in the same section.
 
\section{The solar cycle}
\label{solcy}
The amount of magnetic flux that rises up to the Sun surface varies with time in a cycle called the solar cycle, which also correspond to the periodic change in the Sun\textquoteright s activity (including changes in the levels of solar radiation and ejection of solar material) and appearance (sunspots, flares, etc). This cycle is sometimes referred to as the sunspot cycle which is associated with strong magnetic fields. According to the Sun activity and the appearance of sunspots, the extreme of the solar cycles can be defined as solar maximum or solar minimum. A brief description of these terminology are given below.

\subsection{Magnetic field of the Sun}
\label{sun_mag}
The Sun magnetic field is generated by the motion of conductive plasma inside the Sun. This motion is created through convection. The high temperatures of the Sun cause the positively charged ions and negatively charged electrons that make up its plasma to move around. Such movement of the plasma creates many complicated magnetic fields. Moreover, due to the unequal rotation of the Sun around its axis, the plasma near the poles rotates slower than the plasma at the equator causing twisting and stretching of magnetic fields. The twisted magnetic fields lead to the formation of Sunspots (see Section \ref{sun_spot}), prominences, and an active corona.

\begin{figure}[!ht]
\begin{center}\includegraphics[width=8.5cm]{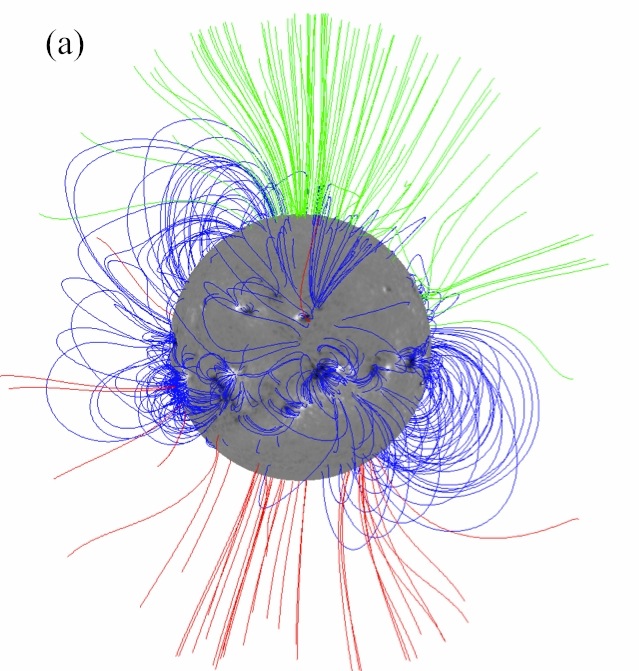}\includegraphics[width=8cm]{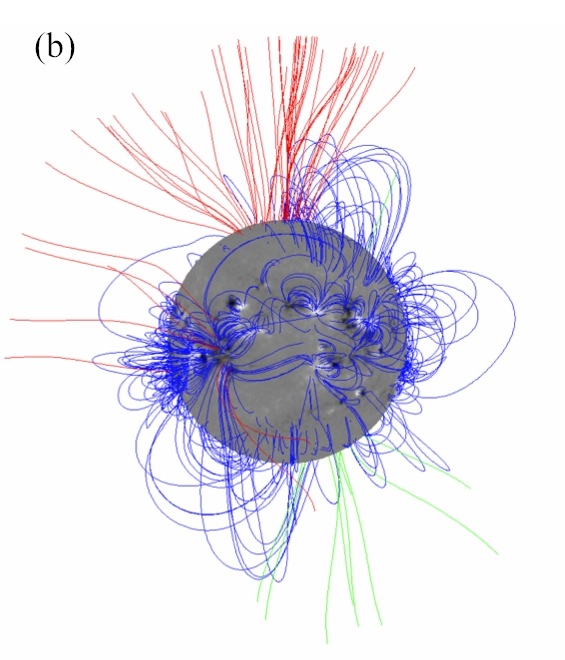}\end{center}
\caption{Approximated global coronal magnetic field structure for the beginnings of years correspond to solar maximum: (a) 1992 and (b) 2002. The photospheric radial field strength is represented by the greyscale, with white/black indicating positive/negative polarity. Green/red field lines represent open fields of positive/negative polarity and blue lines represent closed fields. These figures have been extracted from \cite{Petrie13}. }
\label{magnetic_field}
\end{figure}

The Sun magnetic field is stronger near the poles and weaker at the equator. In addition to being complex, the magnetic lines actually extend far out into space and this distant extension of the magnetic field is called the \gls{IMF}. Indeed, most of the structure of the Sun corona is shaped by the magnetic field which traps and contains the hot gases. As shown in Figure\footnote{This figure has been extracted from the \cite{Petrie13} only for the demonstration purpose.} \ref{magnetic_field}, the basic shape of the Sun magnetic field is like the field of a simple bar magnet that connect opposite polarities. However, this basic field (also called a dipole field) is a much more complex series of local fields that vary over time, scattered all over the surface, and constantly changing their positions and strengths (closed field presented by blue lines on the same figure).
%
 %

On Figure \ref{magnetic_field} one can notice that, the Sun magnetic field has flipped (open field presented by green and red lines) around the time of solar maximum (see Section \ref{solmax}). Eventually after $\sim$11 years, at the peak of the sunspot cycle the magnetic poles exchange places, called {\it polarity reversal}. The winding process then starts over leading to another cycle of $\sim$11 years with another polarity reversal to return to the Sun to its original state. Thus, the solar magnetic field has a $\sim$22 years cycle to return to its original state.
%
 %
 \begin{figure}[!ht]
\begin{center}\includegraphics[width=15cm]{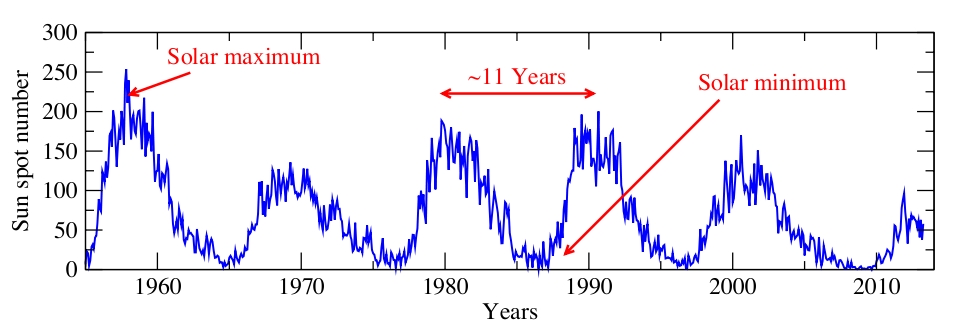}\end{center}
\caption{The solar sunspot cycle\textsuperscript{\ref{sunspot_data}}, since 1955 to present. Figure shows the variation of an average monthly sunspot numbers with time.}
\label{sunspot}
\end{figure}

\subsection{Sunspots} 
\label{sun_spot}
Sunspots are regions where the solar magnetic fields are very strong and twisted. They are temporary phenomena on the photosphere of the Sun that appear visibly as dark spots and typically last for several days, although very large ones may live for several weeks (see \cite{Ringnes64} for age-frequency distribution of sunspot groups). In visible light, sunspots appear darker than their surroundings as they are relatively thousands of degrees cooler than an average temperature of solar surface. The quantity that measures the number of sunspots and the groups of sunspots present on the Sun surface is refereed as sunspot number. The number and location of sunspots change over time. An average monthly distribution of sunspot numbers\footnote{\label{sunspot_data}\url{http://www.sidc.be/sunspot-data/}} are plotted with respect to time in Figure \ref{sunspot}. The peaks in this figure correspond to the highest solar activities and the cycle of these activities last approximately for 11 years on average. 

\subsection{Solar maxima}
\label{solmax}
The solar maxima is the period of greatest solar activity in the $\sim$11 years solar cycle (Figure \ref{sunspot}). During a solar maximum, the solar surface is covered by relatively large active regions \citep{Leon10} and large numbers of sunspots appear on the Sun surface. At solar maximum, the Sun magnetic field lines are the most distorted due to the magnetic field on the solar equator rotating at a slightly faster than at the solar poles and causing more solar activity (see Figure \ref{sol_max_min}). The magnetic field of the Sun approximates that of a dipole at high solar latitudes and reverse polarity during the peak of the sunspot cycle (see Figure \ref{magnetic_field}).

\subsection{Solar minima} 
\label{solmin}
\begin{figure}[!ht]
\begin{center}\includegraphics[width=8cm]{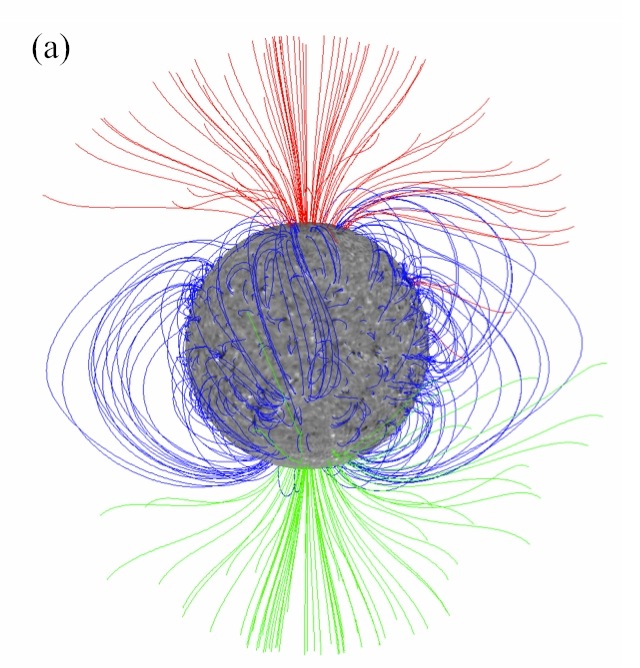}\includegraphics[width=7cm]{CHP3_Plot/2002.jpg}\end{center}
\caption{Approximated global coronal magnetic field structure for the beginnings of years: (a) 2008, solar minimum and (b) 2002, solar maximum. The photospheric radial field strength is represented by the greyscale, with white/black indicating positive/negative polarity. Green/red field lines represent open fields of positive/negative polarity and blue lines represent closed fields. These figures have been extracted from the \cite{Petrie13}.}
\label{sol_max_min}
\end{figure}

In contrast to solar maximum, the solar minimum is the period of least solar activity in the $\sim$11 year solar cycle (Figure \ref{sunspot}). During this time the Sun is in line with its magnetic poles as it has completed an 180 degree reversal and then have less solar activity. Typically, the active regions that are present during this time, occur at high solar latitude (far from the equator) \citep{Leon10}. During a solar minimum the Sun magnetic field, resembles that of an iron bar magnet, with great closed loops near the equator and open field lines near the poles (see Figure \ref{sol_max_min}).
%
 %

\section{The solar wind}
\label{solwd}
The solar wind is a stream of energized, charged particles, primarily electrons and protons, flowing outward from the Sun. The stream of particles varies in temperature and speed over time. Due to high temperature and kinetic energy these particles can escape the Sun gravity. Moreover, as described by \cite{Kojima90}, the large-scale solar wind structure is changing systematically with the phase of solar activity. The solar wind is divided into two components, respectively termed the slow solar wind and the fast solar wind. 
%
 %
 \begin{figure}[ht]
\begin{center}\includegraphics[width=15cm]{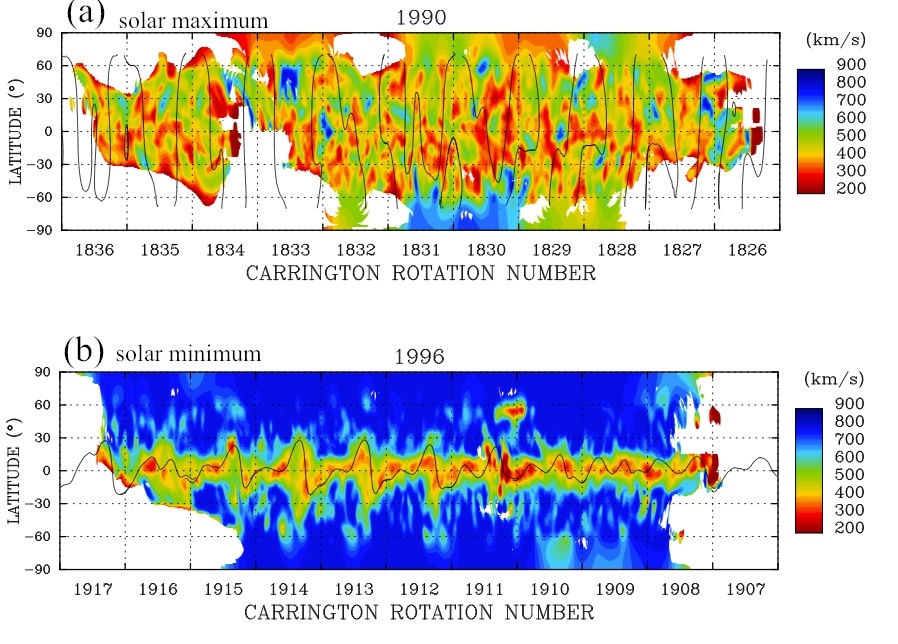}\end{center}
\caption{Synoptic source surface maps of solar wind speeds, in the Carrington rotation number versus latitude for 1990 and 1996, which approximately correspond to the cycle 22 maximum ({\it panel a}) and 22/23 minimum ({\it panel b}), respectively. The dark solid line represents the neutral magnetic line. These figures have been extracted from the \cite{Tokumaru10}.}
\label{sfwind}
\end{figure}
\subsection{Fast solar wind}
\label{fast_wind}
The fast wind has been associated with open field lines extended from coronal holes. It is characterized by a low density, and a low mass flux. The fast solar wind has a typical velocity of $\simeq$750 km/s, a temperature of $\simeq$8$\times$10$^5$ K and it nearly matches the composition of the Sun photosphere \citep{Feldman05}. The fast solar wind is thought to originate from coronal holes, which are the regions of open field lines in the Sun magnetic field \citep{Hassler99}. In these regions the magnetic field lines are open to the interplanetary medium \citep{Zirker77}, see Figure \ref{sol_max_min}. The fast wind areas increase systematically as the solar activity diminishes, reaching the maximum value at the minimum phase \citep{Tokumaru10}.  Figure\footnote{This figure has been extracted from the \cite{Tokumaru10} only for the demonstration purpose.} \ref{sfwind} shows an example of solar wind speed maps extracted from \cite{Tokumaru10}. From panel {\it b} one can see that, during a solar minimum, the high-to-mid latitude regions ($\sim \pm30^\circ$) were occupied with the fast wind. These regions however diminished or disappeared greatly during solar maximum (see panel $a$).

\subsection{Slow solar wind} 
\label{slow_wind}
The slow wind has been associated with the most active coronal regions and mainly with closed magnetic structures \citep{Schwenn83}, which cause its speed to be less than the fast solar wind. It is denser, and has a more complex structure, with turbulent regions and large-scale structures \citep{Kallenrode04}. The slow wind has a velocity of about $\simeq$400 km/s, a temperature of $\simeq$1.4-1.6$\times$10$^6$ K \citep{Feldman05}. The slow solar wind appears to originate from a region around the Sun equatorial belt that is known as the {\it streamer belt}. Coronal streamers extend outward from this region, carrying plasma from the interior along closed magnetic loops \citep{Lang00}. However \cite{Bravo97} suggest that coronal holes may also be the sources of slow winds, which could be emerging from the bordering. During solar maximum, the slow wind area increase systematically over the solar latitude and decrease as solar activity diminishes \citep{Tokumaru10}. Figure \ref{sfwind} shows that, during solar minimum, the low latitude regions were dominated by the slow wind and became ubiquitous at all latitudes during the maximum.

\section{Radio signal perturbation}
\label{rad_sig_per}
The perturbation in the radio signals is one of the consequences of the solar corona. For a deep space probe, when the Sun directly intercepts the radio signals between the spacecraft and the Earth, it induces perturbations in the signals and degrades them enormously. However, such degradations give an opportunity to compute time delay induced by the solar plasma and to study its physical characteristics with the plasma parameter. The time delay indicates the extra time due to the presence of an ionized medium in the propagation path, and the plasma parameter is a measure of the number of particles in a volume (so called electron density) along the line of sight (LOS). 

There could be a number of causes which degrade the radio signal during solar conjunctions. The solar flares and \gls{CME} could be the most important events that occur in the solar corona and cause the degradation of signal. These events are often associated with the solar activity and can occur during both phases of Sun activity, solar maximum and solar minimum. During the low periods of the solar cycle, events are less frequent and generally confined to the Sun equatorial region, when during periods of high solar activity, events are much more frequent and may occur at any place on the Sun surface (see Figure \ref{sfwind}). The brief characteristics of these events are given below:

\begin{itemize}
\item {{\bf Solar flare:}} Sudden energy release in the solar atmosphere is called solar flare \citep{Hudson95}. The enormous explosions of the Sun surface typically last for a few minutes and can release enormous amount of energy. The solar flares are known to be associated with the magnetic field structure of the Sun around the sunspots \citep{Parker63}. If this structure becomes twisted and sheared then magnetic field lines can cross and reconnect with an explosive release of energy. This causes an eruption of gases on the solar surface, and extends hundreds of thousands of kilometers out from the surface of the Sun following the magnetic lines to form a solar flare. During the first stages of the solar flare, high velocity protons are ejected and travel at around a third the speed of light. The ejected material follows the arc of the magnetic lines and then returns to the Sun, although some material is ejected into outer space especially during the larger flares.

\item {\bf Coronal mass ejections:} They are another form of disturbance that can affect radio communications. \gls{CME} are the explosions in the Sun corona that cause huge bubbles of gas that are threaded with magnetic field lines, and the bubbles are ejected over the space during several hours. During the \gls{CME}, the fluctuations of the Sun magnetic fields cause a release of huge quantities of matter and electromagnetic radiation into space above the Sun surface, either near the corona, or farther into the solar system, or even beyond. Such release of matter disrupts the steady flow of the solar wind producing a large increase in the flow. Unlike a solar flare, a \gls{CME} doesn't produce intense light. But it does produce a magnetic shockwave which may interact with the Earth magnetic field. 
\end{itemize}

Both events increase the level of solar radiation in the solar corona regions. During superior solar conjunctions, when radio signals interact with these radiations, then the signals suffer severe degradation and cause a time delay on ranging measurements and a phase advance on Doppler measurements. The magnitudes of these effects are inversely proportional to the square of the signal frequency and the radial distance (outward from the Sun) \citep{Muhleman77,Schwennvol1,Schwennvol2,Guhathakurta94,Bird94,Guhathakurta96}. Thus, solar corona effects on the radio signals are decreasing with increase in frequency and with distance from the Sun. Peaks in the Figures \ref{post-fit-mgs} and \ref{range_bias_mgs} (panel $b$) of Chapter \ref{CHP2} show an example of solar corona effects on Doppler and range measurements. The solar corona model deduce from the range measurements acquired at the time of solar conjunctions allowed us to compute time delay due to the presence of an ionized medium in the propagation path. One can then also estimate an average electron density along the line of sight (LOS). 

\section{Solar corona correction of radio signals and its application to planetary ephemeris}
\label{summary_paper}
As described in sections \ref{solcy} and \ref{solwd}, solar wind area and solar events are frequently changing with the solar activity. Hence, one can expect different distributions of charged particles in slow and fast wind regions \citep{Schwenn06}. Therefore, it is necessary to identify if the region of the radio signal propagation is either affected by the slow wind or by the fast wind. In this section we investigated these regions and derived the characteristics of solar corona models and associated electron density distribution at different phases of solar activity (maximum and minimum) and at different solar wind states (slow and fast). The processed range data obtained after orbit determination have been used to derive the models and to estimate the time delay due to the solar corona. These estimations lead us to remove the solar corona perturbations from the range data. The corrected range data are then used for the construction of planetary ephemeris. This complementary range data noticeably improve the extrapolation capability of the planetary ephemeris and the estimation of the asteroid masses. All results corresponding to this study are published in the Astronomy $\&$ Astrophysics journal. In this section we gave the brief description of the steps used to derive the solar corona models. The outcomes of this study are given in Section \ref{paper} and followed by \cite{verma12}.
%
 %
\begin{figure}[!ht]
\begin{center}
\includegraphics[width=16cm]{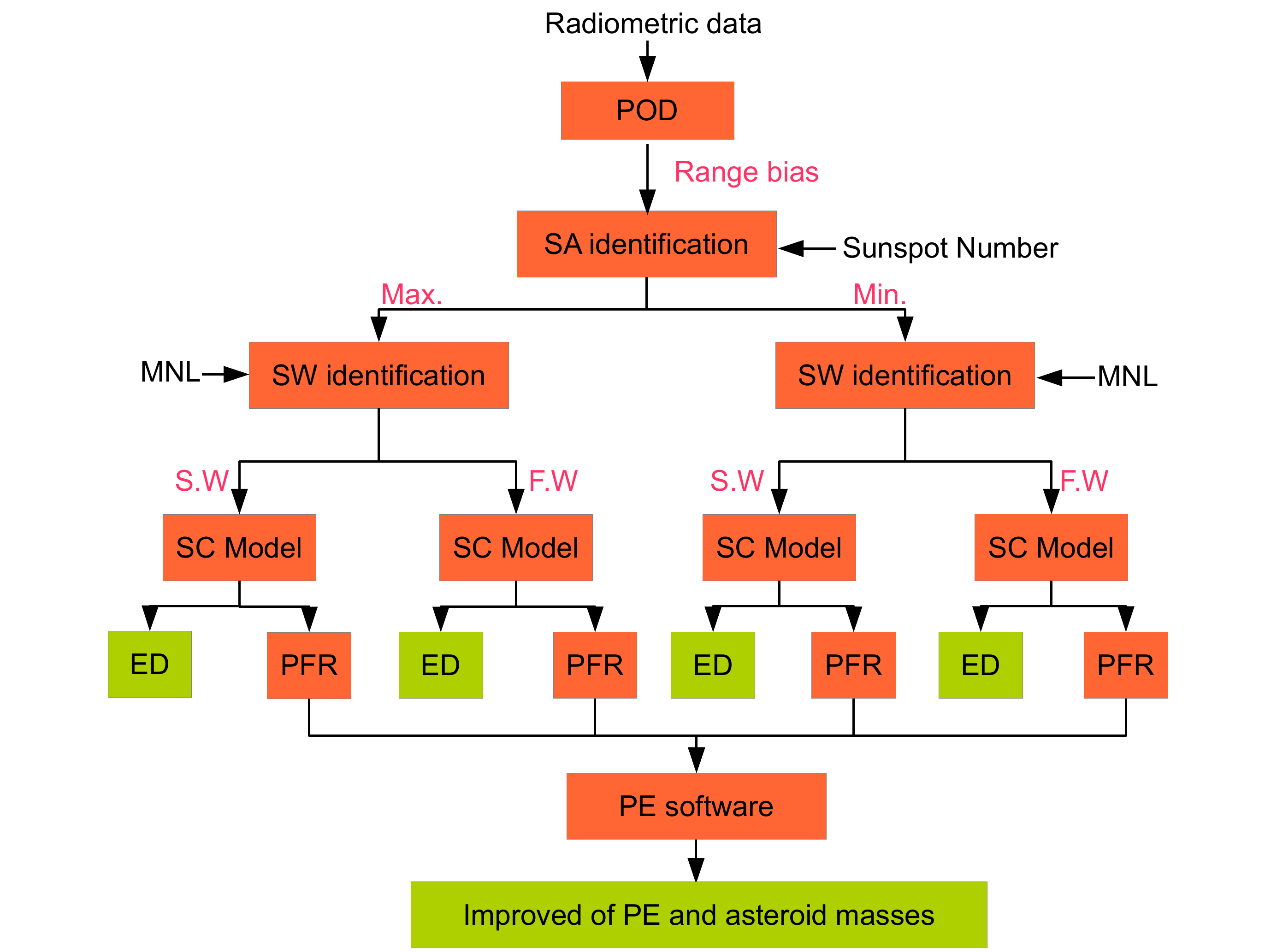}\end{center}
\caption{Flow chart describing the procedure involved in the derivation of solar corona model, electron density computations and further improvement of the planetary ephemeris. The abbreviations used in the flow chart are: \\
\small{POD: Precise Orbit Determination; SA: Solar Activity; SW : Solar Wind \\
 S.W : Slow Wind; F.W : Fast Wind; ED : Electron Density; PFR : PostFit Residuals}}
\label{paper_flowchart}
\end{figure}

Figure \ref{paper_flowchart} described the procedure used to perform this study, that are:
\begin{itemize}

\item {\bf Precise Orbit Determination:} The radiometric data of \gls{MGS}, \gls{MEX}, and \gls{VEX}, consisting of two-way Doppler- and range measurements (see Chapter \ref{CHP1}), have been used to perform this study. These data set are first analyzed by the orbit determination software to compute the spacecraft orbit precisely. Such analysis of the \gls{MGS} radiometric data using \gls{GINS} software have been performed in Chapter \ref{CHP2}. However, for \gls{MEX}, and \gls{VEX} these analysis were performed by \gls{ESA} navigation team. The range bias obtained from these computations were then provided by \gls{ESA}, and we compared them to light-time delays computed with the INPOP10b, and the DE421 ephemerides. The brief description of \gls{MEX} and \gls{VEX} orbit accuracy and their estimation are given in \cite{Fienga2009} and in \cite{verma12} (see Section \ref{paper}).  

\item {\bf Solar activity identifications:} As described in Section \ref{solcy}, unequal rotation of the Sun causes twisting and stretching of the magnetic fields. The twisted magnetic fields leads to the formation of sunspots, prominences, and an active corona. Hence, as shown in Figure \ref{sunspot}, larger (smaller) number of sunspots appeared on the Sun surface represents the maximum (minimum) solar activity. In June 2002, when the solar activity was maximum (see Figure \ref{sunspot}), the MGS \gls{SEP} angle remained below 10$^\circ$ for two months and went at minimum to 3.325$^\circ$. \gls{MEX} experienced superior conjunctions during the minimum phase of solar activities: October 2006, December 2008, and February 2011. During these time \gls{MEX} \gls{SEP} angle remained below 10$^\circ$ for two months. Similarly, the VEX \gls{SEP} angle remained below 8$^\circ$ for two months during October 2006 and June 2008 (solar minima, see Figure \ref{sunspot}). The peaks and gaps shown in Figure 2 of Section \ref{paper} demonstrate the effect of the solar conjunction on the estimated range bias for \gls{MGS}, \gls{MEX}, and \gls{VEX}. 

\item {\bf Solar wind identifications:} As shown in Figure \ref{sfwind}, the distribution of the solar slow-, and fast-winds varies significantly with the solar activities. Hence, as described in \cite{Schwenn06}, the electronic profiles are very different in slow- and fast-wind regions. MGS, and MEX and VEX experienced superior conjunctions during the solar maxima and minima, respectively. It is therefore necessary to identify if the region of the \gls{LOS} is either affected by the slow-wind or by the fast-wind when passes through the solar corona regions. Such identification can be performed by locating the \gls{MDLOS} heliographic longitudes and latitudes with the maps of the solar corona magnetic field as provided by the \gls{WSO}\footnote{\url{http://wso.stanford.edu/}}. \gls{MNL}, where the resultant magnetic field is zero, can be then used to define the limit of slow-, and fast-winds. We took limits of the slow solar wind regions as a belt of 20$^\circ$ above and below the \gls{MNL} during the solar minima and $\pm$70$^\circ$ during solar maxima \citep{You07,You12}. The entire distributions of the \gls{LOS} for all three spacecraft are given in Section \ref{paper}.   

\item {\bf Solar corona model:} From Figure \ref{paper_flowchart} one can see that, the solar corona models have been derived separately for different types of solar winds (fast, slow), and for solar activity phases (minimum, maximum). As mentioned previously, interactions between radio signals and solar corona regions cause severe degradations in the signals and a time delay on ranging measurements. Such effects on the radio signals decrease with an increase in frequency and with distance from the Sun, and can be modeled by integrating the entire ray path from the Earth station ({\itshape L$_{{Earth}_{s/n}}$}) to the spacecraft ({\itshape L$_{s/c}$}) at a given epoch. This model is defined as:
 \begin{equation}
 \label{delay_eq}
 \Delta \tau = \frac{1}{2cn_{cri} (f)} \times \int_{L_{{Earth}_{s/n}}} ^{L_{s/c}}N_{\mathrm{e}}(l) \ dL
 \end{equation}
 
 \begin{displaymath}
   n_{cri} (f) = 1.240 \times 10^4\ \bigg(\frac{f}{1\ MHz}\bigg)^2 \ \ \mathrm{cm^{-3}}  \ \ ,
 \end{displaymath}
 
where $c$ is the speed of light, $n_{cri}$ is the critical plasma density for the radio carrier frequency $f$, and $N_{e}$ is an electron density in the unit of electrons per cm$^{3}$ and is expressed as \citep{Bird96}
 
\begin{equation}
  \label{electron_eq1}
N_e{(l)} =   B\ \bigg(\frac{l}{R_\odot}\bigg)^{-\epsilon}  \ \ \mathrm{cm^{-3}}  \ \ .
 \end{equation}

where $B$ and $\epsilon$ are the real positive parameters to be determined from the data. R$_\odot$ and $l$ are the solar radius and radial distance in AU. The maximum contribution in the electron density occurs when $l$ equals the \gls{MDLOS}, $p$, from the Sun. At a given epoch, \gls{MDLOS} is estimated from the planetary and spacecraft ephemerides. We also defined the impact factor $r$ as the ration between \gls{MDLOS} $p$ and the solar radii (R$_\odot$). The complete analytical solutions for computing the Equation \ref{delay_eq} is given in Appendix A of Section \ref{paper}. Moreover, in addition to \cite{Bird96} corona model, we have also used \cite{Guhathakurta96} corona model that added one or more terms to Equation \ref{electron_eq1}, that is:
 \begin{equation}
   \label{electron_eq2}
N_e{(l)} =  A\bigg(\frac{l}{R_\odot}\bigg)^{-c} +  B\bigg(\frac{l}{R_\odot}\bigg)^{-d}  \ \ \mathrm{cm^{-3}}  
 \end{equation}
with c $\simeq$ 4 and d = 2.

The parameters of Equations \ref{electron_eq1} and \ref{electron_eq2} are calculated using least-squares techniques. These parameters are obtained for various ranges of the \gls{MDLOS}, from 12R$_\odot$ to 215R$_\odot$ for  MGS, 6R$_\odot$ to 152R$_\odot$ for MEX, and from 12R$_\odot$ to 154R$_\odot$ for VEX. The fitting of the parameters were performed, for all available data acquired at the time of the solar conjunctions, for each spacecraft individually, and separately for fast- and slow-wind regions. The summary of estimated parameters for each spacecraft and for each solar wind region are given in Table 2 of Section \ref{paper}).

%
%
\begin{figure}[!ht]
\begin{center}
\includegraphics[width=16cm]{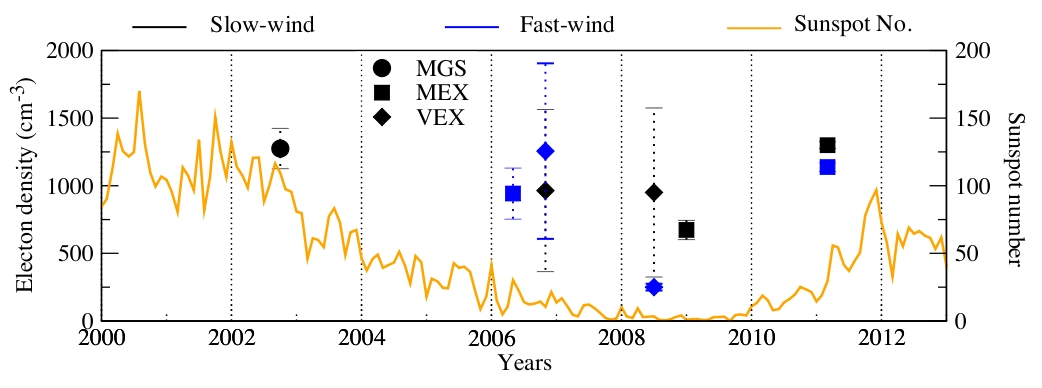}\end{center}
\caption{An average electron density distribution at 20R$_\odot$ during different phases of solar activities and for different states of solar wind. Higher number of sunspots correspond to maximum phase  of solar activity, while smaller number correspond to minimum phase.}
\label{SWnFW}
\end{figure}
\begin{enumerate}[(a)]
\item {\bf Electron density:} The MGS experienced its superior conjunction in 2002 when the solar activity was maximum and the slow-wind region was spread at about $\pm$70$^\circ$ of heliolatitude \citep{Tokumaru10}. Hence, \gls{MDLOS} of the \gls{MGS} exclusively affected by the slow-wind region (see Figure 6 of Section \ref{paper}). In contrast, MEX and VEX experienced their superior conjunctions during solar minima (2006, 2008, and 2011). During these periods, the slow-wind region was spread at about $\pm$30$^\circ$ of heliolatitude \citep{Tokumaru10}. Owing to small slow-wind region, \gls{MDLOS} of the MEX and VEX were affected by both slow-, and fast-wind regions (see Figure 4 of Section \ref{paper}).

For an example, on Figure \ref{SWnFW}, are plotted the average electron densities obtained at 20R$_\odot$ for all three spacecraft, computed separately for fast- and slow-wind regions using Equation \ref{electron_eq1}. These densities are then compared with the various models, as given in Table \ref{comparison}.  

\begin{table}[!ht]
\caption{Electron densities estimated from different models at 20R$_\odot$ and at 215R$_\odot$ (1AU). }
\centering
\begin{center}
\renewcommand{\arraystretch}{1.4}
\small
\begin{threeparttable}
\begin{tabular}{ c c c c c c}
 \hline  
                        &                           &                                            \\
    Authors & Spacecraft  &    Solar  &\gls{MDLOS} & Ne @ 20R$_\odot$ & Ne @ 215R$_\odot$  \\ 
     &                         &          activity&&      (el. cm$^-3$)& (el. cm$^-3$)                      \\

 \hline
\cite{Leblanc} &Wind&  Min&1.3-215 &847&  7.2 \\

\cite{Bougeret}  &  Helios 1 and 2& Min/Max & 65-215&  890& 6.14\\
    
 \cite{Issautier98} & Ulysses & Min & 327-497 & 307\tnote{*} & 2.65$\pm$0.5\tnote{*}\\
 
 This chapter\tnote{**} &  MEX06 &  Min &  6-40& 942$\pm$ 189 & 2.3$\pm$0.9 \\
 
 This chapter\tnote{**} &  MEX08 &  Min &  6-71& 673$\pm$ 72 & 2.3$\pm$1.3 \\
 
 This chapter\tnote{**} &  VEX08 &  Min &  12-154& 950$\pm$ 625 & 3$\pm$2 \\
 
\cite{Muhleman77} & Mariner 6 and 7 &Max. &5-100&1231$\pm$ 64&9$\pm$3\\
 
 This chapter\tnote{**} & MGS &  Max &  12-215& 1275$\pm$ 150 & 11.0$\pm$1.5 \\
     
\cite{Bird94} &  Ulysses &  Max &  5-42& 1700$\pm$ 100 & 4.7$\pm$0.415 \\

 \cite{AndersonV2}&  Voyager 2 &  Max & 10-88& 6650$\pm$ 850 & 38$\pm$4\\
    \hline
    
\end{tabular}
\begin{tablenotes}
\item[*] Mean electron density corresponds to latitude $\geq$40$^\circ$ 
\item[**] The values correspond to Table 2 of Section \ref{paper}, \cite{verma12}
\end{tablenotes}
\end{threeparttable}
\end{center}
\label{comparison}
\end{table}
In Table \ref{comparison}, we provide the average electron density at 20R$_\odot$ and 215R$_\odot$, based on the corresponding model parameters (if not given by the authors). Table \ref{comparison} shows a wide range of the average electron densities, estimated at 20R$_\odot$ and 215R$_\odot$ during different phases of solar activity. As one can notice, our estimates of the average electron density are very close to the previous estimates, especially during solar minimum. The widest variations between our results and the earlier estimates were found during solar maxima and can be explained from the high variability of the solar corona during these periods.

Moreover, as one can notice on Figure \ref{SWnFW} and Table \ref{comparison}, dispersions in the estimation of electron densities using \gls{VEX} range data are relatively large compared to the \gls{MGS} or \gls{MEX}. This can be explained by the limitations in the VEX orbit determination \citep{Fienga2009}, which introduced bias in the estimation of the model parameters and consequently in the electron densities. The detailed analysis of these results are discussed in Section \ref{paper}

\item {\bf Postfit residuals:} In addition to electron density computations, one can also compute the light time delay due to the solar corona. This time delay can then be removed from the range bias to minimize the effect of solar corona. Statistics of the range bias before (prefit) and after (postfit) the correction of the solar corona perturbations are given in Table \ref{STAT}. From this table one can notice that, the estimated dispersions in the postfit range bias are one order of magnitude lower than the dispersions in the prefit range bias. It shows a good agreement between the model estimates and the range radiometric data. However, because of the degraded quality of VEX orbit \citep{Fienga2009}, the statistics of the VEX residuals are not as good as for the MGS and MEX. 

\begin{table}[!ht]
\caption{Statistics of the range bias before and after solar corona corrections.}
\centering
\renewcommand{\arraystretch}{1.4}
\small
\begin{threeparttable}
\begin{tabular}{ccccccc}\Xhline{2\arrayrulewidth}

\multicolumn{1}{c}{\multirow{2}{*}{{\bf S/C}}} &\multicolumn{4}{c}{{\bf Pre-fit}}     & \multicolumn{2}{c}{{\bf Post-fit}} \\ \cline{3-4} \cline{6-7}
&& {\bf mean (m)} &{\bf $\sigma$ (m) }&&{\bf mean (m)} &{\bf $\sigma$ (m)} \\ \Xhline{2\arrayrulewidth}

MGS, 2002&& {6.02} &{10.10}&&{-0.16} &{2.89} \\ 
MEX, 2006&& {42.03} &{39.30}&&{0.85} &{9.06} \\
MEX, 2008&& {16.00} &{20.35}&&{-0.10} &{4.28} \\
MEX, 2011&& {15.44} &{19.20}&&{0.11} &{6.48} \\
VEX, 2006&& {5.47} &{11.48}&&{-0.74} &{6.72} \\
VEX, 2008&& {3.48} &{11.48}&&{-0.87} &{7.97} \\
\Xhline{2\arrayrulewidth}
\end{tabular}
\end{threeparttable}
\label{STAT}
\end{table}
\end{enumerate}

\item {\bf Planetary ephemerides improvements:} As described in Chapter \ref{CHP0}, range bias data are very important for the construction of the planetary ephemerides. Usually, due to high uncertainties, the range bias affected by the solar corona perturbations ($\sim$4-6 months of data) are not included in the construction of the planetary ephemerides. Thanks to the solar corona corrections, it was possible to use for the first time these range bias. To demonstrate the impact of these complementary data, corrected for the solar corona perturbations, we construct two ephemerides, INPOP10c and INPOP10d, both fitted over the same data set as was used for the construction of INPOP10b (see Chapter \ref{CHP0}). INPOP10c was constructed without solar corona corrections while these corrections were included in the INPOP10d construction. These additional data represents $\sim$8$\%$ of whole data set used for the ephemeris construction.

In particular, the Mars orbit is affected by the belt of asteroids. The asteroid masses may cause a degradation in the estimates of the Mars orbit. Therefore by keeping more observations during solar conjunction intervals we have noticed the noticeable improvements in the extrapolation capability of the planetary ephemerides and the estimation of the asteroid masses. Table 5 of Section \ref{paper} gives such estimations of the asteroid masses. On Table 6 of Section \ref{paper}, significantly improved masses, estimated with INPOP10d, are then compared with the one found in the literature and showed 80$\%$ better consistency with respect to INPOP10c. Moreover, MEX extrapolated residuals computed with INPOP10d show a better long-term behavior compared with INPOP10c with 30$\%$ less degraded residuals after two years of extrapolation. The complete analysis of these results are discussed in Section \ref{paper}. 
\end{itemize} 

\section{Conclusion}
\label{chp3_con}
We have analyzed the large-scale structure of the corona electron density, since 2001 to 2011. This analysis has been done with the range bias data of the MGS, MEX, and VEX spacecraft acquired during solar conjunction periods. The parameters of the solar corona models are then deduced from these data. These parameters were estimated separately for each spacecraft at different phases of solar activity (maximum and minimum) and at different solar wind states (slow and fast). We compared our results with the previous estimations that were based on different techniques and different data sets. Our results are consistent, especially during solar minima, with estimations obtained by these different techniques. However, during the solar maxima, electron densities obtained with different methods or different spacecraft show weaker consistencies. 

We have also demonstrated the impact of solar corona correction on the construction of planetary ephemerides. The supplementary data, corrected from the solar corona perturbations, allowed us to gain $\sim$8$\%$ of whole data set. Such corrected data are then used for the first time in the construction of INPOP and induce a noticeable improvement in the estimation of the asteroid masses and a better long-term behavior of the ephemerides. In addition to these improvements, recent Uranus observations and positions of Pluto have beed added to construct the latest INPOP10e ephemerides \citep{Fienga13}. For the further improvement, INPOP10e is then used for the analysis of MESSENGER radioscience data (see Chapter \ref{CHP4}).   

As stated previously, all results described in this chapter are published in the Astronomy $\&$ Astrophysics journal. The Section \ref{paper} therefore represented by the \cite{verma12}, which described the complete analysis of these results. 
\section{\cite{verma12}} 
\label{paper}
\includepdf[pages=1-17]{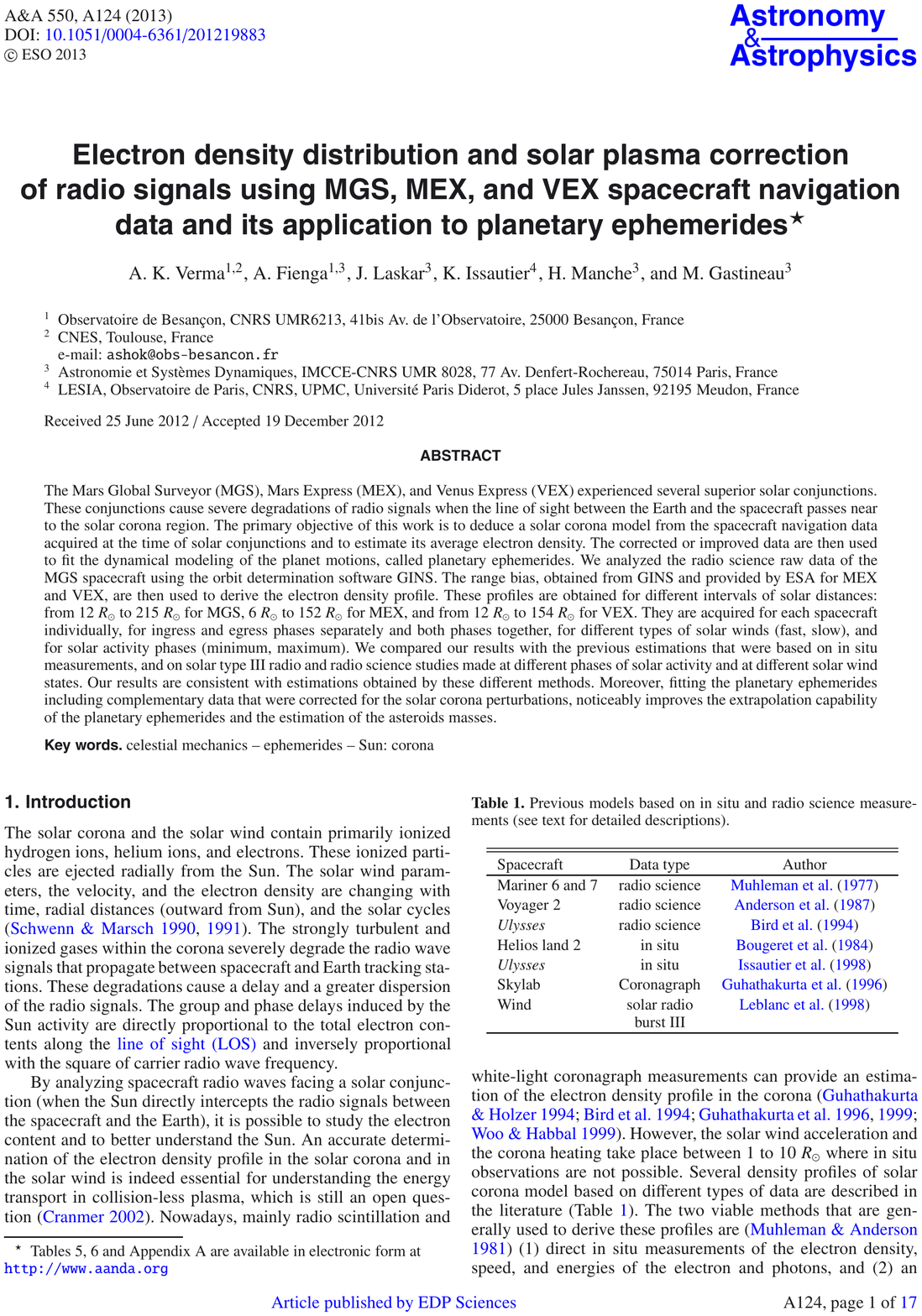}

%% file: CHP4.tex
\chapter{Improvement of the planetary ephemeris and test of general relativity with MESSENGER}    
\label{CHP4}


\section{Introduction}
Mercury is the smallest and least explored terrestrial planet of the solar system. Mariner 10 was the first spacecraft which made three close encounters (two in 1974 and one in 1975) to this planet and provided most of our current knowledge of the planet until early 2008 \citep{Smith10}. In addition to Mariner 10 flyby observations, ground based radar measurements were the only observations which were used to study the gravity field of Mercury and its physical structure (spherical body with slightly flattened at the poles and mildly elongated equator) \citep{Anderson87,Anderson1996}. In 2004, \gls{NASA} launched a dedicate mission, \gls{MGR}, to learn more about this planet. \gls{MGR} made three close encounters (two in 2008 and one in 2009) to Mercury and became the first spacecraft which observed Mercury from its orbit.   

Till now, \gls{MGR} has completed more than two years on orbit at Mercury. During the orbital period, radio tracking of \gls{MGR} routinely measured the Doppler and range observables at \gls{DSN} stations. These observables are important to estimate the spacecraft state vectors (position and velocity) and to improve the knowledge of Mercury\textquoteright s gravity field and its geophysical properties \citep{Srinivasan07}. Using the first six months of radioscience data during the orbital period, \cite{Smith12} computed the gravity field and gave better constraints on the internal structure (density distribution) of Mercury. This updated gravity field becomes crucial for the present computation of \gls{MGR} orbit and to perform precise relativistic tests.    

The primary objectives of this work is to determine the precise orbit of the \gls{MGR} spacecraft around Mercury using radioscience data and then improve the planetary ephemeris INPOP (see Chapter \ref{CHP0}). The updated spacecraft and planetary ephemerides are then used to perform sensitive relativistic tests of the \gls{PPN} parameters ($\gamma$ and $\beta$). 

As described in Chapter \ref{CHP0}, spacecraft range measurements are used for the construction of planetary ephemerides. These measurements approximately cover 56\% of whole INPOP data and impose strong constraints on the planet orbits and on the other solar system parameters including asteroid masses. However, until now, only five flybys (two from Mariner 10 and three from \gls{MGR}) range measurements were available for imposing strong constraints to the Mercury\textquoteright s orbit \citep{Fienga2011}. Therefore, range measurements obtained by \gls{MGR} spacecraft during its mapping period are important to improve our knowledge of the orbit of Mercury.  

Moreover, high precision radioscience observations also gave an opportunity to perform sensitive relativistic tests by estimating possible violation of \gls{GR} of the two relativistic parameters ($\gamma$ and $\beta$) of the \gls{PPN} formalism of general relativity \citep{Will93}. The previous estimations of these parameters, using different techniques and different data set, can be found in \cite{Bertotti03,Muller08,Pitjeva09,Williams09,Manche2010,Konopliv11,Fienga2011}. However, because of Mercury relatively large eccentricity and close proximity to the Sun, its orbital motion provides one of the best solar system tests of general relativity \citep{Anderson97}. In addition, \cite{Fienga10,Fienga2011} also demonstrated that, Mercury observations are far more sensitive to \gls{PPN} modification of \gls{GR} than other data used in the planetary ephemerides construction. We therefore, also performed such tests with the latest \gls{MGR} observations to obtain one of the best value for \gls{PPN} parameters.

In this chapter, we introduce the updated planetary ephemeris INPOP13a and summarize the technique used for the estimation of the \gls{PPN} parameters. The outline of the chapter is as follow: The section \ref{messenger} gives an overview of the \gls{MGR} mission. The radioscience data analysis and the dynamical modeling of \gls{MGR} are also discussed in the same section. In section \ref{orbit}, we present the results obtained during the orbit determination. The evolution of INPOP with the accuracy of MESSENGER orbit, and the brief description of tests performed with the INPOP13a are also discussed in this section.

The results presented in this chapter are gathered in an article published in Astronomy $\&$ Astrophysics. The Section \ref{paper_mgr} is therefore stated as \cite{verma14}. This section deals with the detailed analysis of the tests performed with the high precision Mercury ephemeris INPOP13a.

\section{MESSENGER data analysis}
\label{messenger}
\subsection{Mission design}
\label{design}
Under the NASA\textquoteright s Discovery program, the MErcury Surface, Space ENvironment, GEochemistry, and Ranging (\gls{MGR}) spacecraft is the first probe to orbit the planet Mercury. It launched in August 3, 2004, from Pad B of Space Launch Complex 17 at Cape Canaveral Air Force Station, Florida, aboard a three-stage Boeing Delta II rocket. On March 18, 2011, \gls{MGR} successfully entered Mercury\textquoteright s orbit after completing three flybys of Mercury following two flybys of Venus and one of Earth.

\begin{figure}[]
\begin{center}\includegraphics[width=13cm]{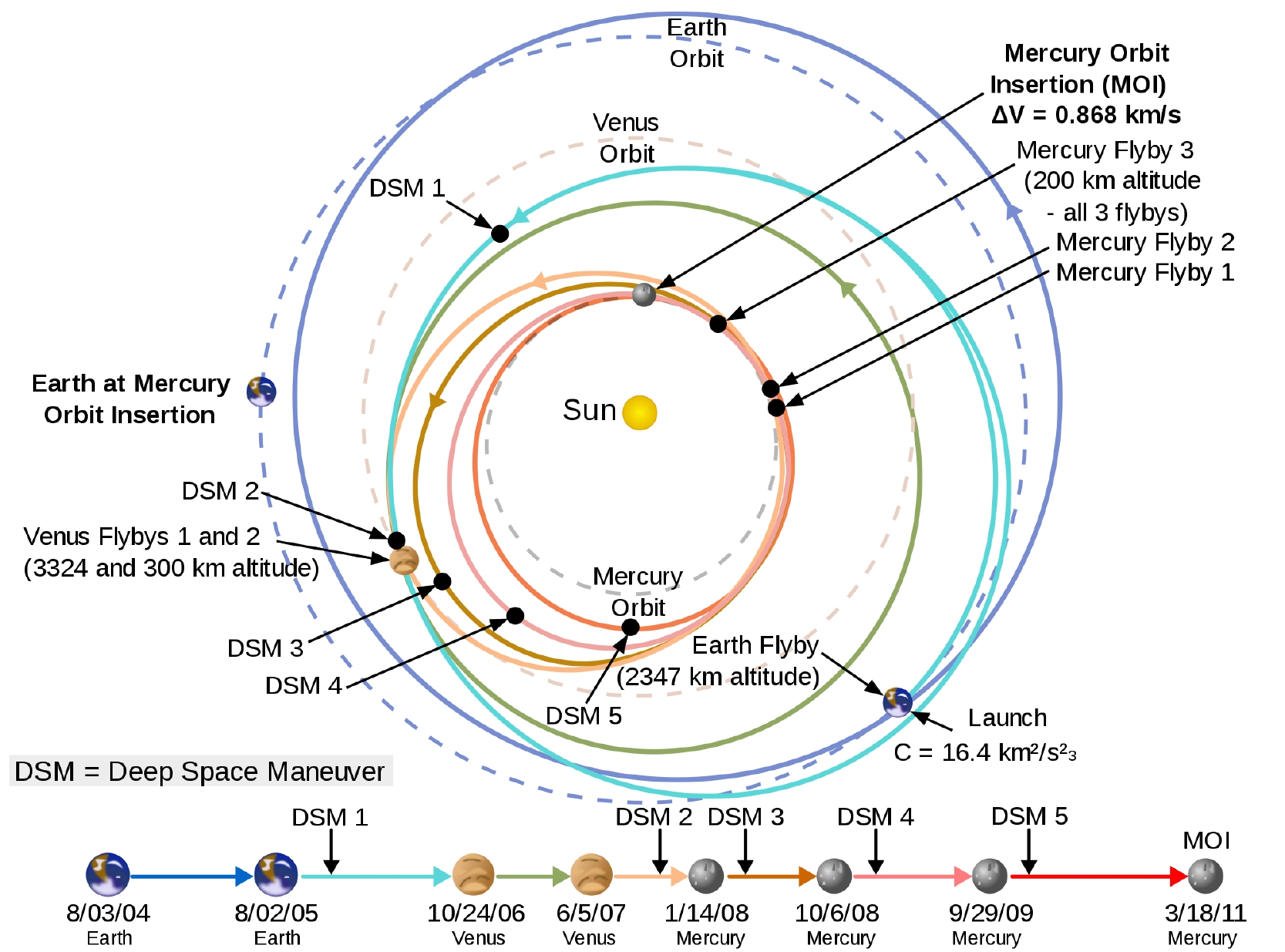}\end{center}
\caption{Summary of MESSENGER\textquoteright s entire trajectory from launch to mapping period \citep{McAdams07}.}
\label{mission}
\end{figure}

The 6.6-year trip from launch to Mercury orbit insertion is one of the longest interplanetary cruise phase options considered for \gls{MGR}. The spacecraft system design lifetime accounted for a seven-year journey to Mercury followed by a prime (one-year) and extended (one- or more-year) Mercury orbit phases. \gls{MGR} used gravity assists from Earth, Venus and Mercury to lower its speed relative to Mercury at orbit insertion. Several \gls{TCM}, including five large \gls{DSM} were also used to adjust its path to Mercury \citep{McAdams07}. As a summary of the entire trajectory from launch to mapping period is shown in Fig. \ref{mission}. 

As shown in this figure, approximately one year after launch, \gls{MGR} made its first flyby of Earth on August 2, 2005 and then headed toward the Venus flybys using largest DSM. The \gls{MGR} made two Venus flybys, first occurred on October 24, 2006 and second on June 5, 2007. These flybys headed \gls{MGR} towards Mercury by additionally using the DSM performed after second flyby of Venus. In January 14, 2008 \gls{MGR} became first spacecraft which provided first close-up look of Mercury in more than 30 years. The Mercury flyby 1-to-Mercury orbit insertion transfer trajectory, shown in Fig. \ref{mission}, includes three Mercury flyby-DSM segments that lower the spacecraft speed relative to Mercury and on March 18, 2011, \gls{MGR} successfully entered Mercury\textquoteright s orbit.

The \gls{MGR} spacecraft was initially inserted into a $\sim$12-hour, near-polar orbit around Mercury, with an initial periapsis altitude of 200 km, initial periapsis longitude of 60\textdegree N, and apoapsis at $\sim$15,200 km altitude in the southern hemisphere. After successful first year flight in the orbit, mission was extended to one or more year which began on 18 March 2012. During first extended mission, two orbit-correction maneuvers were executed, four days apart, in April 2012 to reduce MESSENGER\textquoteright s orbital period from $\sim$12 to $\sim$8 hours \citep{Flanigan13}

\subsection{Spacecraft geometry}
\label{sc_geo}
 \begin{figure}[]
\begin{center}\includegraphics[width=16cm]{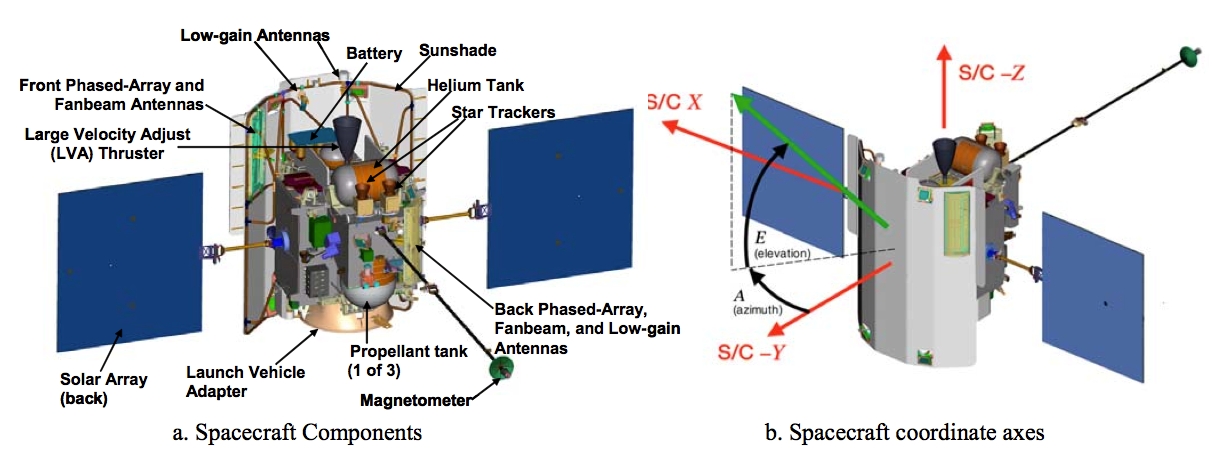}\end{center}
\caption{MESSENGER spacecraft geometry \citep{Vaughan06}.}
\label{sc}
\end{figure}
The \gls{MGR} spacecraft was designed and constructed to withstand the harsh environments associated with achieving and operating in Mercury orbit. The spacecraft structure is constructed primarily of lightweight composite material housing a complex dual-mode propulsion system. Figure \ref{sc} shows the complex geometry of the \gls{MGR} spacecraft, where, panel (a) shows the spacecraft components, and panel (b) represents the coordinate system. Table \ref{macro_mgr} gives the approximated characteristics of the spacecraft components, extracted from the \cite{Vaughan02}. 

The whole system of the \gls{MGR} spacecraft can be divided into eight subsystems. The briefly introduction of these subsystems is given below and detailed information can be found in \cite{Leary07}: 
\begin{itemize}
\item {\bf Structures and mechanisms:} the primary spacecraft structures are the core, the adapter ring, the sunshade, the solar panels, and the magnetometer boom. The core of the spacecraft tightly integrates support panels with the propulsion system. Three mechanical assemblies were deployed during operation, the two solar panels and the 3.6-m magnetometer boom. The solar array hinges are located at each end of the arms connecting the panels to the core structure. Whereas, the magnetometer boom is separated into two segments with one hinge between the spacecraft structure and the first segment and the other between the two segments.
\item {\bf Propulsion:} the \gls{MGR} propulsion system is a pressurized bipropellant dual-mode system. As shown in Figure \ref{sc}, they are three main propellant tanks, a refillable auxiliary fuel tank, and a helium pressurant tank provide propellant and pressurant storage. MESSENGER carries seventeen thrusters. Three thruster types, arranged in five different thruster module configurations, provide the required spacecraft forces and torques.
\item {\bf Thermal:} the thermal design of the MESSENGER spacecraft relies upon a ceramic-cloth sunshade to protect the vehicle from the intense solar environment. Sunshade can experienced maximum temperature of 350\textdegree C at Mercury but creates a benign thermal environment for the main spacecraft bus, allowing the use of essentially standard electronics, components, and thermal blanketing materials.
\item {\bf Power:} the power system is designed to support about 390 W of load power near Earth and 640 W during Mercury orbit. The power is primarily provided by two solar panels that are mounted on small booms extendable beyond the sunshade and rotating to track the Sun.
\item {\bf Avionics:} MESSENGER is equipped with redundant integrated electronics modules (IEM). The IEM implements command and data handling, guidance and control, and fault protection functions. A primary driver of the IEM architecture was to simplify spacecraft fault protection.
\item {\bf Software:} it provides the main processor-supported code that performs commanding, data handling, and spacecraft control.
\item {\bf Guidance and control:} this subsystem maintains spacecraft attitude and executes propulsive maneuvers for spacecraft trajectory control. It also controls the solar panel orientation to maintain a Sun offset angle providing sufficient power at moderate panel temperatures. It is also responsible for keeping the sunshade pointed towards the Sun to protect the spacecraft bus from extreme heat and radiation.
\item {\bf Radio frequency telecommunications:} this subsystem consists of small deep space transponders, solid-state power amplifiers, phased-array antennas, and medium- and low-gain antennas. The goals of this subsystem is, to provide the highest quality and quantity of scientific data, and to provide highly accurate Doppler and range data for navigation and science.
\end{itemize}
  
\begin{table*}[ht]
\caption{MESSENGER spacecraft \textit{macro-model} characteristics \citep{Vaughan02}.}
\centering
\begin{center}
\renewcommand{\arraystretch}{1.2}
\small
\begin{tabular}{ccccc}\Xhline{2\arrayrulewidth}
\hline \hline
S/C body & Components & App. area (m$^2$) & Diffuse Ref. & Specular Ref.\\
\Xhline{2\arrayrulewidth}
& $\pm$X side of sunshade & 2.057 & 0.35 & 0.15 \\
Spacecraft Bus &Center of sunshade, along -Y & 1.132 & 0.35 & 0.15 \\  
& +Y side of the spacecraft & 4.933 & 0.35 & 0.15 \\
\Xhline{2\arrayrulewidth}
& Front side of $\pm$X solar panel & 2.5 & 0.07 & 0.52 \\
Solar Arrays & Back side of $\pm$X solar panel & 2.5 & 0.07 & 0.52 \\ 
\Xhline{2\arrayrulewidth}
\end{tabular}
\end{center}
\label{macro_mgr}
\end{table*}

\subsection{Radioscience data}
\label{data}
The \gls{MGR} spacecraft was tracked by the NASA\textquoteright s \gls{DSN} stations at X-band frequency, 7.2 GHz for uplink from the ground stations and 8.4 GHz for downlink from the spacecraft. Communications are accomplished via the 34-m and 70-m antennas of \gls{DSN} stations in Goldstone, USA; Madrid, Spain; and Canberra, Australia. The MESSENGER X-band tracking consists in measuring the round-trip time delay (two-way range), and the two- and three-way ramped Doppler shift of the carrier frequency of the radio link between the spacecraft and the \gls{DSN} stations on Earth. The precision of the Doppler measurement for the radio frequency subsystem is within $\pm$0.1 mm/s over 10s to several minutes of integration time \citep{Srinivasan07}.
  
\subsection{Dynamical modeling and orbit determination processes}
\label{analysis}
We have analyzed one-and-half year of tracking data collected by the \gls{DSN} during the \gls{MGR} orbital period. This data sample corresponds to one year of prime mission and six months of first extended mission. The complete data set that were used for the analysis are available on the Geoscience node\footnote{\label{pds}\url{http://pds-geosciences.wustl.edu/messenger/}} of the NASA\textquoteright s \gls{PDS}. For precise orbit determination, all available observations were analyzed with the help of the \gls{GINS} software (see Chapter \ref{CHP1}).

The precise orbit determination is based on a full dynamical approach. The dynamical modeling includes gravitational (gravitational attraction of Mercury, Eq. \ref{potential}, third-body gravity perturbations from the Sun and other planets, Eq. \ref{newton}, and relativistic corrections, Eq. \ref{relativity}) and non-gravitational (solar radiation pressure, Eq. \ref{solar_rad}, Mercury radiation pressure, Eqs. \ref{albedo} and \ref{infra}) forces that are acting on the spacecraft. These forces have been taken into account in the force budget of \gls{MGR}. Because of the thin atmosphere of Mercury, we have assumed that \gls{MGR} experienced negligible resistance due to the atmosphere, hence atmospheric drag force was not included in the force budget. Moreover, \gls{MGR} fires small thrusters (nominally on each Tuesday) to perform \gls{MDM}\footnote{\label{anc}\url{http://pds-geosciences.wustl.edu/messenger/mess-v_h-rss-1-edr-rawdata-v1/messrs_0xxx/ancillary/}}  for reducing the spacecraft angular momentum to a safe level. In addition to \gls{MDM}, \gls{MGR} also performed \gls{OCM}\textsuperscript{\ref{anc}} typically once every Mercury year ($\sim$88 Earth days) to maintain minimum altitude below 500 kilometers. Due to insufficient information of these maneuvers, we therefore did not include the epoch of each maneuver during the orbit computation. Hence, empirical delta accelerations, radial, along-track, and cross-track, at the epoch of maneuvers were not included in the force budget.

The measurement (Doppler and range) models (see Sec. \ref{drobs}) and the light time corrections (see Sec. \ref{lidel}), that are modeled in \gls{GINS}, are described in chapter \ref{CHP1}. During the computations, \gls{DSN} station coordinates were corrected from the Earth\textquoteright s polar motion, solid-Earth tides, and from the ocean loading, based on the formulation given in \cite{Moyer}. In addition to these corrections, radiometric data also have been corrected from tropospheric propagation through the meteorological data\textsuperscript{\ref{anc}} (pressure, temperature and humidity) of the stations.

The complex geometry of the \gls{MGR} spacecraft was treated as a combination of flat plates arranged in the shape of box, with attached solar arrays, so called \textit{Box-Wing} macro-model.  The approximated characteristics of this macro-model, which includes cross-section area and specular and diffuse reflectivity coefficients of the components, extracted from \cite{Vaughan02} and are given in Table \ref{macro_mgr}. In addition to the  macro-model characteristics, orientations of the spacecraft were also taken in account. The attitude of the spacecraft, and of its articulated panels in an inertial frame are usually defined in terms of quaternions. The approximate value of these quaternions were extracted from the SPICE \gls{NAIF} software. The macro-model and its orientation have allowed to calculate the non-gravitational accelerations that are acting on the \gls{MGR} spacecraft due to the radiation pressure from the Sun and Mercury (albedo and thermal infra-red emission).

For orbit computations and for parameters estimations, a multi-arc approach was used to get an independent estimate of the \gls{MGR} accelerations. In this method, we integrated the equations of motion using the time-step of 50s and then, orbital fits were obtained from short data-arcs fitted over the observations span of one-day using an iterative process. The short data-arcs of one-day have been chosen to account for the model imperfections (see section \ref{comps} of Chapter \ref{CHP2}).

\begin{table*} 
\caption{Summary of the Doppler and range tracking data used for orbit determination.}
\centering
\begin{center}
\renewcommand{\arraystretch}{1.4}
\small
\begin{tabular}{ c c c c c c c}
 \hline
 \hline  
    Mission       &    Begin date      &    End date      &  Number of        & Number of        & Number of   \\
         phase    &     dd-mm-yyyy    &     dd-mm-yyyy    & 2-way Doppler   & 3-way Doppler  & range           \\
 \hline    
  Prime & 17-05-2011  & 18-03-2012  &  2108980 & 184138 & 11540 \\
  Extended & 26-03-2012 & 18-09-2012  & 1142974 & 23211 & 5709 \\  
  \hline    
\end{tabular}
\end{center}
\label{data_sum}
\end{table*}
An iterative least-square fit was performed on the complete set of Doppler- and range-tracking data-arcs corresponding to the orbital phase of the mission. The summary of these tracking data are given in Table \ref{data_sum}. Several parameters have been estimated during the orbit computation. They are similar to the one estimated for \gls{MGS} spacecraft (see Sec. \ref{solve_4_para_mgs} of Chapter \ref{CHP2}) except scale factor for atmospheric drag and empirical delta accelerations. 

  
\section{Orbit determination}
\label{orbit}  
\subsection{Acceleration budget}
\label{acc_bug_mgr}
\begin{figure}
\begin{center}\includegraphics[width=15cm]{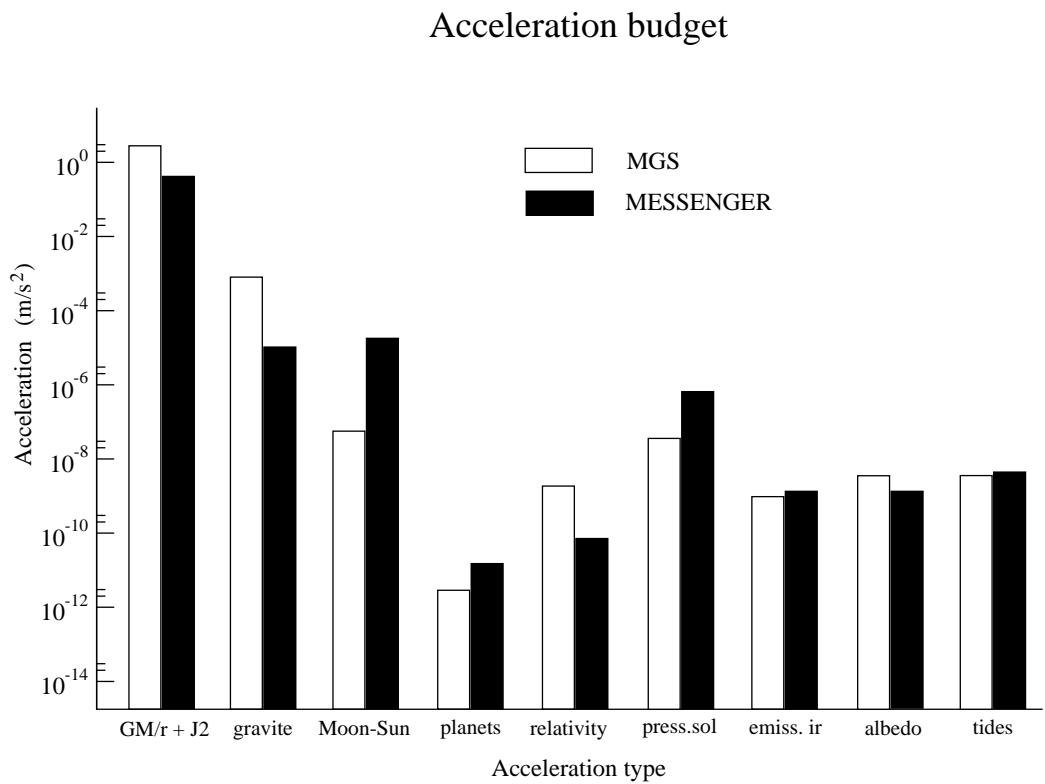}\end{center}
\caption{Gravitational and non-gravitational accelerations acting on the MESSENGER spacecraft. The empirical accelerations correspond to maneuvers and accelerations due to atmospheric drag were not computed for MESSENGER (see Section \ref{analysis}). }
\label{acc_mgr}
\end{figure}

As mentioned in the Chapter \ref{CHP2}, the accurate orbit determination of a planetary spacecraft requires a good knowledge of gravitational and non-gravitational forces which are acting on the spacecraft. The model of these forces is described in the Chapter \ref{CHP1}. The Figure \ref{acc_mgr} illustrates an average of various accelerations that are acting on the \gls{MGR} spacecraft during the prime and extended phases of the orbital periods. The comparison between \gls{MGS} (see Chapter \ref{CHP2}) and \gls{MGR} accelerations that are taken in account in the force budget is also shown in Figure \ref{acc_mgr}.



%
 %

Mercury and Mars have different positions in the solar system and have different physical properties (such as internal mass distribution). These physical properties of the planets, and the shape of the spacecraft orbit, largely affect the force budget of \gls{MGR} and \gls{MGS} spacecraft, respectively. The comparison between \gls{MGR} and \gls{MGS} accelerations based on Figure \ref{acc_mgr} can lead to the following comments:

\begin{itemize}
\item {\bf Accelerations due to gravitational potential:} Mars and Mercury, both planets have very different mass distributions, and \gls{MGS} and \gls{MGR}, both spacecraft have very different shapes of their orbits (\gls{MGS}: low altitude near circular; \gls{MGR}: highly eccentric). As a consequence, the contribution of the mean and zonal coefficients ($GM/r + J_2$) in the accelerations for \gls{MGR} spacecraft is $\sim$8 times smaller than for the \gls{MGS} spacecraft. Similarly, the contribution of the tesseral ($l$ $\ne$ $m$) and sectoral ($l$ = $m$) coefficients in the accelerations is $\sim$90 times smaller in \gls{MGR} spacecraft. 

\item {\bf Accelerations due to third body attractions:} Mercury is the closest planet to the Sun and hence, experiences the largest gravitational attraction from the Sun as compared to other planets of the solar system. As a result, accelerations due to the Sun attraction on \gls{MGR} is $\sim$172 time greater than the one experienced by \gls{MGS} spacecraft. 

\item {\bf Accelerations due to general relativity} from Eq. \ref{relativity} one can see that, the acceleration due to \gls{GR} is a function of the gravitational constant (GM) of the orbiting body and the position and velocity vectors of the spacecraft relative to the center of orbiting body. Due to the smaller GM of the Mercury than the Mars one, and to the highly eccentric \gls{MGR} orbit, the acceleration due to \gls{GR} experienced by \gls{MGR} is $\sim$14 time smaller than \gls{MGS}. However, beside this situation, Mercury is the planet the most affected by \gls{GR} as its advance of perihelia induced by the Sun gravity is about 30 times bigger than the advance of the Mars orbit (43 arcsecond/cy and 1.3 arcsecond/cy for Mars).
 
\item {\bf Accelerations due to the solar radiation pressure:} It is the largest non-gravitational acceleration acting on the both spacecraft. Due to the close proximity of \gls{MGR} to the Sun, it experienced $\sim$13 time more solar radiation pressure than \gls{MGS}. 

\item {\bf Accelerations due to planet radiation:} The Infra-Red radiation and the Albedo of the planets cause small accelerations in the spacecraft motion, respectively. These are the smallest non-gravitational accelerations that are acting on both spacecraft. An average value of these accelerations are relatively similar for both missions. 

\item {\bf Accelerations due to solid planetary tides:} During prime mission of the \gls{MGR} spacecraft, an average value of computed accelerations due to solid planetary tides is similar for both spacecraft, while during extended mission, \gls{MGR} experienced relatively larger acceleration (see Table \ref{ave_acc}). 
\end{itemize}

\begin{figure}[!ht]
\begin{center}\includegraphics[width=8cm]{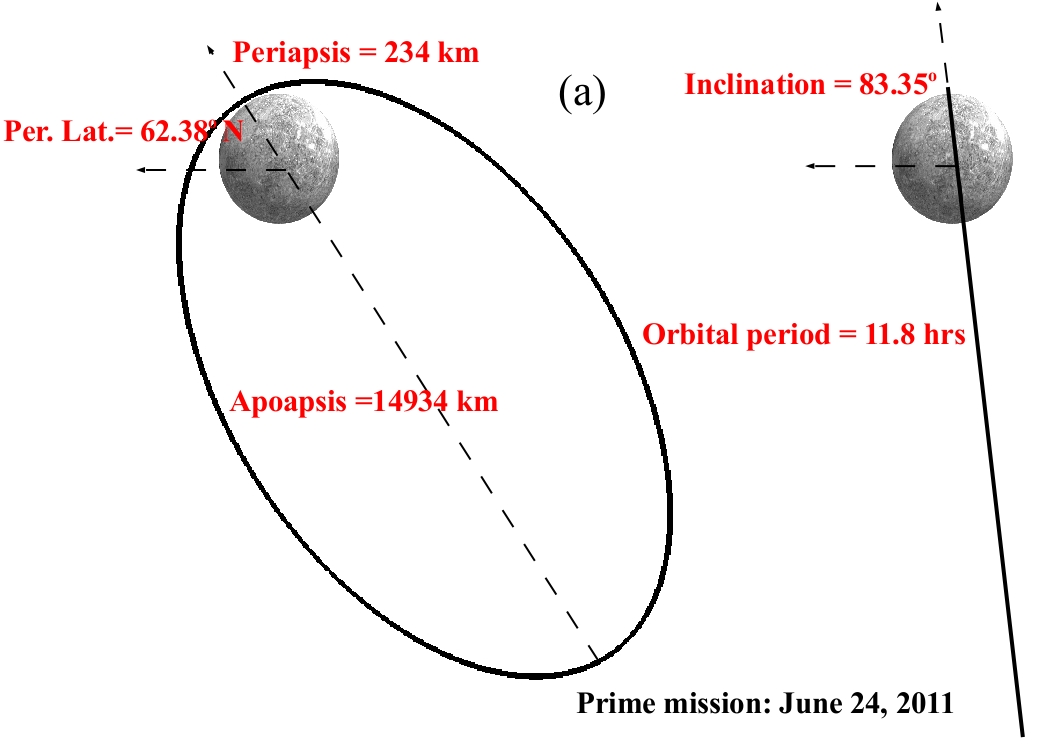}\includegraphics[width=8cm]{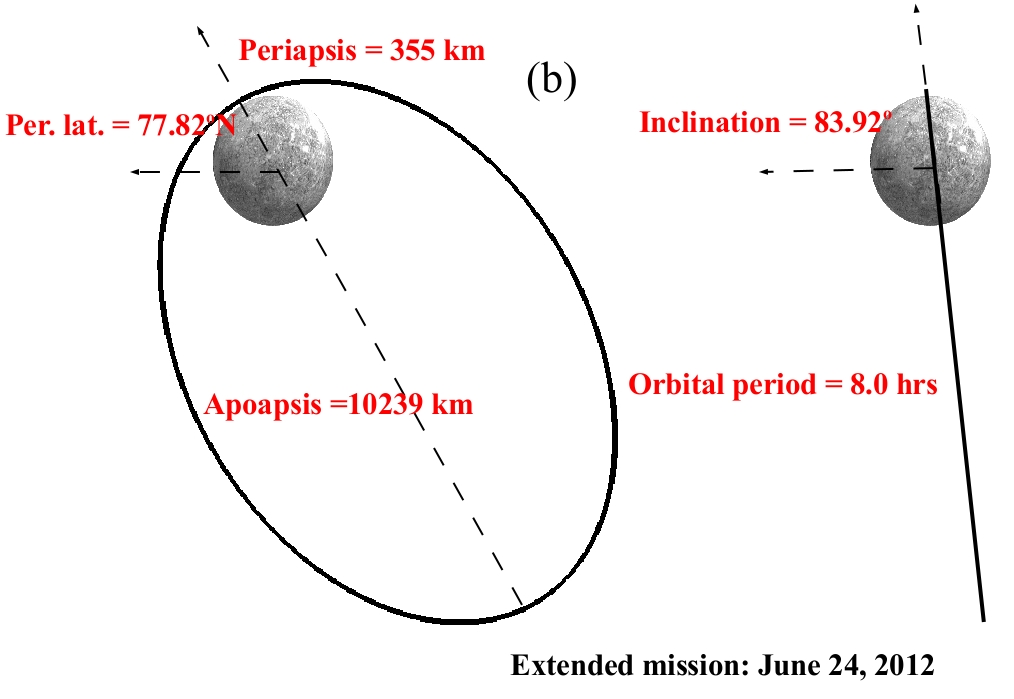}\end{center}
\caption{MESSENGER orbit: (a) prime phase (June 24, 2011) (b) extended phase (June 24, 2012).}
\label{mgr_orb}
\end{figure}

As stated previously, during the extended phase of the mission apoapsis altitude of \gls{MGR} was significantly tuned up to 5000km. Because of its relatively shorter orbit compared to the prime phase (for example, see Figure \ref{mgr_orb}), one can expect a different distribution of the accelerations during both phases. A summary of an average magnitude of these accelerations are given on Table \ref{ave_acc}. As expected during the extended phase, \gls{MGR} experienced $\sim$40$\%$ greater accelerations due to the gravitational potential of Mercury and approximately the same percentage of increment has been estimated for the acceleration due to \gls{GR}. These increments in the accelerations can be explained from the close approach of \gls{MGR} to Mercury, as shown in Figure \ref{mgr_orb}. However, gravitational accelerations due to the third body (including the Moon and the Sun) attraction are relatively smaller during extended phase. In addition to gravitational accelerations, non-gravitational accelerations due to Mercury radiations, Infra-Red radiation and Albedo, were also enhanced by $\sim$42$\%$ and $\sim$11$\%$, respectively, whereas solar radiation pressure remains approximately similar during both phases. 


\begin{table}
\caption{An average magnitude of MESSENGER accelerations estimated during prime and extended phase of the mission.}
\centering
\renewcommand{\arraystretch}{1.4}
\small
\begin{threeparttable}
\begin{tabular}{cccccc}\Xhline{2\arrayrulewidth} 
\multicolumn{1}{c}{\multirow{2}{*}{{\bf Acceleration}}} &\multicolumn{4}{c}{{\bf Mission Phase}}  \\ \cline{3-5}
& & {\bf Prime (m/s$^2$)} &{\bf Extended (m/s$^2$) }&  \\ \Xhline{2\arrayrulewidth}
GM/r + J2 && 3.25 $\times$ \ 10$^{-1}$ & 4.62 $\times$ \ 10$^{-1}$ & \\ 
Gravity  && 0.86 $\times$ \ 10$^{-5}$ & 1.19 $\times$ \ 10$^{-5}$& \\
Moon-Sun && 1.12 $\times$ \ 10$^{-5}$ & 0.87 $\times$ \ 10$^{-5}$ &\\
Planets   && 1.53 $\times$ \ 10$^{-11}$ & 0.87 $\times$ \ 10$^{-11}$ &\\
Relativity   && 6.78 $\times$ \ 10$^{-11}$ & 9.44 $\times$ \ 10$^{-11}$&\\
Solar rad. press.&& 5.47 $\times$ \ 10$^{-7}$ & 5.87 $\times$ \ 10$^{-7}$ &\\
IR emissi. && 1.24 $\times$ \ 10$^{-9}$ & 1.77 $\times$ \ 10$^{-9}$ &\\
Albedo && 1.03 $\times$ \ 10$^{-9}$ & 1.15 $\times$ \ 10$^{-9}$ &\\
Solid tides && 4.32 $\times$ \ 10$^{-9}$ & 6.31 $\times$ \ 10$^{-9}$ &\\
\Xhline{2\arrayrulewidth}
\end{tabular}
\end{threeparttable}
\label{ave_acc}
\end{table}


%
 %

\subsection{Significance of MESSENGER observation for INPOP}
\label{inpop}
As discussed in Chapter \ref{CHP0}, the INPOP planetary ephemerides are built on a regular basis and are provided to users thought the IMCCE website \url{www.imcce.fr/inpop}. The INPOP10e ephemerides were the latest release \citep{Fienga13} and were delivered as the official Gaia mission planetary ephemerides used for the navigation of the satellite as well as for the analysis of the data. Specific developments and analysis were done for the Gaia release such as TCB time-scale version or an accurate estimation of the INPOP link to ICRF. 

With the delivery of the MESSENGER radioscience data, a new opportunity was offered to improve drastically our knowledge of the Mercury orbit and to perform tests of gravity at a close distance from the Sun. In order to perform such tests with a decisive accuracy, one should first reduce the uncertainty of the Mercury orbit. Indeed, as it was stated previously, only five positions of Mercury were deduced from spacecraft flybys over 40 years: 2 positions in the 70's from the Mariner flybys and 3 in 2008 and 2009 from the MESSENGER flybys. These positions gave very accurate positions of Mercury of about several tens of meters compared to direct radar observations of the surface of planet obtained with an accuracy of about 1 kilometer from the 70's to the late 90's. Using the 1.5 year range measurements (see Section \ref{graviF}) is then a crucial chance for obtaining a better tangle over the $\sim$0.3 year Mercury orbit. The few meter accuracy of the MESSENGER range data will give big constraints over short period perturbations on the Mercury orbit when the 5 flyby positions obtained with Mariner and the MESSENGER flybys will still be significant for the measurements of long term perturbations. 

\subsection{Evolution of INPOP with the accuracy of MESSENGER orbit}
\label{postfit}
The \gls{rms} values of the postfit Doppler and range residuals give some indications about the quality of the orbit fit and the quality of the estimated parameters. Moreover, the quality of the used parameters associated with the physical model can also be estimated from these residuals. However, Mercury is the least explored terrestrial planet and the poor knowledge of the parameters associated with the Mercury physical model highly influenced the \gls{MGR} orbit. Therefore, numbers of tests have been performed to compute precise orbit of \gls{MGR} with a gradual improvement of Mercury physical model, such as: spherical harmonic model of the gravity field, rotational model, and the Mercury ephemeris. 
%
 %
 \begin{figure}[!ht]
\begin{center}\includegraphics[width=13.5cm]{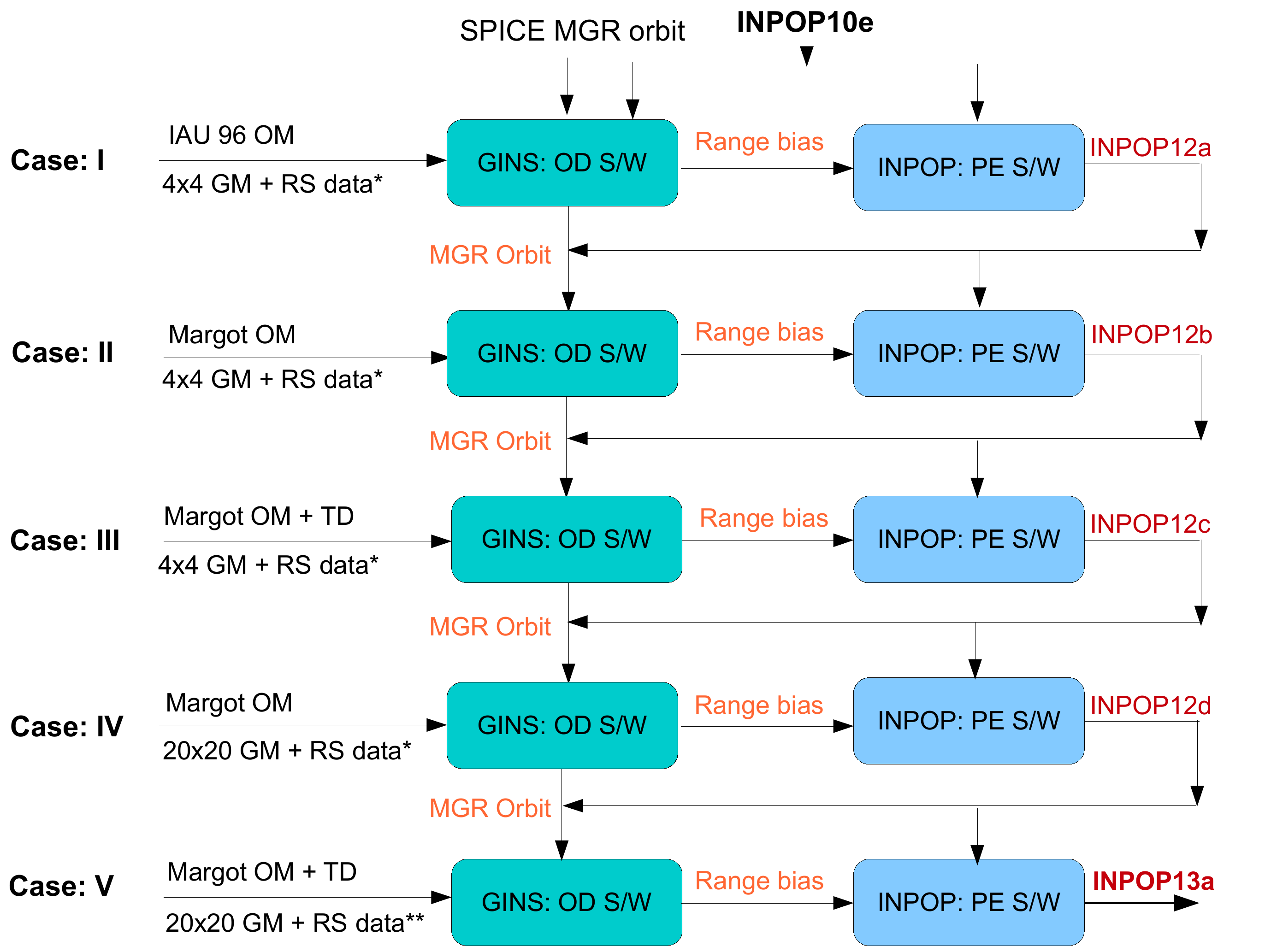}\end{center}
\caption{Schematic diagram of the evolution of planetary ephemeris from INPOP10e to INPOP13a with the improvement of \gls{MGR} orbit.  \\
\footnotesize{OM: Orientation Model; GM: Gravity Model; TD : Transponder Delay; RS : Radioscience \\
 OD S/W : Orbit Determination software;  PE S/W : Planetary Ephemeris software \\
 (*) Prime mission; (**) prime+extended mission.}}
\label{flow_chart_mgr}
\end{figure}

The evolution of planetary ephemeris from INPOP10e to INPOP13a with these gradual improvements is shown in Figure \ref{flow_chart_mgr}. From this figure one can see that, improved inputs have been implemented to GINS software in the subsequent tests. Using these inputs and the one described in Sections \ref{analysis}, GINS constructs the \gls{MGR} orbit precisely. The estimated range bias from GINS are then used to built the new planetary ephemeris. The constants and dynamical modeling used for the construction of the new ephemerides are similar to INPOP10e (see Chapter \ref{CHP0}). A global adjustment of the planet initial conditions including Pluto, the mass of the Sun, the oblateness of the Sun, the ratio between the mass of the Moon and 140 asteroid masses. The improved ephemeris and the \gls{MGR} orbit are then used as a input for subsequent tests to reconstruct the \gls{MGR} orbit and planetary ephemeris. The detailed analysis of these tests are given in the following sections:

\subsubsection{Case I: First guess orbit for Messenger and INPOP12a}
\label{iau96}
\myparagraph{Description}
Before the \gls{MGR} mission, Mariner 10 flybys observations and ground based radar measurements were the only observations which were used to study the gravity field of Mercury and its physical structure (spherical body with slightly flattened at the poles and mildly elongated equator) \citep{Anderson87,Anderson1996}. Later, using the altimetric and radio tracking observations from \gls{MGR} first two flybys of Mercury \cite{Smith10} derived the spherical harmonic model of Mercury gravity field HgM001 developed up to degree and order 4. In this test, we have analyzed radioscience data of the \gls{MGR} spacecraft acquired at the time of prime phase of the mission (see Table \ref{data_sum}) using HgM001 gravity model. The \gls{MGR} orbital fits were obtained from one-day data-arcs using an iterative process (see Section \ref{analysis} for more details). To initialize the iteration, initial position and velocity vectors of \gls{MGR} were taken from the SPICE \gls{NAIF} kernels\footnote{\url{ftp://naif.jpl.nasa.gov/pub/naif/pds/data/mess-e_v_h-spice-6-v1.0/}}. The positions and velocities of the planets were accessed through the latest INPOP10e planetary ephemeris.  Moreover, the orientation of Mercury was defined as recommended by the \gls{IAU}\footnote{It is a default orientation of Mercury defined in the GINS software} (see Table \ref{iau96_Mer}). 
  
\begin{figure}[!ht]
\begin{center}\includegraphics[width=14cm]{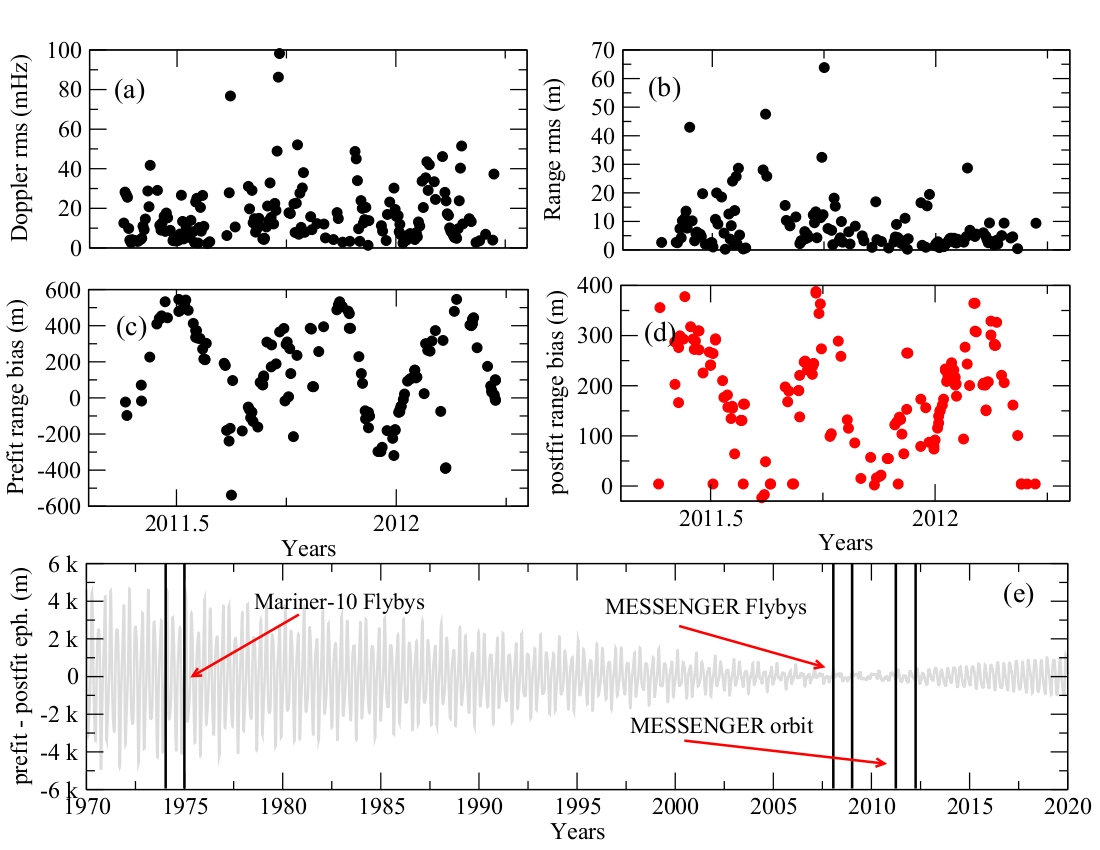} \end{center}
\caption{{Case 1: (a) rms values of the postfit two- and three-way Doppler residuals for each one-day data-arc, (b) rms values of the postfit two-way range residuals for each one-day data-arc, (c) range bias (prefit) correspond to INPOP10e, (d) range bias (postfit) correspond to newly fitted INPOP12a ephemeris, and (e) difference in the Mercury-Earth geometric distances between INPOP10e and INPOP12a ephemerides. The indicated area are intervals of time corresponding to Mariner 10 and MESSENGER observations.}}
\label{iau_rot}
\end{figure}

\begin{table}[!ht] 
\caption{Recommended values for the direction of the north pole of rotation and the prime meridian of the Mercury, 1996 \citep{Davies96}.}
\renewcommand{\arraystretch}{1.8}
\small
\begin{minipage}[t]{0.4\textwidth}
\vspace{0pt}
\begin{tabular}{ c c}
 \hline
 \hline  
   Parameter    &    value       \\
 \hline    
 $\alpha_0$ \ = & 281.01 - 0.033T \\
$\delta_0$ \ = & 61.45 - 0.005T \\
  $W$ \ = &329.68 + 6.1385025d \\
  \hline    
\end{tabular}
\end{minipage}
\begin{minipage}[t]{0.6\textwidth}
\vspace{0pt}\raggedright
$\alpha_0$, $\delta_0$: are right ascension and declination respectively, which define the spin axis with equinox J2000 at epoch J2000. \\
$W$: is the rotational phase \\
 T: is the interval in Julian centuries (of 36525 days) from the standard epoch \\
 d: is the interval in days (of 86400 SI seconds) from the standard epoch, with epochs defined in \gls{TDB}
\end{minipage}
\label{iau96_Mer}
\end{table}
\myparagraph{Results} 
As stated before, \gls{rms} values of the postfit Doppler and range residuals give some indications about the quality of the orbit fit and the quality of the estimated parameters. On panel $a$ of Figure \ref{iau_rot}, are plotted the \gls{rms} values of two- and three-way Doppler residuals, obtained for each data-arc and expressed in millihertz (mHz). The range measurements were also used to assist in fitting the Doppler data for a precise orbit determination. The panel $b$ of Figure \ref{iau_rot} presents the \gls{rms} values of two-way range residuals, obtained for each data-arc. On panel $c$, are plotted the range bias (error in the Earth-Mercury distances) estimated with INPOP10e. These range bias are then used to fit the planetary ephemeris software (see Figure \ref{flow_chart_mgr}). The postfit range bias of newly fitted INPOP12a ephemeris are plotted in panel $d$. The differences in the Earth-Mercury distances over 50 years between these two ephemeris are plotted in panel $e$ of the same figure.  

The statistics of these results are given in Table \ref{case1_tab} and are compared with the required accuracy. From this table and Figure \ref{iau_rot} one can see that, the rms values of the postfit Doppler and range residuals are widely dispersed, and experienced $\sim$3 times more dispersion compared to required accuracy with a mean value of about $\sim$-0.3mHz and $\sim$0.89m, respectively. Such high discrepancies in the residuals are likely due to insufficient spherical harmonic coefficients of the gravity model, that are not able to capture small spatial scales. Moreover, Mercury has shorter orbit than other planets and hence, it experiences several (approximately three) superior conjunctions in one Earth year. During the conjunction period, one can expect severe degradations in the signals (see Chapter \ref{CHP3}). However, because of the high dispersion in the residuals, the corona impact on the residuals is not clearly visible. 

Nevertheless, it is worth to note that computed range bias (one per arc for ranging measurements) for INPOP10e were still useful for improving the planetary ephemeris. As one can see on Figure \ref{iau_rot} and Table \ref{case1_tab}, the ephemeris INPOP12a newly fitted over these observations, improved the Earth-Mercury distance by a factor 2 during the observational period. However, the dispersion in the range bias is still 10 times greater than the required accuracy. The differences between INPOP10e and INPOP12a ephemerides in terms of Earth-Mercury distances are plotted on panel $e$ of the same figure. One can see that, \gls{MGR} observations (which is the only difference between these two ephemerides construction) can cause up to 4km differences in the Earth-Mercury distances over the time period of $\sim$50 years.

\begin{table}[!ht] 
\caption{Statistics of the residuals obtained for Case I, i) postfit Doppler and range residuals, ii) prefit (INPOP10e) and postfit (INPOP12a) range bias. }
\renewcommand{\arraystretch}{1.4}
\small
\centering
\begin{center}
\begin{threeparttable}
\begin{tabular}{ c c c c c c}
 \hline
 \hline  
   Residuals    &  & Value\tnote{**}                      & & Required accuracy   \\  
      type          &       &      & & \citep{Srinivasan07} \\
 \hline    
 2-, and 3-way Doppler  &       & -0.3$\pm$16.3 mHz & & $\textless$ 0.1 \ \ mm/s ($\sim$5.7 \ \ mHz\tnote{*} \ )  \\
 2-way Range      &    & -0.002$\pm$8 m  & &  $\textless$ 3 m\\
 1-way range bias, INPOP10e & &  161$\pm$249 m  & & $\textless$ 10 m\\
 1-way range bias, INPOP12a  && 181$\pm$102 m& & $\textless$ 10 m \\
 \hline    
\end{tabular}
\begin{tablenotes}
\item[*] 2/3-way: 1 mHz = 0.0178 mm/s =  0.5$\times$speed of light / X-band frequency.
\item[**] mean$\pm$1-$\sigma$ dispersion of the rms values
\end{tablenotes}
\end{threeparttable}
\end{center}
\label{case1_tab}
\end{table}
  
\subsubsection{Case II: New Mercury orientation model and INPOP12b}
\label{margoRot}
\myparagraph{Description}
 \begin{figure}[!ht]
\begin{center}\includegraphics[width=16cm]{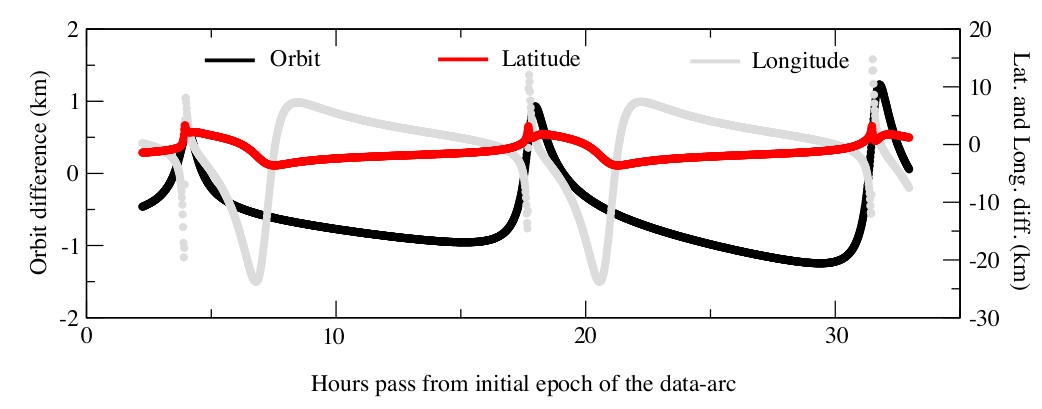}\end{center}
\caption{An example for the change of \gls{MGR} orbit characteristics due to \cite{Margot09} Mercury orientation model. The differences are plotted with respect to IAU 1996 Mercury orientation model.}
\label{diff_lat_lon}
\end{figure}
\begin{table}[!ht] 
\caption{Recommended model for the orientation of Mercury \citep{Margot09}.}
\renewcommand{\arraystretch}{1.2}
\small
\begin{minipage}[t]{0.6\textwidth}
\hspace{2cm}
\begin{tabular}{ c c}
 \hline
 \hline  
   Parameter    &    value       \\
 \hline    
 $\alpha_0$ \ = & 281.0097 - 0.0328T \\
$\delta_0$ \ = & 61.4143 - 0.0049T \\
  $W$ \ = &329.75 + 6.1385025d \\
  & + 0.00993822 sin(M1) \\
  & -0.00104581 sin(M2) \\
  & -0.00010280 sin(M3) \\
  & -0.00002364 sin(M4) \\ 
  &-0.00000532 sin(M5)\\
  where & \\
  & M1 = 174.791086 + 4.092335d \\ 
  & M2 = 349.582171 + 8.184670d \\ 
  & M3 = 164.373257 + 12.277005d \\ 
  & M4 = 339.164343 + 16.369340d \\ 
  & M5 = 153.955429 + 20.461675d\\
  \hline    
\end{tabular}
\end{minipage}
\begin{minipage}[t]{0.4\textwidth}
Angles are expressed in degrees, and T and d are defined as in Table \ref{iau96_Mer}.
\end{minipage}
\label{margot_model}
\end{table}
In this case, to reconstruct the \gls{MGR} orbit and to initialize the iteration, initial position and velocity vectors of \gls{MGR} were deduced from the previous solution and the positions and velocities of the planets were accessed through the newly fitted ephemeris INPOP12a (see Figure \ref{flow_chart_mgr}). In addition to these changes, a new model of Mercury orientation \citep{Margot09} has been implemented in the \gls{GINS}. The previous orientation model for Mercury was inadequate because it uses an obsolete spin orientation, neglects oscillations in the spin rate called longitude librations, and relies on a prime meridian that no longer reflects its intended dynamical significance \citep{Margot09}. These effects induce positional errors on the surface of $\sim$2.5 km in latitude and up to several km in longitude (e.g, see Figure \ref{diff_lat_lon}). The \cite{Margot09} updated orientation model incorporates modern values of the spin orientation, includes non-zero obliquity and librations, and restores the dynamical significance to the prime meridian. The characteristics of this model is given in Table \ref{margot_model}. The gravity model used for this computations is the one used for previous solution (i.e, HgM001). 

\begin{figure}[!ht]
\begin{center}\includegraphics[width=14cm]{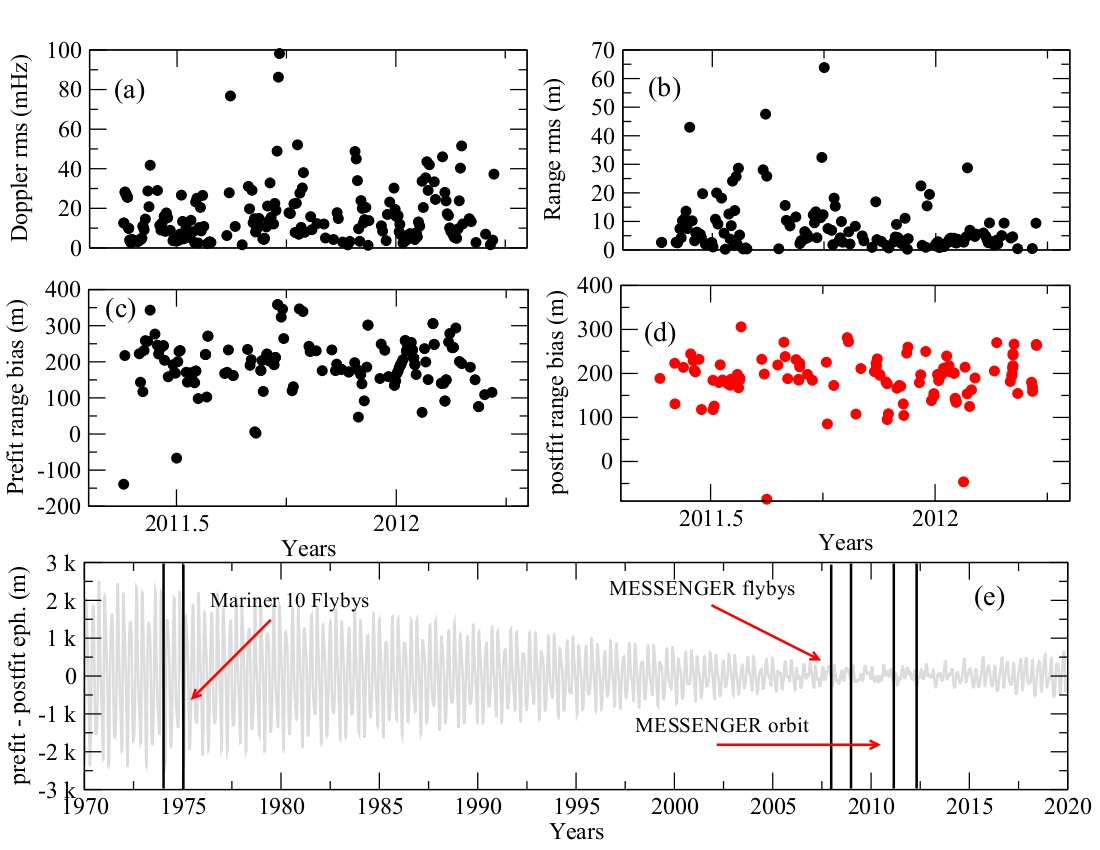}\end{center}
\caption{{Case II: (a) rms values of the postfit two- and three-way Doppler residuals for each one-day data-arc, (b) rms values of the postfit two-way range residuals for each one-day data-arc, (c) range bias (prefit) correspond to INPOP12a, (d) range bias (postfit) correspond to newly fitted INPOP12b ephemeris, and (e) difference in the Mercury-Earth geometric distances between INPOP12a and INPOP12b ephemerides. The indicated area are intervals of time corresponding to Mariner 10 and MESSENGER observations.}}
\label{mar_rot}
\end{figure}

\myparagraph{Results} 
On Table \ref{case2_tab}, statistics and comparison of the obtained results are given. From this table and the Figure \ref{mar_rot} (similar to  \ref{iau_rot}) one can see that, because of the same gravity model, discrepancies in the residuals are similar to the Case I. These discrepancies are about $\sim$3, $\sim$2.5, and $\sim$6 times larger compared to the required values of Doppler, range, and range bias residuals, respectively. However, \cite{Margot09} orientation model of Mercury removed the systematic trend in the range bias (see Figure \ref{iau_rot}) and reduced the dispersion up to 29 m in the INPOP12a range bias. 

Moreover, on panel $d$ of Figure \ref{mar_rot}, one can noticed that, the newly fitted ephemeris INPOP12b shows a clear offset of about 190 m with a dispersion of about 56 m. This offset can be explained from the transponder delay, which was not taken in account during the orbit construction of \gls{MGR}. This offset with 1$\sigma$ of dispersion is however compatible with the transponder delay that was measured on the ground (see Section \ref{paper_mgr} for more detailed analysis), calibrated from 1,356.89 ns ($\sim$407 m) to 1,383.74 ns ( $\sim$415 m) depending on the radio frequency configuration (transponder, solid-state power amplifiers, and antenna configuration) \citep{Srinivasan07}. Thus, such compatibility of the ephemeris offset with the measured transponder delay suggests that, there is not a large error included in the spacecraft and in the planetary orbit fit procedure. Moreover, on panel $e$ of Figure \ref{mar_rot}, are plotted the Earth-Mercury distance differences between INPOP12a and INPOP12b. The change in the Mercury orientation model brought up to 2km of differences in the Earth-Mercury distances over the time period of $\sim$50 years.
 
\begin{table}[!ht] 
\caption{Statistics of the residuals obtained for Case II, i) postfit Doppler and range residuals, ii) prefit (INPOP12a) and postfit (INPOP12b) range bias. }
\renewcommand{\arraystretch}{1.4}
\small
\centering
\begin{center}
\begin{threeparttable}
\begin{tabular}{ c c c c c }
 \hline
 \hline  
   Residuals    &  & Value\tnote{**}                      & & Required accuracy   \\  
      type          &       &   (mean$\pm$rms)   & & \citep{Srinivasan07} \\
 \hline    
 2-, and 3-way Doppler  &       & -0.01$\pm$15.6 mHz & & $\textless$ 0.1 \ \ mm/s ($\sim$5.7 \ \ mHz\tnote{*} \ )  \\
 2-way Range      &    & -0.04$\pm$7.6 m  & &  $\textless$ 3 m\\
 1-way range bias, INPOP12a & &  195$\pm$73 m  & & $\textless$ 10 m\\
 1-way range bias, INPOP12b  && 189$\pm$56 m& & $\textless$ 10 m \\
 \hline    
\end{tabular}
\begin{tablenotes}
\item[*] 2/3-way: 1 mHz = 0.0178 mm/s =  0.5$\times$speed of light / X-band frequency.
\item[**] mean$\pm$1-$\sigma$ dispersion of the rms values
\end{tablenotes}
\end{threeparttable}
\end{center}
\label{case2_tab}
\end{table}
  
\subsubsection{Case III: Group delay and INPOP12c}
\label{transINP12b}
\myparagraph{Description}
Similarly to the previous case, to reconstruct the \gls{MGR} orbit and to initialize the iteration, initial position and velocity vectors of \gls{MGR} were deduced from the old solution (Case II) and the positions and velocities of the planets were accessed through the newly fitted ephemeris INPOP12b (see Figure \ref{flow_chart_mgr}). The gravity model and the Mercury orientation model used for this computation are the one used for the previous solution. However, a mean value of 1,371 ns \citep{Srinivasan07} has been implemented in \gls{GINS} as a transponder delay for range measurements.   

\myparagraph{Results} 
\begin{table}[!ht] 
\caption{Statistics of the residuals obtained for Case III, i) postfit Doppler and range residuals, ii) prefit (INPOP12b) and postfit (INPOP12c) range bias. }
\renewcommand{\arraystretch}{1.4}
\small
\centering
\begin{center}
\begin{threeparttable}
\begin{tabular}{ c c c c c }
 \hline
 \hline  
   Residuals    &  & Value\tnote{**}                      & & Required accuracy   \\  
      type          &       &   (mean$\pm$rms)   & & \citep{Srinivasan07} \\
 \hline    
 2-, and 3-way Doppler  &       & -0.01$\pm$16 mHz & & $\textless$ 0.1 \ \ mm/s ($\sim$5.7 \ \ mHz\tnote{*} \ )  \\
 2-way Range      &    & 0.02$\pm$6.9 m  & &  $\textless$ 3 m\\
 1-way range bias, INPOP12b & &  8$\pm$50 m  & & $\textless$ 10 m\\
 1-way range bias, INPOP12c  && 9$\pm$48 m& & $\textless$ 10 m \\
 \hline    
\end{tabular}
\begin{tablenotes}
\item[*] 2/3-way: 1 mHz = 0.0178 mm/s =  0.5$\times$speed of light / X-band frequency.
\item[**] mean$\pm$1-$\sigma$ dispersion of the rms values
\end{tablenotes}
\end{threeparttable}
\end{center}
\label{case3_tab}
\end{table}
 \begin{figure}[!ht]
\begin{center}\includegraphics[width=14cm]{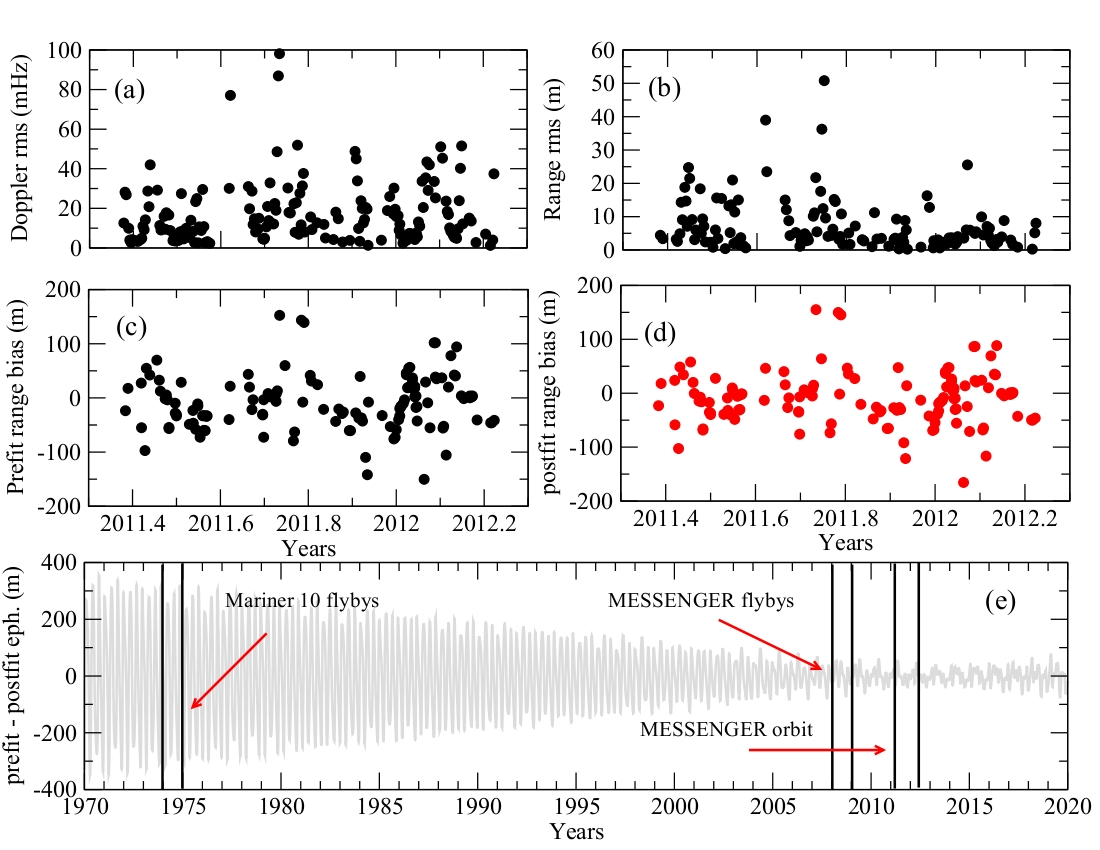}\end{center}
\caption{{Case III: (a) rms values of the postfit two- and three-way Doppler residuals for each one-day data-arc, (b) rms values of the postfit two-way range residuals for each one-day data-arc, (c) range bias (prefit) correspond to INPOP12b, (d) range bias (postfit) correspond to newly fitted INPOP12c ephemeris, and (e) difference in the Mercury-Earth geometric distances between INPOP12b and INPOP12c ephemerides. The indicated area are intervals of time corresponding to Mariner 10 and MESSENGER observations.}}
\label{trans}
\end{figure}

\gls{MGR} orbit is mainly constrained by the Doppler observations which describing the range rate between the spacecraft and the ground station. Therefore, as expected, a constant transponder delay did not affect the \gls{MGR} orbit, thus there is no significant change in the Doppler postfit residuals as compared to Case II (see Table \ref{case3_tab}). In contrast, the time delay due to the ranging transponder directly affects the light time and adds additional bias in the range observables. 

The range bias estimation relies on all the range observations of the data-arc. This range bias is related to the onboard range devices themselves (transponder in this case) and to the expected error of ephemerides. Therefore, one can noticed that on Figure \ref{trans} and on Table \ref{case3_tab}, new estimations of range bias are dramatically different from the Case II. The offset in the range bias of about 190 m found for INPOP12b (see Section \ref{margoRot}) is thus almost removed in the new estimation of range bias. Forthwith, the new ephemeris INPOP12c fitted over these range bias has an offset of about 9 m with the dispersion of about 50 m. Moreover, on panel $e$ of Figure \ref{trans}, are plotted the Earth-Mercury distance differences between INPOP12b and INPOP12c. The implementation of the transponder delay brought up to 300 m of differences in the Earth-Mercury distances over the time period of $\sim$50 years.

\subsubsection{Case IV: New gravity field HgM002 and INPOP12d}
\label{graviH}
\myparagraph{Description}
The radioscience data are significantly important to improve our knowledge of Mercury gravity field and its geophysical properties. Using the first six months of radioscience data of the orbital period, \citep{Smith12} computed the gravity field and the internal structure (density distribution) of Mercury. This updated gravity field solution is crucial for the precise computation of MESSENGER orbit and also to perform precise relativistic tests. In this case, the spherical harmonic model \citep{Smith12} of Mercury gravity field HgM002\footnote{\url{http://pds-geosciences.wustl.edu/missions/messenger/rs.htm}} developed up to degree and order 20 has been implemented in \gls{GINS}. The initial conditions for \gls{MGR}, and the positions and velocities of the planets were taken from the case III (see Figure \ref{flow_chart_mgr}). 
\begin{table}[!ht] 
\caption{Statistics of the residuals obtained for Case IV, i) postfit Doppler and range residuals, ii) prefit (INPOP12c) and postfit (INPOP12d) range bias. }
\renewcommand{\arraystretch}{1.3}
\small
\centering
\begin{center}
\begin{threeparttable}
\begin{tabular}{ c c c c c }
 \hline
 \hline  
   Residuals    &  & Value\tnote{**}                      & & Required accuracy   \\  
      type          &       &   (mean$\pm$rms)   & & \citep{Srinivasan07} \\
 \hline    
2-, and 3-way Doppler  &       & -0.002$\pm$5.0 mHz & & $\textless$ 0.1 \ \ mm/s ($\sim$5.7 \ \ mHz\tnote{*} \ )  \\
 2-way Range      &    & -0.002$\pm$1.7 m  & &  $\textless$ 3 m\\
 1-way range bias, INPOP12c & &  1.8$\pm$17 m  & & $\textless$ 10 m\\
 1-way range bias, INPOP12d  && 0.7$\pm$7.5 m& & $\textless$ 10 m \\
 \hline    
\end{tabular}
\begin{tablenotes}
\item[*] 2/3-way: 1 mHz = 0.0178 mm/s =  0.5$\times$speed of light / X-band frequency.
\item[**] mean$\pm$1-$\sigma$ dispersion of the rms values, excluding solar corona 
\end{tablenotes}
\end{threeparttable}
\end{center}
\label{case4_tab}
\end{table}
 \begin{figure}[!ht]
\begin{center}\includegraphics[width=14cm]{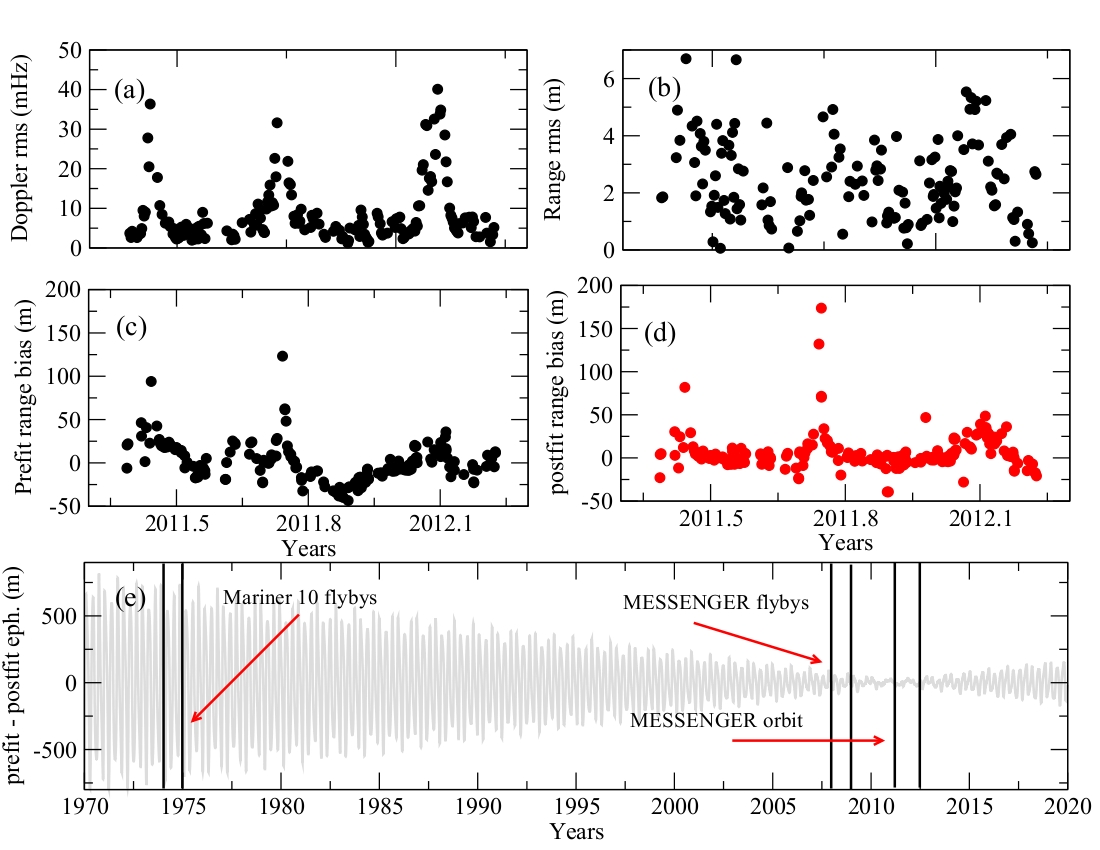}\end{center}
\caption{{Case IV: (a) rms values of the postfit two- and three-way Doppler residuals for each one-day data-arc, (b) rms values of the postfit two-way range residuals for each one-day data-arc, (c) range bias (prefit) correspond to INPOP12c, (d) range bias (postfit) correspond to newly fitted INPOP12d ephemeris, and (e) difference in the Mercury-Earth geometric distances between INPOP12c and INPOP12d ephemerides. The indicated area are intervals of time corresponding to Mariner 10 and MESSENGER observations.}}
\label{gravi_1yr}
\end{figure}
\myparagraph{Results}
The spherical harmonic model HgM002 of degree and order 20, gave a strong constraint to the \gls{MGR} orbit. Thanks to HgM002 model, we are then able to construct a very precise orbit of \gls{MGR} and then of Mercury. As one can see on Figure \ref{gravi_1yr} and Table \ref{case4_tab}, the rms values of postfit Doppler and range residuals are $\sim$3 times smaller with almost zero mean values compared to the one found with HgM001 gravity model. Because of such accuracy in the orbit computation, one can easily disentangle the impact of the solar corona on the observations (peaks on Figure \ref{gravi_1yr}). As one can noticed from Table \ref{case4_tab}, the estimated rms values of the Doppler and range residuals, excluding superior solar conjunction periods, are compatible with the required accuracy of about 0.1 mm/s and 3 m, respectively \citep{Srinivasan07}. Moreover, we have also adjusted Doppler offsets per arc and per DSN station (participating in the data-arc) accounting for the systematic errors generated by the devices at each tracking station. As expected, the fitted values for each \gls{DSN} station tracking pass is of the order of a few tenths of mHz, which is lower than the Doppler postfit residuals for each data-arc. No large offset was then detected in the modeling of the Doppler shift measurements at each tracking station.

As a consequence of the precise \gls{MGR} orbit, the dispersion in the range bias estimated for INPOP12c is almost 30 m less than the one found with previous estimations (see Figure \ref{gravi_1yr} and Table \ref{case4_tab}). The new ephemeris INPOP12d fitted over these range bias has an offset of about 7.5 m with a mean value less than meter. Such accuracy in the range bias also shows the compatibility with the required accuracy of about 10 m. Further improvements in the range bias however may be limited by the uncalibrated transponder time delays due to either temperature variations or electronic perturbations from other devices. On panel $e$ of Figure \ref{gravi_1yr}, are plotted the Earth-Mercury distance differences between INPOP12c and INPOP12d. The improvement of the \gls{MGR} orbit can lead up to 500 m of changes in the Earth-Mercury distances. The accuracy of Mercury orbit depends then upon the quality of the range bias. 

\subsubsection{Case V: Extension of the mission and INPOP13a}
\label{graviF}
\myparagraph{Description}
As stated before, on March 2012, the \gls{MGR} mission was extended to one or more years. Therefore, in the construction of the latest INPOP13a ephemeris, we have analyzed all available \gls{MGR} radioscience data, including one year of prime phase and six months of extended phase. The analysis was performed in the same manner as for previous cases.   
 \begin{figure}[!ht]
\begin{center}\includegraphics[width=14cm]{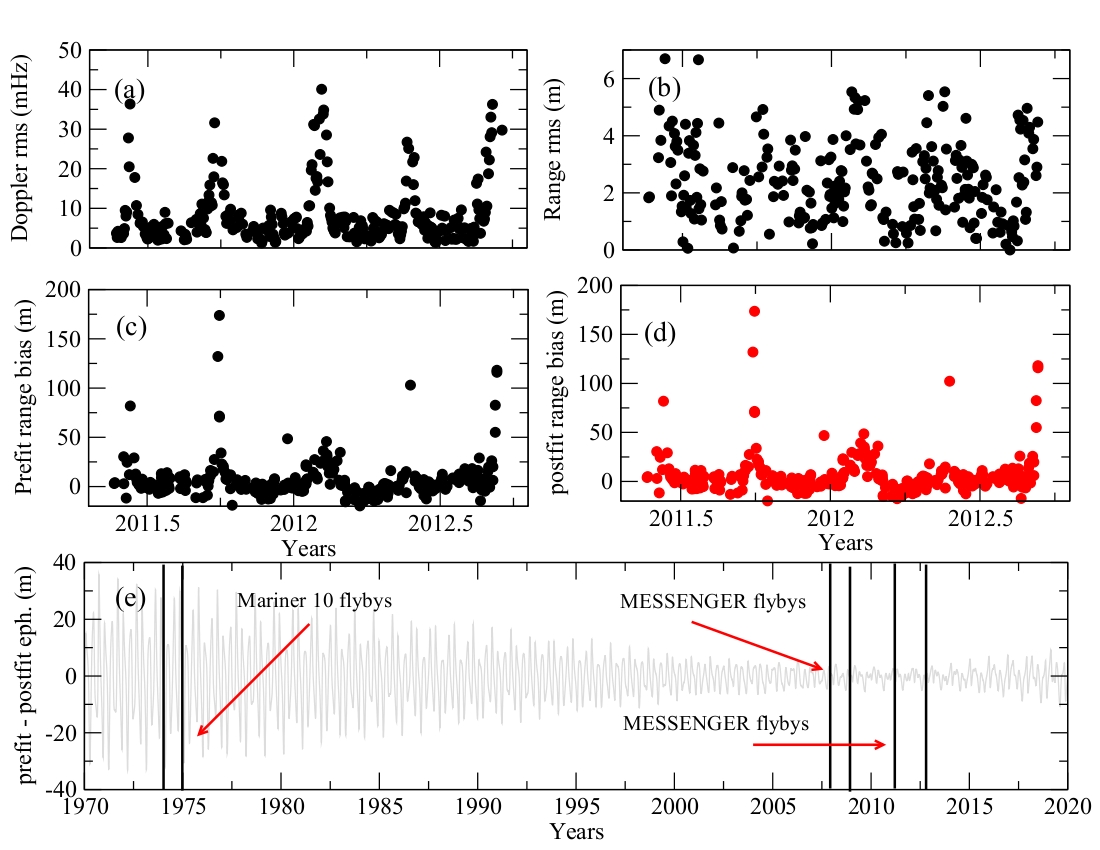}\end{center}
\caption{{Case V: (a) rms values of the postfit two- and three-way Doppler residuals for each one-day data-arc, (b) rms values of the postfit two-way range residuals for each one-day data-arc, (c) range bias (prefit) correspond to INPOP12d, (d) range bias (postfit) correspond to newly fitted INPOP13a ephemeris, and (e) difference in the Mercury-Earth geometric distances between INPOP12d and INPOP13a ephemerides. The indicated area are intervals of time corresponding to Mariner 10 and MESSENGER observations.}}
\label{gravi_full}
\end{figure}

\begin{table}[!ht] 
\caption{Statistics of the residuals obtained for Case V , i) postfit Doppler and range residuals, ii) prefit (INPOP12d) and postfit (INPOP13a) range bias. }
\renewcommand{\arraystretch}{1.3}
\small
\centering
\begin{center}
\begin{threeparttable}
\begin{tabular}{ c c c c c }
 \hline
 \hline  
    Residuals    &  & Value\tnote{**}                      & & Required accuracy   \\  
      type          &       &   (mean$\pm$rms)   & & \citep{Srinivasan07} \\
 \hline    
 2-, and 3-way Doppler  &       & -0.00063$\pm$4.8 mHz & & $\textless$ 0.1 \ \ mm/s ($\sim$5.7 \ \ mHz\tnote{*} \ )  \\
 2-way Range      &    & -0.003$\pm$1.5 m  & &  $\textless$ 3 m\\
 1-way range bias, INPOP12d  && 0.6$\pm$8.0 m& & $\textless$ 10 m \\
  1-way range bias, INPOP13a & &  -0.4$\pm$8.4 m  & & $\textless$ 10 m\\
 \hline    
\end{tabular}
\begin{tablenotes}
\item[*] 2/3-way: 1 mHz = 0.0178 mm/s =  0.5$\times$speed of light / X-band frequency.
\item[**] mean$\pm$1-$\sigma$ dispersion of the rms values, excluding solar corona 
\end{tablenotes}
\end{threeparttable}
\end{center}
\label{case5_tab}
\end{table}
\myparagraph{Results}
During the extended phase of the mission, \gls{MGR} was placed in an $\sim$8-hour orbit to conduct further scans of Mercury by significantly tuned the apoapsis altitude of \gls{MGR} (e.g, see Figure \ref{mgr_orb}). As described in Section \ref{acc_bug_mgr}, this change in the orbit significantly altered the acceleration budget of the \gls{MGR} spacecraft (see Table \ref{ave_acc}). On figure \ref{gravi_full}, are plotted the rms values of the postfit Doppler and range residuals for both phases. Statistics of these residuals are similar to the Case IV and given on Table \ref{case5_tab}. These residuals and estimated range bias are comparable with the required accuracy for \gls{MGR}, hence confirmed that there is no large error included in the spacecraft dynamical modeling, and in the planetary orbit fit procedure. 

\begin{figure}[!ht]
\begin{center}\includegraphics[width=14cm]{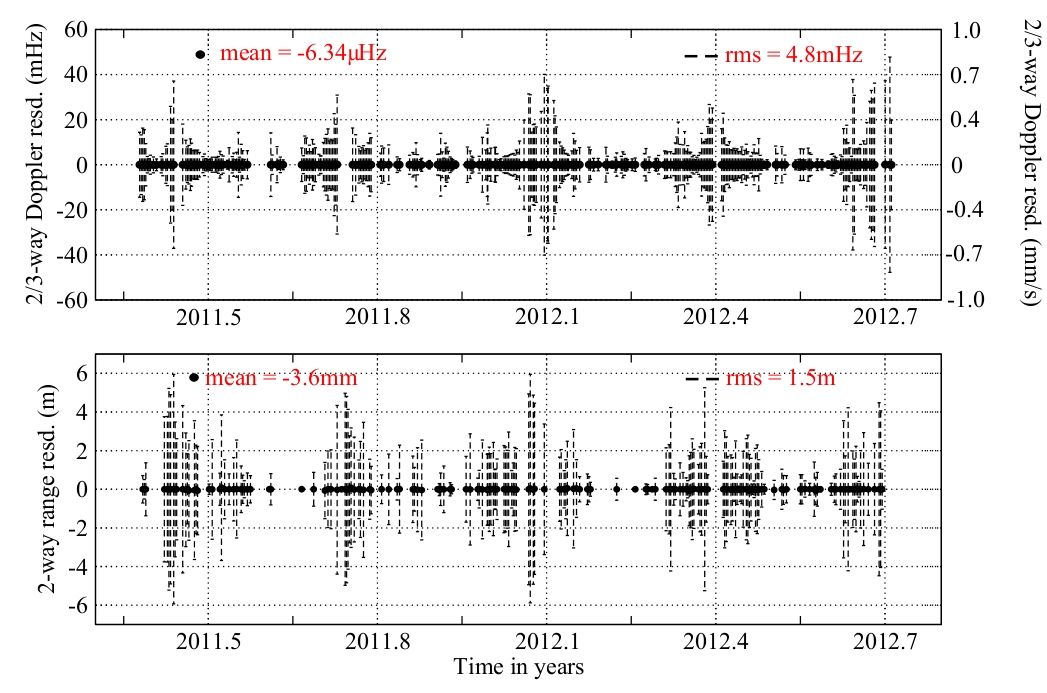}\end{center}
\caption{Mean and rms values of the postfit Doppler and range residuals, estimated for each data-arc.}
\label{mean_sd}
\end{figure}

\myparagraph{Comparisons}
To check the quality of \gls{MGR} orbit in terms of postfit Doppler and range residuals, we compared our estimation with the ones found in the literature \citep{Genova13,Smith12}. On Figure \ref{mean_sd}, are plotted the estimated mean and 1-sigma values of postfit residuals for each measurement types. These values are obtained separately for each data arc. Typical mean and rms values of the postfit Doppler, and two-way range residuals are estimated of about -0.00063$\pm$4.8 mHz\footnote{2/3-way: 1 mHz = 0.0178 mm/s =  0.5$\times$speed of light / X-band frequency}, and -0.003$\pm$1.5 m, respectively.

\begin{table}[!ht] 
\caption{Comparisons of postfit residuals between different authors.}
\renewcommand{\arraystretch}{1.3}
\small
\centering
\begin{center}
\begin{threeparttable}
\begin{tabular}{ c c c c c }
 \hline
 \hline  
    Author    &  &    Doppler                   & & Range   \\  
 \hline    
 This chapter  &       & -0.00063$\pm$4.8 mHz & & -0.003$\pm$1.5 m  \\
 \cite{Genova13}      &    & -0.00088$\pm$3.6 mHz  & &  -0.06$\pm$1.87 m\\
 \cite{Smith12}  && 0.4$\pm$2.0\tnote{*} \ \ mm/s & & - \\
 \hline    
\end{tabular}
\begin{tablenotes}
\item[*] 2/3-way: 1 mHz = 0.0178 mm/s =  0.5$\times$speed of light / X-band frequency.
\end{tablenotes}
\end{threeparttable}
\end{center}
\label{comp_tab}
\end{table}

On Table \ref{comp_tab}, are presented the comparisons. Our estimations are comparable with \cite{Smith12} and \cite{Genova13}. These authors, however only used the first six months of orbital data for their computations. Our estimations are very close to the one found by \cite{Genova13}. However, comparatively high uncertainties in our Doppler residuals (of about 1.2 mHz) can be explained from the adjusted parameters. In addition to our adjustment, \cite{Genova13} also adjusted the Mercury gravity field up to degree 20, scale factors for albedo and infrared radiation pressure, and empirical delta accelerations for accounting the \gls{OCM}. Such improved dynamical model could explain the better uncertainties in the \cite{Genova13} residuals.

\myparagraph{INPOP13a ephemeris}
 \begin{figure}[!ht]
\begin{center}\includegraphics[width=15cm]{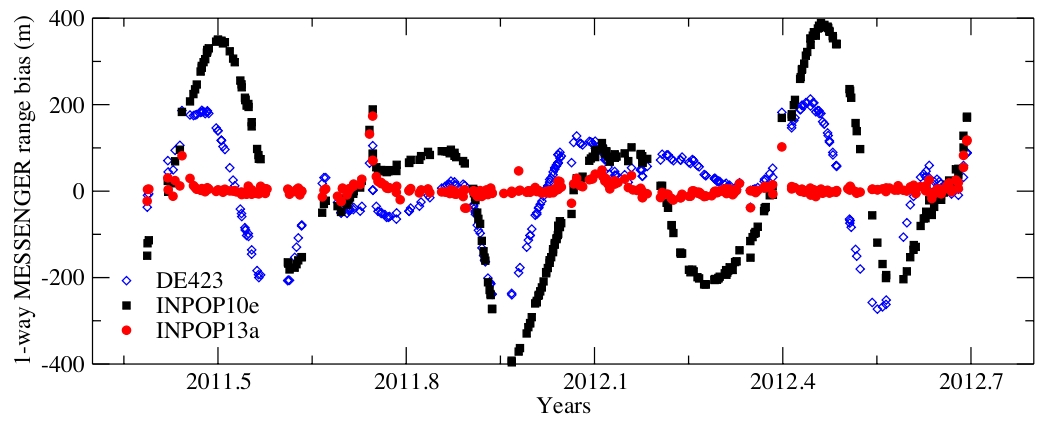}\end{center}
\caption{MESSENGER one-way range residuals obtained with INPOP13a, INPOP10a and DE423.}
\label{fortmessD}
\end{figure}
One of the main objective of this work is to improve the planetary ephemerides. Range bias computed during the Case V are then used to further refinement of planetary ephemeris, especially of the Mercury orbit. The newly fitted planetary ephemeris INPOP13a showed postfit residuals in the Earth-Mercury distances of -0.4$\pm$8.4 m excluding solar corona period, which is two order of improvement than any recent planetary ephemerides (e.g INPOP10e (21$\pm$187 m) and DE423 (15$\pm$105 m), see Figure \ref{fortmessD} and Section \ref{paper_mgr}). Moreover, as discussed in Section \ref{paper_mgr}, over longer intervals of time, INPOP13a is consistent with the DE423 ephemerides and improved the internal accuracy of INPOP by a factor two on the geocentric distances. Furthermore, to understand the impact of six months complementary data of the extended phase over the Mercury orbit, on panel $e$ of Figure \ref{gravi_full}, are plotted the Earth-Mercury distance differences between INPOP12d and INPOP13a ephemerides. The additional data of the extended phase indeed gave small constraints to $\sim$0.3 Earth year of Mercury orbit, and brought up to 20 m of differences in the Earth-Mercury distances over the time period of $\sim$50 years. 

As a scientific exploitation of these results, such high precision planetary ephemeris INPOP13a allowed us to perform several tests. The detailed analysis of these tests are described in the Section \ref{paper_mgr}. An example of such tests are: (a) Impact of planetary ephemeris over the MESSENGER orbit, and (b) \gls{GR} tests of \gls{PPN}-formalism and its impact over the \gls{MGR} orbit. \\

\begin{enumerate}[{\bf (a)}]

\item The geometric distances between the Earth and Mercury are $\sim$16 times ameliorated in INPOP13a compared to INPOP10e. To analyze the impact of the improvement of the planetary ephemeris over the spacecraft orbit, we reanalyzed the entire one and half year of radioscience data using INPOP10e ephemeris. The dynamical modeling and the orbit determination process accounted for performing this analysis are the same as used for Case V. The improvement in the Mercury ephemeris however brought negligible variations in the \gls{MGR} orbit. The differences between the two solutions, one obtained with INPOP10e and other with INPOP13a, in the rms postfit Doppler and range residuals were estimated as 0.008$\pm$0.04 mHz, and 0.05$\pm$0.3 m, respectively. These values are however far below compared to the estimated accuracy of 4.8 mHz, and 1.5 m respectively. 


\item \gls{GR} tests of \gls{PPN}-formalism were also performed with INPOP13a. Because of the high precision Mercury ephemeris, our estimations of the PPN parameters are most stringent than previous results. We considered the 5, 10 and 25$\%$ of changes in the postfit residuals compared to INPOP13a. These changes in the postfit residuals are then used to estimate the possible violation of two relativistic parameters ($\gamma$ and $\beta$) of the PPN formalism of \gls{GR}. This analysis shows ten times smaller uncertainty in the estimation of $\beta$ and $\gamma$ than our previous results with INPOP10a or INPOP08. Moreover, one of the best estimation of the parameter $\gamma$ by \cite{Bertotti03} is compatible with our 25$\%$ of estimation.
\end{enumerate}

\section{\cite{verma14}} 
\label{paper_mgr}
\includepdf[pages=1-13]{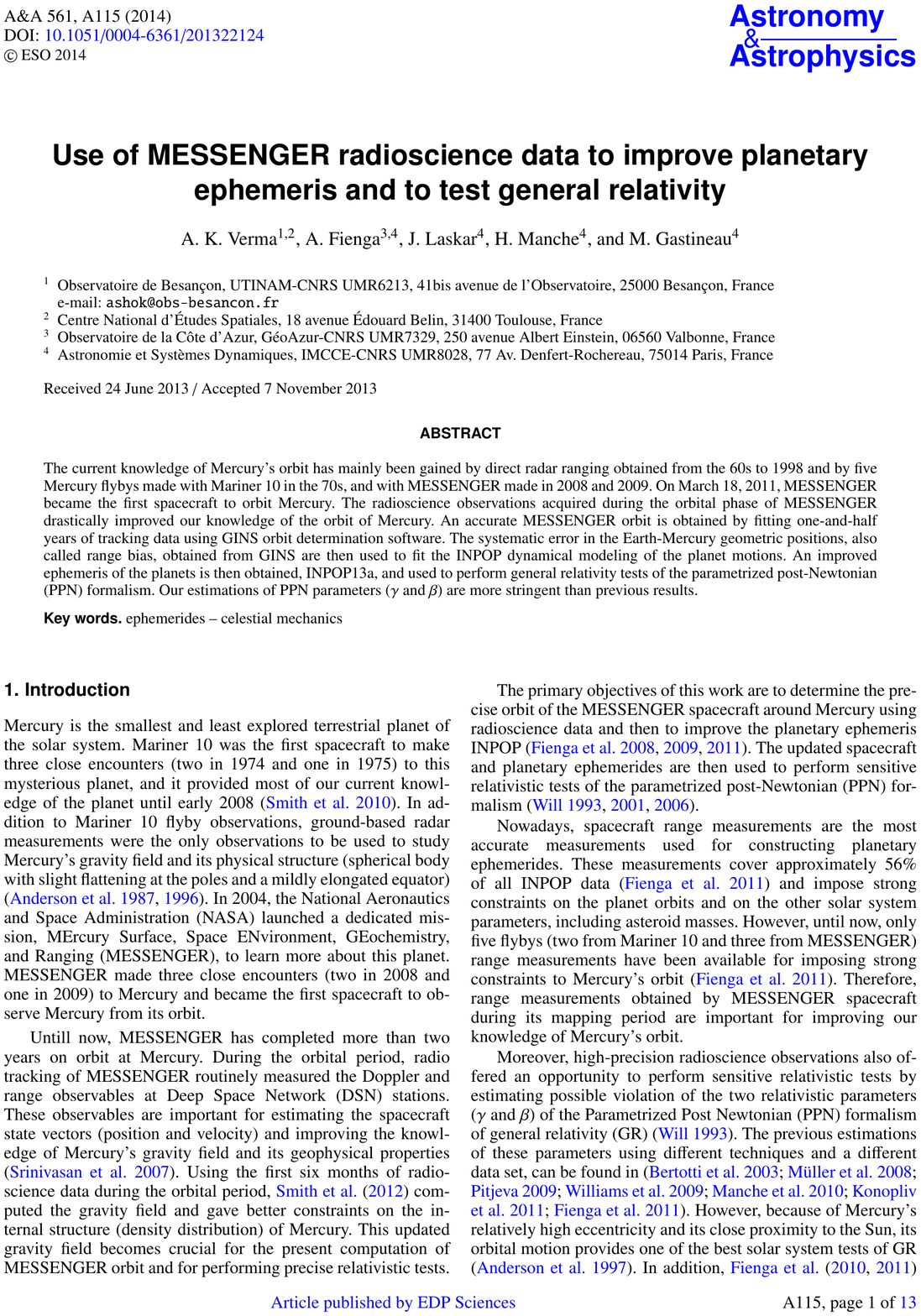}

%% file: CHP5.tex
\chapter{General conclusions}    
\label{CHP5}

This thesis has been essentially motivated by the independent analysis of the past and present space mission radiometric data, and to provide data analysis tools for the improvement of the planetary ephemerides INPOP, as well as to use improved ephemerides to perform tests of physics such as general relativity, solar corona studies, etc.

The thesis has presented the research results obtained from the direct analysis of the \gls{MGS} and \gls{MGR} radiometric raw data. To start the analysis, we have first developed an independent software to extract the contents of the Orbit Data Files (ODFs). These contents are then used to reconstruct the precise orbits of the \gls{MGS} and \gls{MGR} spacecraft using \gls{GINS} software developed by \gls{CNES}. In addition, based on the algorithms given in Chapter \ref{CHP1}, we have also developed an independent light time solution software in order to treat the JPL light time data.

The first part of the thesis deals with the analysis of the \gls{MGS} radiometric data as an academic case to test our understanding of the raw radiometric data and their analysis with GINS by comparing our results with the literature. We have analyzed the entire radioscience data of \gls{MGS} since 1999 to 2006 using \textit{Box-Wing} macro-model consisting ten-plate and a parabolic high gain antenna. On average, an estimated root mean square (rms) value of the post-fit Doppler- and two-way range residuals are less than 5 mHz and 1 m respectively, excluding the residuals at the time of solar conjunctions. Such accuracy of the orbit and of the estimated parameters (see Chapter \ref{CHP2}) are consistent with the results found in the literature (\cite{Yuan01,Lemoine01,Konopliv06,JMarty}, see Table \ref{comp_mgs}). Moreover, we also compared range bias that were computed from GINS with the reduced light time data provided by JPL. These range bias are consistent with each other, and hence confirm the validity of our analysis with respect to JPL \gls{ODP} software.

 Furthermore, the study has been also performed to understand the impact of the macro-model over the orbit perturbations and the estimated parameters. Instead to a \textit{Box-Wing} macro-model, a \textit{Spherical} macro-model was used for a new analysis of the entire radiometric data. By comparing the outcomes of these models (see Chapter \ref{CHP2}), we confirmed that, in the absence of precise knowledge of the spacecraft characteristics, short data-arc can be preferable to account for the mis-modeling in the spacecraft model without costing the orbit accuracy. However, this analysis may be not preferable for extracting an accurate geophysical signals from the Doppler measurements, such as the estimation of gravity field coefficients.

As a supplementary exploitation of MGS, we also performed solar physic studies for the first time using range bias data acquired at the time of solar conjunction periods. In addition to MGS, range bias data of MEX and VEX missions were also used and allowed us to analyze the large-scale structure of the electron density of the solar corona since 2001 to 2011. The parameters of the solar corona models, estimated separately for each spacecraft at different phases of solar activity (maximum and minimum) and at different solar wind states (slow and fast), are then deduced from these data using least square techniques. We compared our estimates with earlier results obtained with different methods. These estimates were found to be consistent during the same solar activities, especially during solar minima. However, during the solar maxima, electron densities obtained with different methods or different spacecraft show weaker consistencies.

We have also demonstrated the impact of solar corona correction on the construction of planetary ephemerides. Data acquired during the period of solar conjunctions show a severe degradation in the radio signals and consequently in the range bias. The observations obtained two months before and
after the conjunctions were therefore usually removed from the fitted data sample of planetary ephemerides. The supplementary data, corrected from the solar corona perturbations, allowed us to gain $\sim$8$\%$ of whole data set. Such corrected data are then used for the first time in the construction of INPOP and induce a noticeable improvement in the estimation of the asteroid masses and a better long-term behavior of the ephemerides. These results are published in the Astronomy $\&$ Astrophysics journal, \cite{verma12}. 

After the successful analysis of the \gls{MGS} radiometric data, the second part of the thesis deals with the complete analysis of the MESSENGER tracking data. MESSENGER is the first spacecraft orbiting Mercury. It therefore gives an ample opportunity to improve our knowledge of the Mercury orbit and also to perform one of the most sensitive \gls{GR} tests of PPN-formalism. We analyzed one and half year of radioscience data using GINS software. The {\it Box-Wing} macro-model of the MESSENGER spacecraft was seen as a combination of flat plates arranged in the shape of a box, with attached solar arrays. For orbit computation and for parameters estimation, a multi-arc approach was used to get an independent estimate of the MESSENGER accelerations. In this method, we integrated the equations of motion using the time step of 50s and then, orbital fits were obtained from short data-arcs fitted over the observation span of one-day using an iterative process. Excluding solar corona period, the estimated typical mean and \gls{rms} value of postfit Doppler, and two-way range residuals of about -0.00063$\pm$4.8 mHz, and -0.003$\pm$1.5 m respectively. Such accuracies are comparable with \cite{Genova13,Smith12,Stanbridge11,Srinivasan07}.

The range bias computed from the GINS are then used for further refinement in the Mercury ephemeris. Using these range bias, we built the planetary ephemerides INPOP13a. This ephemeris showed an accuracy of -0.4$\pm$8.4 m in the geocentric Mercury residuals excluding solar corona period, which is two order of improvement than any latest planetary ephemerides ({\it e.g} INPOP10e (21$\pm$187 m) and DE423 (15$\pm$105 m)). The \gls{GR} tests of \gls{PPN}-formalism were then performed with such high precision Mercury ephemerides. To determine the acceptable intervals of the violation of \gls{GR} thought the \gls{PPN} parameters ($\beta$, $\gamma$), small variations of these two parameters near unity were imposed in the construction of alternative planetary ephemerides fitted over the whole data sets. The percentage difference between these ephemerides to INPOP13a are then used to define the intervals of \gls{PPN} parameters $\beta$ and $\gamma$.

As expected, our estimations of PPN parameters are most stringent than previous results. We considered 5, 10 and 25$\%$ of changes in the postfit residuals. The \gls{PPN} intervals corresponding to these changes are compatible with \gls{GR} with an uncertainty at least ten times smaller than our previous results with INPOP10a or INPOP08. Moreover, so far one of the best estimation of parameter $\gamma$ has been estimated from the Cassini observations by \cite{Bertotti03}, which is compatible with our 25$\%$ of estimation. Furthermore, our 5 and 10$\%$ estimations give no indication of possible asymmetry. We also demonstrated that, despite the less quantity and quality of the Mariner 10 data, these latest are very important to consider for long term effects of the \gls{PPN} parameters. The results of this study are published in Astronomy $\&$ Astrophysics, \cite{verma14}.

In 2015, the BepiColombo mission, a joint mission of the European Space Agency (ESA) and the Japan Aerospace and eXploration Agency (JAXA) to the planet Mercury, will be launch. The main future work will be therefore to analyze the radioscience data obtained from this mission to further improvement in the accuracy of the \gls{PPN} parameters. Moreover, the tools developed during these thesis will also be useful for the analysis of past, present, and future space missions, such as  Odyssey, MRO etc..., and to deliver most up-to-date high accurate ephemerides to the users. This will definitely propel INPOP at the forefront of planetary ephemerides and will make team INPOP independent from the navigation team of the space agencies.

%% file: Verma_Thesis.bbl
\begin{thebibliography}{116}
\expandafter\ifx\csname natexlab\endcsname\relax\def\natexlab#1{#1}\fi
\expandafter\ifx\csname url\endcsname\relax
  \def\url#1{\texttt{#1}}\fi
\expandafter\ifx\csname urlprefix\endcsname\relax\def\urlprefix{URL }\fi

\bibitem[{{Afonso} et~al.(1989){Afonso}, {Barlier}, {Mignard}, {Carpino}, and
  {Farinella}}]{Afonso89}
{Afonso}, G., {Barlier}, F., {Mignard}, F., {Carpino}, M., {Farinella}, P.,
  Oct. 1989. {Orbital effects of LAGEOS seasons and eclipses}. Annales
  Geophysicae 7, 501--514.

\bibitem[{{Albee} et~al.(2001){Albee}, {Arvidson}, {Palluconi}, and
  {Thorpe}}]{albee01}
{Albee}, A.~L., {Arvidson}, R.~E., {Palluconi}, F., {Thorpe}, T., Oct. 2001.
  {Overview of the Mars Global Surveyor mission}. Journal of Geophysical
  Research 106, 23291--23316.

\bibitem[{{Anderson} et~al.(1987{\natexlab{a}}){Anderson}, {Colombo},
  {Espsitio}, {Lau}, and {Trager}}]{Anderson87}
{Anderson}, J.~D., {Colombo}, G., {Espsitio}, P.~B., {Lau}, E.~L., {Trager},
  G.~B., Sep. 1987{\natexlab{a}}. {The mass, gravity field, and ephemeris of
  Mercury}. \icarus 71, 337--349.

\bibitem[{{Anderson} et~al.(1996){Anderson}, {Jurgens}, {Lau}, {Slade}, and
  {Schubert}}]{Anderson1996}
{Anderson}, J.~D., {Jurgens}, R.~F., {Lau}, E.~L., {Slade}, III, M.~A.,
  {Schubert}, G., Dec. 1996. {Shape and Orientation of Mercury from Radar
  Ranging Data}. \icarus 124, 690--697.

\bibitem[{{Anderson} et~al.(1978){Anderson}, {Keesey}, {Lau}, {Standish}, and
  {Newhall}}]{Anderson78}
{Anderson}, J.~D., {Keesey}, M.~S.~W., {Lau}, E.~L., {Standish}, Jr., E.~M.,
  {Newhall}, X.~X., 1978. Tests of general relativity using astrometric and
  radio metric observations of the planets. Acta Astronautica 5, 43.

\bibitem[{{Anderson} et~al.(1976){Anderson}, {Keesey}, {Lau}, {Standish},
  {Newhall}, and {V.~N.~Novikov \& D.~N.~Cheverov}}]{Anderson76}
{Anderson}, J.~D., {Keesey}, M.~S.~W., {Lau}, E.~L., {Standish}, Jr., E.~M.,
  {Newhall}, X.~X., {V.~N.~Novikov \& D.~N.~Cheverov} (Eds.), Oct. 1976. {Tests
  of general relativity using astrometric and radiometric observations of the
  planets}.

\bibitem[{{Anderson} et~al.(1987{\natexlab{b}}){Anderson}, {Krisher},
  {Borutzki}, {Connally}, {Eshe}, {Hotz}, {Kinslow}, {Kursinski}, {Light},
  {Matousek}, {Moyd}, {Roth}, {Sweetnam}, {Taylor}, {Tyler}, {Gresh}, and
  {Rosen}}]{AndersonV2}
{Anderson}, J.~D., {Krisher}, T.~P., {Borutzki}, S.~E., {Connally}, M.~J.,
  {Eshe}, P.~M., {Hotz}, H.~B., {Kinslow}, S., {Kursinski}, E.~R., {Light},
  L.~B., {Matousek}, S.~E., {Moyd}, K.~I., {Roth}, D.~C., {Sweetnam}, D.~N.,
  {Taylor}, A.~H., {Tyler}, G.~L., {Gresh}, D.~L., {Rosen}, P.~A., Dec.
  1987{\natexlab{b}}. {Radio range measurements of coronal electron densities
  at 13 and 3.6 centimeter wavelengths during the 1985 solar conjunction of
  Voyager 2}. The Astrophysical Journal 323, L141--L143.

\bibitem[{{Anderson} et~al.(1997){Anderson}, {Turyshev}, {Asmar}, {Bird},
  {Konopliv}, {Krisher}, {Lau}, {Schubert}, and {Sjogren}}]{Anderson97}
{Anderson}, J.~D., {Turyshev}, S.~G., {Asmar}, S.~W., {Bird}, M.~K.,
  {Konopliv}, A.~S., {Krisher}, T.~P., {Lau}, E.~L., {Schubert}, G., {Sjogren},
  W.~L., Jan. 1997. {Radio-science investigation on a Mercury Orbiter mission}.
  \planss 45, 21--29.

\bibitem[{{Ash} et~al.(1967){Ash}, {Shapiro}, and {Smith}}]{Ash67}
{Ash}, M.~E., {Shapiro}, I.~I., {Smith}, W.~B., Apr. 1967. {Astronomical
  constants and planetary ephemerides deduced from radar and optical
  observations}. \aj 72, 338--+.

\bibitem[{{Barlier} et~al.(1978){Barlier}, {Berger}, {Falin}, {Kockarts}, and
  {Thuillier}}]{Barlier78}
{Barlier}, F., {Berger}, C., {Falin}, J.~L., {Kockarts}, G., {Thuillier}, G.,
  Mar. 1978. {A thermospheric model based on satellite drag data}. Annales de
  Geophysique 34, 9--24.

\bibitem[{{Bertotti} et~al.(1987){Bertotti}, {Ciufolini}, and
  {Bender}}]{Bertotti87}
{Bertotti}, B., {Ciufolini}, I., {Bender}, P.~L., Mar. 1987. {New test of
  general relativity - Measurement of de Sitter geodetic precession rate for
  lunar perigee}. Physical Review Letters 58, 1062--1065.

\bibitem[{{Bertotti} et~al.(2003){Bertotti}, {Iess}, and
  {Tortora}}]{Bertotti03}
{Bertotti}, B., {Iess}, L., {Tortora}, P., Sep. 2003. {A test of general
  relativity using radio links with the Cassini spacecraft}. \nat 425,
  374--376.

\bibitem[{{Bird} et~al.(1996){Bird}, {Paetzold}, {Edenhofer}, {Asmar}, and
  {McElrath}}]{Bird96}
{Bird}, M.~K., {Paetzold}, M., {Edenhofer}, P., {Asmar}, S.~W., {McElrath},
  T.~P., Dec. 1996. {Coronal radio sounding with Ulysses: solar wind electron
  density near 0.1AU during the 1995 conjunction.} Astronomy \& Astrophysics
  316, 441--448.

\bibitem[{{Bird} et~al.(1994){Bird}, {Volland}, {Paetzold}, {Edenhofer},
  {Asmar}, and {Brenkle}}]{Bird94}
{Bird}, M.~K., {Volland}, H., {Paetzold}, M., {Edenhofer}, P., {Asmar}, S.~W.,
  {Brenkle}, J.~P., May 1994. {The coronal electron density distribution
  determined from dual-frequency ranging measurements during the 1991 solar
  conjunction of the ULYSSES spacecraft}. The Astrophysical Journal 426,
  373--381.

\bibitem[{{Bougeret} et~al.(1984){Bougeret}, {King}, and {Schwenn}}]{Bougeret}
{Bougeret}, J.-L., {King}, J.~H., {Schwenn}, R., Feb. 1984. {Solar radio burst
  and in situ determination of interplanetary electron density}. Solar Physics
  90, 401--412.

\bibitem[{{Bravo} and {Stewart}(1997)}]{Bravo97}
{Bravo}, S., {Stewart}, G.~A., Nov. 1997. {Fast and Slow Wind from Solar
  Coronal Holes}. \apj 489, 992.

\bibitem[{{Cash} et~al.(2008){Cash}, {Emmons}, and {Welgemoed}}]{Cash08}
{Cash}, P., {Emmons}, D., {Welgemoed}, J., Dec. 2008. { Ultra-stable
  oscillators for space application}. In: 40th Annual Precise Time and Time
  Interval (PTTI) Meeting. pp. 51--56.

\bibitem[{{Chao}(1971)}]{chao}
{Chao}, C.~C., Oct. 1971. {New Tropospheric Range Corrections with Seasonal
  Adjustment}. New Tropospheric Range Corrections with Seasonal Adjustment,
  Deep Space Network Progress Report, 31-1526, VI, 67-82 31.

\bibitem[{{Ciufolini}(1986)}]{Ciufolini86}
{Ciufolini}, I., Jan. 1986. {Measurement of the Lense-Thirring drag on
  high-altitude, laser-ranged artificial satellites}. Physical Review Letters
  56, 278--281.

\bibitem[{{Davies} et~al.(1996){Davies}, {Abalakin}, {Bursa}, {Lieske},
  {Morando}, {Morrison}, {Seidelmann}, {Sinclair}, {Yallop}, and
  {Tjuflin}}]{Davies96}
{Davies}, M.~E., {Abalakin}, V.~K., {Bursa}, M., {Lieske}, J.~H., {Morando},
  B., {Morrison}, D., {Seidelmann}, P.~K., {Sinclair}, A.~T., {Yallop}, B.,
  {Tjuflin}, Y.~S., Jan. 1996. {Report of the IAU/IAG/COSPAR Working Group on
  Cartographic Coordinates and Rotational Elements of the Planets and
  Satellites: 1994}. Celestial Mechanics and Dynamical Astronomy 63, 127--148.

\bibitem[{{Dow}(1988)}]{Dow88}
{Dow}, J.~M., 1988. {An improved tidal model for high accuracy satellite orbit
  determination}. In: 16th International Symposium on Space Technology and
  Science, Vol. 1. Vol.~1. pp. 701--705.

\bibitem[{{Eshleman} et~al.(1977){Eshleman}, {Tyler}, {Anderson}, {Fjeldbo},
  {Levy}, {Wood}, and {Croft}}]{Eshleman77}
{Eshleman}, V.~R., {Tyler}, G.~L., {Anderson}, J.~D., {Fjeldbo}, G., {Levy},
  G.~S., {Wood}, G.~E., {Croft}, T.~A., Nov. 1977. {Radio science
  investigations with Voyager}. Space Science Reviews 21, 207--232.

\bibitem[{{Feldman} et~al.(2005){Feldman}, {Landi}, and
  {Schwadron}}]{Feldman05}
{Feldman}, U., {Landi}, E., {Schwadron}, N.~A., Jul. 2005. {On the sources of
  fast and slow solar wind}. Journal of Geophysical Research (Space Physics)
  110, 7109.

\bibitem[{{Fienga}(2011)}]{FiengaHDR}
{Fienga}, A., May 2011. {Ephemerides planetaires et systemes de reference}.
  Habilitation a Diriger des Recherches, Universeite de Franche-Comte -
  Observatoire de Besancon, in French.

\bibitem[{{Fienga} et~al.(2011b){Fienga}, {Kuchynka}, {Laskar}, {Manche}, , and
  {Gastineau}}]{Fienga2011b}
{Fienga}, A., {Kuchynka}, P., {Laskar}, J., {Manche}, H., , {Gastineau}, M.,
  Oct. 2011b. Asteroid mass determinations with the inpop planetary
  ephemerides. EPSC-DPS Join Meeting 2011. p. 1879.

\bibitem[{{Fienga} et~al.(2010){Fienga}, {Laskar}, {Kuchynka}, {Le
  Poncin-Lafitte}, {Manche}, and {Gastineau}}]{Fienga10}
{Fienga}, A., {Laskar}, J., {Kuchynka}, P., {Le Poncin-Lafitte}, C., {Manche},
  H., {Gastineau}, M., Jan. 2010. {Gravity tests with INPOP planetary
  ephemerides}. In: {Klioner}, S.~A., {Seidelmann}, P.~K., {Soffel}, M.~H.
  (Eds.), IAU Symposium. Vol. 261 of IAU Symposium. pp. 159--169.

\bibitem[{{Fienga} et~al.(2011a){Fienga}, {Laskar}, {Manche}, {Kuchynka},
  {Desvignes}, {Gastineau}, and {Cognard}}]{Fienga2011}
{Fienga}, A., {Laskar}, J., {Manche}, H., {Kuchynka}, P., {Desvignes}, G.,
  {Gastineau}, M, ., {Cognard}, I., a.~G., 2011a. The planetary ephemerides
  inpop10a and its applications in fundamental physics. Celestial Mechanics and
  Dynamical Astronomy.

\bibitem[{{Fienga} et~al.(2009){Fienga}, {Laskar}, {Morley}, {Manche},
  {Kuchynka}, {Le Poncin-Lafitte}, {Budnik}, {Gastineau}, and
  {Somenzi}}]{Fienga2009}
{Fienga}, A., {Laskar}, J., {Morley}, T., {Manche}, H., {Kuchynka}, P., {Le
  Poncin-Lafitte}, C., {Budnik}, F., {Gastineau}, M., {Somenzi}, L., Dec. 2009.
  {INPOP08, a 4-D planetary ephemeris: from asteroid and time-scale
  computations to ESA Mars Express and Venus Express contributions}. Astronomy
  \& Astrophysics 507, 1675--1686.

\bibitem[{{Fienga} et~al.(2008){Fienga}, {Manche}, and
  {Gastineau}}]{Fienga2008}
{Fienga}, A., {Manche}, H., a. L.~J., {Gastineau}, M., 2008. Inpop06: a new
  numerical planetary ephemeris. Astronomy \& Astrophysics 477, 315--327.

\bibitem[{{Fienga} et~al.(2013){Fienga}, {Manche}, {Laskar}, {Gastineau}, and
  {Verma}}]{Fienga13}
{Fienga}, A., {Manche}, H., {Laskar}, J., {Gastineau}, M., {Verma}, A., Jan.
  2013. {INPOP new release: INPOP10e}. ArXiv e-prints.

\bibitem[{{Fjeldbo} et~al.(1971){Fjeldbo}, {Kliore}, and
  {Eshleman}}]{Fjeldbo71}
{Fjeldbo}, G., {Kliore}, A.~J., {Eshleman}, V.~R., Mar. 1971. {The Neutral
  Atmosphere of Venus as Studied with the Mariner V Radio Occultation
  Experiments}. Astronomical Journal 76, 123.

\bibitem[{{Flanigan} et~al.(2013){Flanigan}, {OShaughnessy}, {Wilson}, and
  {Hill}}]{Flanigan13}
{Flanigan}, S.~H., {OShaughnessy}, D.~J., {Wilson}, M.~N., {Hill}, T.~A., 2013.
  {MESSENGER maneuvers to reduce orbital period during the extended mission:
  Ensuring maximum use of the bi-propellant propulsion system}. In: 23rd Space
  Flight Mechanics Meeting, American Astronomical Society. pp. 13--382.

\bibitem[{{Folkner}(2010)}]{DE423}
{Folkner}, W.~M., Feb. 2010. {Planetary ephemeris DE423 fit to Messenger
  encounters with Mercury}. JPL, interoffice memorandum IOM 343R-10-001.

\bibitem[{{Folkner} et~al.(2008){Folkner}, {Williams}, and {Boggs}}]{DE421}
{Folkner}, W.~M., {Williams}, J.~G., {Boggs}, D.~H., Mar. 2008. {JPL planetary
  and lunar ephemerides DE421,} IOM 343R-08-003.

\bibitem[{{Genova} et~al.(2013){Genova}, {Iess}, and {Marabucci}}]{Genova13}
{Genova}, A., {Iess}, L., {Marabucci}, M., Jun. 2013. {Mercury's gravity field
  from the first six months of MESSENGER data}. \planss 81, 55--64.

\bibitem[{{Guhathakurta} and {Holzer}(1994)}]{Guhathakurta94}
{Guhathakurta}, M., {Holzer}, T.~E., May 1994. {Density structure inside a
  polar coronal hole}. The Astrophysical Journal 426, 782--786.

\bibitem[{{Guhathakurta} et~al.(1996){Guhathakurta}, {Holzer}, and
  {MacQueen}}]{Guhathakurta96}
{Guhathakurta}, M., {Holzer}, T.~E., {MacQueen}, R.~M., Feb. 1996. {The
  Large-Scale Density Structure of the Solar Corona and the Heliospheric
  Current Sheet}. The Astrophysical Journal 458, 817.

\bibitem[{{Hairer} et~al.(1987){Hairer}, {Norsett}, and {Wanner}}]{Hairer87}
{Hairer}, E., {Norsett}, S.~P., {Wanner}, G. (Eds.), 1987. Solving Ordinary
  Differential Equations I: Nonstiff Problems. Springer series in Computational
  Mathematics. Springer.

\bibitem[{{Hassler} et~al.(1999){Hassler}, {Dammasch}, {Lemaire}, {Brekke},
  {Curdt}, {Mason}, {Vial}, and {Wilhelm}}]{Hassler99}
{Hassler}, D.~M., {Dammasch}, I.~E., {Lemaire}, P., {Brekke}, P., {Curdt}, W.,
  {Mason}, H.~E., {Vial}, J.-C., {Wilhelm}, K., Feb. 1999. {Solar Wind Outflow
  and the Chromospheric Magnetic Network}. Science 283, 810.

\bibitem[{{Hedin} et~al.(1996){Hedin}, {Fleming}, {Manson}, {Schmidlin},
  {Avery}, {Clark}, {Franke}, {Fraser}, {Tsuda}, {Vial}, and
  {Vincent}}]{Hedin96}
{Hedin}, A.~E., {Fleming}, E.~L., {Manson}, A.~H., {Schmidlin}, F.~J., {Avery},
  S.~K., {Clark}, R.~R., {Franke}, S.~J., {Fraser}, G.~J., {Tsuda}, T., {Vial},
  F., {Vincent}, R.~A., Sep. 1996. {Empirical wind model for the upper, middle
  and lower atmosphere.} Journal of Atmospheric and Terrestrial Physics 58,
  1421--1447.

\bibitem[{{Howard} et~al.(1974){Howard}, {Tyler}, {Fjeldbo}, {Kliore}, {Levy},
  {Brunn}, {Dickinson}, {Edelson}, {Martin}, {Postal}, {Seidel}, {Sesplaukis},
  {Shirley}, {Stelzried}, {Sweetnam}, {Zygielbaum}, {Esposito}, {Anderson},
  {Shapiro}, and {Reasenberg}}]{Howard74}
{Howard}, H.~T., {Tyler}, G.~L., {Fjeldbo}, G., {Kliore}, A.~J., {Levy}, G.~S.,
  {Brunn}, D.~L., {Dickinson}, R., {Edelson}, R.~E., {Martin}, W.~L., {Postal},
  R.~B., {Seidel}, B., {Sesplaukis}, T.~T., {Shirley}, D.~L., {Stelzried},
  C.~T., {Sweetnam}, D.~N., {Zygielbaum}, A.~I., {Esposito}, P.~B., {Anderson},
  J.~D., {Shapiro}, I.~I., {Reasenberg}, R.~D., Mar. 1974. {Venus: Mass,
  Gravity Field, Atmosphere, and Ionosphere as Measured by the Mariner 10
  Dual-Frequency Radio System}. Science 183, 1297--1301.

\bibitem[{{Huang} et~al.(1990){Huang}, {Ries}, {Tapley}, and
  {Watkins}}]{Huang90}
{Huang}, C., {Ries}, J.~C., {Tapley}, B.~D., {Watkins}, M.~M., Jun. 1990.
  {Relativistic effects for near-earth satellite orbit determination}.
  Celestial Mechanics and Dynamical Astronomy 48, 167--185.

\bibitem[{{Hudson} et~al.(1995){Hudson}, {Haisch}, and {Strong}}]{Hudson95}
{Hudson}, H., {Haisch}, B., {Strong}, K.~T., Mar. 1995. {Comment on 'The solar
  flare myth' by J. T. Gosling}. \jgr 100, 3473--3477.

\bibitem[{{Issautier} et~al.(1998){Issautier}, {Meyer-Vernet}, {Moncuquet}, and
  {Hoang}}]{Issautier98}
{Issautier}, K., {Meyer-Vernet}, N., {Moncuquet}, M., {Hoang}, S., Feb. 1998.
  {Solar wind radial and latitudinal structure - Electron density and core
  temperature from ULYSSES thermal noise spectroscopy}. Journal of Geophysical
  Research 103, 1969.

\bibitem[{{Jacchia}(1977)}]{Jacchia77}
{Jacchia}, L.~G., Mar. 1977. {Thermospheric Temperature, Density, and
  Composition: New Models}. SAO Special Report 375.

\bibitem[{{Kallenrode}(2004)}]{Kallenrode04}
{Kallenrode}, M.-B., 2004. {Space physics : an introduction to plasmas and
  particles in the heliosphere and magnetospheres}.

\bibitem[{{Kaula}(1966)}]{Kaula66}
{Kaula}, W.~M. (Ed.), 1966. {Theory of satellite geodesy: Applications of
  satellites to geodesy}. Dover Publications.

\bibitem[{{Klioner}(2008)}]{Klioner08}
{Klioner}, S.~A., Feb. 2008. {Relativistic scaling of astronomical quantities
  and the system of astronomical units}. \aap 478, 951--958.

\bibitem[{{Kojima} and {Kakinuma}(1990)}]{Kojima90}
{Kojima}, M., {Kakinuma}, T., Aug. 1990. {Solar cycle dependence of global
  distribution of solar wind speed}. Space Sci Rev, 53, 173--222.

\bibitem[{{Konopliv} et~al.(2011){Konopliv}, {Asmar}, {Folkner}, {Karatekin},
  {Nunes}, {Smrekar}, {Yoder}, and {Zuber}}]{Konopliv11}
{Konopliv}, A.~S., {Asmar}, S.~W., {Folkner}, W.~M., {Karatekin}, {\"O}.,
  {Nunes}, D.~C., {Smrekar}, S.~E., {Yoder}, C.~F., {Zuber}, M.~T., Jan. 2011.
  {Mars high resolution gravity fields from MRO, Mars seasonal gravity, and
  other dynamical parameters}. icarus 211, 401--428.

\bibitem[{{Konopliv} et~al.(2006){Konopliv}, {Yoder}, {Standish}, {Yuan}, and
  {Sjogren}}]{Konopliv06}
{Konopliv}, A.~S., {Yoder}, C.~F., {Standish}, E.~M., {Yuan}, D.-N., {Sjogren},
  W.~L., May 2006. {A global solution for the Mars static and seasonal gravity,
  Mars orientation, Phobos and Deimos masses, and Mars ephemeris}. Icarus 182,
  23--50.

\bibitem[{{Kuchynka}(2010)}]{KuchynkaPHD}
{Kuchynka}, P., 2010. Etude des perturbations induites par les asteroides sur
  les mouvements des planetes et des sondes spatiales autour du point de
  lagrange l2. {PhD} in astronomy, Observatoire de Paris.

\bibitem[{{Kwok}(2000)}]{trk}
{Kwok}, A., 2000. {TRK-2-18 Tracking System Interfaces Orbit Data File
  Interface}. TRK-2-18 Tracking System Interfaces Orbit Data File Interface,
  Deep Space Mission System, JPL D-16765, 820-013.

\bibitem[{{Lainey} et~al.(2007){Lainey}, {Dehant}, and
  {P{\"a}tzold}}]{Lainey07}
{Lainey}, V., {Dehant}, V., {P{\"a}tzold}, M., Apr. 2007. {First numerical
  ephemerides of the Martian moons}. Astronomy and Astrophysics 465,
  1075--1084.

\bibitem[{{Lang}(2000)}]{Lang00}
{Lang}, K.~R., Sep. 2000. {The Sun From Space}. \apss 273, 1--6.

\bibitem[{Lawson and Hanson(1995)}]{Lawson95}
Lawson, C.~L., Hanson, R.~J., 1995. Solving Least Squares Problems. SIAM,
  Philadelphia, PA.

\bibitem[{{Leary} et~al.(2007){Leary}, {Conde}, {Dakermanji}, {Engelbrecht},
  {Ercol}, {Fielhauer}, {Grant}, {Hartka}, {Hill}, {Jaskulek}, {Mirantes},
  {Mosher}, {Paul}, {Persons}, {Rodberg}, {Srinivasan}, {Vaughan}, and
  {Wiley}}]{Leary07}
{Leary}, J.~C., {Conde}, R.~F., {Dakermanji}, G., {Engelbrecht}, C.~S.,
  {Ercol}, C.~J., {Fielhauer}, K.~B., {Grant}, D.~G., {Hartka}, T.~J., {Hill},
  T.~A., {Jaskulek}, S.~E., {Mirantes}, M.~A., {Mosher}, L.~E., {Paul}, M.~V.,
  {Persons}, D.~F., {Rodberg}, E.~H., {Srinivasan}, D.~K., {Vaughan}, R.~M.,
  {Wiley}, S.~R., Aug. 2007. {The MESSENGER Spacecraft}. Space Sci Rev, 131,
  187--217.

\bibitem[{{Leblanc} et~al.(1998){Leblanc}, {Dulk}, and {Bougeret}}]{Leblanc}
{Leblanc}, Y., {Dulk}, G.~A., {Bougeret}, J.-L., Nov. 1998. {Tracing the
  Electron Density from the Corona to 1au}. Solar Physics 183, 165--180.

\bibitem[{{Lemoine} et~al.(1999){Lemoine}, {Rowlands}, {Smith}, {Chinn},
  {Pavlis}, {Luthcke}, {Neumann}, and {Zuber}}]{lemoine99}
{Lemoine}, F.~G., {Rowlands}, D.~D., {Smith}, D.~E., {Chinn}, D.~S., {Pavlis},
  D.~E., {Luthcke}, S.~B., {Neumann}, G.~A., {Zuber}, M.~T., Aug. 1999. {Orbit
  determination for Mars Global Surveyor during mapping}. In: AAS/AIAA
  Astrodynamics Specialist Conference, 16-19 August. pp. 99--328.

\bibitem[{{Lemoine} et~al.(2001){Lemoine}, {Smith}, {Rowlands}, {Zuber},
  {Neumann}, {Chinn}, and {Pavlis}}]{Lemoine01}
{Lemoine}, F.~G., {Smith}, D.~E., {Rowlands}, D.~D., {Zuber}, M.~T., {Neumann},
  G.~A., {Chinn}, D.~S., {Pavlis}, D.~E., Oct. 2001. {An improved solution of
  the gravity field of Mars (GMM-2B) from Mars Global Surveyor}. \jgr 106,
  23359--23376.

\bibitem[{{Leon} and {Jay}(2010)}]{Leon10}
{Leon}, G., {Jay}, M.~P., 2010. {The solar corona; 2nd ed.} Cambridge Univ.
  Press.

\bibitem[{{Manche}(2011)}]{Manche2011}
{Manche}, H., 2011. Elaboration des ephemeride inpop : modele dynamique et
  ajustements aux donnees de telemetrie laser lune. Ph.D. thesis, in french,
  Observatoire de Paris.

\bibitem[{{Manche} et~al.(2010){Manche}, {Fienga}, {Laskar}, {Gastineau},
  {Bouquillon}, {Francou}, and {Kuchynka}}]{Manche2010}
{Manche}, H., {Fienga}, A., {Laskar}, J., {Gastineau}, M., {Bouquillon}, S.,
  {Francou}, G., {Kuchynka}, P., Nov. 2010. {LLR residuals of the latest INPOP
  solution and constraints on post-Newtonian parameters}. In: Journ{\'e}es
  Syst{\`e}mes de R{\'e}f{\'e}rence Spatio-temporels 2010. Journees Systemes de
  references.

\bibitem[{{Margot}(2009)}]{Margot09}
{Margot}, J.-L., Dec. 2009. {A Mercury orientation model including non-zero
  obliquity and librations}. Celestial Mechanics and Dynamical Astronomy 105,
  329--336.

\bibitem[{{Marshall}(1992)}]{Marshall92}
{Marshall}, J.~A., 1992. {Modeling radiation forces acting on TOPEX/Poseidon
  for precision orbit determination}.

\bibitem[{{Marty}(2011)}]{Jcmarty10}
{Marty}, J.~C., 2011. {Characteristics of the MGS macro-model}. Private
  communication.

\bibitem[{{Marty} et~al.(2009){Marty}, {Balmino}, {Duron}, {Rosenblatt}, {Le
  Maistre}, {Rivoldini}, {Dehant}, and {van Hoolst}}]{JMarty}
{Marty}, J.~C., {Balmino}, G., {Duron}, J., {Rosenblatt}, P., {Le Maistre}, S.,
  {Rivoldini}, A., {Dehant}, V., {van Hoolst}, T., Mar. 2009. {Martian gravity
  field model and its time variations from MGS and Odyssey data}. Planetary and
  Space Science 57, 350--363.

\bibitem[{{McAdams} et~al.(2007){McAdams}, {Farquhar}, {Taylor}, and
  {Williams}}]{McAdams07}
{McAdams}, J.~V., {Farquhar}, R.~W., {Taylor}, A.~H., {Williams}, B.~G., Aug.
  2007. {MESSENGER Mission Design and Navigation}. Space Sci Rev, 131,
  219--246.

\bibitem[{McCarthy and Petit(2003)}]{McCarthyIERS32}
McCarthy, D.~D., Petit, G., 2003. {IERS} technical note no 32. Tech. rep.,
  {IERS} Convention Centre, http://www.iers.org/iers/publications/tn/tn32/.
\newline\urlprefix\url{http://www.iers.org/iers/publications/tn/tn32/}

\bibitem[{{McCarthy} and {Petit}(2004)}]{McCarthy04}
{McCarthy}, D.~D., {Petit}, G., 2004. {IERS Conventions (2003)}. IERS Technical
  Note 32, 1.

\bibitem[{{Milani} et~al.(1987){Milani}, {Nobili}, and {Farinella}}]{Milani87}
{Milani}, A., {Nobili}, A.~M., {Farinella}, P., 1987. {Non-gravitational
  perturbations and satellite geodesy.}

\bibitem[{{Moyer}(1971)}]{Moyer1971}
{Moyer}, T., 1971. Dpodp manual. IOM 3215-37, JPL.

\bibitem[{{Moyer}(2003)}]{Moyer}
{Moyer}, T.~D., 2003. {Formulation for Observed and Computed Values of Deep
  Space Network Data Types for Navigation}. Vol.~2. John Wiley \& Sons.

\bibitem[{{Muhleman} and {Anderson}(1981)}]{Muhleman81}
{Muhleman}, D.~O., {Anderson}, J.~D., Aug. 1981. {Solar wind electron densities
  from Viking dual-frequency radio measurements}. The Astrophysical Journal
  247, 1093--1101.

\bibitem[{{Muhleman} et~al.(1977){Muhleman}, {Esposito}, and
  {Anderson}}]{Muhleman77}
{Muhleman}, D.~O., {Esposito}, P.~B., {Anderson}, J.~D., Feb. 1977. {The
  electron density profile of the outer corona and the interplanetary medium
  from Mariner-6 and Mariner-7 time-delay measurements}. Astrophysics Journal
  211, 943--957.

\bibitem[{{M{\"u}ller} et~al.(2008){M{\"u}ller}, {Soffel}, and
  {Klioner}}]{Muller08}
{M{\"u}ller}, J., {Soffel}, M., {Klioner}, S.~A., Mar. 2008. {Geodesy and
  relativity}. Journal of Geodesy 82, 133--145.

\bibitem[{{Newhall}(1989)}]{Newhall89}
{Newhall}, X.~X., 1989. {Numerical Representation of Planetary Ephemerides}.
  Celestial Mechanics 45, 305.

\bibitem[{{Newhall} et~al.(1983){Newhall}, {Standish}, and
  {Williams}}]{Newhall83}
{Newhall}, X.~X., {Standish}, E.~M., {Williams}, J.~G., Aug. 1983. {DE 102 - A
  numerically integrated ephemeris of the moon and planets spanning forty-four
  centuries}. Astronomy \& Astrophysics 125, 150--167.

\bibitem[{{Parker}(1963)}]{Parker63}
{Parker}, E.~N., Jul. 1963. {The Solar-Flare Phenomenon and the Theory of
  Reconnection and Annihiliation of Magnetic Fields.} \apjs 8, 177.

\bibitem[{{P{\"a}tzold} et~al.(1995){P{\"a}tzold}, {Bird}, {Edenhofer},
  {Asmar}, and {McElrath}}]{Patzold95}
{P{\"a}tzold}, M., {Bird}, M.~K., {Edenhofer}, P., {Asmar}, S.~W., {McElrath},
  T.~P., 1995. {Dual-frequency radio sounding of the solar corona during the
  1995 conjunction of the Ulysses spacecraft}. Geographical research letters
  22, 3313--3316.

\bibitem[{{P{\"a}tzold} et~al.(2004){P{\"a}tzold}, {Neubauer}, {Carone},
  {Hagermann}, {Stanzel}, {H{\"a}usler}, {Remus}, {Selle}, {Hagl}, {Hinson},
  {Simpson}, {Tyler}, {Asmar}, {Axford}, {Hagfors}, {Barriot}, {Cerisier},
  {Imamura}, {Oyama}, {Janle}, {Kirchengast}, and {Dehant}}]{patzold04}
{P{\"a}tzold}, M., {Neubauer}, F.~M., {Carone}, L., {Hagermann}, A., {Stanzel},
  C., {H{\"a}usler}, B., {Remus}, S., {Selle}, J., {Hagl}, D., {Hinson}, D.~P.,
  {Simpson}, R.~A., {Tyler}, G.~L., {Asmar}, S.~W., {Axford}, W.~I., {Hagfors},
  T., {Barriot}, J.-P., {Cerisier}, J.-C., {Imamura}, T., {Oyama}, K.-I.,
  {Janle}, P., {Kirchengast}, G., {Dehant}, V., Aug. 2004. {MaRS: Mars Express
  Orbiter Radio Science}. In: {Wilson}, A., {Chicarro}, A. (Eds.), Mars
  Express: the Scientific Payload. Vol. 1240 of ESA Special Publication. pp.
  141--163.

\bibitem[{{Petrie}(2013)}]{Petrie13}
{Petrie}, G.~J.~D., May 2013. {Solar Magnetic Activity Cycles, Coronal
  Potential Field Models and Eruption Rates}. \apj 768, 162.

\bibitem[{{Pitjeva}(2005)}]{Pitjeva05}
{Pitjeva}, E.~V., May 2005. {High-Precision Ephemerides of Planets-EPM and
  Determination of Some Astronomical Constants}. Solar System Research 39,
  176--186.

\bibitem[{{Pitjeva}(2009)}]{Pitjeva09}
{Pitjeva}, E.~V., May 2009. {EPM Ephemerides and Relativity}. In: IAU
  Symposium, American Astronomical Society. Vol. 261. p. 603.

\bibitem[{{Pitjeva}(2010)}]{Pitjeva10}
{Pitjeva}, E.~V., Jan. 2010. {EPM ephemerides and relativity}. In: {Klioner},
  S.~A., {Seidelmann}, P.~K., {Soffel}, M.~H. (Eds.), IAU Symposium. Vol. 261
  of IAU Symposium. pp. 170--178.

\bibitem[{{Pitjeva} and {Pitjev}(2013)}]{Pitjeva13}
{Pitjeva}, E.~V., {Pitjev}, N.~P., Jul. 2013. {Relativistic effects and dark
  matter in the Solar system from observations of planets and spacecraft}.
  \mnras 432, 3431--3437.

\bibitem[{{Ringnes}(1964)}]{Ringnes64}
{Ringnes}, T.~S., 1964. {On the lifetime of sunspot groups}. Astrophysica
  Norvegica 9, 95.

\bibitem[{{Schwenn}(1983)}]{Schwenn83}
{Schwenn}, R., Nov. 1983. {The average solar wind in the inner heliosphere:
  Structures and slow variations}. In: NASA Conference Publication. Vol. 228 of
  NASA Conference Publication. pp. 489--507.

\bibitem[{{Schwenn}(2006)}]{Schwenn06}
{Schwenn}, R., Jun. 2006. {Solar Wind Sources and Their Variations Over the
  Solar Cycle}. Space Science Reviews 124, 51--76.

\bibitem[{Schwenn and Marsch(1990)}]{Schwennvol1}
Schwenn, R., Marsch, E. (Eds.), 1990. Physics of the Inner Heliosphere, Vol. I:
  Large-Scale Phenomena. Vol.~20 of Physics and Chemistry in Space. Springer.

\bibitem[{Schwenn and Marsch(1991)}]{Schwennvol2}
Schwenn, R., Marsch, E. (Eds.), 1991. Physics of the Inner Heliosphere, Vol.
  II: Particles, Waves and Turbulence. Vol.~21 of Physics and Chemistry in
  Space. Springer.

\bibitem[{{Shapiro}(1964)}]{Shapiro64}
{Shapiro}, I.~I., Dec. 1964. {Fourth Test of General Relativity}. Physical
  Review Letters 13, 789--791.

\bibitem[{{Smith} et~al.(2012){Smith}, {Zuber}, {Phillips}, {Solomon}, {Hauck},
  {Lemoine}, {Mazarico}, {Neumann}, {Peale}, {Margot}, {Johnson}, {Torrence},
  {Perry}, {Rowlands}, {Goossens}, {Head}, and {Taylor}}]{Smith12}
{Smith}, D.~E., {Zuber}, M.~T., {Phillips}, R.~J., {Solomon}, S.~C., {Hauck},
  S.~A., {Lemoine}, F.~G., {Mazarico}, E., {Neumann}, G.~A., {Peale}, S.~J.,
  {Margot}, J.-L., {Johnson}, C.~L., {Torrence}, M.~H., {Perry}, M.~E.,
  {Rowlands}, D.~D., {Goossens}, S., {Head}, J.~W., {Taylor}, A.~H., Apr. 2012.
  {Gravity Field and Internal Structure of Mercury from MESSENGER}. Science
  336, 214--.

\bibitem[{{Smith} et~al.(2010){Smith}, {Zuber}, {Phillips}, {Solomon},
  {Neumann}, {Lemoine}, {Peale}, {Margot}, {Torrence}, {Talpe}, {Head},
  {Hauck}, {Johnson}, {Perry}, {Barnouin}, {McNutt}, and {Oberst}}]{Smith10}
{Smith}, D.~E., {Zuber}, M.~T., {Phillips}, R.~J., {Solomon}, S.~C., {Neumann},
  G.~A., {Lemoine}, F.~G., {Peale}, S.~J., {Margot}, J., {Torrence}, M.~H.,
  {Talpe}, M.~J., {Head}, J.~W., {Hauck}, S.~A., {Johnson}, C.~L., {Perry},
  M.~E., {Barnouin}, O.~S., {McNutt}, R.~L., {Oberst}, J., Sep. 2010. {The
  equatorial shape and gravity field of Mercury from MESSENGER flybys 1 and 2}.
  icarus 209, 88--100.

\bibitem[{{Srinivasan} et~al.(2007){Srinivasan}, {Perry}, {Fielhauer}, {Smith},
  and {Zuber}}]{Srinivasan07}
{Srinivasan}, D.~K., {Perry}, M.~E., {Fielhauer}, K.~B., {Smith}, D.~E.,
  {Zuber}, M.~T., Aug. 2007. {The Radio Frequency Subsystem and Radio Science
  on the MESSENGER Mission}. Space Science Reviews 131, 557--571.

\bibitem[{{Stanbridge} et~al.(2011){Stanbridge}, {Williams}, {Taylor}, {Page},
  {Bryan}, {Dunham}, {Wolff}, and {Williams}}]{Stanbridge11}
{Stanbridge}, D.~R., {Williams}, K.~E., {Taylor}, A.~H., {Page}, B.~R.,
  {Bryan}, C.~G., {Dunham}, D.~W., {Wolff}, P., {Williams}, B.~G., Aug. 2011.
  Achievable force model accuracies for messenger in mercury orbit. In:
  AAS/AIAA Astrodynamics Specialist Conference. AAS 11-548.

\bibitem[{{Standish}(1998)}]{DE405}
{Standish}, E.~M., Aug. 1998. {JPL planetary and lunar ephemerides,
  DE405/LE405,} IOM 312F-98-483.

\bibitem[{{Standish} et~al.(1976){Standish}, {Keesey}, and
  {Newhall}}]{Standish76}
{Standish}, Jr., E.~M., {Keesey}, M.~S.~W., {Newhall}, X.~X., Feb. 1976. {JPL
  Development Ephemeris number 96}.

\bibitem[{{Standish}(1990)}]{Standish90}
{Standish}, E.~M., J., Jul. 1990. {The observational basis for JPL's DE 200,
  the planetary ephemerides of the Astronomical Almanac}. Astronomy \&
  Astrophysics 233, 252--271.

\bibitem[{{Stark} and {Parker}(1995)}]{Stark&Parker95}
{Stark}, P., {Parker}, R., 1995. {Bounded-variable least-squares: an algorithm
  and applications}. Computational Statistics 10, 129\u2013141.

\bibitem[{{Tokumaru} et~al.(2010){Tokumaru}, {Kojima}, and
  {Fujiki}}]{Tokumaru10}
{Tokumaru}, M., {Kojima}, M., {Fujiki}, K., Apr. 2010. {Solar cycle evolution
  of the solar wind speed distribution from 1985 to 2008}. Journal of
  Geophysical Research (Space Physics) 115, 4102.

\bibitem[{{Tyler} et~al.(2001){Tyler}, {Balmino}, {Hinson}, {Sjogren}, {Smith},
  {Simpson}, {Asmar}, {Priest}, and {Twicken}}]{Tyler01}
{Tyler}, G.~L., {Balmino}, G., {Hinson}, D.~P., {Sjogren}, W.~L., {Smith},
  D.~E., {Simpson}, R.~A., {Asmar}, S.~W., {Priest}, P., {Twicken}, J.~D., Oct.
  2001. {Radio science observations with Mars Global Surveyor: Orbit insertion
  through one Mars year in mapping orbit}. Journal of geophysical research 106,
  23327--23348.

\bibitem[{{Tyler} et~al.(1981){Tyler}, {Eshleman}, {Anderson}, {Levy},
  {Lindal}, {Wood}, and {Croft}}]{Tyler81}
{Tyler}, G.~L., {Eshleman}, V.~R., {Anderson}, J.~D., {Levy}, G.~S., {Lindal},
  G.~F., {Wood}, G.~E., {Croft}, T.~A., Apr. 1981. {Radio science
  investigations of the Saturn system with Voyager 1 - Preliminary results}.
  Science 212, 201--206.

\bibitem[{{Tyler} et~al.(1986){Tyler}, {Eshleman}, {Hinson}, {Marouf},
  {Simpson}, {Sweetnam}, {Anderson}, {Campbell}, {Levy}, and
  {Lindal}}]{Tyler86}
{Tyler}, G.~L., {Eshleman}, V.~R., {Hinson}, D.~P., {Marouf}, E.~A., {Simpson},
  R.~A., {Sweetnam}, D.~N., {Anderson}, J.~D., {Campbell}, J.~K., {Levy},
  G.~S., {Lindal}, G.~F., Jul. 1986. {Voyager 2 radio science observations of
  the Uranian system Atmosphere, rings, and satellites}. Science 233, 79--84.

\bibitem[{{van der Ha} and {Modi}(1977)}]{vanderHa77}
{van der Ha}, J.~C., {Modi}, V.~J., Dec. 1977. {Analytical evaluation of solar
  radiation induced orbital perturbations of space structures}. Journal of the
  Astronautical Sciences 25, 283--306.

\bibitem[{{Vaughan} et~al.(2002){Vaughan}, {Haley}, {OShaughnessy}, and
  {Shapiro}}]{Vaughan02}
{Vaughan}, R.~M., {Haley}, D.~R., {OShaughnessy}, D.~J., {Shapiro}, H.~S.,
  2002. {Momentum management for the MESSENGER mission}. Advances in the
  Astronautical Sciences 109, 1139--1158.

\bibitem[{{Vaughan} et~al.(2006){Vaughan}, {Leary}, {Conde}, {Dakermanji},
  {Ercol}, {Fielhauer}, {Grant}, {Hartka}, {Hill}, {Jaskulek}, {McAdams},
  {Mirantes}, {Persons}, and {Srinivasan}}]{Vaughan06}
{Vaughan}, R.~M., {Leary}, J.~C., {Conde}, R.~F., {Dakermanji}, G., {Ercol},
  C.~J., {Fielhauer}, K.~B., {Grant}, D.~G., {Hartka}, T.~J., {Hill}, T.~A.,
  {Jaskulek}, S.~E., {McAdams}, J.~V., {Mirantes}, M.~A., {Persons}, D.~F.,
  {Srinivasan}, D.~K., Jan. 2006. {Return to Mercury: The MESSENGER spacecraft
  and mission}. In: Institute of Electrical and Electronics Engineers (IEEE)
  Aerospace Conference. IEEEAC paper 1562,15 pp., 2006.

\bibitem[{{Verma} et~al.(2013){Verma}, {Fienga}, {Laskar}, {Issautier},
  {Manche}, and {Gastineau}}]{verma12}
{Verma}, A.~K., {Fienga}, A., {Laskar}, J., {Issautier}, K., {Manche}, H.,
  {Gastineau}, M., Feb. 2013. {Electron density distribution and solar plasma
  correction of radio signals using MGS, MEX, and VEX spacecraft navigation
  data and its application to planetary ephemerides}. Astronomy \& Astrophysics
  550, A124.

\bibitem[{{Verma} et~al.(2014){Verma}, {Fienga}, {Laskar}, {Manche}, and
  {Gastineau}}]{verma14}
{Verma}, A.~K., {Fienga}, A., {Laskar}, J., {Manche}, H., {Gastineau}, M., Jan.
  2014. {Use of MESSENGER radioscience data to improve planetary ephemeris and
  to test general relativity}. \aap 561, A115.

\bibitem[{{Wahr}(1981)}]{Wahr81}
{Wahr}, J.~M., Mar. 1981. {Body tides on an elliptical, rotating, elastic and
  oceanless earth.} Geophysical Journal International 64, 677--703.

\bibitem[{{Will}(1993)}]{Will93}
{Will}, C.~M. (Ed.), 1993. pp.~396.~ISBN 0521439736. Theory and Experiment in
  Gravitational Physics. Cambridge University Press.

\bibitem[{{Williams} et~al.(2009){Williams}, {Turyshev}, and
  {Boggs}}]{Williams09}
{Williams}, J.~G., {Turyshev}, S.~G., {Boggs}, D.~H., 2009. {Lunar Laser
  Ranging Tests of the Equivalence Principle with the Earth and Moon}.
  International Journal of Modern Physics D 18, 1129--1175.

\bibitem[{{You} et~al.(2012){You}, {Coles}, {Hobbs}, and {Manchester}}]{You12}
{You}, X.~P., {Coles}, W.~A., {Hobbs}, G.~B., {Manchester}, R.~N., May 2012.
  {Measurement of the electron density and magnetic field of the solar wind
  using millisecond pulsars}. \mnras 422, 1160--1165.

\bibitem[{{You} et~al.(2007){You}, {Hobbs}, {Coles}, {Manchester}, and
  {Han}}]{You07}
{You}, X.~P., {Hobbs}, G.~B., {Coles}, W.~A., {Manchester}, R.~N., {Han},
  J.~L., Dec. 2007. {An Improved Solar Wind Electron Density Model for Pulsar
  Timing}. The Astrophysical Journal 671, 907--911.

\bibitem[{{Yuan} et~al.(2001){Yuan}, {Sjogren}, {Konopliv}, and
  {Kucinskas}}]{Yuan01}
{Yuan}, D.-N., {Sjogren}, W.~L., {Konopliv}, A.~S., {Kucinskas}, A.~B., Oct.
  2001. {Gravity field of Mars: A 75th Degree and Order Model}. \jgr 106,
  23377--23402.

\bibitem[{{Zirker}(1977)}]{Zirker77}
{Zirker}, J.~B., Aug. 1977. {Coronal holes and high-speed wind streams}.
  Reviews of Geophysics and Space Physics 15, 257--269.

\end{thebibliography}
